%% file: main.tex
\documentclass[
  reprint,
  aps,
  prx,
  amsmath,amssymb,
  superscriptaddress,
  longbibliography
]{revtex4-2}

\usepackage{style/setup}
\usepackage{style/macros}

\newcommand{\sharedcorrmark}{%
  \protect\textsuperscript{\protect\makebox[0pt][l]{\protect\hspace{0.55em},*}}%
}

\makeatletter
\newcommand{\localfootnote}[1]{%
  \begingroup
    \let\ltx@footmark\ltx@footmark@latex
    \let\ltx@foottext\ltx@foottext@latex
    \let\thempfn\thempfn@latex
    \footnote{#1}%
  \endgroup
}
\makeatother

\makeatletter
\newcommand{\localdaggerfootnote}[1]{%
  \begingroup
    \renewcommand{\thefootnote}{\fnsymbol{footnote}}%
    \let\ltx@footmark\ltx@footmark@latex
    \let\ltx@foottext\ltx@foottext@latex
    \let\thempfn\thempfn@latex
    \footnotemark[2]%
    \footnotetext[2]{#1}%
  \endgroup
}
\makeatother

\begin{document}
\setlength{\skip\footins}{6pt plus 1pt minus 1pt}
\DisableTOC

\title{Learning Arbitrary Lindbladians with Quantum Error Correction}

\author{Nikita Romanov}
\thanks{emails:
\href{mailto:nromanov@g.harvard.edu}{nromanov@g.harvard.edu};\,\,
\href{mailto:syelin@g.harvard.edu}{syelin@g.harvard.edu}.}
\affiliation{Department of Physics, Harvard University, Cambridge, MA 02138, USA}
\author{Petr Ivashkov}
\affiliation{Department of Information Technology and Electrical Engineering, ETH Z\"urich, Z\"urich, Switzerland}
\affiliation{Department of Physics, Harvard University, Cambridge, MA 02138, USA}
\author{Weiyuan Gong}
\affiliation{School of Engineering and Applied Sciences, Harvard University, Allston, MA 02134, USA}

\author{\protect\\[-0.2em]Ishaan~Kannan}
\affiliation{Department of Physics, Harvard University, Cambridge, MA 02138, USA}
\author{Andi Gu}
\affiliation{Department of Physics, Harvard University, Cambridge, MA 02138, USA}
\author{Hong-Ye Hu}
\affiliation{Department of Physics, Harvard University, Cambridge, MA 02138, USA}
\author{Susanne F. Yelin\sharedcorrmark}
\affiliation{Department of Physics, Harvard University, Cambridge, MA 02138, USA}

\begin{abstract}
    We study ansatz-free Lindbladian learning, the problem of reconstructing the generator of an open quantum system without prior knowledge of its Hamiltonian or dissipator structures. This problem exhibits two fundamental precision limits. Hamiltonian components not obscured by dissipation are Heisenberg-limited, while full Lindbladian reconstruction is subject to the quadratically worse standard quantum limit. This creates an apparent algorithmic impasse: achieving the Heisenberg-limited Hamiltonian learning requires suppressing unknown dissipation, which seemingly demands prior reconstruction of the noise and thereby incurs the standard-quantum-limit cost. In this work, we resolve this obstruction by giving an algorithm that learns the Hamiltonian disjoint from dissipator (HDD) at the Heisenberg limit without prior knowledge of either the Hamiltonian or dissipator supports. Our main technical ingredient is a randomized recursive stabilizer-code construction that progressively identifies and suppresses the dominant dissipative terms without reconstructing the full dissipator. Building on this framework, we also introduce an efficient end-to-end algorithm that learns the entire sparse Lindbladian at the standard quantum limit. Finally, we prove that in the ansatz-free setting, the HDD terms constitute the maximal set of Hamiltonian coefficients uniformly learnable at the Heisenberg limit. We show that every term outside HDD is fundamentally standard-quantum limited, extending prior no-go results to the ansatz-free setting. Together, our protocols provide a scalable framework for characterizing open quantum systems, with quantum error correction serving as a key learning primitive.
\end{abstract}
\maketitle

\twocolumngrid

\section{Introduction}\label{sec:introduction}
\vspace{-0.8em}
Reconstructing the generator of a quantum system from observations of its dynamics is a foundational task across quantum science. It underlies the benchmarking and diagnosis of noisy quantum processors~\cite{chuang1997prescription,poyatos1997complete,emerson2005scalable,knill2008randomized,erhard2019characterizing,magesan2011scalable,majidy2024building}, metrology in the presence of decoherence,~\cite{huelga1997improvement,escher2011general,demkowicz2012elusive}, the design of error-correcting codes tailored to a device's dominant noise~\cite{fletcher2008channel,aliferis2008fault,bonilla2021xzzx,tuckett2019tailoring,chuang1997bosonic,wu2025bias,kuehnke2025hardware, valenti2019hamiltonian}, and the validation of quantum simulators~\cite{wiebe2014hamiltonian,hangleiter2024robustly,kraft2025bounded}. Under Markovian noise, quantum dynamics are generated by a Lindbladian, which combines coherent Hamiltonian evolution with dissipation~\cite{gorini1976completely, lindblad1976generators}. This gives rise to two complementary learning tasks. Hamiltonian learning targets the coherent part of the generator, treating dissipation as noise that can obscure or bias estimates of the Hamiltonian parameters~\cite{da2011practical,wiebe2014hamiltonian}. Lindbladian learning is the strictly harder task of reconstructing the full generator, including its dissipative components~\cite{boulant2003robust}. In this work, we address both tasks: Hamiltonian learning in the presence of unknown Markovian dissipation, and full Lindbladian learning.\localdaggerfootnote{For a more extensive review of related work, see Appendix~\ref{sec:related_work}.}

Recent years have seen substantial progress on Hamiltonian learning from dynamics under the idealized assumption of noiseless evolution~\cite{da2011practical,bairey2019learning,zubida2021optimal,stilck2024efficient,gu2024practical,caro2024learning,haah2024learning,huang2023learning,mirani2024learning,ni2024quantum,dutkiewicz2024advantage,bakshi2024structure,ma2024learning,zhao2025learning,hu2025ansatz,sinha2025improvedhamiltonianlearningsparsity,abbas2025nearly}. A key figure of merit in this problem is the total black-box evolution time required to estimate Hamiltonian parameters to accuracy $\varepsilon$. The best possible scaling, known as the Heisenberg limit (HL), is $1/\varepsilon$, while the standard quantum limit (SQL) corresponds to the quadratically worse  $1/\varepsilon^2$. Attaining the HL is a long-standing goal in quantum sensing and Hamiltonian learning~\cite{giovannetti2006quantum,higgins2007entanglement,kitaev1995quantum,huang2023learning,he2026optimal}. Although the HL scaling has been achieved under unitary dynamics~\cite{huang2023learning,mirani2024learning,ni2024quantum,dutkiewicz2024advantage,bakshi2024structure,ma2024learning,zhao2025learning,hu2025ansatz,sinha2025improvedhamiltonianlearningsparsity,abbas2025nearly}, realistic devices inevitably experience dissipation and noise. Such noise can obscure the long-time coherent evolution these algorithms rely on and degrade the achievable scaling toward the SQL~\cite{huelga1997improvement,demkowicz2012elusive,escher2011general,demkowicz2014using,smirne2016noisyfreqest, zhou2018achieving,demkowicz2017qecmetrology,gorecki2020optimal}. Therefore, recovering the Heisenberg limit in Hamiltonian learning in the presence of noise remains a major challenge.

In metrology, quantum error correction (QEC) has emerged as a central tool for restoring HL scaling in local parameter estimation under Markovian noise~\cite{dur2014improved,kessler2014quantum,arrad2014sensingqec,marti2015qecenhanced,zhou2018achieving,demkowicz2017qecmetrology,gorecki2020optimal,zhou2020optimal,reiter2017dissipative,layden2018spatial,shettell2021practical,layden2019ancillafree,zhou2024ancillafree,antu2025stabilizercodes,qiao2026distributed,kapourniotis2019faulttolerant,sahu2026achieving,marrero2026encoded}. Existing constructions strongly rely on prior knowledge of the Hamiltonian and dissipator structures, using it to design QEC codes that suppress the specified dissipative dynamics while preserving the target Hamiltonian terms as nontrivial logical operators~\cite{zhou2018achieving,demkowicz2017qecmetrology,gorecki2020optimal}. For quantum-system characterization, however, the relevant control imperfections (Hamiltonian terms) and error mechanisms may be unknown and nonlocal~\cite{kraft2025bounded}, making such structure-aware code design unavailable. Moreover, learning the dissipator is fundamentally a standard-quantum-limited task~\cite{brady2026precision}, raising the possibility that the cost of discovering the appropriate QEC code may overwhelm any subsequent Heisenberg-limited advantage. This motivates a fundamental question: can one attain Heisenberg-limited Hamiltonian learning in the \emph{ansatz-free} setting, where neither the Hamiltonian nor dissipator supports are known a priori?

In this work, we answer this question affirmatively by constructing an algorithm that learns the Hamiltonian part disjoint from the dissipator (HDD), defined formally below, at the Heisenberg limit without prior access to Hamiltonian or dissipator structures. Our main technical contribution is a recursive QEC-based learning framework that progressively identifies and suppresses the strongest Lindbladian terms while preserving access to weaker unknown ones. This framework bypasses the standard quantum limit by identifying only the heavy dissipator support needed for QEC codes, rather than reconstructing the full dissipator.

To establish the optimality of the Heisenberg-limited coverage of our algorithm, we prove that the HDD sector is the maximal Hamiltonian target uniformly learnable at the Heisenberg limit. Every coefficient outside HDD is fundamentally standard-quantum limited in the ansatz-free setting, extending prior no-go results~\cite{zhou2018achieving,demkowicz2017qecmetrology,gorecki2020optimal,abbasgholinejad2026error}.

We next turn from noisy Hamiltonian learning to the more demanding task of full Lindbladian learning, which characterizes dissipative dynamics alongside the coherent evolution. For this task, HL scaling is known to be impossible in general~\cite{zhou2018achieving,demkowicz2017qecmetrology,brady2026precision}, so the best achievable \(\varepsilon\)-dependence is SQL. In the ansatz-free setting, the difficulty is further compounded by the exponentially many potential Lindbladian terms. Prior approaches focused on local models or otherwise restricted structures~\cite{stilck2024efficient,franca2025learning,montana2025efficiently,pastori2022characterization,olsacher2025hamiltonian,berg2025large,kraft2025bounded}, while our previous ansatz-free protocol incurred a structure-identification cost scaling as \(1/\varepsilon^4\)~\cite{ivashkov2026ansatzfreelindblad}. In this work, we introduce the first efficient end-to-end SQL algorithm for learning arbitrary sparse Lindbladians without prior knowledge of their structure. Crucially, our protocol's resources scale polynomially in system size, sparsity, and \(1/\varepsilon\), even when the Lindbladian terms have unbounded locality\localdaggerfootnote{Several concurrent and subsequent works have developed complementary efficient Lindbladian-learning protocols; see Sec.~\ref{sec:outlook} for comparison.}.

We achieve this by combining the aforementioned QEC-based learning framework with a new coefficient-estimation primitive. The recursive framework first identifies the dissipator structure and learns the Hamiltonian coefficients disjoint from the dissipator. To estimate the dissipative and the remaining Hamiltonian coefficients, we introduce a family of low-rank observables on the Choi state of the dynamics that selectively isolate Lindbladian coefficients. This direct access avoids constructing and solving a large linear system, as in prior derivative-based approaches~\cite{stilck2024efficient, franca2025learning}, thereby eliminating the need to control a potentially unfavorable conditioning factor.

Together, our protocols achieve HL scaling for HDD Hamiltonian learning and efficient SQL scaling for full Lindbladian learning in the ansatz-free setting. They require only simple Pauli or Bell-state preparations and measurements, Clifford gates, and polynomial-time classical processing, providing a scalable framework for noisy-system calibration and characterization.

\begin{figure*}
\centering
\includegraphics[width=\linewidth]{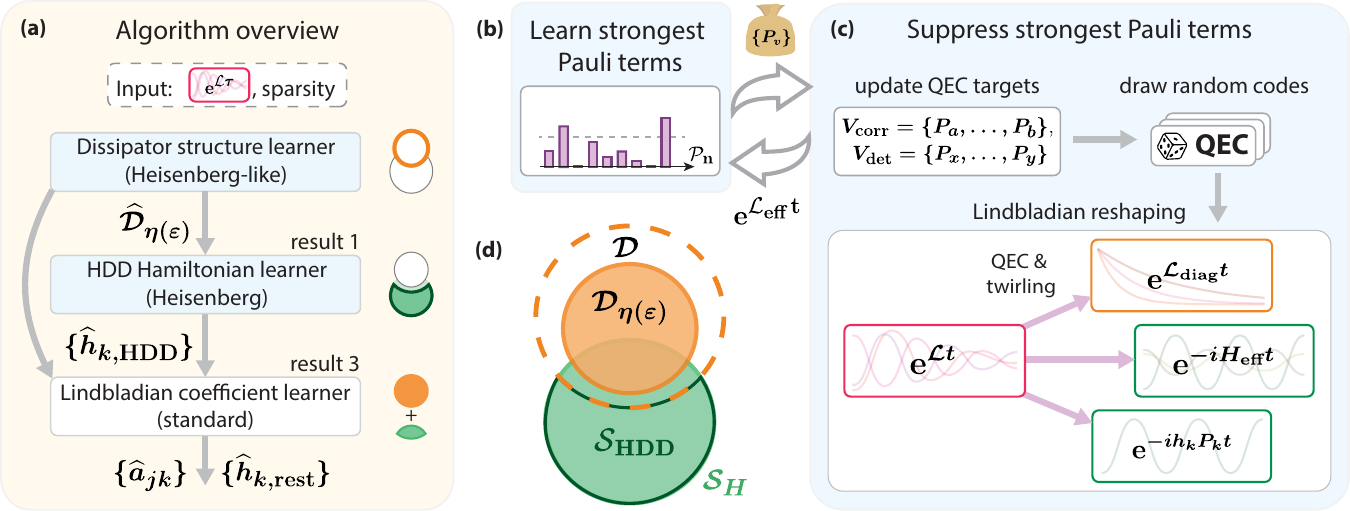}
\caption{
\textbf{Overview of ansatz-free Hamiltonian and Lindbladian learning algorithms.}
(a)~Given black-box access to $e^{\Lindblad t}$ and an upper bound $M$ on the Lindbladian sparsity, the algorithm first learns a tolerant estimate
$\EstEtaEpsFullDissStruct$ of the $\eta(\varepsilon)$-heavy dissipator footprint in total evolution time
$\SoftBigO{1/\eta}$.
The choice of $\eta(\varepsilon)$ depends on the final target.
For \HND{} (\HNDAbbrev{}) learning, choosing
$\eta=\Theta(\varepsilon/M)$ suffices to estimate every coefficient in the full
\HNDAbbrev{} sector
$\HNDStruct=\HamiltStruct\setminus\FullDissStruct$
to accuracy $\varepsilon$ in Heisenberg-limited total evolution time
$\SoftBigO{M^2/\varepsilon}$.
For full Lindbladian learning, we instead choose
$\eta=\Theta(\varepsilon^2/M)$ and append a standard-quantum-limited
coefficient-learning stage, yielding total evolution time
$\SoftBigO{M/\varepsilon^2+M^2/\varepsilon}$. (b)~The key primitive is a recursive construction that learns the strongest
Pauli terms $\{P_v\}$, suppresses them using QEC reshaping, and repeats on
the residual effective Lindbladian until the desired accuracy is reached. (c)~Selective suppression is implemented by adaptive stabilizer QEC.
Previously learned Hamiltonian terms are placed in $\VDetect$, for which
we enforce detectability, while previously identified dissipator terms are
placed in $\VCorrect$, for which we enforce correctability.
Random stabilizer codes deterministically suppress the targeted Pauli terms
while preserving sensitivity to unknown weaker terms with non-negligible
probability. The resulting error-correction scheme reshapes the original
Lindbladian into effective dynamics suitable for dissipator learning,
Hamiltonian learning, or single-coefficient estimation. (d)~Venn diagram of the learning targets. The dashed region $\FullDissStruct$ is the full dissipator footprint, while $\EtaEpsFullDissStruct\subseteq\FullDissStruct$ is its $\eta$-heavy subset learned by the dissipator stage. The Hamiltonian learner identifies and estimates all coefficients in the dark green HDD sector $\HNDStruct=\HamiltStruct\setminus\FullDissStruct$ and may additionally access the weak-footprint terms $\HamiltStruct\cap \bigl(\FullDissStruct\setminus\EtaEpsFullDissStruct\bigr)$ at finite resolution. The final coefficient learner estimates the remaining Hamiltonian and dissipator coefficients associated with the learned dissipator footprint.
}
\label{fig:alg-overview}
\end{figure*}

\vspace{-0.8em}
\section{Overview of Results}\label{sec:main_results}
\vspace{-0.6em}

We consider the problem of learning the coefficients of a time-independent $n$-qubit Lindbladian $\Lindblad$ given access to its dynamics $e^{t\Lindblad}$. We express $\Lindblad$ in the Pauli basis:
\begin{equation} \label{eq:lindbladian-intro}
    \begin{aligned}
        \Lindblad(\rho) = -i&\underbrace{\sum_{P_k\in \HamiltStruct} h_k \comm{P_k}{\rho}}_{\mathrm{Hamiltonian}} + \\ &+\underbrace{\sum_{P_k,P_j \in \DiagDissStruct} a_{kj}\!\left(P_k \rho P_j - \tfrac12\acomm{P_j P_k}{\rho}\right)}_{\mathrm{Dissipator}},
    \end{aligned}
    \vspace{-0.5em}
\end{equation}
where $P_i,P_j$ are $n$-qubit traceless Pauli operators \footnote{Although Lindbladian generators admit a unitary gauge freedom in the choice of jump operators, the Pauli-basis expansion Eq.~\eqref{eq:lindbladian-intro} is unique.}, $\HamiltStruct$ and $\DiagDissStruct$ denote the unknown Hamiltonian and dissipator supports, and $\{h_k,a_{kj}\}$ are the corresponding unknown coefficients. We also take as an input a sparsity parameter $M \ge |\HamiltStruct| + |\DiagDissStruct|^2$, which upper bounds the total number of nonzero coefficients. The goal is to output estimates $\{\widehat h_k\}\cup\{\widehat a_{kj}\}$ of nonzero coefficients to within additive error $\varepsilon$, with high success probability. As the Lindbladian global scale can be absorbed into the evolution time, we assume without loss of generality that $|h_k|\le 1$ and $|a_{kj}|\le 1$ for all $k,j$. We refer to the positive semidefinite coefficient matrix $a$ as the Kossakowski matrix~\cite{gorini1976completely,Chruscinski2017history}.

Throughout the main analysis, we assume noiseless auxiliary computation and control, with noise entering only through the black-box evolution $e^{t\Lindblad}$. In Appendix \ref{sec:interface-noise}, we quantify the admissible interface noise strength and control speed under which our guarantees remain valid.

\noindent\textbf{Heisenberg-limited learning.}
Our main result is that, despite not knowing either the Hamiltonian or
dissipator supports in advance, one can learn the Hamiltonian component outside the dissipator footprint with Heisenberg-limited total evolution time. To make this precise, let us define the full dissipator footprint:
\vspace{-0.2em}
\begin{equation}
\label{eq:full-dissipator-footprint}
\FullDissStruct
\coloneqq
\DiagDissStruct
\cup
\bigl\{
P_\ell P_m:
P_\ell,P_m\in\DiagDissStruct,\;
P_\ell\neq P_m
\bigr\},
\vspace{-0.3em}
\end{equation}
where pairwise Pauli products are taken modulo phase. Correspondingly, we define the Hamiltonian-disjoint-from-dissipator
(\HNDAbbrev{}) sector as
\begin{equation}
\label{eq:hdd-def-main}
\HNDStruct
\coloneqq
\HamiltStruct\setminus\FullDissStruct .
\vspace{-0.3em}
\end{equation}

\begin{result}[Ansatz-free \HNDAbbrev{} Hamiltonian Learning]
\label{res:informal-main}
    There exists a quantum algorithm that outputs an estimator of the true
    \HND{} coefficients such that with high probability:
    \begin{equation*}
        |\widehat{h_k}-h_k| \le \varepsilon,
        \qquad \forall P_k \in \HNDStruct .
    \end{equation*}
    The total evolution time under $e^{\Lindblad t}$ is Heisenberg-limited:
    \begin{equation}
        t_{\mathrm{tot}} = \SoftBigO{M^2/\varepsilon}.
    \end{equation}
    The algorithm employs only Clifford operations, uses
    $n + \BigO{\log M}$ noiseless ancillary qubits, and has classical
    processing cost $\SoftBigO{\Poly{n,M,1/\varepsilon}}$.
\end{result}

To attain the Heisenberg limit in \HNDAbbrev{} Hamiltonian learning, we design in \Cref{sec:lindbladian_reshaping} a Lindbladian reshaping procedure based on stabilizer QEC. This procedure can be used to suppress selected Lindbladian terms, thereby enabling the long coherent evolutions needed for attaining the HL. If the structures of both the Hamiltonian and the dissipator were known a priori, one could apply the reshaping procedure directly: choose a code that suppresses the dissipative terms while allowing the relevant Hamiltonian terms to act as logical operators~\cite{zhou2018achieving,demkowicz2017qecmetrology,gorecki2020optimal}. This would reduce the task to Hamiltonian learning within a logical subspace, where one could apply their favorite Heisenberg-limited Hamiltonian learning method~\cite{huang2023learning,bakshi2024structure,hu2025ansatz,ma2024learning}.

To remove the known-structure assumption, we introduce a QEC-based recursive framework that underlies both the dissipator-structure learner and the \HNDAbbrev{} Hamiltonian learner highlighted in \cref{fig:alg-overview}~(a). At each stage, the algorithm identifies the strongest remaining Lindbladian terms, suppresses them by adding them to the target error-correction or error-detection set, and then proceeds to weaker terms, as illustrated in \cref{fig:alg-overview}~(b)--(c). By randomizing the stabilizer-code construction, we ensure that the unknown weaker terms survive as non-trivial logical operators with non-negligible probability, while the targeted terms are deterministically suppressed. We show that this can be achieved with only $\BigO{\log M}$ ancillas. 

Crucially, the dissipator stage avoids SQL scaling by identifying only
the heavy dissipator support needed for the subsequent QEC-code
construction, rather than estimating dissipator coefficients. For a
resolution scale $\eta>0$, define the $\eta$-heavy dissipator structure
\begin{equation*}
    \EtaDiagDissStruct
    \coloneqq
    \{P_k\in\DiagDissStruct:a_{kk}\ge\eta\},
\end{equation*}
and let $\EtaFullDissStruct$ denote its corresponding footprint, obtained
by replacing $\DiagDissStruct$ with $\EtaDiagDissStruct$ in
Eq.~\eqref{eq:full-dissipator-footprint}. In
\Cref{sec:dissipator_struct_learning}, we show that this
$\eta$-heavy structure can be identified in total evolution time
$\SoftBigO{1/\eta}$. Operationally, at target accuracy $\varepsilon$, choosing $\eta=\BigO{\varepsilon/M}$ suffices for the QEC construction underlying full \HNDAbbrev{} learning, with the unresolved weaker dissipative terms contributing only a controlled residual error. Thus, combining the
$\eta$-heavy dissipator learner from
\Cref{sec:dissipator_struct_learning} with the logical Hamiltonian learner
summarized in \Cref{sec:hamiltonian_learning} yields an end-to-end,
Heisenberg-limited, ansatz-free algorithm for \HNDAbbrev{} Hamiltonian
learning. Formal resource guarantees are given in
\Cref{thm:full-hdd-learned-dissipator} and
\Cref{cor:ansatz-free-hnd}.

In Appendix~\ref{sec:appendix-reducing-the-ancilla-cost}, we also provide an ancilla-reduced variant of \Cref{res:informal-main} that improves the total ancilla count from  $n+\BigO{\log M}$ to $\BigO{\log M}$ following~\cite{hu2025ansatz} at the cost of increased black-box evolution time $t_{\mathrm{tot}} = \SoftBigO{M^{4}\log{(n)}/\varepsilon}$.

\noindent\textbf{Optimality of the \HNDAbbrev{} boundary.}
Furthermore, we prove that the Heisenberg-limited coverage of \Cref{res:informal-main} is optimal in the ansatz-free setting.

\begin{result}[Optimality of the \HNDAbbrev{} boundary]
\label{res:hdd-optimality}
For every nonidentity Pauli $P_r\in\FullDissStruct$, learning its
Hamiltonian coefficient $h_r$ to additive accuracy $\varepsilon$ in
the ansatz-free setting requires, in the worst case,
\begin{equation}
    t_{\mathrm{tot}}
    =
    \Omega(1/\varepsilon^2).
\end{equation}
This holds even under known dissipator support and arbitrary adaptive
quantum protocols with noiseless ancillas and quantum memory. Thus,
$\HNDStruct$ is the maximal Hamiltonian sector admitting a uniform
Heisenberg-limited learning guarantee.
\end{result}

To prove this lower bound, we consider the simpler task of distinguishing
a given Lindbladian $\Lindblad_0$ from a nearby hypothesis
$\Lindblad_\theta$ along a one-parameter joint Hamiltonian--dissipator
deformation. For every $P_r\in\FullDissStruct$, we construct the deformation such that the target Hamiltonian coefficient
changes linearly,
$h_r(\theta)=h_r(0)+\theta$, while the distinguishability generated by
the dynamics is only second order in $\theta$. Concretely, a query of
duration $t$ can decrease the root fidelity between the two hypotheses
by only $\mathcal O(\theta^2 t)$. We then use the learning-tree formalism~\cite{chen2022exponential} and telescope
this fidelity bound over the complete adaptive measurement history,
showing that no adaptive protocol can distinguish the two hypotheses
with constant bias unless
$t_{\mathrm{tot}}=\Omega(1/\theta^2)$. Taking
$\theta=4\varepsilon$, any $\varepsilon$-accurate learner would provide
such a distinguisher, yielding
$t_{\mathrm{tot}}=\Omega(1/\varepsilon^2)$.
Details of the proof and formal guarantees, together with a comparison to the
Hamiltonian-not-in-Lindblad-span (HNLS) criterion~\cite{zhou2018achieving,demkowicz2017qecmetrology},
are provided in Appendix~\ref{sec:hdd-boundary}.

\noindent\textbf{Finite-resolution extension.}
At finite target accuracy $\varepsilon$, our Hamiltonian learner may access
terms beyond the asymptotic \HNDAbbrev{} sector in the weak dissipator footprint
\begin{equation*}
    \HamiltStruct\cap
    \bigl(\FullDissStruct\setminus\EtaEpsFullDissStruct\bigr),
\end{equation*}
illustrated
in \cref{fig:alg-overview}~(d). For any fixed Lindbladian, this additional sector becomes empty as
$\varepsilon\to0$. In Sec.~\ref{sec:finite-resolution-hamilt-learning}, we characterize when these additional coefficients can still be learned at the
Heisenberg limit and explain how this finite-resolution extension is
consistent with our lower bound \Cref{res:hdd-optimality}.

\noindent\textbf{Standard-quantum-limited learning.}
Our third result is an ansatz-free standard-quantum-limited algorithm for
learning all coefficients $\{h_k\}\cup\{a_{kj}\}$ of a Lindbladian as
in Eq.~\eqref{eq:lindbladian-intro}. The algorithm remains efficient
even when the Lindbladian terms have unbounded locality. Its structure-identification step follows the \HNDAbbrev{} Hamiltonian learner above, but uses the more stringent threshold 
$\eta=\mathcal O(\varepsilon^2/M)$, which limits how strongly
dissipative coherences can bias Hamiltonian terms in the weak dissipator
footprint $\FullDissStruct\setminus\EtaEpsFullDissStruct$. After this step,
we have identified the relevant dissipator structure and learned all
Hamiltonian coefficients outside the learned dissipator footprint to
accuracy $\varepsilon$ (see \cref{fig:alg-overview}~(a)). The only
remaining unknowns are the dissipator coefficients and the Hamiltonian
coefficients inside the learned dissipator footprint.

To learn these remaining coefficients, we use the Choi state of the short-time channel, obtained by applying $e^{\Lindblad t}\otimes I_n$ to the maximally entangled $2n$-qubit state. In the Choi representation, the first-order expansion separates into orthogonal directions corresponding to individual Hamiltonian and dissipative coefficients. We exploit this property by explicitly constructing \(2n\)-qubit low-rank observables whose short-time derivatives provide direct access to the desired coefficients. Thus, rather than assembling a large linear system, each remaining coefficient is isolated by the derivative of a single Choi expectation value. These derivatives are estimated using shadow process tomography~\cite{caro2024learning,levy2024qpt,kunjummen2023qpt} in combination with Chebyshev polynomial interpolation. This direct-observable construction is the main novel ingredient: compared with the black-box learner of~\cite{caro2024learning}, it improves the evolution-time scaling from \(\SoftBigO{nM^3/\varepsilon^4}\) to \(\SoftBigO{M/\varepsilon^2}\), and compared with general shadow process tomography approaches~\cite{levy2024qpt,kunjummen2023qpt}, it avoids condition-number-dependent error amplification from linear-system inversion. We summarize the protocol in \Cref{sec:sql-coeff-learning}, with formal guarantees given in Appendix~\ref{sec:appendix-dissipator-hid-coeff-learning}.

\begin{result}[Ansatz-free Lindbladian Learning] \label{res:ansatz-free-lindblad}
    There exists a quantum algorithm for learning  a Lindbladian as in Eq.~\eqref{eq:lindbladian-intro} that outputs an estimator $\hat{\bm{\lambda}} = (\hat{\bm{h}},\hat{\bm{a}})$ of the true Hamiltonian and dissipator coefficient vector $\bm{\lambda} = (\bm{h},\bm{a})$ such that:
    \vspace{-0.4em}
    \begin{equation}
        \|\hat{\bm{\lambda}} - \bm{\lambda}\|_\infty \le \varepsilon
    \vspace{-0.3em}
    \end{equation}
    with high probability. The total evolution time under $e^{\Lindblad t}$ is
    \begin{equation}
        t_{\mathrm{tot}} = \widetilde{\mathcal O}\left(\frac{M}{\varepsilon^{2}} + \frac{M^{2}}{\varepsilon}\right).
    \end{equation}
    The algorithm employs only Clifford operations, uses $n + \BigO{\log M}$  noiseless ancillary qubits, and has classical processing cost $\Poly{n,M,1/\varepsilon}$.
\end{result}

\vspace{-0.4em}
\section{QEC Lindbladian Reshaping}
\label{sec:lindbladian_reshaping}
\vspace{0.0em}

Both our Heisenberg- and standard-quantum-limited results are enabled by selective suppression of previously identified terms in the Lindbladian expansion Eq.~\eqref{eq:lindbladian-intro}. We refer to these suppression techniques broadly as
\emph{Lindbladian reshaping}, illustrated schematically in
\cref{fig:alg-overview}~(c). In Hamiltonian learning, such reshaping is typically achieved via Trotterization or Pauli twirling~\cite{huang2023learning,bakshi2024structure,hu2025ansatz,ma2024learning}. For Lindbladian dynamics, however, suppressing dissipative terms requires going beyond unitary control. 

We thus use stabilizer quantum error correction as the main reshaping mechanism: repeated syndrome measurements suppress detectable Hamiltonian terms via the projective Zeno effect~\cite{Mobus2019QZEGeneralized,Becker2021QZEOpen,Burgarth2020QZD,PazSilva2012ZenoQC,franceschetto2025hamiltonian}, while active recovery suppresses the dissipative terms ~\cite{zhou2018achieving,demkowicz2017qecmetrology,kessler2014quantum,atalaya2021continuous}. By randomly constructing the code, we make the specified Pauli operators detectable or correctable while preserving the remaining unknown terms as identifiable logical operators with high probability. Since our learning framework identifies the relevant error-generating mechanisms explicitly, and there are at most $M$ of them, we use a direct lookup-table decoder.

As a toy example, consider constructing a deterministic QEC reshaping scheme for the two-qubit Lindbladian
\begin{equation} \label{eq:reshaping-ex}
    \Lindblad(\rho)
    =    -i\bigl[\underbrace{\alpha_1 XI+\alpha_2 IX}_{\mathrm{Hamiltonian}},\rho\bigr]
    +
    \gamma(ZZ\rho ZZ-\rho),
\end{equation}
and suppose that we want to isolate the dynamics generated by $\alpha_1 XI$ in order to learn $\alpha_1$. Recall briefly that an $n$-qubit stabilizer code is specified by commuting Pauli checks $G_1,\ldots,G_m$ generating an abelian subgroup $S\subseteq\PnHerm$ with $-I\notin S$. The code space $\mathcal H_c$ is their joint $+1$ eigenspace, with projector $\Pi_0=\prod_{\ell=1}^m(I+G_\ell)/2$. A Pauli $Q$ that anticommutes with at least one check is detectable and satisfies $\Pi_0 Q\Pi_0=0$, while a Pauli $Q$ that commutes with every check preserves the code space and induces the logical operator $\overline Q\coloneqq \Pi_0 Q\Pi_0$ on $\mathcal H_c$. Errors are classified through the syndrome map $r:\mathbb P_n\to\{0,1\}^m$, defined by setting $r_\ell(P)=1$ if $P$ anticommutes with $G_\ell$, and $r_\ell(P)=0$ otherwise.

For the example in  Eq.~\eqref{eq:reshaping-ex}, a natural choice of stabilizer generating set is $\{IY\}$. This check anticommutes with the unwanted terms $IX$ and $ZZ$ and commutes with the desired term $XI$. This makes $IX$ and $ZZ$ detectable, whereas $XI$ survives as a nontrivial logical operator. Let $\Pi_0=(I+IY)/2$ and $\Pi_1=(I-IY)/2$ be the corresponding syndrome projectors. Upon observing the nontrivial syndrome, apply the recovery $R_1=ZZ$ to correct the dissipative jump, and set $R_0=I$ for the trivial syndrome. The resulting recovery channel is
\begin{equation}
    \RMap(\rho)
    =
    \sum_{s\in\{0,1\}}R_s\Pi_s\rho\Pi_sR_s^\dagger .
\end{equation}
Given this channel and for any code-space state $\rho_c=\Pi_0\rho_c\Pi_0$, a single QEC round yields
\begin{equation}
    \RMap\circ e^{\tau\Lindblad}(\rho_c)
    =
    \rho_c-i\alpha_1\tau[XI,\rho_c]+\BigO{\tau^2},
\end{equation}
which, as desired, approximates the evolution under $\LindbladEff(\rho) = -i\alpha_1[XI, \rho]$. Notably, the unwanted Hamiltonian term $IX$ is suppressed by error detection alone through the projective Zeno effect. Thus, in this example, a single stabilizer check suffices both to suppress the Hamiltonian error and to correct the dissipative jump. 

\begin{figure*}
    \centering
    \includegraphics[width=\linewidth]{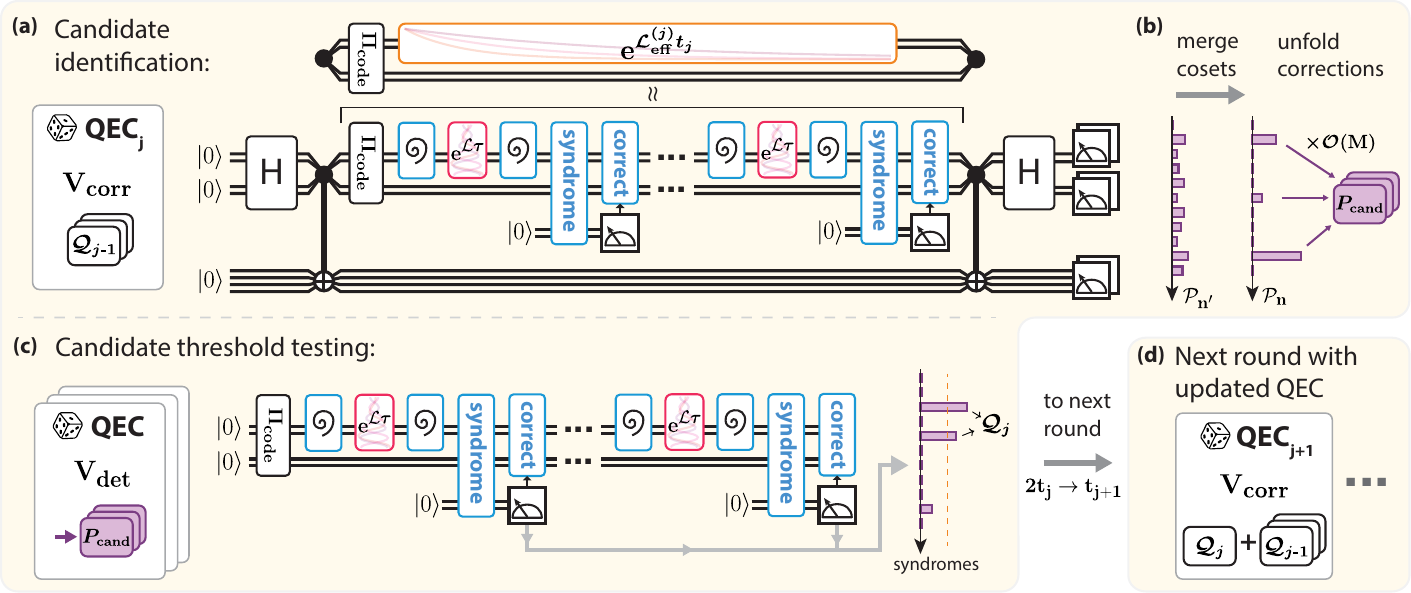}
    \caption{
    \textbf{Hierarchical dissipator structure learner.}
    This subroutine constructs a tolerant estimator $\EstEtaDiagDissStruct$ of the $\eta$-heavy dissipator by recursively learning and suppressing progressively weaker terms. Suppression isolates the remaining weaker terms through Lindbladian reshaping, implemented by repeated QEC combined with Pauli twirling, as indicated by the twirl gates.
    (a)~In each hierarchical round, the first step identifies the strongest as-yet unknown terms from the effective logical evolution, using either Bell sampling or population recovery~\cite{flammia2021pauli}. The figure depicts the Bell-sampling approach. In round $j$, the unknown terms of interest appear with probabilities proportional to $t_j a_{kk}$, which are $\Omega(1/M_D)$ by construction, where $M_D \le \sqrt{M}$ upper bounds the dissipator sparsity $|\DiagDissStruct|$. These terms can therefore be identified using $\SoftBigO{M_D}$ repetitions of the Bell-sampling experiment.
    (b)~Because sampling is performed after error correction, the observed logical terms do not directly specify the original physical Pauli candidates and require post-processing. First, since logical operators are equal up to stabilizers, we merge samples belonging to the same stabilizer cosets. Second, each observed logical Pauli can arise from $\BigO{M}$ possible physical terms through the $\BigO{M}$ possible recovery operators; we therefore add all such preimages to the candidate superset. (c)~The resulting candidates are threshold-tested in parallel using the syndrome-measurement histogram to remove false positives. Because different physical Pauli terms can share the same syndrome, we draw sufficiently many random codes so that each tested Pauli is collision-free for at least one code. (d)~Once the round-$j$ terms have been identified and validated, they are added to the QEC target set $\VCorrect$ for suppression in subsequent rounds. The updated code initializes round \(j+1\), where the evolution time is doubled to expose weaker dissipator terms.
    }
    \label{fig:diss-struct-learner}
\end{figure*}

In the general Lindbladian-learning setting, we implement QEC-based suppression of previously learned terms by maintaining two sets: $\VDetect \subseteq \HamiltStruct$ for Hamiltonian terms targeted for detection, and $\VCorrect \subseteq \DiagDissStruct$ for dissipator terms targeted for correction.  Detectability of a Pauli $P$ is guaranteed by a nonzero syndrome, $r(P)\neq 0$. Correctability of $\VCorrect$ means that all Paulis in $\VCorrect$ have distinct nonzero syndromes. Using the stabilizer-syndrome identity $r(P_aP_b)=r(P_a)\oplus r(P_b)$, this is equivalent to requiring all pairwise products of Paulis in $\VCorrect$ to be detectable. Therefore, the target suppression conditions are guaranteed by making every element of the induced footprint
\begin{equation}
    \mathcal F \coloneqq
    \VDetect
    \cup
    \VCorrect
    \cup
    \{P_\ell P_m : P_\ell,P_m\in \VCorrect,\ P_\ell\neq P_m\}
\end{equation}
detectable, where $|\mathcal F| \le M$. Since any fixed nonidentity Pauli anticommutes with a uniformly random Pauli with probability $1/2$, drawing $m=\BigO{\log M}$ random stabilizer checks suffices to guarantee the suppression of target terms with constant probability. Repeating the random construction boosts the success probability arbitrarily close to one. The same construction also preserves access to yet-unidentified terms. Conditioned on the target suppression, any Pauli $P_r \notin \mathcal F$ outside of the target footprint survives as an unmodified logical with probability $\propto 2^{-m}\propto 1/M$. Finally, to ensure that the checks commute and hence define a valid stabilizer code, we add at most $\BigO{\log M}$ ancillary qubits. See Appendix~\ref{sec:qec-construction} for details on the code construction.

\vspace{-0.4em}
\section{Dissipator Structure Learning}\label{sec:dissipator_struct_learning}
\vspace{0.0em}
Dissipator-structure identification is a crucial first step in both our \HNDAbbrev{} Hamiltonian and general Lindbladian learners. Recall that the aim of this step is to recover the $\eta$-heavy dissipator structure $\EtaDiagDissStruct \coloneqq \{P_k\in \DiagDissStruct : a_{kk}\ge \eta\}$. Our approach proceeds by hierarchically learning progressively weaker bins, as schematically depicted in \cref{fig:alg-overview}~(b)-(c). We first identify all Paulis with rates above $1/2$, suppress them by QEC reshaping, then double the evolution time and repeat the procedure at threshold $1/4$, continuing recursively until the target threshold $\eta$ is reached.

More specifically, in the $j$-th round, as illustrated in \cref{fig:diss-struct-learner}~(a), the protocol begins with Bell sampling to identify a candidate superset $\widehat{\DHierSet}_j \supseteq \DHierSet_j$ of the round-$j$ dissipator bin
\begin{equation}
    \DHierSet_j
    \coloneqq
    \bigl\{P_k\in \DiagDissStruct : 2^{-(j+1)} < a_{kk} \le 2^{-j}\bigr\},
\end{equation}
for $j\in \{0,\ldots,\lceil \log_2(1/\eta)\rceil-1\}$ (\Cref{subsec:diss-candidate-identification}). It then tests these candidates using syndrome measurements to remove false positives (\Cref{subsec:diss-candidate-testing}). The newly verified terms are added to the correction target set $\VCorrect$ and subsequently suppressed via QEC-based Lindbladian reshaping (\Cref{subsec:diss-dynamics-reshaping}). Eventually, $\VCorrect$ contains all bins $\DHierSet_j$ above the desired threshold $\eta$, thereby providing an estimate of $\EtaDiagDissStruct$. The total black-box evolution time of dissipator structure identification across the hierarchical rounds is:
\vspace{-0.4em}
\begin{equation*}
    t_\mathrm{tot} = \SoftBigO{1/\eta}
\end{equation*}
Importantly, within this hierarchical approach, Bell sampling can be replaced by population recovery~\cite{flammia2021pauli}, reducing the ancilla cost from $n+\BigO{\log M}$ to $\BigO{\log M}$, similarly to~\cite{hu2025ansatz}. The details of this alternative approach are provided in Appendix~\ref{sec:appendix-reducing-the-ancilla-cost}.

\vspace{-0.4em}
\subsection{Reshaping the dynamics} \label{subsec:diss-dynamics-reshaping}
\vspace{0.0em}

We first simplify the learning problem by Pauli twirling~\cite{emerson2007symmetrized} the Lindbladian channel, which we depict by the twirl gates in \cref{fig:diss-struct-learner}~(a)-(c). For a Pauli $P$, let $\mathcal U_P(\rho)\coloneqq P\rho P^\dagger$. At each time step, we draw $P\in\PnHerm$ uniformly at random and conjugate the short-time evolution $e^{\tau\Lindblad}$ by $\mathcal U_P$. Averaging over these random Pauli conjugations yields an effective generator with only the diagonal dissipator remaining:
\begin{equation} \label{eq:lindblad-twirled}
     \Lindblad_\mathrm{twrl} = \underset{P \sim \PnHerm}{\mathbb E} \left\{
    \mathcal U_P \circ
    \Lindblad \circ
    \mathcal U_P (\rho)\right\} = \sum_{P_k \in \DiagDissStruct}a_{kk}(P_k \rho P_k -\rho).
\end{equation}

The round-$j$ QEC-based Lindbladian reshaping is implemented according to \Cref{sec:lindbladian_reshaping} via a random $n'$-qubit error-correcting code with recovery map $\RMap_j$. Applying it on top of the $n$-qubit twirled Lindbladian dynamics yields the following effective generator when acting on code-states $\rho_c = \Pi_0 \rho_c \Pi_0$:
\begin{equation} \label{eq:reshaped-logical-lindblad}
    \begin{aligned}
       & \LindbladEff^{(j)}(\rho_c)
    = \RMap_j \circ (\Lindblad_\mathrm{twrl}\otimes I_{n'-n}) (\rho_c) =  \\
    &=
    \sum_{\substack{P_k\in \DiagDissStruct, \\ P_k \notin \VCorrect}}
    a_{kk}
    \Bigl(
        (Q_{P_k}\otimes I_{n'-n})\rho_c(Q_{P_k}^{\dagger}\otimes I_{n'-n})
        -\rho_c
    \Bigr),
    \end{aligned}
\end{equation}
where $Q_P\coloneqq R_{r(P)}P\in\mathbb P_n$ denotes the reshaped Pauli obtained by composing $P$ with the recovery associated to its syndrome $r(P)$. Thus, previously identified terms are suppressed, while each unknown term $P_k\in\DHierSet_j\setminus\VCorrect$ enters the reshaped dynamics as the logical Pauli $Q_{P_k}$.

\vspace{-0.4em}
\subsection{Identifying dissipator candidates} \label{subsec:diss-candidate-identification}
\vspace{0.0em}

To identify the unknown candidates from $\DHierSet_j$, we use Bell sampling of the reshaped logical Lindbladian evolution $e^{\LindbladEff^{(j)}t}$. The resulting procedure requires total evolution time $t_{\mathrm{sample},j}=\SoftBigO{2^j}\le \SoftBigO{1/\eta}$.

To illustrate the principle, consider first evolution under a twirled Lindbladian from Eq.~\eqref{eq:lindblad-twirled} without hierarchical error correction. We first prepare a $2n$-qubit maximally entangled state $\rho_0 \coloneq \ket{\Phi_0}\bra{\Phi_0}$ where $\ket{\Phi_0} \coloneq \ket{\Phi_+}^{\otimes n}$ and $\ket{\Phi_+} = (\ket{00} + \ket{11})/\sqrt{2}.$ Applying $e^{\Lindblad_{\mathrm{twrl}} \tau}$ to the first qubit of each Bell pair yields:
\begin{equation}
    \begin{aligned}
        (e^{\Lindblad_{\mathrm{twrl}} \tau}\otimes I_n)(\ket{\Phi_0}\bra{\Phi_0}) = (1-\Gamma_\mathrm{tot}\tau)\ket{\Phi_0}\bra{\Phi_0} +\\+ \sum_{P_k \in \DiagDissStruct}\tau a_{kk} \ket{\Phi_k}\bra{\Phi_k} + \BigO{\tau^2},
    \end{aligned}
\end{equation}
where $\Gamma_\mathrm{tot} \coloneq \sum_{P_k \in \DiagDissStruct}a_{kk}$ denotes the total decay rate and $\ket{\Phi_k} \coloneq P_k \otimes I_n \ket{\Phi_0}$. Performing a Bell-basis measurement then yields outcomes with probabilities proportional to $\{a_{kk}\}$. We have thus reduced identification of sizable dissipator terms to the standard coupon collector problem. Let $\alpha_{\min}$ be the smallest nonzero rate we wish to identify. Repeating the experiment $\SoftBigO{1/(\alpha_{\min}\tau)}$ times then suffices to sample each Pauli term of interest at least once with high probability \footnote{
Combined with $\tau=\Theta(\eta/M^2)$ and thresholding of the resulting histogram, this immediately yields a direct procedure for identifying the $\eta$-heavy dissipator structure using total black-box evolution time $t=\SoftBigO{1/\eta}$ and $\SoftBigO{M^2/\eta^2}$ Bell-sampling experiments. We instead proceed with the hierarchical approach, reducing the number of Bell-sampling experiments to $\SoftBigO{M}$. Moreover, as discussed at the beginning of this section, the hierarchical approach makes it possible to replace Bell sampling with population recovery, reducing the number of ancillas from $n+\BigO{\log M}$ to $\BigO{\log M}$.}.

Bell sampling of the logical Lindbladian evolution $e^{\LindbladEff^{(j)}t}$ proceeds essentially identically, with two technical details.
First, since logical operators are equivalent up to multiplication by stabilizers, we merge samples belonging to the same stabilizer coset in post-processing. 
Second, the effective logical Paulis appearing in Eq.~\eqref{eq:reshaped-logical-lindblad} coincide with the unknown physical Paulis only up to multiplication by correction operators. To recover the original physical candidates, we therefore unfold each sampled logical Pauli by multiplying it by all $\BigO{M}$ correction operators of the code, as illustrated in \cref{fig:diss-struct-learner}~(b). 

Overall, by setting the per-experiment evolution time in round $j$ to $t_j = 2^j/M_D$, where $M_D \le \sqrt{M}$ is an upper bound on the dissipator sparsity $|\DiagDissStruct|$, we guarantee that for any $P_k \in \DHierSet_j$,
\begin{equation*}
    \Pr[P_k \text{ is covered}] \ge \Omega(a_{kk}t_j) = \Omega(1/M_D).
\end{equation*}
By repeating the Bell-sampling experiment $\SoftBigO{M_D}$ times, we ensure that each term in $\DHierSet_j$ is sampled at least once. Consequently, the logical Bell-sampling routine outputs a candidate superset $\widehat{\DHierSet}_j \supseteq \DHierSet_j$ of size at most $|\widehat{\DHierSet}_j|=\SoftBigO{MM_D}=\SoftBigO{M^{3/2}}$. The factor $M$ arises from lifting each sampled logical Pauli to its physical candidates. See Appendix~\ref{sec:diss-structure-sampling} for details on hierarchical candidate identification and the de-sparsified variant of the algorithm.

\vspace{-0.4em}
\subsection{Threshold-testing dissipator candidates} \label{subsec:diss-candidate-testing}
\vspace{0.0em}

The goal of this step is to remove false positives from the candidate set $\widehat{\DHierSet}_j$ via tolerant-testing. For robustness, we accept all terms with rates at least $\eta$ and reject all terms with rates below $\eta/2$. We show that this can be achieved with high probability using total evolution time $t_{\mathrm{test},j} = \mathcal{O}\bigl(\log{}(|\widehat{\DHierSet}_j|)/\eta\bigr)$ and $\mathcal{O}\bigl(\log{} (|\widehat{\DHierSet}_j| \cdot M_D)\bigr) = \BigO{\log M}$ ancillas. The resulting sub-routine based on syndrome measurements is shown in \Cref{fig:diss-struct-learner} (c).

The guiding observation here is that for a single round of QEC and any $P_k\in\mathcal P_{n'}$, the probability of observing the binary syndrome $r(P_k)$ is given by
\begin{equation} \label{eq:prob-syndrome-mass}
\Pr\!\bigl[\text{syndrome}=r(P_k)\bigr] =
\!\!\!\!\!\sum_{\substack{P_v\in\DiagDissStruct\\ r(P_v)=r(P_k)}}
\!\!\!\!\!a_{vv}\tau
+\mathcal O(M^2\tau^2),
\end{equation}\vspace{-0.3em}
which highlights the well-known identifiability issue~\cite{wagner2021optimal,wagner2022pauli} that a given syndrome rate receives contributions from multiple errors. While we can deterministically guarantee unique syndromes for the candidate terms $\widehat{\DHierSet}_j$, syndrome collisions with unknown nonzero terms may still occur. We circumvent this issue by drawing sufficiently many random stabilizer codes to ensure that each tested Pauli is syndrome-isolated from heavy dissipator terms at least once. Choosing $\tau = \Theta(\eta/M^2)$ guarantees that the second-order error in  Eq.~\eqref{eq:prob-syndrome-mass} is bounded. Thus, in the absence of collision,
\begin{align} \label{eq:syndrome-prob-gap}
    \Pr\!\bigl[\text{syndrome}=r(P_k) \mid a_{kk} \ge \eta\bigr] \ge \frac{7}{8}\eta\tau, \\ \label{eq:syndrome-prob-gap-2}
    \Pr\!\bigl[\text{syndrome}=r(P_k) \mid a_{kk} \le \frac{\eta}{2}\bigr] \le \frac{5}{8}\eta\tau.
\end{align}
Running the twirled Lindbladian dynamics for
$N_{\mathrm{QEC}}=\BigO{\log{}(|\widehat{\DHierSet}_j|)/(\eta\tau)}$
consecutive error-correction rounds suffices to resolve the gap between Eq.~\eqref{eq:syndrome-prob-gap} and Eq.~\eqref{eq:syndrome-prob-gap-2} for all candidate terms in parallel. This is guaranteed with high probability by the multiplicative Azuma's inequality~\cite{freedman1975martingales,kuszmaul2025multiplicative}. Altogether, this yields a tester with total black-box evolution time of $t_{\mathrm{test},j}=\BigO{\log{}(|\widehat{\DHierSet}_j|)/\eta}$. See Appendix~\ref{sec:diss-struct-testing} for details and the de-sparsified version of the algorithm. 

By combining the candidate identification and testing recursively, we eventually obtain $\EstEtaDiagDissStruct \gets \VCorrect$ that approximates the true $\eta$-heavy dissipator structure:
\begin{equation}
    \EtaDiagDissStruct \subseteq \EstEtaDiagDissStruct \subseteq \HalfEtaDiagDissStruct,
\end{equation}
using $t_\mathrm{tot} = \SoftBigO{1/\eta}$ evolution time. See Appendix~\ref{sec:appendix_dissipator_structure_learning} for formal guarantees on the full hierarchical learning algorithm.

\vspace{0.0em}
\section{Hamiltonian learning}\label{sec:hamiltonian_learning}
\vspace{0.0em}

Once the $\eta(\varepsilon)$-heavy dissipator structure has been identified, we proceed to learn the Hamiltonian outside the identified dissipator footprint. At the desired accuracy $\varepsilon$, define the corresponding finite-resolution target as
\begin{equation}
\label{eq:hid-hnid-defs}
\EpsHNDStruct
\coloneqq
\HamiltStruct\setminus\EtaEpsFullDissStruct .
\end{equation}
By enforcing error correction of the identified dissipator via QEC
reshaping, we approximately reduce the problem to ansatz-free logical
Hamiltonian learning:
\begin{equation}
\label{eq:logical-hamilt-generator}
    \LindbladEff(\rho_c)
    \approx
    -i \sum_{\substack{
    P_k \in \EpsHNDStruct,\\
    r(P_k)=0}}
    h_k[P_k\otimes I_{n'-n},\rho_c].
\end{equation}
To solve this problem, we adapt the hierarchical Heisenberg-limited
algorithm of~\cite{hu2025ansatz,sinha2025improvedhamiltonianlearningsparsity}
to the logical setting. At a high level, the resulting procedure closely
parallels the dissipator structure learner, recursively identifying
progressively weaker terms by alternating between candidate identification
and candidate testing. We therefore defer the full description to
Appendix~\ref{sec:appendix-hamilt-learning} and highlight only the key
technical differences relative to the dissipator structure learner of
\Cref{sec:dissipator_struct_learning}.

The first major difference is that, in the Hamiltonian setting, the candidate-testing step can be performed at the Heisenberg limit by directly estimating the coefficients via Robust Frequency Estimation~\cite{kimmel2015robust}. Consequently, the \HNDAbbrev{} Hamiltonian protocol not only identifies the heavy structure but also learns the corresponding coefficients to accuracy $\varepsilon$ in the process.

Another major difference lies in the Lindbladian reshaping subroutine. First, to avoid the suppression of Hamiltonian evolution, we no longer apply physical twirling during structure identification. Instead, we rely on QEC-based reshaping, which suppresses both the heavy dissipator and the Hamiltonian terms identified in previous rounds. More importantly, by Eq.~\eqref{eq:logical-hamilt-generator}, only zero-syndrome Hamiltonian terms survive Zeno suppression. We therefore draw $\SoftBigO{M}$ random QEC codes so that each target Hamiltonian term survives QEC reshaping for at least one code. This requirement introduces an additional factor of $M$ in the total evolution time, yielding $t_{\mathrm{tot}}=\SoftBigO{M^{2}/\varepsilon}$ compared to the $\SoftBigO{M/\varepsilon}$ scaling achieved for noiseless ansatz-free Hamiltonian learning~\cite{sinha2025improvedhamiltonianlearningsparsity}.

\vspace{0.0em}
\subsection{Hamiltonian learning beyond HDD at finite resolution}
\label{sec:finite-resolution-hamilt-learning}
\vspace{0.0em}
We now discuss the learning guarantees for the Hamiltonian terms that enter the finite-resolution target \(\EpsHNDStruct\) beyond the asymptotic \HNDAbbrev{} sector \(\HNDStruct\). Since
$\EtaEpsFullDissStruct\subseteq\FullDissStruct$, we can write
\begin{equation}
    \EpsHNDStruct
    =
    \HNDStruct
    \cup
    \left[
    \HamiltStruct\cap
    \bigl(\FullDissStruct\setminus\EtaEpsFullDissStruct\bigr)
    \right].
\end{equation}
Thus, operationally, in addition to the full \HNDAbbrev{} sector, our learner
may access Hamiltonian terms in the weak dissipator footprint, as illustrated
in \cref{fig:alg-overview}~(d). While \Cref{res:informal-main} guarantees Heisenberg-limited
learning of all coefficients in $\HNDStruct$,
we now ask when the same guarantee extends to these additional terms.
Note that this enlargement is intrinsically finite-resolution: for any fixed Lindbladian, as $\varepsilon\to0$ and $\eta(\varepsilon)\to0$, eventually $\EtaEpsFullDissStruct = \FullDissStruct$, and the additional sector vanishes.

The key difficulty in this finite-resolution regime is a phenomenon we call
weak--strong Kossakowski coupling. The following minimal example illustrates
the effect. Consider a single imbalanced jump operator
\begin{equation}
    J=X+\sqrt{\frac{\varepsilon}{2}}Y
\end{equation}
superimposed on a Hamiltonian term $H=h_z Z$. The corresponding Lindbladian is
\begin{equation}
    \begin{aligned}
        \Lindblad(\rho)
        &=
        -ih_z[Z,\rho]
        +\sqrt{\frac{\varepsilon}{2}}(Y\rho X+X\rho Y)+\\
        &\quad +(X\rho X-\rho)
        +\frac{\varepsilon}{2}(Y\rho Y-\rho).
    \end{aligned}
\end{equation}
Note the off-diagonal entry of order $\sqrt{\varepsilon}$ that saturates the positivity bound $|a_{XY}|\le \sqrt{a_{XX}a_{YY}}$. If the dissipator-structure learner is
run at threshold $\eta\simeq\varepsilon$, it identifies the strong $X$
direction but not the weak $Y$ direction. After QEC reshaping suppresses the identified $X$ error, the weak--strong $X/Y$ coupling can induce an effective Hamiltonian $Z$ bias of order $\sqrt{\varepsilon}$, too large for an $\varepsilon$-accurate Hamiltonian estimate:
\begin{equation}
    \LindbladEff(\rho)
    =
    -i\left(h_z-\sqrt{\frac{\varepsilon}{2}}\right)[Z,\rho]
    +\frac{\varepsilon}{2}(Z\rho Z-\rho).
\end{equation}
The obstruction is intermediate in scale: the diagonal $Y$ component of the original Lindbladian is too weak to be learned, but its off-diagonal Kossakowski coupling with $X$ is too strong to be negligible. This example illustrates why, without additional information about the
dissipator, Hamiltonian terms inside the weak dissipator footprint do not in
general inherit the Heisenberg-limited guarantee of the full
\HNDAbbrev{} sector $\HNDStruct$, as formalized by
\Cref{res:hdd-optimality}.

To exclude precisely this imbalanced weak--strong regime while otherwise allowing off-diagonal Kossakowski structure, we introduce the balanced
Kossakowski-tail condition
\begin{equation}
\label{eq:main-balanced-tail-cond}
|a_{kj}|
\le
|\DiagDissStruct|\min\{a_{kk},a_{jj}\},
\qquad
P_k,P_j\in\DiagDissStruct .
\end{equation}
Importantly, the condition remains compatible with many common noise models. For example, amplitude damping is allowed: a jump proportional
to $X+iY$ produces a balanced $X/Y$ block, whose two directions are either
learned and corrected together or remain jointly below the final
$\varepsilon$-accuracy budget. Standard depolarizing and Pauli-dephasing
noise satisfy the condition trivially because their Kossakowski matrices are
diagonal. Collective dephasing or drive fluctuations~\cite{kraft2025bounded}
generated by jumps such as $\sum_i \alpha_i Z_i$ are also allowed, provided
the coupled amplitudes are not too imbalanced. We therefore get the following extension of \Cref{res:informal-main}:

\begin{result}[Heisenberg-limited finite-resolution Hamiltonian learning]
\label{res:finite-resolution-hamilt-learning}
Under the balanced Kossakowski-tail condition in
Eq.~\eqref{eq:main-balanced-tail-cond}, choosing
$\eta(\varepsilon)=\Theta(\varepsilon/M)$, the Hamiltonian learner estimates
all coefficients outside the learned dissipator footprint to accuracy
$\varepsilon$ with high probability:
\begin{equation*}
    |\widehat h_k-h_k|\le\varepsilon,
    \qquad
    \forall P_k\in
    \HamiltStruct\setminus\EstEtaEpsFullDissStruct ,
\end{equation*}
where $\EtaFullDissStruct \subseteq \EstEtaFullDissStruct \subseteq \HalfEtaFullDissStruct$. The total evolution time is Heisenberg-limited:
$t_{\mathrm{tot}}=\SoftBigO{M^2/\varepsilon}$.
\end{result}

Note that this balanced-tail condition is mild in a precise sense. Positivity gives $|a_{kj}|\le \sqrt{a_{kk}a_{jj}}$, so $|a_{kj}| \le |\DiagDissStruct|\min\{a_{kk},a_{jj}\}$ holds automatically whenever
\begin{equation}
\frac{\max\{a_{kk},a_{jj}\}}{\min\{a_{kk},a_{jj}\}}
\le
|\DiagDissStruct|^2 .
\end{equation}
Thus the condition only limits off-diagonal coupling between Pauli terms whose diagonal dissipative weights are imbalanced by more than a quadratic factor in sparsity. A more general version with an explicit imbalance parameter replacing $|\DiagDissStruct|$ is discussed in Appendix~\ref{sec:appendix-removing-known-dissipator-assumption}.

\vspace{0.0em}

\section{Learning remaining Lindbladian coefficients}\label{sec:sql-coeff-learning}
\vspace{0.0em}
The preceding sections identify the dissipator structure and learn the \HNDAbbrev{} Hamiltonian terms. It remains to estimate the dissipator coefficients \(\{a_{lm}\}\), together with the corresponding Hamiltonian-overlapping-dissipator coefficients \(\{h_k:P_k\in\HamiltStruct\cap\EstEtaEpsFullDissStruct\}\). At this stage, the relevant Pauli labels are already known, so the task is only to estimate the remaining numerical coefficients. We do this using the Choi state of the short-time channel. The advantage of the Choi representation is that the first-order action of the Lindbladian separates into simple, orthogonal directions corresponding to distinct Hamiltonian and dissipative terms.

Let $\rho_0$ be the $2n$-qubit maximally entangled state and define the Lindbladian Choi state
\begin{equation*}
    \rho(t)
    \coloneq
    (e^{t\mathcal L}\otimes I_n)(\rho_0),
\end{equation*}
where \(e^{t\mathcal L}\) acts only on one half of the underlying Bell pairs. For the Pauli terms whose coefficients remain unknown, we introduce a family of Choi observables $\{O_\lambda\}$. Each label $\lambda$ corresponds to one of the four cases: $\lambda 
\in \{h_k, a_{mm}, \Re(a_{lm}), \Im(a_{lm})\}$. As shown in Appendix~\ref{sec:appendix-dissipator-hid-coeff-learning}, expectation values of observables $\{O_\lambda\}$ take the common form
\begin{equation*}
\tr\left(O_\lambda\rho(t)\right)
=
\lambda t + O(t^2).
\end{equation*}
Thus each unknown coefficient is isolated by the derivative of one or two Choi expectation values at \(t=0\).

It remains to estimate these derivatives from finite-time experiments. For each target observable \(O\), define
\begin{equation*}
    f_O(t)
    \coloneq
    \tr\!\left(O\rho(t)\right).
\end{equation*}
We evaluate \(f_O(t)\) at Gauss--Chebyshev times \(t_q\in(0,\tau_{\max})\), fit a low-degree Chebyshev interpolant, and take the derivative of the interpolant at \(t=0\). At each time node, all target observables can be estimated in parallel using global Clifford classical shadows~\citep{huang2020shadows}. This parallelization is efficient because the observables $\{O_\lambda\}$ are low-rank and Hilbert--Schmidt normalized. Consequently, the number of shadow snapshots depends only logarithmically on the number of coefficients being estimated. Combining the shadow estimates across the Chebyshev nodes gives simultaneous derivative estimates for all remaining coefficients.

Putting these ingredients together yields the standard-quantum-limited Lindbladian coefficient learner proved in \Cref{thm:appendix-lindblad-coefficients_learning_shadows}. Applied to the Hamiltonian-overlapping-dissipator terms and to the learned dissipator footprint, the algorithm estimates all remaining coefficients to accuracy \(\varepsilon\) with high probability in total evolution time
\begin{equation}
    t_{\mathrm{tot}}
    =
    \widetilde{\mathcal O}
    \!\left(
        \frac{M}{\varepsilon^2}
    \right).
\end{equation}

\vspace{0.0em}
\section{Conclusion and Outlook}\label{sec:outlook}
\vspace{0.0em}
In this work, we developed an ansatz-free framework for learning arbitrary
sparse Lindbladians using quantum error correction. Our algorithms learn the \HNDAbbrev{} Hamiltonian sector at the Heisenberg limit without prior knowledge of either the Hamiltonian or dissipator structures, while providing an efficient SQL algorithm for learning the full Lindbladian. We further show that the \HNDAbbrev{} sector is the maximal Hamiltonian sector admitting a uniform Heisenberg-limited guarantee in the unrestricted ansatz-free setting. These results add to a growing viewpoint that QEC can serve not only as a tool for protecting quantum information, but also as an active diagnostic primitive for probing microscopic dynamics~\cite{fowler2014scalable,blume2025estimating,
takou2025estimating,remm2026experimentally,wagner2021optimal,
wagner2022pauli,wagner2023learning,xiao2026insitu,zheng2026efficient}.
Looking ahead, we highlight several open questions.

\noindent\textbf{Reducing short-time requirements.}
Our algorithms rely on frequently applying quantum controls to implement
Lindbladian reshaping and attain the Heisenberg- and
standard-quantum-limited scalings. The resulting short-time control requirements pose a central challenge for experimental and fault-tolerant implementations~\cite{cotler2026noisylearning,kannan2026exponential}.
As such, a valuable direction lies in designing Lindbladian-learning
algorithms that admit longer evolutions, similar
to~\cite{shin2026mintime,depradenne2026longtimes}.
Several concurrent and subsequent works have made substantial progress
toward optimal minimum evolution times in various
settings~\cite{arad2026nearoptimallearninglocallindbladians,
mobus2026robuststructurelearningklocal,
lewis2026learningstructureopenquantum}, but the known guarantees can still
carry strong dependence on interaction strength, locality, or large
constant prefactors. Achieving a practically accessible time resolution
therefore remains an important open problem for experimental implementations of Lindbladian learning.

\noindent\textbf{Exploiting low-rank dissipative structure.}
Our approach strongly relies on a compact Pauli-basis description of the Lindbladian. While this covers many common interactions and noise models, a single jump operator can have exponentially broad Pauli support. Motivated by low-rank quantum tomography methods that exploit low state or
Kraus rank~\cite{flammia2012quantum,Kliesch2019guaranteedrecovery}, it is natural to ask whether a Kossakowski-rank promise can similarly be exploited in ansatz-free Lindbladian learning. On the \HNDAbbrev{} Hamiltonian-learning side, such an extension would require suppressing general learned error subspaces rather than individual Pauli errors, naturally pointing toward QEC constructions beyond the stabilizer framework.

\noindent\textbf{Beyond exact sparsity.}
Our algorithms assume an $M$-sparse Lindbladian, but in practice one may
instead expect decay or norm bounds on the coefficient vector rather than
exact sparsity. Indeed, a Lindbladian may contain exponentially many very
small Pauli coefficients whose aggregate contribution is controlled in an
appropriate norm, and such tails need not affect learning at the target
accuracy. This motivates desparsifying the framework. We partially address
this by desparsifying the heavy dissipator structure learner in
Appendix~\ref{sec:appendix_dissipator_structure_learning}. While concurrent works have relaxed exact-sparsity
assumptions~\cite{chen2026learning,lewis2026learningstructureopenquantum,
mobus2026robuststructurelearningklocal}, extending the Heisenberg-limited
\HNDAbbrev{} guarantee to effective-sparsity or norm-tail assumptions remains open.

\noindent\textit{Note.}
During the completion of this manuscript, we became aware of several concurrent and independent works~\cite{arad2026nearoptimallearninglocallindbladians,mobus2026robuststructurelearningklocal,lewis2026learningstructureopenquantum,sinha2026lindblad,song2026efficient,chen2026learning,zhou2026characterizingarbitrarylindbladiandynamics} that address Lindbladian learning. The works of \citet{arad2026nearoptimallearninglocallindbladians,mobus2026robuststructurelearningklocal,lewis2026learningstructureopenquantum} focus on the $k$-local setting and are therefore complementary to our sparse, arbitrary-locality setting. The work of \citet{sinha2026lindblad} is likewise complementary: it considers a similar sparse Lindbladian setting, but focuses on SPAM robustness and identifiability rather than optimal scaling with the target accuracy $\varepsilon$. Most closely related to our sparse full-Lindbladian result are
\citet{chen2026learning,zhou2026characterizingarbitrarylindbladiandynamics}, which achieve efficient SQL Lindbladian learning without locality assumptions, ancillas, or interleaved control, thereby substantially relaxing the experimental requirements. Our distinct contribution is Heisenberg-limited \HNDAbbrev{} Hamiltonian learning without prior knowledge of either the Hamiltonian or dissipator structures, together with the corresponding optimality boundary.

\section{Acknowledgment}
We are thankful for the insightful discussions with Richard R. Allen, Maria Elovenkova, Antonio Anna Mele, Jordi A. Monta\~{n}\`{a} L\'{o}pez, Liyuan Chen, Nik O. Gjonbalaj, John M. Martyn, Juan Castaneda. We acknowledge the support from DOE through
the QUACQ program (DE-SC0025572) and from the NSF through the CUA PFC (PHY-2317134) and QuSEC (OMA-2326787). N.R. acknowledges support from the NTT research fellowship. W.G. acknowledges support from NSF Grant CCF-2430375 and the Von Neumann Award from Harvard Computer Science.

\section{AI statement}
The authors developed the learning algorithms and identified preliminary hard instances underlying the lower-bound construction. GPT-6 was subsequently used to prove the adaptive lower bound and to assist with the organization, presentation, and polishing of several algorithmic proofs. All AI-assisted arguments and text were independently checked, revised, and approved by the authors, who take full responsibility for the content of the manuscript.

\clearpage
\bibliography{bib/main}

\clearpage
\onecolumngrid

\makeatletter
\markboth{\MakeUppercase{\@shorttitle}}{\MakeUppercase{\@shorttitle}}
\makeatother
\clearpage

\appendix

\EnableTOC

\begingroup
\setlength{\parskip}{0pt}  

\makeatletter
\def\l@subsubsection#1#2{} 
\renewcommand{\tocname}{Contents of Appendix}
\tableofcontents
\endgroup

\clearpage


\input{appendix/related_work}

\input{appendix/background}

\input{appendix/Lindbladian_reshaping}

\input{appendix/dissipator_structure_learning}
\input{appendix/hamiltonian-HDD-learning}
\input{appendix/HDD-barrier}
\input{appendix/dissipator_coefficients_learning}

\input{appendix/interface-noise}

\input{appendix/reducing_ancillas}

\end{document}

%% file: appendix/related_work.tex
\section{Related Work}\label{sec:related_work}

\noindent\textbf{QEC-enhanced quantum metrology.}
The problem of parameter estimation has a long history in quantum metrology and sensing. In its simplest form, one considers a one-parameter Hamiltonian $H(\omega)=\omega G$ and seeks to estimate $\omega$ to accuracy $\varepsilon$ by preparing probes, letting them interact with the system, and measuring the resulting states. The natural resource is the total evolution time $t_{\mathrm{tot}}$. With independent probes and classical averaging, the estimation error decreases only as $1/\sqrt{t_{\mathrm{tot}}}$, giving the standard quantum limit $t_{\mathrm{tot}}=\mathcal{O}(1/\varepsilon^2)$~\citep{braunstein1994statistical}. In noiseless dynamics, this scaling can be improved to the Heisenberg limit $t_{\mathrm{tot}}=\mathcal{O}(1/\varepsilon)$ by coherently accumulating phase, either in parallel using entangled probe states or sequentially through phase-estimation-type protocols~\citep{giovannetti2006quantum,higgins2007entanglement,kitaev1995quantum}. However, this Heisenberg scaling is fragile: generic Markovian dissipation destroys the coherent phase accumulation needed for Heisenberg-limited estimation~\citep{huelga1997improvement,demkowicz2012elusive,escher2011general,demkowicz2014using,smirne2016noisyfreqest}. More precisely, it has been established that Hamiltonian terms that lie in the noise span are constrained to standard-quantum-limited scaling, while Heisenberg scaling can be recovered only for directions outside this dissipative span~\citep{zhou2018achieving,demkowicz2017qecmetrology,gorecki2020optimal}. For such identifiable coherent directions, quantum error correction emerged as a central tool to suppress the dissipative dynamics while preserving the signal Hamiltonian as a nontrivial logical, allowing the encoded state to evolve for long times and accumulate phase coherently~\citep{dur2014improved,kessler2014quantum,arrad2014sensingqec,marti2015qecenhanced,zhou2018achieving,demkowicz2017qecmetrology,gorecki2020optimal,zhou2020optimal,reiter2017dissipative,layden2018spatial,shettell2021practical}. 

A rich family of QEC-metrology protocols has since been developed.
Ancilla-free constructions show that, under additional commutation or code-design conditions, noiseless ancillas are not always necessary~\citep{layden2019ancillafree,zhou2024ancillafree}.
Approximate-QEC frameworks balance residual noise against signal retention~\citep{zhou2020optimal}, while multiparameter QEC-metrology addresses the joint estimation of several Hamiltonian directions under Markovian dissipation~\citep{gorecki2020optimal}.
Stabilizer-code constructions further specialize these ideas to many-body Hamiltonian sensing, including surface-code, Reed–Muller, Shor, and repetition-code protocols that protect nontrivial many-body logical signals~\citep{antu2025stabilizercodes,qiao2026distributed,marrero2026encoded}.
Recent fault-tolerant variants have begun to incorporate noisy state preparation, syndrome extraction, and logical readout, establishing thresholds below which Heisenberg scaling can still be recovered~\citep{kapourniotis2019faulttolerant,sahu2026achieving}. Despite this breadth, these protocols rely crucially on prior knowledge of the dissipative and Hamiltonian structures: the code is designed specifically to correct the known noise operators while retaining sensitivity to the target Hamiltonian. It has therefore remained an open problem whether Heisenberg-limited Hamiltonian learning is possible when neither the Hamiltonian nor the dissipator are known in advance.
\medskip

\noindent\textbf{Quantum metrology of noise.}
A complementary line of work treats noise itself as the signal to be estimated.
Early examples include quantum estimation of loss rates and dissipative couplings, where the goal is to optimize probes and measurements for a specified Markovian open-system model~\citep{monras2007optimal,benatti2014dissipative,beau2017nonlinear}.
For environments with non-negligible temporal correlations or memory effects, quantum noise spectroscopy methods have been developed, using controlled probe dynamics, often through dynamical-decoupling filter functions, to reconstruct noise spectra and higher-order correlations~\citep{alvarez2011measuring,norris2016qubit,pazsilva2017multiqubit,szankowski2017environmental,montana2025efficiently}.
Returning to the Markovian setting, recent work has shown that entanglement can give large advantages for sensing collective or all-to-all Lindbladian noise~\citep{wang2024exponential,brady2026correlated}.
Related estimation results characterize known Lindblad rates and multiparameter Markovian noise~\citep{gardner2025lindblad,brady2026precision}. These works typically assume a known noise model and a narrow local parameter regime around a calibrated operating point, then use quantum Fisher information and Cramér–Rao-type bounds to optimize high-precision estimators for the pre-specified structure.
The setting considered in this work is complementary: we neither assume the structure of the Lindbladian nor restrict to a narrow parameter neighborhood, so the resulting ansatz-free Lindbladian learner can serve as a coarse calibration and structure-discovery stage feeding into subsequent high-precision metrology protocols.
\medskip

\noindent\textbf{Hamiltonian learning.}
Hamiltonian learning asks one to reconstruct the coefficients of an unknown Hamiltonian, often written in a Pauli expansion $H=\sum_i h_i P_i$. Although a completely unstructured $n$-qubit Hamiltonian has exponentially many Pauli coefficients, many physically relevant systems are local or sparse, and recent Hamiltonian-learning protocols exploit these properties to achieve sample and runtime guarantees with only polynomial dependence on the system size. Progress has come from both static quantum data and real-time dynamics. One line of work learns Hamiltonians from static objects, such as eigenstates or Gibbs states~\citep{qi2019determining, evans2019scalable, gu2024practical, anshu2021sample, rouze2024learning, haah2024learning, bakshi2024learning}. These approaches avoid coherent real-time evolution, but require access to sufficiently informative states, whose preparation can itself be computationally or experimentally demanding~\citep{kempe2004complexity, brandao2019finite}.

A parallel line of work learns Hamiltonians directly from dynamics~\citep{da2011practical,wiebe2014hamiltonian,wiebe2014quantum,wang2017experimental,bairey2019learning,zubida2021optimal,haah2024learning,stilck2024efficient,gu2024practical,castaneda2025hamiltonian}. The central idea is that short-time expectation values of suitable observables contain linear information about the Hamiltonian coefficients. In particular, if the relevant Pauli term $P_i$ is known, one can always choose a probe state $\rho$ and observable $O$ such that $\langle O(t)\rangle=2h_i t+\mathcal{O}(t^2)$, or equivalently $\frac{d}{dt}\langle O(t)\rangle|_{t=0}=2h_i$~\citep{haah2024learning}. Estimating such derivatives and assembling the resulting equations gives scalable coefficient-learning algorithms, with standard-quantum-limited sample complexity $\mathcal{O}(1/\varepsilon^2)$ achievable under appropriate locality or sparsity assumptions~\citep{haah2024learning,stilck2024efficient,gu2024practical}. 

More recently, a major effort has focused on achieving Heisenberg-limited scaling for Hamiltonian learning~\citep{huang2023learning, li2024heisenberg, mirani2024learning, ni2024quantum, dutkiewicz2024advantage, bakshi2024structure, ma2024learning, zhao2025learning, hu2025ansatz, sinha2025improvedhamiltonianlearningsparsity, abbas2025nearly, shin2026mintime, he2026optimal}.
The first such protocols in the many-body setting used Hamiltonian reshaping, which coherently amplifies the signal of selected coefficients and enables estimation with total evolution time $\mathcal{O}(1/\varepsilon)$~\citep{huang2023learning}.
Subsequent works developed related Heisenberg-limited approaches using additional quantum control, iterative bootstrapping, and extensions to bosonic and fermionic systems~\citep{dutkiewicz2024advantage,li2024heisenberg,mirani2024learning,ni2024quantum}.
A central challenge in this line is removing the need to know the Hamiltonian structure in advance, since an unknown Hamiltonian may be supported on an exponentially large set of possible Pauli terms.
This challenge has been addressed in local and bounded-interaction settings using structure-learning procedures based on Pauli-channel estimation~\citep{yu2023robust,harper2021fast} and Goldreich--Levin-type search with bootstrapping~\citep{bakshi2024structure}.
More recently, \citet{hu2025ansatz} developed a generic ansatz-free Hamiltonian-learning framework for arbitrary sparse Pauli Hamiltonians, combining support discovery with Heisenberg-limited coefficient estimation.

We also highlight two practically motivated lines of work in Hamiltonian learning.
Since many Heisenberg-limited protocols rely on frequent quantum control and arbitrarily short-time queries, recent work has studied how to retain learning guarantees when each query has a nonvanishing minimum duration, when only long-time evolution is available, and even when the total evolution time is uncontrollable and unknown~\citep{shin2026mintime,bakshi2024structure,depradenne2026longtimes}.
Another important constraint is limited spatial access: quantum probe tomography shows that, for structured many-body Hamiltonian families, one can reconstruct Hamiltonian parameters from a single local probe coupled to a small subsystem of a thermal state, with query complexity polynomial in \(1/\varepsilon\) and classical post-processing time polylogarithmic in \(1/\varepsilon\)~\citep{chen2025quantumprobetomography}. These directions begin to circumvent the need for linear-in-time response functions that are only accessible when one can assume short-time access.

These developments leave open the regime most relevant to noisy devices: Hamiltonian learning from Lindbladian dynamics, where dissipation can obscure the long-time coherent evolution needed for Heisenberg-limited scaling~\citep{huelga1997improvement,demkowicz2012elusive,escher2011general,demkowicz2014using,smirne2016noisyfreqest,zhou2018achieving,demkowicz2017qecmetrology,gorecki2020optimal,cotler2026noisylearning}.
The present work addresses this ansatz-free noisy setting by learning the Hamiltonian component disjoint from the dissipator at the Heisenberg limit, without prior knowledge of either the Hamiltonian or dissipator supports.
\medskip

\noindent\textbf{Lindbladian learning.}
Lindbladian learning seeks to identify the generator of open-system dynamics and has recently attracted growing attention as a natural extension of Hamiltonian learning. A prominent line of works reconstructs the channels $e^{t\mathcal L}$ at one or several times and then infers the generator $\mathcal L$ from these tomographic snapshots~\citep{boulant2003robust,samach2022lindblad,dobrynin2024compressed,onorati2023fitting1,onorati2023fitting2,liu2025robust}. Unlike in the Hamiltonian case, however, taking a logarithm of $e^{t\mathcal L}$ does not generally yield a unique or valid Lindbladian: one must decide whether some branch of the complex logarithm corresponds to a legitimate time-independent generator, a version of the \emph{Markovianity problem}, which is \textbf{NP}-hard in general~\citep{wolf2008assessing,cubitt2012complexity,liu2025robust}. Even when this ambiguity is absent, such methods inherit the exponential cost of full process tomography and are therefore mostly limited to few-qubit systems.

Another line of work learns Lindbladians from steady states~\citep{bairey2020learning}. This can be more scalable in favorable settings, but steady states generally do not determine the generator uniquely. For example, many distinct Lindbladians with Hermitian jump operators share the maximally mixed state as a steady state~\citep{baumgartner2008analysis}. Moreover, even when the steady state fixes the generator structure, it cannot determine the overall rate, since $\mathcal L$ and $\gamma\mathcal L$ have the same fixed points. Preparing the relevant steady states can also be experimentally demanding~\citep{temme2010chi}.

Scalable guarantees have also been obtained under additional structural assumptions, most commonly locality, sparsity, or a known operator ansatz. 
One approach extends short-time Hamiltonian-learning ideas to open-system dynamics by estimating derivatives of observables under \(e^{t\mathcal L}\) and using them to form linear systems for the Lindbladian coefficients~\citep{stilck2024efficient,franca2025learning,montana2025efficiently}. 
For local Lindbladians, robust interpolation gives standard-quantum-limited sample complexity, \(\mathcal{O}(1/\varepsilon^2)\), and related ideas have been extended to time-dependent Lindbladians and structured non-Markovian dynamics~\citep{stilck2024efficient,franca2025learning,montana2025efficiently}. 
Testing problems, such as detecting dissipative contributions, have also been considered~\citep{cai2026optimal}. 
In parallel, numerical and model-fitting approaches have demonstrated promising Lindbladian reconstruction in medium- and large-scale digital and analog systems, typically under model-specific structural assumptions or without worst-case performance guarantees~\citep{pastori2022characterization,olsacher2025hamiltonian,kraft2025bounded,berg2025large,birke2026demonstrating}.

A related multiparameter-sensing approach uses quantum scrambling, in the style of classical shadows, to estimate many coherent and incoherent Pauli signals in parallel~\citep{gong2026multiparameter}. 
While sample-efficient for large pre-specified families of Hamiltonian and dissipative rates, an ansatz-free application would require searching over exponentially many Pauli terms and does not directly learn general off-diagonal Kossakowski coherences. 
This is complementary to our goal of learning arbitrary sparse Lindbladians with unknown support using polynomial quantum and classical resources.

Most recently, ansatz-free Lindbladian learning was shown to be possible in situ by identifying candidate Hamiltonian and dissipative supports directly from short-time channel data~\citep{ivashkov2026ansatzfreelindblad}. 
This removes the need for prior knowledge of the operator structure, but the structure-identification step requires \(\mathcal{O}(1/\varepsilon^4)\) samples, bottlenecked by Hamiltonian structure identification from short-time Pauli error-rate derivatives. 
Thus, this ansatz-free in situ approach does not achieve the standard-quantum-limited \(\mathcal{O}(1/\varepsilon^2)\) scaling. 
Whether end-to-end ansatz-free Lindbladian learning can achieve the standard quantum limit has remained open.
\medskip

\noindent\textbf{Learning from syndromes.}
Syndrome data provide a natural source of information about noise in error-corrected devices, revealing faults without directly measuring the encoded state.
Early works extracted error models from the output of fixed error-detection circuits, while more recent experimental and phenomenological approaches use syndrome correlations to learn detector error models, decoder priors, or circuit-level noise features for a specified code or decoding experiment~\citep{fowler2014scalable,combes2014insitu,spitz2018adaptive,blume2025estimating,takou2025estimating,remm2026experimentally}.
A parallel theoretical line has characterized which Pauli noise parameters are identifiable from syndrome statistics alone for fixed stabilizer and data-syndrome codes~\citep{wagner2021optimal,wagner2022pauli}.
In this setting, the basic obstruction is that all errors with the same syndrome contribute to the same observed syndrome rate, so a fixed syndrome map only reveals certain combinations of physical error probabilities.
Going beyond physical operators, \citet{wagner2023learning} showed that syndrome data can also be used to learn the induced logical Pauli channel of a stabilizer, subsystem, or data-syndrome code under appropriate correctability assumptions.
Rather than estimating rare logical failures directly, the syndrome distribution can reveal the logical noise induced by the physical Pauli channel modulo stabilizers.
Recent work extends this syndrome-learning perspective to fault-tolerant Clifford circuits through spacetime-code formalisms, characterizing physical or logical Pauli noise from the syndrome data produced by a fixed circuit family~\citep{xiao2026insitu,zheng2026efficient}.

These works take the code or syndrome-extraction circuit as part of the object being characterized: they ask what physical, detector-level, or logical noise information is learnable from the syndrome stream of a chosen error-correction experiment.
This perspective is well suited to refining decoder weights, detector error models, and syndrome-extraction circuits for a given code.
Our goal is complementary: we seek the underlying physical generator itself, independent of any fixed code, so that the learned noise model can inform the design of tailored QEC procedures or guide hardware-level improvements.
In particular, our overall task goes beyond Pauli-channel learning: we aim to learn a sparse Lindbladian, including both Hamiltonian and dissipative components.
Within our algorithm, syndrome data are used to identify the strongest diagonal dissipative terms needed for QEC-based Hamiltonian learning.
By going beyond a fixed code, we circumvent the associated identifiability obstruction and ensure, through random stabilizer-code redraws, that each candidate Pauli is identifiable at least once with high probability.
This yields tolerant threshold tests for the $\eta$-heavy dissipator structure in total evolution time $\SoftBigO{1/\eta}$, in contrast with rate estimation, which remains standard-quantum-limited and requires $\Omega(1/\eta^2)$ total evolution time to achieve accuracy $\eta$.

\noindent\textbf{Noisy learning theory.}
Beyond noisy Hamiltonian and Lindbladian learning, recent work has studied the role of noise in learning from quantum experiments more generally. The core message of Ref.~\citep{cotler2026noisylearning} is that while future quantum processors may be fault tolerant, their interactions with quantum states or dynamics will be subject to noise, and that this consideration can eliminate several of the known advantages of quantum-enhanced learning protocols. In Appendix~\ref{sec:interface-noise} we analyze our Lindbladian learning algorithm under their NBQP noisy-oracle model and demonstrate that our performance guarantees persist up to an error floor determined by the quality of the query interface. Moreover, \citet{kannan2026exponential} propose a method of ``quantum uploading," demonstrating that an unknown experimental quantum system can be embedded into encoded memory such that quantum learning speedups can persist through a noisy processor-experiment interface. These results can be extended to encode the output state of unknown quantum dynamics as in Lindbladian learning, so that intermediate quantum processing can be executed effectively noiselessly. Therefore, optimizing our algorithm to lengthen query times and reduce reshaping steps would enable it to interface with quantum uploading and improve its practical noise-robustness. Complementarily, our progress on learning coherent and noisy dynamics simultaneously may suggest strategies to optimize the performance of other quantum learning primitives in structured noisy environments.

%% file: appendix/background.tex
\section{Preliminaries}

\subsection{Notation and global conventions}
\label{sec:appendix-notation}

\noindent\textbf{Basic spaces and superoperators.}
We consider an $n$-qubit probe system with Hilbert space $\mathcal H_n \coloneq (\mathbb C^2)^{\otimes n}$ with $d\coloneq 2^n$, and write $\mathcal B(\mathcal H_n)$ for the space of linear operators on $\mathcal H_n$. A linear map
$\Phi:\mathcal B(\mathcal H_n)\to\mathcal B(\mathcal H_n)$ is called a superoperator. We use the Hilbert--Schmidt inner product
\[
    \langle A,B\rangle_{\mathrm{HS}}
    \coloneq
    \tr(A^\dagger B).
\]
The adjoint $\Phi^\dagger$ of a superoperator $\Phi$ is defined with respect to this inner product by $\langle \Phi(A),B\rangle_{\mathrm{HS}} = \langle A,\Phi^\dagger(B)\rangle_{\mathrm{HS}}$, with $A,B\in\mathcal B(\mathcal H_n)$. For an observable $O$ and a state $\rho$, we write $\langle O\rangle_\rho \coloneq \tr(O\rho)$. For a unitary $U$, we write $\mathcal U_U$ for the conjugation channel $\mathcal U_U(X)\coloneq UXU^\dagger$.

Several parts of the algorithm append noiseless ancillas. When $n'-n$ ancilla qubits are added, the enlarged Hilbert space is
\[
    \mathcal H_{n'}\cong \mathcal H_n\otimes \mathcal H_{n'-n}.
\]
An $n$-qubit operator $A$ is then identified with $A\otimes I_{n'-n}$ unless stated otherwise.

For Choi-state calculations, we use left and right registers $L,R$ and define
\[
    \ket{\Phi_+}\coloneq \frac{1}{\sqrt2}(\ket{00}+\ket{11}),
    \qquad
    \ket{\Phi_0}\coloneq \ket{\Phi_+}^{\otimes n},
    \qquad
    \rho_0\coloneq \ket{\Phi_0}\bra{\Phi_0}.
\]
The Lindbladian evolution acts on the left register: $\rho(t) \coloneq (e^{t\Lindblad}\otimes I_R)(\rho_0)$. Unless stated otherwise in the Choi-state sections, an $n$-qubit operator $A$ is understood as $A\otimes I_R$.

\medskip
\noindent\textbf{Norms.}
For $x\in\mathbb C^m$ and $p\in[1,\infty)$, we use
\[
    \|x\|_p
    \coloneq
    \left(\sum_{i=1}^m |x_i|^p\right)^{1/p},
    \qquad
    \|x\|_\infty
    \coloneq
    \max_i |x_i|.
\]
The vector $\ell_\infty$ norm is used for entrywise coefficient and estimator errors. For an operator $X\in\mathcal B(\mathcal H_n)$ with singular values $\{s_i(X)\}_{i=1}^d$, the Schatten-$p$ norm is
\[
    \|X\|_p
    \coloneq
    \left(\sum_{i=1}^d s_i(X)^p\right)^{1/p},
    \qquad
    1\le p<\infty,
    \qquad
    \|X\|_\infty
    \coloneq
    \max_i s_i(X).
\]
Thus $\|X\|_1$ is the trace norm, $\|X\|_2$ is the Hilbert--Schmidt norm, and $\|X\|_\infty$ is the operator norm. We use
\[
    \|X\|_\infty\le \|X\|_2\le \|X\|_1,
    \qquad
    \|X\|_1\le \sqrt d\,\|X\|_2,
    \qquad
    \|X\|_2\le \sqrt d\,\|X\|_\infty.
\]
We also use Hölder's inequality for Schatten norms,
\[
    |\tr(A^\dagger B)|
    \le
    \|A\|_p\,\|B\|_q,
    \qquad
    \frac1p+\frac1q=1.
\]
For a superoperator $\Phi$, its diamond norm is
\begin{equation}
    \label{eq:diamond-norm-def}
    \|\Phi\|_\diamond
    \coloneq
    \sup_{r\ge 1}
    \sup_{X\neq0}
    \frac{\|(\Phi\otimes I_r)(X)\|_1}{\|X\|_1},
\end{equation}
where $I_r$ denotes the identity map on $\mathcal B(\mathbb C^r)$. Diamond norms are submultiplicative:
\[
    \|\Phi\circ\Psi\|_\diamond
    \le
    \|\Phi\|_\diamond\,\|\Psi\|_\diamond.
\]
If $\Phi$ is a quantum channel, then $\|\Phi\|_\diamond=1$. For a scalar function $f:I\to\mathbb C$ on an interval $I\subseteq\mathbb R$, we write
\[
    \|f\|_{\infty,I}
    \coloneq
    \sup_{t\in I}|f(t)|.
\]

\medskip
\noindent\textbf{Pauli conventions.}
We use $I, X, Y, Z$ to denote the identity and the three Pauli matrices:
\[
    I=\begin{pmatrix}1&0\\0&1\end{pmatrix},\quad
    X=\begin{pmatrix}0&1\\1&0\end{pmatrix},\quad
    Y=\begin{pmatrix}0&-i\\ i&0\end{pmatrix},\quad
    Z=\begin{pmatrix}1&0\\0&-1\end{pmatrix}.
\]
The single-qubit Pauli group $\mathbb{P}_1$ is the group generated by the three Pauli matrices along with the identity:
\[
    \mathbb{P}_1 = \langle I, X,Y,Z \rangle = \{\pm I,\pm iI,\pm X,\pm iX,\pm Y,\pm iY,\pm Z,\pm iZ \}.
\]
The $n$-qubit Pauli group $\mathbb{P}_n$ of order $|\mathbb{P}_n| = 4\cdot4^{n}$ is generated by single-qubit Pauli operators acting on each of the $n$ qubits:
\[
    \mathbb{P}_n = \langle \sigma_1\otimes\cdots\otimes \sigma_n \;:\; \sigma_i \in \{I,X,Y,Z \} \rangle.
\]
For computational purposes, a convenient Hermitian operator basis for the space of linear operators $\mathcal{B}(\mathcal{H})$ on an $n$-qubit Hilbert space $\mathcal{H} = (\mathbb{C}^2)^{\otimes n}$ is given by the set of all $n$-qubit Pauli strings:
\[
  \PnHerm = \{I, X,Y,Z\}^{\otimes n}.
\]
In other words, $\PnHerm$ is obtained by stripping off the overall phases from $\mathbb{P}_n$, such that any two elements $P_i,P_j \in \PnHerm$ are orthogonal to each other under the Hilbert–Schmidt inner product: $\tr{P_i^\dagger P_j} = 2^n\delta_{ij}$. The set $\PnHerm$ has cardinality $|\PnHerm| = 4^n$. When multiplying two Pauli strings $P_i,P_j\in\PnHerm$, the product is another Pauli up to a phase:
\[
    P_i P_j = \omega_{ij}\, P_{i\oplus j}, \qquad \omega_{ij}\in\{\pm1,\pm i\}.
\]
These phases record commutation relations of Paulis and satisfy $\omega_{ij}=\omega_{ji}^*$, so that $(\omega_{ij})$ is Hermitian. Keeping track of $\omega_{ij}$ allows us to work with the phase-stripped basis $\PnHerm$, which is, unlike $\mathbb P_n$, is not closed under multiplication. We also use a related definition of the commutation sign:
\begin{equation} \label{eq:pauli_commutation_sign}
    \operatorname{sgn}(P_i,P_j):=\omega_{ij}^2\in\{\pm1\},\quad\text{so that } \;
    P_i P_j=\operatorname{sgn}(P_i,P_j)\,P_j P_i.
\end{equation}

For any Pauli-group element \(Q=\omega P\), with
\(P\in\PnHerm\) and \(\omega\in\{\pm1,\pm i\}\), we write
\(\Herm(Q):=P\) for its phase-stripped Hermitian representative.

We adopt the convention of labeling $P_0 = I^{\otimes n}$ as the $n$-qubit identity operator. Every other element $P_i \in \PnHerm$ with $i \neq 0$ is traceless: $\tr{P_i} = 2^n \delta_{i0}$. The orthogonality and completeness of $\PnHerm$ means that any operator $X \in \mathcal{B}(\mathcal{H})$ can be uniquely expanded in the Pauli basis as:
\[
X = \sum_{i=0}^{4^n-1} x_i P_i,
\]
where the expansion coefficients are given by $x_i = \frac{1}{2^n}\tr{P_i X}$.

\medskip
\noindent\textbf{Lindbladian Pauli--Kossakowski form.}
Throughout the paper, the unknown time-independent $n$-qubit Lindbladian is written in the Pauli--Kossakowski form
\[
    \Lindblad(\rho)
    =
    -i\sum_{P_k\in\HamiltStruct} h_k \comm{P_k}{\rho}
    +
    \sum_{P_k,P_j\in\DiagDissStruct}
    a_{kj}
    \left(
        P_k\rho P_j
        -
        \frac12\acomm{P_jP_k}{\rho}
    \right),
\]
where $a\succeq0$ and $h_k \in \mathbb R$ and
\[
    \HamiltStruct
    \coloneq
    \{P_k\in\PnHerm\setminus\{I\}:h_k\neq0\}
\]
is the Hamiltonian structure. The Kossakowski matrix $a\succeq0$ is indexed by nonidentity Paulis. Its diagonal support is
\[
    \DiagDissStruct
    \coloneq
    \{P_k\in\PnHerm\setminus\{I\}:a_{kk}\neq0\}.
\]
Since $a\succeq0$, the condition $a_{kk}=0$ implies $a_{kj}=a_{jk}=0$ for every $j$. Hence all dissipative coefficients are supported on $\DiagDissStruct\times\DiagDissStruct$. We write
\[
    M_H\coloneq |\HamiltStruct|,
    \qquad
    M_D\coloneq |\DiagDissStruct|,
    \qquad
    M\coloneq M_H+M_D^2.
\]
Thus $M$ is the natural real-parameter count for the Hamiltonian coefficients together with the Hermitian Kossakowski submatrix on $\DiagDissStruct$.

The Lindbladian global scale can be absorbed into the evolution time. Therefore, we can assume the normalization
\[
    |h_k|\le1,
    \qquad
    |a_{kj}|\le1.
\]

\medskip
\noindent\textbf{Dissipator footprints and Heisenberg-learnable Hamiltonian targets.}
The dissipator structure $\DiagDissStruct$ induces the full dissipator footprint
\begin{equation*}
    \FullDissStruct
    \coloneq
    \DiagDissStruct
    \cup
    \bigl\{
        \Herm(P_\ell P_m):
        P_\ell,P_m\in\DiagDissStruct,\,
        P_\ell\neq P_m
    \bigr\}.
\end{equation*}
That is, $\FullDissStruct$ contains both the Pauli operators appearing directly in the dissipator and all Pauli operators generated, up to phase, by pairwise products of dissipator Paulis. We define the \HND{} (\HNDAbbrev{}) Pauli set as
\begin{equation*}
    \HNDStruct
    \coloneq
    \HamiltStruct\setminus\FullDissStruct,
\end{equation*}
with the complementary \HD{} set
\begin{equation*}
    \HDStruct
    \coloneq
    \HamiltStruct\cap\FullDissStruct.
\end{equation*}
In \cref{sec:hdd-boundary}, we show that, in the black-box learning model considered here, $\HNDStruct$ is the maximal set of Hamiltonian Pauli targets for which Heisenberg-limited coefficient learning can be guaranteed uniformly as $\varepsilon\to0$.

At finite target accuracy, however, sufficiently weak dissipative directions need not obstruct Heisenberg-limited learning. We therefore introduce finite-resolution versions of these structures. For a resolution scale $\eta>0$, define the $\eta$-heavy dissipator structure by
\begin{equation*}
    \EtaDiagDissStruct
    \coloneq
    \{P_k\in\DiagDissStruct:a_{kk}\ge\eta\}.
\end{equation*}
The corresponding $\eta$-heavy dissipator footprint is
\begin{equation*}
    \EtaFullDissStruct
    \coloneq
    \EtaDiagDissStruct
    \cup
    \bigl\{
        \Herm(P_\ell P_m):
        P_\ell,P_m\in\EtaDiagDissStruct,\,
        P_\ell\neq P_m
    \bigr\}.
\end{equation*}
The \HNDAbbrev{} and \HDAbbrev{} sets at scale $\eta$ are respectively
\begin{equation*}
    \EtaHNDStruct
    \coloneq
    \HamiltStruct\setminus\EtaFullDissStruct,
    \qquad
    \EtaHDStruct
    \coloneq
    \HamiltStruct\cap\EtaFullDissStruct.
\end{equation*}
Since $\EtaFullDissStruct\subseteq\FullDissStruct$, one has
$\HNDStruct\subseteq\EtaHNDStruct$; thus, at finite resolution, the Heisenberg-learnable Hamiltonian sector can be strictly larger than its asymptotic counterpart.

When the threshold is chosen as a function of the target accuracy, $\eta=\eta(\varepsilon)$, we write
\begin{equation*}
    \EtaEpsFullDissStruct
    \coloneq
    \EtaFullDissStruct\big|_{\eta=\eta(\varepsilon)},
    \qquad
    \EpsHNDStruct
    \coloneq
    \EtaHNDStruct\big|_{\eta=\eta(\varepsilon)}.
\end{equation*}


\medskip
\noindent\textbf{QEC notation.}
The stabilizer-code conventions are introduced in Appendix~\ref{app:stabilizer-prelims}. The recovery-map and Lindbladian-reshaping conventions are introduced in Appendix~\ref{sec:appendix-lindbladian-reshaping}.

\medskip
\noindent\textbf{Accuracy parameters and asymptotics.}
Hats denote estimators. We use $\delta$ for failure probabilities, $\eta$ for dissipator-structure thresholds, and $\varepsilon$ for final coefficient-learning accuracy. The symbol $\epsilon_s$ is reserved for nodewise sampling accuracy in Chebyshev or shadow-estimation routines. We write $N_{\exp}$ for the number of experimental repetitions, $t_{\mathrm{tot}}$ for total black-box evolution time, and $t_{\mathrm{min}}$ for the smallest evolution-time scale that the protocol must support.

We use standard Landau notation. The notation $\widetilde{\mathcal O}(\cdot)$ suppresses polylogarithmic factors in the relevant parameters, such as $n$, $M$, $1/\varepsilon$, $1/\eta$, and $1/\delta$.

\subsection{Lindbladian Taylor Tail}
\label{sec:appendix-lindbladian-taylor-tail}
Because our algorithms repeatedly invoke reshaping primitives---including physical Pauli twirling~\cite{emerson2007symmetrized} in Appendix~\ref{sec:appendix_dissipator_structure_learning} and the Lindbladian reshaping introduced in Appendix~\ref{sec:appendix-lindbladian-reshaping}---we will repeatedly need short-time expansions with controlled remainder terms. We therefore record \Cref{lem:lindblad-taylor-tail}, which we use throughout to bound evolution errors.

\begin{lemma}[Lindbladian Taylor tail]\label{lem:lindblad-taylor-tail}
Given a GKSL generator $\Lindblad$ and $\tau\ge 0$, the following is true:
\begin{equation*}
    e^{\tau\Lindblad} = I + \tau\Lindblad + \int_{0}^{\tau}(\tau -s)\Lindblad^2e^{s\Lindblad}\,ds.
\end{equation*}
Moreover, defining the remainder
\[
F(\tau)\;:=\;\int_{0}^{\tau}(\tau -s)\Lindblad^2e^{s\Lindblad}\,ds,
\]
we have the diamond-norm bound
\[
\|F(\tau)\|_\diamond \;\le\; \tfrac12\,\|\Lindblad\|_\diamond^{2}\,\tau^{2}.
\]
\begin{proof}
From the power series of $e^{s\Lindblad}$, it follows that
\begin{equation*}
    \frac{d}{ds}e^{s\Lindblad} = \Lindblad e^{s\Lindblad}.
\end{equation*}
By the fundamental theorem of calculus,
\begin{equation}\label{eq:lindblad-exponent-minus-identity}
    e^{\tau\Lindblad} - I
    = \int_{0}^{\tau}\frac{d}{ds}e^{s\Lindblad} \, ds
    = \int_{0}^{\tau}\Lindblad e^{s\Lindblad}\,ds .
\end{equation}
Integrate by parts with $u(s):=\tau-s$ and $v(s):=-\Lindblad e^{s\Lindblad}$ to obtain
\begin{equation*}
    \int_{0}^{\tau}\Lindblad e^{s\Lindblad}\,ds
    = -(\tau -s)\Lindblad e^{s\Lindblad}\Big|_{0}^\tau
      + \int_{0}^{\tau}(\tau -s)\Lindblad^2e^{s\Lindblad}\,ds
    = \tau \Lindblad + \int_{0}^{\tau}(\tau -s)\Lindblad^2e^{s\Lindblad}\,ds,
\end{equation*}
and plugging this back into Eq.~\eqref{eq:lindblad-exponent-minus-identity} gives the identity.

For the remainder bound, use submultiplicativity of $\|\cdot\|_\diamond$ and that
$e^{s\Lindblad}$ is CPTP (hence $\|e^{s\Lindblad}\|_\diamond=1$) for $s\ge 0$:
\begin{align*}
\|F(\tau)\|_\diamond
&\le \int_{0}^{\tau}(\tau-s)\,\|\Lindblad^2e^{s\Lindblad}\|_\diamond\,ds \\
&\le \int_{0}^{\tau}(\tau-s)\,\|\Lindblad\|_\diamond^2\,\|e^{s\Lindblad}\|_\diamond\,ds \\
&\le \|\Lindblad\|_\diamond^2 \int_{0}^{\tau}(\tau-s)\,ds
= \tfrac12\,\|\Lindblad\|_\diamond^{2}\,\tau^{2}.
\end{align*}
\end{proof}
\end{lemma}

\subsection{Lindbladian Diamond Norm Bound}
\label{sec:appendix-lindbladian-norm-bound}
Here, we present a simple Lindbladian diamond norm bound that would be useful for evaluating effective evolution errors in Appendix~\ref{sec:qec-effective-evolution} as well as the measurement error in \Cref{lem:syndrome-hit-prob}.

\begin{lemma}[Diamond norm of a Pauli-sparse Lindbladian]\label{lem:diamond-sparse-lindblad}
    Let $\mathcal H$ be a $d$-dimensional Hilbert space and let the Lindbladian $\Lindblad: \mathcal{B}(\mathcal{H}) \to \mathcal{B}(\mathcal{H})$ be expressed in the Pauli basis:
    \[
        \Lindblad(\rho)
        = -i \sum_{i} h_i [P_i,\rho]
          + \sum_{i,j} a_{ij}\!\left(P_i \rho P_j - \tfrac12\{P_j P_i,\rho\}\right),
    \]
    where each $P_i$ is a (multi-qubit) Pauli operator. Assume there are at most $M_H$ non-zero $h_i$ and at most $M_D^2$ non-zero $a_{ij}$, with $|h_i|\le 1$ and $|a_{ij}|\le 1$ for all non-zero coefficients. Let $M := M_H + M_D^2$. Then
    \begin{equation}
        \norm{\Lindblad}_\diamond \;\le\; 2M,
    \end{equation}
    where $\norm{\cdot}_\diamond$ denotes the diamond norm on density matrices from \cref{eq:diamond-norm-def}.
\end{lemma}

\begin{proof}
    For unitaries $U,V$, the map $\mathcal{S}_{U,V}(\rho) := U\rho V$ satisfies
    \[
        (\mathcal{S}_{U,V}\otimes I)(X) = (U\otimes I)X(V\otimes I),
    \]
    Since unitary matrices don't change the singular values, $\|(\mathcal{S}_{U,V}\otimes I)(X)\|_1 = \|X\|_1$, hence $\|\mathcal{S}_{U,V}\|_\diamond = 1$.

    \emph{Hamiltonian part.} For a single non-zero $h_i$,
    \[
        \Lindblad^{(H)}_i(\rho) := -i h_i [P_i,\rho]
        = -i h_i\big(P_i\rho - \rho P_i\big)
    \]
    is a linear combination of two maps of diamond norm $1$, so
    $\|\Lindblad^{(H)}_i\|_\diamond \le 2|h_i| \le 2$. Summing and using the triangle inequality,
    \[
        \|\Lindblad_H\|_\diamond
        := \Big\|\sum_i \Lindblad^{(H)}_i\Big\|_\diamond
        \le \sum_i \|\Lindblad^{(H)}_i\|_\diamond
        \le 2M_H.
    \]

    \emph{Dissipative part.} For a single non-zero $a_{ij}$,
    \[
        \Lindblad^{(D)}_{ij}(\rho)
        := a_{ij}\Big(P_i \rho P_j - \tfrac12 P_j P_i \rho - \tfrac12 \rho P_j P_i\Big)
    \]
    is a linear combination of three maps of diamond norm $1$, so
    $\|\Lindblad^{(D)}_{ij}\|_\diamond \le 2|a_{ij}| \le 2$. Thus
    \begin{equation}\label{eq:dissipator-diamond-norm}\|\Lindblad_D\|_\diamond
        := \Big\|\sum_{i,j} \Lindblad^{(D)}_{ij}\Big\|_\diamond
        \le \sum_{i,j} \|\Lindblad^{(D)}_{ij}\|_\diamond
        \le 2M_D^2.
    \end{equation}

    Finally, $\Lindblad = \Lindblad_H + \Lindblad_D$ implies
    \[
        \norm{\Lindblad}_\diamond
        \le \|\Lindblad_H\|_\diamond + \|\Lindblad_D\|_\diamond
        \le 2M_H + 2M_D^2 = 2M. \qedhere
    \]
\end{proof}
\begin{corollary}[Diamond norm bound for a diagonal Pauli dissipator]\label{cor:diamond-diag-pauli}
Assume $\Lindblad$ has no Hamiltonian part ($h_i=0$ for all $i$) and has only
diagonal Kossakowski entries, i.e.\ $a_{ij}=0$ for all $i\neq j$, with at most
$M_D$ nonzero diagonal coefficients $a_{ii}$ satisfying $|a_{ii}|\le 1$. Then
\begin{equation}
\label{eq:diamond-diag-pauli-bound}
\|\Lindblad\|_\diamond \;\le\; 2M_D.
\end{equation}
\end{corollary}

\begin{proof}
This is Lemma~\ref{lem:diamond-sparse-lindblad} with $M_H=0$ and with at most
$M_D$ nonzero dissipative terms (rather than $M_D^2$) since only the diagonal
coefficients $a_{ii}$ are present. Hence $\|\Lindblad\|_\diamond\le 2M_D$.
\end{proof}

\subsection{Stabilizer QEC Preliminaries}
\label{app:stabilizer-prelims}

\subsubsection{Stabilizers, code space, and projector}
A stabilizer group $S\subseteq \mathbb P_n$ is an abelian subgroup with $-I\notin S$.
Fix a (possibly redundant) generating set $G=\{G_\ell\}_{\ell=1}^m$.
The code space $\mathcal H_c$ is the joint $+1$ eigenspace of all $G_\ell$, with orthogonal projector
\begin{equation}
\Pi=\frac{1}{|S|}\sum_{\sigma\in S}\sigma
=\prod_{\ell=1}^m \frac{I+G_\ell}{2},\qquad \sigma\Pi=\Pi \sigma=\Pi\ \ \forall \sigma\in S.
\label{eq:stabilizer-projector}
\end{equation}

We will repeatedly work on the corner algebra:
\begin{equation}
\mathfrak A_c:=\Pi\,\mathcal B(\mathcal H)\,\Pi\cong\mathcal B(\mathcal H_c),
\quad\text{so that }\ \Pi X\Pi=X\ \text{ for }X\in\mathfrak A_c .
\label{eq:corner-algebra-def}
\end{equation}

Define the corresponding corner projection map $\mathcal C_\Pi: \mathcal B(\mathcal H) \to \mathfrak A_c$:
\begin{equation} \label{eq:def-projective-conjugation}
    \mathcal C_\Pi(X) \coloneq  \Pi X\Pi, \qquad \forall X\in \mathcal B(\mathcal H)
\end{equation}

\begin{definition}[Detectability]
\label{def:detectability}
A Pauli $Q\in \mathbb P_n$ is \emph{detectable} by the stabilizer code if it
anticommutes with at least one generator in $G$ (equivalently, with some $\sigma\in S$).
\end{definition}

\begin{proposition}[Detectability $\Longleftrightarrow$ compression to zero]
\label{prop:stab-detectability}
For a Pauli $Q\in \mathbb P_n$ and stabilizer projector $\Pi$, the following are equivalent:
\begin{enumerate}[label=(\roman*)]
\item $Q$ is detectable, i.e., there exists $g\in G$ with $\{g,Q\}=0$;
\item $\Pi Q \Pi = 0$.
\end{enumerate}
Consequently, for all $X\in\mathfrak A_c$, $\Pi Q X \Pi=0$ and $\Pi\{Q,X\}\Pi=0$.
\begin{proof}
(i)$\Rightarrow$(ii): Pick $g\in G$ with $\{g,Q\}=0$. Then
\[
g(\Pi Q\Pi)=(\Pi g)Q\Pi=\Pi(gQ)\Pi=-\Pi(Qg)\Pi=-(\Pi Q\Pi)g.
\]
Since $g$ acts as the identity on $\mathcal H_c$, the only operator on the code
that anticommutes with $g$ is $0$, hence $\Pi Q \Pi=0$.

(ii)$\Rightarrow$(i): Suppose $\Pi Q \Pi=0$ and, for contradiction, that $Q$ commutes
with every generator $G_\ell$. Then $Q$ preserves the code space:
for any $\psi\in\mathcal H_c$ we have $G_\ell(Q\psi)=Q(G_\ell\psi)=Q\psi$, so $Q\psi\in\mathcal H_c$.
Therefore $\Pi Q \Pi\,\psi=\Pi(Q\psi)=Q\psi$. But $\Pi Q \Pi=0$ forces $Q\psi=0$
for all $\psi\in\mathcal H_c$, which is impossible since $Q$ is unitary and
$\mathcal H_c\neq\{0\}$. Hence some generator must anticommute with $Q$.
\end{proof}
\end{proposition}

\subsubsection{Normalizer and logical Paulis}
The \emph{normalizer} (same as centralizer in $\mathbb P_n$) of \(S\) in the Pauli group is
\(N(S):=\{P\in \mathbb P_n:\ [P,\sigma]=0\ \forall \,\sigma\in S\}\).
For \(Q\in N(S)\) we define its logical action on the code by
\begin{equation}
\overline Q \;:=\; \Pi Q \Pi \in \mathfrak A_c .
\label{eq:logical-action}
\end{equation}
If \(Q\in S\) then \(\overline Q=\Pi\) (identity on the code).
More generally, \(Q\in N(S)\) acts trivially on the code up to an
overall phase iff \(Q\) is proportional to an element of \(S\);
otherwise \(\overline Q\) is a nontrivial logical Pauli up to global phase.

\subsubsection{Syndromes and syndrome projectors}

\paragraph{Syndrome (commutation) map.}
For a Pauli $P\in\mathbb P_n$ define its \emph{syndrome} with respect to the
generators $G=\{G_\ell\}_{\ell=1}^m$ by
\begin{equation}
r(P)\in\{0,1\}^m,\qquad
r_\ell(P)=
\begin{cases}
0,& [P,G_\ell]=0,\\[2pt]
1,& \{P,G_\ell\}=0.
\end{cases}
\label{eq:syndrome-map}
\end{equation}
Equivalently, $G_\ell P=(-1)^{r_\ell(P)}\,P G_\ell$.
Then:
\begin{enumerate}[label=(\alph*)]
\item $r(PQ)=r(P)\oplus r(Q)$ for all $P,Q\in\mathbb P_n$.
\item $r(P)=\mathbf 0$ iff $P\in N(S)$ (i.e.\ $P$ commutes with the stabilizer group).
\end{enumerate}

\paragraph{Syndrome projectors.}
Given \(G=\{G_\ell\}_{\ell=1}^m\), define syndrome projectors
\begin{equation}
\Pi_s \;:=\; \prod_{\ell=1}^m \frac{I+(-1)^{s_\ell}\, G_\ell}{2},
\qquad s\in\{0,1\}^m,
\qquad \sum_s \Pi_s = I,\ \ \Pi_s\Pi_{s'}=\delta_{s,s'}\Pi_s,
\label{eq:syndrome-projectors}
\end{equation}
with \(\Pi_{(0,\ldots,0)}=\Pi\).

\paragraph{Shift / block identities.}
For every $s,t\in\{0,1\}^m$ and $P\in\mathbb P_n$,
\begin{align}
\Pi_s\,P &= P\,\Pi_{\,s\oplus r(P)}, \qquad\qquad
P\,\Pi_s = \Pi_{\,s\oplus r(P)}\,P,
\label{eq:syndrome-shift-rule}\\[4pt]
\Pi_s\,P\,\Pi_t &= 
\begin{cases}
0, & t\neq s\oplus r(P),\\[2pt]
\Pi_s\,P = P\,\Pi_t, & t=s\oplus r(P).
\end{cases}
\label{eq:syndrome-block-structure}
\end{align}
\emph{Proof sketch.} Using \eqref{eq:syndrome-projectors},
\(
\Pi_s=\prod_{\ell=1}^m \frac{I+(-1)^{s_\ell}G_\ell}{2}.
\)
Move $P$ through each factor with $G_\ell P=(-1)^{r_\ell(P)}PG_\ell$; each factor gains
the bit flip $s_\ell\mapsto s_\ell\oplus r_\ell(P)$, yielding \eqref{eq:syndrome-shift-rule}.
Multiplying on the right by $\Pi_t$ and using orthogonality
$\Pi_u\Pi_v=\delta_{u,v}\Pi_u$ gives \eqref{eq:syndrome-block-structure}.
\qed

\paragraph{Detectability in terms of $r(\cdot)$.}
A Pauli $Q$ is detectable iff $r(Q)\neq \mathbf 0$ (rephrasing of \Cref{prop:stab-detectability}).
Conversely, if $r(Q)=\mathbf 0$ then $Q$ preserves the code and
$\overline Q=\Pi Q\Pi$ is its logical action (see \eqref{eq:logical-action}).

\subsubsection{Binary symplectic representation}
\label{app:binary-symplectic}

We use the standard binary symplectic picture
\cite{gottesman1997stabilizer, nielsen2010quantum, gottesman2024surviving}.
In particular, we use the standard identification of phase-stripped
Pauli labels with binary symplectic vectors:

\[
\PnHerm \ \longleftrightarrow\ \mathbb F_2^{2n},\qquad
P \ \longleftrightarrow\ v_P=(x_P\,|\,z_P),\ \ 
P\propto\bigotimes_{j=1}^n X_j^{x_{P,j}} Z_j^{z_{P,j}} ,
\]
with the symplectic product
\[
\langle v,w\rangle_s \;:=\; x\!\cdot z' \,+\, z\!\cdot x' \ \in\ \mathbb F_2
\quad\text{for } v=(x|z),\ w=(x'|z').
\]
Then \( [P,Q]=0 \) iff \( \langle v_P, v_Q\rangle_s=0 \), and
\( \{P,Q\}=0 \) iff \( \langle v_P, v_Q\rangle_s=1 \).

\paragraph{Syndrome matrix and normalizer as a nullspace.}
Let \(S\) be generated by \(G=\{G_\ell\}_{\ell=1}^s\) with binary rows
\(h_\ell:=v_{G_\ell}\in\mathbb F_2^{2n}\). Form the \(s\times 2n\) matrix
\(H\) with rows \(h_\ell\). Let
\[
\Lambda \;:=\;
\begin{pmatrix}
0 & I_n\\
I_n & 0
\end{pmatrix},
\qquad
\langle u,v\rangle_s \;:=\; u^\top \Lambda v
\]
be the standard symplectic form on \(\mathbb F_2^{2n}\).
For any \(P\in\PnHerm\) with binary vector \(v_P\), the syndrome is
\begin{equation}
\label{eq:symp-syndrome}
r(P) \;=\; H\,\Lambda\,v_P \pmod 2,
\end{equation}
so \(r(P)=\mathbf 0\) iff \(v_P\in\ker(H\Lambda)\).
Hence
\[
N(S)\ \cong\ \{\,v\in\mathbb F_2^{2n}: H\Lambda v=0\,\}
\;=\;(\mathrm{rowspan}(H))^{\perp_s},
\]
where \((\cdot)^{\perp_s}\) denotes orthogonality with respect to
\(\langle\cdot,\cdot\rangle_s\).
If the \(s\) generators are independent, then \(\mathrm{rank}(H)=\mathrm{rank}(H\Lambda)=s\) and
\(\dim_{\mathbb F_2} N(S)=2n-s\).

\paragraph{Logical Paulis and their count.}
Two normalizer Paulis \(P,Q\in N(S)\) induce the same logical Pauli up
to global phase iff \(P^\dagger Q\) is proportional to an element of
\(S\), i.e.\ iff \(v_P-v_Q\in\mathrm{rowspan}(H)\).
Thus, after quotienting out global phases, the space of logical Paulis
(including identity) is
\[
N(S)/(\langle iI\rangle S)
\ \cong\
(\ker(H\Lambda))/(\mathrm{rowspan}(H)),
\]
which has
\[
\dim_{\mathbb F_2} \bigl(N(S)/S\bigr) \;=\; (2n-s)-s \;=\; 2(n-s).
\]
Equivalently, if \(k:=n-s\) is the number of logical qubits, then there are
\(2^{2k}=4^k=4^{\,n-s}\) distinct logical Paulis (including the logical identity).

\begin{lemma}[Linear algebra lemma (Gottesman~{\cite[Lem.~3.15]{gottesman2024surviving}})]
\label{lem:gottesman-315}
Let \(\{P_1,\dots,P_m\}\subset\PnHerm\) map to linearly independent
\(v_{P_i}\in\mathbb F_2^{2n}\). For any choice \(s\in\{0,1\}^m\) there exists
\(Q\in\PnHerm\) such that
\(
\langle v_{P_i}, v_Q\rangle_s = s_i
\)
for all \(i\). Moreover, the set of such \(Q\) has size \(2^{\,2n-m}\).
\end{lemma}

\begin{corollary}[Every nontrivial logical has an anticommuting partner]
\label{cor:anticommute-partner}
Let \(P\in N(S)\setminus(\langle iI\rangle S)\) be a nontrivial logical Pauli. Then there exists
\(Q\in N(S)\setminus S\) such that \(\{P,Q\}=0\).
\end{corollary}
\begin{proof}
Apply Lemma~\ref{lem:gottesman-315} to the independent set consisting of the
\(s\) generators \(G_1,\dots,G_s\) together with \(P\). Choose the commutation
pattern \(s=(0,\dots,0,1)\), i.e.\ commuting with each \(G_\ell\) and
anticommuting with \(P\). The resulting \(Q\) commutes with all \(G_\ell\),
hence \(Q\in N(S)\), and anticommutes with \(P\). Since every \(S\)-element
commutes with \(P\), this \(Q\notin S\).
\end{proof}

\begin{corollary}[Half of all logical Paulis anticommute with a fixed nontrivial logical]
\label{cor:half-anticommute}
Let \(P\in N(S)\setminus S\) be a nontrivial logical Pauli.
Then among the distinct logical Paulis \(N(S)/S\) 
exactly half anticommute with \(\overline P\) and
exactly half commute with \(\overline P\). 
\end{corollary}

\begin{proof}
Apply Lemma~\ref{lem:gottesman-315} to the independent set consisting of the
\(m\) generators \(G_1,\dots,G_m\) together with \(P\).
Choose the commutation pattern
\[
s=(\underbrace{0,\ldots,0}_{m\ \text{entries}},1),
\]
i.e., commute with each \(G_\ell\) and anticommute with \(P\).
By Lemma~\ref{lem:gottesman-315} there are exactly
\(2^{\,2n-(m+1)}=2^{\,2n-m-1}\) physical Paulis \(Q\) satisfying these
constraints. Every stabilizer element commutes with \(P\), so none of these
\(Q\) lies in \(S\).

Passing to logical operators amounts to quotienting by \(S\):
each logical class in \(N(S)/S\) has exactly \(|S|=2^{\,m}\) representatives
in \(N(S)\). Hence the number of \emph{distinct} logical Paulis that
anticommute with \(\overline P\) is
\[
\frac{2^{\,2n-m-1}}{2^{\,m}} \;=\; 2^{\,2n-2m-1}\;=\;2^{\,2(n-m)-1}.
\]
Repeating the argument with the pattern
\((0,\ldots,0,0)\) (commuting also with \(P\)) yields the same count for the
logical Paulis that commute with \(\overline P\).
Since the total number of logical Paulis is \(2^{\,2(n-m)}\),
exactly half anticommute with \(\overline P\) and half commute.
\end{proof}

Below, we cite a crucial lemma for Lindbladian reshaping (Appendix~\ref{sec:appendix-lindbladian-reshaping}):
\begin{lemma}[Logical Pauli twirling \cite{emerson2007symmetrized}]\label{lem:logical-pauli-twirling}
Let $\overline{P}$ be a logical Pauli string  of some code and let $\overline{\mathcal K}_{\overline P_a}$ be the commuting logical set of a target non-identity logical Pauli ${\overline P_a}$:
\begin{equation*}
\overline{\mathcal K}_{\overline P_a}
\;:=\;\{\overline Q\in\overline{\mathcal P}:\ [\overline Q,\overline P_a]=0\}.
\end{equation*}

Then
\[
\frac{1}{|\overline{\mathcal K}_{\overline P_a}|}
\sum_{\overline Q\in \overline{\mathcal K}_{\overline P_a}}
\overline Q\,\overline P\,\overline Q
\;=\;
\begin{cases}
\;\overline P,& \overline P \in \{ \overline P_a, I\},\\[2pt]
\;0,& \text{otherwise}.
\end{cases}
\]
\end{lemma}

\begin{proof}
If $\overline P \in \{ \overline P_a, I\}$, then by construction, $\overline P$ commutes with all $\overline Q$. Hence, $\overline{Q} \,\overline{P} \,\overline{Q} = \overline P$ and the entire sum is $\overline P$. 

If $\overline P \notin \{ \overline P_a, I\}$, then we prove that half of logical Paulis in $\overline{\mathcal K}_{\overline P_a}$ anti-commute with $\overline P$ and half commute, hence averaging to zero. For that, apply Lemma~\ref{lem:gottesman-315} to the independent set consisting of the
\(m\) Stabilizer generators \(G_1,\dots,G_m\) together with \(\overline P_a, \overline P\). The anti-commuting logical Paulis have the following commutation pattern:
\[
s=(\underbrace{0,\ldots,0}_{m\ \text{entries}},0,1),
\]
By Lemma~\ref{lem:gottesman-315} there are exactly
$2^{\,2n-(m+2)}=2^{\,2n-m-2}$ physical Paulis, and hence $2^{\,2n-m-2}/2^m$ logical Paulis (taking quotient by $S$) satisfying these constraints. This constitutes exactly one half of $\overline{\mathcal K}_{\overline P_a}$ (by \Cref{cor:half-anticommute}). Since Paulis either commute or anti-commute, the other half of $\overline{\mathcal K}_{\overline P_a}$ commutes with $\overline P$, concluding the proof.
\end{proof}

\subsection{Polynomial interpolation}

Here we introduce the polynomial-interpolation primitive used for derivative estimation. This primitive is used in Appendix~\ref{sec:appendix-ancilla_assisted_coeff_learning}, where Lindbladian coefficients are obtained from first derivatives at \(t=0\) of scalar functions of the form $f(t) \coloneq \tr\!\left(O\,(e^{t\Lindblad}\otimes I_R)(\rho_0) \right)$. In this subsection, however, we keep the discussion general.  

Let \(f:[0,\infty)\to\mathbb R\) be a sufficiently smooth scalar function. Suppose we have access to noisy estimates $\widehat f(t)$ of $f(t)$. The idea is to esitmate \(f\) at a set of carefully chosen nodes on the interval \([0,\tau_{\max}]\), fit a degree-\(r\) polynomial interpolant through those values, and then take the analytical derivative of the fitted polynomial to obtain an estimate for the derivative of $f(t)$. This is a standard approach in numerical analysis, and multiple choices of nodes are possible. We use Chebyshev–Gauss nodes because they provide a stable interpolation rule with controlled derivative bias and controlled amplification of nodewise sampling noise. The following numerical guarantees are adapted from \cite[Appendix~C]{ivashkov2026ansatzfreelindblad}, which contains a more detailed treatment of Chebyshev interpolation and also treats second-derivative estimation.

Fix an interpolation degree \(r\ge2\) and a maximum evolution time \(\tau_{\max}>0\).  Let \(T_j\) denote the Chebyshev polynomial of the first kind, defined equivalently by
\begin{equation}
    T_j(\cos\theta)=\cos(j\theta).
\end{equation}
We use the \(r+1\) Chebyshev--Gauss nodes
\begin{equation}
    z_q
    =
    -\cos\!\left(\frac{(2q-1)\pi}{2(r+1)}\right),
    \qquad
    q=1,\ldots,r+1,
    \label{eq:appendix-cheb-gauss-nodes}
\end{equation}
which are the roots of \(T_{r+1}\).  Mapping the standard interval \([-1,1]\) to the physical time interval \([0,\tau_{\max}]\), we set
\begin{equation}
    t_q
    =
    \frac{\tau_{\max}}{2}(z_q+1),
    \qquad
    z(t)
    =
    \frac{2t}{\tau_{\max}}-1.
    \label{eq:appendix-cheb-time-map}
\end{equation}
Given noisy estimates \(\{\widehat f(t_q)\}_{q=1}^{r+1}\), define the degree-\(r\) Chebyshev interpolant
\begin{equation}
    \widehat p(t)
    =
    \sum_{j=0}^{r}\widehat c_j T_j(z(t)).
    \label{eq:appendix-cheb-interpolant}
\end{equation}
The coefficients are computed from the discrete orthogonality relations at the Chebyshev--Gauss nodes:
\begin{equation}
    \widehat c_0
    =
    \frac{1}{r+1}\sum_{q=1}^{r+1}\widehat f(t_q),
    \qquad
    \widehat c_j
    =
    \frac{2}{r+1}\sum_{q=1}^{r+1}\widehat f(t_q)T_j(z_q),
    \quad j=1,\ldots,r .
    \label{eq:appendix-cheb-coefficients}
\end{equation}
The first-derivative estimator is obtained by differentiating \(\widehat p\) at \(t=0\):
\begin{equation}
    \widehat p^{\,\prime}(0)
    =
    \frac{2}{\tau_{\max}}
    \sum_{j=1}^{r}(-1)^{j+1}j^2\widehat c_j .
    \label{eq:chebyshev_derivative_estimator_first}
\end{equation}
Equivalently, substituting Eq.~\eqref{eq:appendix-cheb-coefficients}, the estimator is a fixed linear combination of the (noisy) function values:
\begin{equation}
    \widehat p^{\,\prime}(0)
    =
    \sum_{q=1}^{r+1}\alpha_q\,\widehat f(t_q),
    \qquad
    \alpha_q
    =
    \frac{4}{\tau_{\max}(r+1)}
    \sum_{j=1}^{r}(-1)^{j+1}j^2T_j(z_q).
    \label{eq:appendix-cheb-first-derivative-weights}
\end{equation}
The weights \(\{\alpha_q\}_{q=1}^{r+1}\) depend only on \(r\) and \(\tau_{\max}\), and can therefore be precomputed once and reused for every function sampled on the same grid.

\begin{theorem}[Noisy first-derivative Chebyshev estimator; Theorem~C.1 in \cite{ivashkov2026ansatzfreelindblad}]
\label{thm:total-error-noisy}
Let \(\tau_{\max}>0\), and let \(f\) be \((r+1)\)-times continuously differentiable on \([0,\tau_{\max}]\).  Suppose the available noisy samples \(\{\widehat f(t_q)\}_{q=1}^{r+1}\) at the Chebyshev--Gauss nodes from Eq.~\eqref{eq:appendix-cheb-time-map} satisfy the nodewise noise bound
\begin{equation}
    |\widehat f(t_q)-f(t_q)|\le \epsilon_s,
    \qquad
    q=1,\ldots,r+1 .
    \label{eq:appendix-cheb-nodewise-noise}
\end{equation}
Then the estimator \(\widehat p^{\,\prime}(0)\) defined in Eq.~\eqref{eq:chebyshev_derivative_estimator_first} satisfies
\begin{equation}
    \bigl|\widehat p^{\,\prime}(0)-f'(0)\bigr|
    \le
    \underbrace{\frac{5r^3}{2\tau_{\max}}\,\epsilon_s}_{\text{\em noise term}}
    +
    \underbrace{
    2(r+1)^2\tau_{\max}^r
    \frac{\norm{f^{(r+1)}}_{\infty,[0,\tau_{\max}]}}{r!}
    }_{\text{\em bias term}},
    \label{eq:chebyshev_total_error_first}
\end{equation}
where
\begin{equation}
    \norm{f^{(r+1)}}_{\infty,[0,\tau_{\max}]}
    \coloneq
    \sup_{t\in[0,\tau_{\max}]} |f^{(r+1)}(t)|.
\end{equation}
\end{theorem}
The theorem separates the error into two contributions: the noise term, which quantifies the cost of differentiating noisy node values, and the bias term, which is the deterministic interpolation error and is controlled by the \((r+1)\)-st derivative of \(f\) on the sampled interval.  The following corollary provides a convenient parameter choice when the derivatives of \(f\) grow at most exponentially with order.

\begin{corollary}[First derivative: sufficient \(\tau_{\max}\), \(r\), and \(\epsilon_s\); Corollary~C.3 in \cite{ivashkov2026ansatzfreelindblad}]
\label{cor:first_chebyshev_derivative_parameters}
Suppose that for some \(B,\Lambda>0\), the derivatives of \(f\) obey
\begin{equation}
    \norm{f^{(k)}}_{\infty,[0,1/(2\Lambda)]}
    \le
    B\Lambda^k,
    \qquad
    k\ge1 .
    \label{eq:appendix-cheb-derivative-growth-assumption}
\end{equation}
Fix a target accuracy \(\epsilon>0\), and set
\begin{equation}
    \tau_{\max}
    =
    \frac{1}{2\Lambda},
    \qquad
    r
    =
    \max\left\{
        2,
        \left\lceil
            \log_2\!\left(\frac{18B\Lambda}{\epsilon}\right)
        \right\rceil
    \right\},
    \qquad
    \epsilon_s
    =
    \frac{\epsilon}{10\Lambda r^3}.
    \label{eq:appendix-cheb-first-derivative-parameter-choice}
\end{equation}
Then the derivative estimator \(\widehat p^{\,\prime}(0)\) in Eq.~\eqref{eq:chebyshev_derivative_estimator_first} satisfies
\begin{equation}
    \bigl|\widehat p^{\,\prime}(0)-f'(0)\bigr|\le \epsilon .
\end{equation}
\end{corollary}

Finally, we note that the spacing between the Gauss--Chebyshev is not too small, which guarantees that the interpolation procedure does not require arbitrarily fine time control as the target accuracy increases.

\begin{lemma}[Time resolution of Chebyshev--Gauss nodes; Lemma~C.6 in \cite{ivashkov2026ansatzfreelindblad}]
\label{lem:appendix-cheb-time-resolution}
Let \(r\ge1\), and let \(\{t_q\}_{q=1}^{r+1}\) be the sampling times defined in Eq.~\eqref{eq:appendix-cheb-time-map}.  Then
\begin{equation}
    \min_{1\le q\le r+1} t_q
    \ge
    \frac{\tau_{\max}}{4(r+1)^2},
    \qquad
    \min_{1\le q\le r}|t_{q+1}-t_q|
    \ge
    \frac{2\tau_{\max}}{(r+1)^2}.
    \label{eq:appendix-cheb-time-resolution}
\end{equation}
In particular, if \(\tau_{\max}=\Theta(1/\Lambda)\) and \(r=\Theta(\log(\Lambda/\epsilon))\), then the smallest relevant time scale is
\begin{equation}
    t_{\mathrm{min}}
    =
    \Omega\!\left(
        \frac{1}{\Lambda\log^2(\Lambda/\epsilon)}
    \right),
    \label{eq:appendix-cheb-time-resolution-corollary}
\end{equation}
up to absolute constants.
\end{lemma}

%% file: appendix/Lindbladian_reshaping.tex
\section{QEC Lindbladian Reshaping}
\label{sec:appendix-lindbladian-reshaping}

In this section, we introduce Lindbladian reshaping, the main subroutine behind our dissipator and Hamiltonian learning algorithms. Our scheme combines the quantum Zeno effect \cite{Mobus2019QZEGeneralized, Becker2021QZEOpen, Burgarth2020QZD, PazSilva2012ZenoQC} with stabilizer quantum error correction (QEC) \cite{zhou2018achieving,demkowicz2017qecmetrology, kessler2014quantum} to selectively suppress specified Lindblad terms. Under the appropriate detectability and correctability conditions, Lindbladian reshaping suppresses the targeted Lindblad terms and thereby enables longer effective evolution times. This primitive is instrumental to our Heisenberg-limited and standard-quantum-limited algorithms for learning Lindbladian dynamics in the setting of unknown structure.

We now summarize the notation used in this section. The $n$-qubit probe system has Hilbert space $\mathcal H_n \cong (\mathbb C^2)^{\otimes n}$, and its dynamics is generated by a Lindbladian $\Lindblad:\mathcal B(\mathcal H_n)\to \mathcal B(\mathcal H_n)$, which we write in the Pauli basis as
\begin{equation}\label{eq:lindblad-eq-reshaping}
    \Lindblad(\rho) = -i\sum_{P_k\in \HamiltStruct} h_k \comm{P_k}{\rho} + \sum_{P_k,P_j \in \DiagDissStruct} a_{kj}\left(P_k \rho P_j - \tfrac12\acomm{P_j P_k}{\rho}\right).
\end{equation}

To realize a stabilizer code, we append $n'-n$ ancilla qubits, thereby enlarging the Hilbert space to $\mathcal H_{n'} \cong \mathcal H_n \otimes \mathcal H_{n'-n}$. Let $\Pi$ denote the projector onto the code space $\mathcal H_c \coloneq \Pi \mathcal H_{n'}$. The associated corner algebra (defined in Eq.~\eqref{eq:corner-algebra-def}) is
\begin{equation} \label{eq:corner-algebra-def-lindbladian-reshaping}
    \mathfrak A_c \coloneq \Pi \mathcal B(\mathcal H_{n'}) \Pi \cong \mathcal B(\mathcal H_c).
\end{equation}
The corresponding corner projection is $\mathcal C_\Pi(X) \coloneq \Pi X\Pi$ for any $X \in \mathcal B(\mathcal H_{n'})$. We also fix a completely positive and trace-preserving (CPTP) recovery channel $\RMap:\mathcal B(\mathcal H_{n'}) \to \mathfrak A_c$ which implements syndrome extraction followed by a corrective operation and returns states to the code space. A precise definition of $\RMap$ is given in Appendix~\ref{sec:qec-recovery-map}.

To perform targeted QEC, we maintain the following two sets of Paulis:
\begin{equation} \label{eq:detect-and-correct-sets}
    \VDetect \subset \PnHerm,
    \qquad
    \VCorrect \subset \PnHerm.
\end{equation}
Operationally, $\VDetect$ keeps track of previously learned Hamiltonian Pauli terms, for which we enforce detectability, and $\VCorrect$ contains previously learned dissipator Pauli terms corresponding to nonzero diagonal Kossakowski matrix elements, for which we enforce correctability. 

The sets $\VDetect$ and $\VCorrect$ consist of probe-system Paulis in $\mathcal P_n$. Whenever such a Pauli is used together with code-space objects on $\mathcal H_{n'}$, we implicitly view it via the canonical embedding $P \mapsto P \otimes I_{n'-n}$. In particular, we write $r(P)$ as shorthand for $r(P \otimes I_{n'-n})$, and $\Pi P \Pi$ as shorthand for $\Pi (P \otimes I_{n'-n}) \Pi$.

Since the Lindbladian structure is not known a priori, we use randomly chosen quantum error-correcting codes. The codes are selected to suppress all previously learned terms via detectability and correctability conditions, while ensuring that the remaining unknown terms survive and induce nontrivial logical dynamics with non-negligible probability. We formally treat the random QEC construction in Appendix~\ref{sec:qec-construction}.

In addition to QEC reshaping, we employ randomized Pauli twirling \cite{emerson2007symmetrized}, namely, we interleave the evolution with conjugation by randomly chosen Pauli operators drawn from a specified set. We use twirling in two forms: physical twirling over $n$-qubit Paulis acting on the probe subsystem, and partial logical twirling over a subset of logical Paulis.

\begin{definition}[Unitary conjugation and twirl of a superoperator]
\label{def:twirl-superoperator}
For a unitary $U$, define the conjugation channel $\mathcal U_U(X):=UXU^\dagger$. Given a finite set $\mathcal G$ of unitaries, define the uniform twirl of a linear map $\mathcal M:\mathcal B(\mathcal H_{n'})\to \mathcal B(\mathcal H_{n'})$ by
\begin{equation}
\label{eq:twirl-superoperator}
    \mathbb T_{\mathcal G}(\mathcal M)
    \coloneq
    \frac{1}{|\mathcal G|}
    \sum_{U\in\mathcal G}
    \mathcal U_U \circ \mathcal M \circ \mathcal U_{U^\dagger} .
\end{equation}
\end{definition}

\begin{definition}[Partial logical twirl set]
\label{def:logical-twirl-set}
Let $\overline{\mathcal P}$ denote the logical Pauli group on the code space $\Pi\mathcal H_{n'}$ (for example, $N(S)/S$ for a stabilizer code with stabilizer group $S$), and fix a nonidentity target logical Pauli $\overline P_a\in\overline{\mathcal P}$. Define the commuting logical subset
\begin{equation}
\label{eq:logical-commuting-set}
    \overline{\mathcal K}_{\overline P_a}
    \coloneq
    \{\overline Q\in\overline{\mathcal P}:\ [\overline Q,\overline P_a]=0\}.
\end{equation}
For $\overline Q\in\overline{\mathcal P}$, let $\mathcal U_{\overline Q}$ denote conjugation on the corner algebra $\mathfrak A_c$ induced by any stabilizer coset representative of $\overline Q$.
\end{definition}

By slight abuse of notation, throughout this section we sometimes write $e^{\Lindblad \tau}\otimes I_{n'-n}$ and $\Lindblad\otimes I_{n'-n}$ for the channel and generator acting as $e^{\Lindblad \tau}$ and $\Lindblad$ on the probe subsystem and trivially on the ancillas.

Having set up the notation, we now define the three single-step maps used, via repeated interleaving, to achieve Lindbladian reshaping. Note that the initial application of $\mathcal C_\Pi$ is added to all of them for simplified notation and analysis. Since $\mathcal C_\Pi$ acts as identity on $\mathfrak A_c$, once we get into the code-space, the physically realized reshaping maps can just omit this projection. 
\begin{itemize}
    \item \textbf{QEC reshaping map} (used in \HNDAbbrev{} Hamiltonian structure learning):
    \begin{equation}
    \label{eq:QEC-reshaping}
        A_{\tau, H}^{I}
        \coloneq
        \RMap \circ \bigl(e^{\Lindblad \tau}\otimes I_{n'-n}\bigr) \circ \mathcal C_\Pi.
    \end{equation}
    Repeated QEC is used to suppress most of the dissipator (specified via $\VCorrect$) and some Hamiltonian (specified via $\VDetect$) terms after they have been learned.

    \item \textbf{Physical twirling + partial QEC reshaping} (used in dissipator structure learning):
    \begin{equation}
    \label{eq:physical-twirling-QEC-reshaping}
        A_{\tau, D}
        \coloneq
        \RMap \circ \frac{1}{4^n}\sum_{P\in \PnHerm}
        \bigl(\mathcal U_P\otimes I_{n'-n}\bigr)\circ
        \bigl(e^{\Lindblad \tau}\otimes I_{n'-n}\bigr)\circ
        \bigl(\mathcal U_P\otimes I_{n'-n}\bigr) \circ \mathcal C_\Pi.
    \end{equation}
    This map is used in a setting where Hamiltonian is fully unknown and the dissipator is partially known. Physical Pauli twirling suppresses Hamiltonian contributions and off-diagonal Kossakowski terms, while QEC reshaping suppresses the previously learned diagonal dissipator terms contained in $\VCorrect$.

    \item \textbf{QEC reshaping + partial logical twirling} (used in \HNDAbbrev{} Hamiltonian coefficient learning):
    \begin{equation}
    \label{eq:logical-twirling-QEC-reshaping}
        A_{\tau, H}^{II}
        \coloneq
        \frac{1}{|\overline{\mathcal K}_{\overline P_a}|}
        \sum_{\overline Q\in \overline{\mathcal K}_{\overline P_a}}
        \Bigl(\mathcal U_{\overline Q}\circ \RMap \circ \bigl(e^{\Lindblad \tau}\otimes I_{n'-n}\bigr)\circ \mathcal U_{\overline Q}\Bigr) \circ \mathcal C_\Pi.
    \end{equation}
    Here QEC reshaping suppresses the previously learned part of the dissipator, while the partial logical twirl isolates a specified surviving logical Hamiltonian term $\overline P_a \in \mathfrak A_c$.
\end{itemize}

This section proceeds as follows:
\begin{itemize}
    \item \textbf{QEC recovery channel.} In Appendix~\ref{sec:qec-recovery-map}, we summarize a standard quantum error correction recovery channel $\RMap:\mathcal B(\mathcal H_{n'})\to \mathfrak A_c$ that implements syndrome measurement, decoding, and correction.

    \item \textbf{Effective generators.} In Appendix~\ref{sec:effective-generators}, we derive the effective generators corresponding to the three reshaping maps Eq.~\eqref{eq:QEC-reshaping}, Eq.~\eqref{eq:physical-twirling-QEC-reshaping}, and Eq.~\eqref{eq:logical-twirling-QEC-reshaping} under suitable detectability and correctability conditions.

    \item \textbf{Random QEC construction.} In Appendix~\ref{sec:qec-construction}, we present the random stabilizer code construction tailored for Lindbladian reshaping. This construction ensures that the previously learned terms are suppressed, while the unknown terms survive as distinct nontrivial logical operators with non-negligible probability. 

    \item \textbf{Effective evolution under QEC reshaping.} In Appendix~\ref{sec:qec-effective-evolution}, we compare the evolution under repeated applications of the one-step maps from Eq.~\eqref{eq:QEC-reshaping}, Eq.~\eqref{eq:physical-twirling-QEC-reshaping}, and Eq.~\eqref{eq:logical-twirling-QEC-reshaping} with the desired effective evolution $e^{t\LindbladEff}$ and quantify the resulting diamond-norm errors.

    \item \textbf{Deferred proofs.}
    In Apppendix~\ref{sec:deferred-proofs-effective-QEC-generators}, we provide the deferred proofs related to the effective generators of QEC reshaping.
    
\end{itemize}

\subsection{QEC recovery channel (decoding + correction)} \label{sec:qec-recovery-map}
We assume a stabilizer code $C$ that is non-degenerate with respect to the relevant collection of Pauli errors. We construct an explicit completely positive and trace-preserving (CPTP) recovery channel $\RMap$ implementing syndrome measurement, decoding, and correction via a syndrome lookup table. The construction follows the standard recovery map for non-degenerate stabilizer codes~\cite{nielsen2010quantum}, except that we allow certain Pauli errors to be treated in a \emph{detection-only} mode.

Concretely, let $G=\{G_\ell\}_{\ell=1}^m$ be a generating set of an abelian stabilizer
group $S$ of code $C$, and let $\{\Pi_s\}_{s\in\{0,1\}^m}$ be the associated syndrome projectors in $\mathcal H_{n'}$ from
Eq.~\eqref{eq:syndrome-projectors}. Let $\Pi \coloneq \Pi_0$. Assume \emph{non-degeneracy} on $\VCorrect$:
all $P\in \VCorrect$ have mutually distinguishable nonzero syndromes $r(P)\in\{0,1\}^m$. Additionally, assume \emph{detectability} on $\VDetect$: each $P\in \VDetect$ has nonzero syndrome $r(P)\in\{0,1\}^m$. 

\emph{Pauli–KL relations on $\VCorrect$.}
Under the above assumption, one can check that the Knill–Laflamme equalities
hold for the target set of Paulis (\cite{knill1997theory,gottesman1997stabilizer,nielsen2010quantum}):
for all $P_k,P_j\in \VCorrect$,
\begin{equation}
\label{eq:pauli-KL}
\Pi (P_k) \Pi = 0,
\qquad
\Pi (P_k P_j) \Pi = \delta_{kj}\,\Pi .
\end{equation}
The first equation follows from the nonzero-syndrome assumption of terms in $\VCorrect$. The second equation follows from the requirement of non-degeneracy on $\VCorrect$ (see \Cref{lem:qec-reshaping-generator}). We only invoke Eq.~\eqref{eq:pauli-KL} as a sanity check: the core of our analysis is that with frequent correction rounds, the
stroboscopic evolution $\mathcal A_\tau$ (referring to any of Eq.~\eqref{eq:QEC-reshaping}, Eq.~\eqref{eq:physical-twirling-QEC-reshaping} and Eq.~\eqref{eq:logical-twirling-QEC-reshaping})
removes all specified Pauli terms to first order
and approximates $e^{t\LindbladEff}$ with an explicit norm error bound as shown in \Cref{lem:N-step-reshaping-error}.

We now construct the recovery channel. For each possible error syndrome
\(s\in\{0,1\}^m\), choose a Hermitian Pauli correction
\(R_s\in\mathcal P_{n'}\) as follows:
\begin{equation}
\label{eq:syndrome-correction-unitaries}
R_{r(P\otimes I_{n'-n})} \coloneqq P\otimes I_{n'-n}
\quad (P\in\VCorrect),\qquad
R_0 \coloneqq I_{n'} .
\end{equation}
For every remaining syndrome \(s\notin r(\VCorrect\otimes I_{n'-n})\cup\{0\}\),
choose any Hermitian Pauli \(R_s\in\mathcal P_{n'}\) with syndrome \(r(R_s)=s\).
Whenever such a choice exists with support only on the probe subsystem, we take
\(R_s=\widetilde R_s\otimes I_{n'-n}\) for some Hermitian \(\widetilde R_s\in\mathcal P_n\). For syndromes
for which no probe-supported representative exists, the correction may act on the
ancilla qubits.

Given a density operator $\rho$ and a realized syndrome $s$, the corrected
state is
\[
\rho_{\mathrm{out}\mid s}
\;=\;
R_s\,\frac{\Pi_s\rho\Pi_s}{\tr(\Pi_s\rho)}\,R_s.
\]
Averaging over all outcomes (the operational channel actually implemented)
yields
\[
\sum_{s}\tr(\Pi_s\rho)\,\rho_{\mathrm{out}\mid s}
\;=\;\sum_{s} R_s \Pi_s \rho \Pi_s R_s.
\]

\begin{definition}[Recovery channel]
\label{def:recovery-channel}
With the recovery operators from Eq.~\eqref{eq:syndrome-correction-unitaries} and with the corresponding Kraus operators $K_s\coloneq R_s\Pi_s$, define the recovery channel $\RMap: \mathcal B(\mathcal H_{n'})\to \mathfrak A_c$:
\begin{equation*}
\RMap(\rho)\coloneq \sum_{s\in\{0,1\}^m} K_s\,\rho\,K_s^\dagger
\;=\;\sum_{s} R_s \Pi_s \rho \Pi_s R_s .
\end{equation*}
\end{definition}

\begin{remark}[Well-definedness, range, and identity on the code space]
\label{rem:recovery-welldefined}
The recovery channel $\RMap$ is completely positive and trace preserving (CPTP) and has range in the corner algebra $\mathfrak A_c$ (Eq.~\eqref{eq:corner-algebra-def-lindbladian-reshaping}). Moreover, $\RMap$ acts as the identity on $\mathfrak A_c$, i.e., $\RMap(X)=X$ for all $X\in \mathfrak A_c$.
\end{remark}

\subsection{Effective generators of QEC reshaping}
\label{sec:effective-generators}

In this section, we derive the effective generators corresponding to the three reshaping maps Eq.~\eqref{eq:QEC-reshaping}, Eq.~\eqref{eq:physical-twirling-QEC-reshaping}, and Eq.~\eqref{eq:logical-twirling-QEC-reshaping}. First, consider the QEC reshaping map defined in Eq.~\eqref{eq:QEC-reshaping}:
\begin{equation*}
    A_{\tau,H}^{I}
    \coloneq
    \RMap \circ (e^{\Lindblad \tau} \otimes I_{n'-n}) \circ \mathcal C_\Pi
    =
    \RMap \circ \mathcal C_\Pi
    +
    \tau \underbrace{\RMap \circ (\Lindblad \otimes I_{n'-n}) }_{\to \LindbladEff} \circ \,\mathcal C_\Pi
    +
    \mathcal O(\tau^2).
\end{equation*}

In \Cref{lem:qec-reshaping-generator}, we derive the effective generator of this map under the detectability and correctability assumptions on $\VDetect$ and $\VCorrect$. Intuitively, when $\VCorrect$ contains most of the dissipator, this approximately reduces the problem to logical Hamiltonian learning. Hence, this map is used for the \HND{} structure learning in Appendix~\ref{sec:appendix-hamilt-struct-sampling}.

\begin{lemma}[Effective generator under QEC reshaping]
\label{lem:qec-reshaping-generator}
Let $\Lindblad$ be an $n$-qubit GKSL generator as in Eq.~\eqref{eq:lindblad-eq-reshaping}, and let $\Lindblad \otimes I_{n'-n}$ denote its trivial extension to the ancilla register. Let $C$ be a stabilizer code on $n' \ge n$ qubits with code projector $\Pi$, corner algebra $\mathfrak A_c \coloneq \Pi \mathcal B(\mathcal H_{n'}) \Pi$ and syndrome map $r(\cdot)$ (as in Eq.~\eqref{eq:syndrome-map}). Suppose $\VDetect \subseteq \PnHerm$ and $\VCorrect \subseteq \DiagDissStruct \subseteq \PnHerm$ are specified sets, and that all Paulis in $\VDetect$ have non-zero syndromes. Assume furthermore that all elements of $\VCorrect$, have pairwise distinct nonzero syndromes. Let $\RMap: \mathcal B(\mathcal H_{n'}) \to \mathfrak A_c$ be the associated recovery map from \Cref{def:recovery-channel} based on the specified sets $\VDetect$ and $\VCorrect$. 

Then, for any $X\in\mathfrak A_c$, the effective generator of the QEC reshaping map from Eq.~\eqref{eq:QEC-reshaping} is
\begin{equation}
\label{eq:Leff-qec-reshaping}
    \begin{aligned}
        &\LindbladEff(X) \coloneq \RMap \circ (\Lindblad \otimes I_{n'-n})(X)  =  \sum_{\substack{P_k,P_j\in \DiagDissStruct \setminus \VCorrect,\\ r(P_k) = r(P_j)}} a_{kj}
    \Bigl((Q_{P_k} \otimes I_{n'-n}) X (Q_{P_j}^\dagger \otimes I_{n'-n}) - \frac{1}{2}\{Q_{P_j}^\dagger Q_{P_k}\otimes I_{n'-n} ,X\}\Bigr) \\
    &-i\underbrace{\sum_{\substack{
    P_k \in \HamiltStruct, \\ r(P_k) = 0}} h_k[P_k  \otimes I_{n'-n},X]}_{[H_\mathrm{tar}, X]}
    \,-\,i \underbrace{\sum_{\substack{
    P_k \in \VCorrect, \\ P_j \in \DiagDissStruct \setminus \VCorrect, \\r(P_k) = r(P_j)}} \left({\operatorname{Re}(a_{kj})\left[\frac{i[P_k,P_j]}{2}  \otimes I_{n'-n}\,,\,X\right] + \operatorname{Im}(a_{kj})\left[\frac{\{P_k,P_j\}}{2}  \otimes I_{n'-n}\,,\,X\right]}\right)}_{[H_\mathrm{bias},X]}
    \end{aligned} 
\end{equation}
where $Q_P \coloneqq \widetilde R_{r(P)}P \in \mathbb P_n$ are Pauli operators composed of $P$ and the corresponding recovery actions $\widetilde R_{r(P)}$ as specified in
Eq.~\eqref{eq:syndrome-correction-unitaries} with $R_{r(P)}=\widetilde R_{r(P)}\otimes I_{n'-n}$. Notably, the resulting Hamiltonian part of $\LindbladEff$ contains only zero-syndrome Paulis. In particular, no Pauli term from
\begin{equation*}
\VDetect \cup \VCorrect \cup \{\Herm(PP') : P,P' \in \VCorrect,\ P\neq P'\}
\end{equation*}
is present in the Pauli decompositions of $H_{\mathrm{tar}}$ and $H_{\mathrm{bias}}$.
\end{lemma}
\begin{proof}
    See Apppendix~\ref{sec:deferred-proofs-effective-QEC-generators}.
\end{proof}

\begin{corollary}[Effective QEC-reshaped generator under distinct dissipator syndromes]\label{cor:qec-reshaping-generator-distinct-syndromes}
    Assuming additionally that all Paulis $P \in \DiagDissStruct$ have pairwise distinct nonzero syndromes, one can significantly simplify the $\LindbladEff$ from Eq.~\eqref{eq:Leff-qec-reshaping} to:
    \begin{equation}
    \label{eq:Leff-qec-reshaping-simplified}
            \LindbladEff(X)  =  \sum_{P_k\in \DiagDissStruct \setminus \VCorrect} a_{kk}
        \Bigl((Q_{P_k} \otimes I_{n'-n}) X (Q_{P_k}^\dagger \otimes I_{n'-n}) - X\Bigr) 
        -i\sum_{\substack{
        P_k \in \HamiltStruct, \\ r(P_k) = 0}} h_k[P_k  \otimes I_{n'-n},X]
    \end{equation}
    where the resulting Hamiltonian part contains no Paulis from
    \begin{equation*}
    \VDetect \cup \DiagDissStruct \cup \{\Herm(PP') : P,P' \in \DiagDissStruct,\ P\neq P'\}.
    \end{equation*}
\end{corollary}

The second relevant map is the physical twirling combined with partial QEC reshaping (Eq.~\eqref{eq:physical-twirling-QEC-reshaping}):
\begin{equation*}
    \begin{aligned}
        A_{\tau,D}
        \coloneq
        &\RMap \circ \frac{1}{4^n}\sum_{P\in \PnHerm}
        (\mathcal U_P \otimes I_{n'-n}) \circ
        (e^{\Lindblad \tau} \otimes I_{n'-n}) \circ
        (\mathcal U_P \otimes I_{n'-n}) \circ \mathcal C_\Pi \\
        =
        &\RMap\circ \mathcal C_\Pi
        +
        \tau \underbrace{\RMap \circ \frac{1}{4^n}\sum_{P\in \PnHerm}
        (\mathcal U_P \otimes I_{n'-n}) \circ
        (\Lindblad \otimes I_{n'-n}) \circ
        (\mathcal U_P \otimes I_{n'-n})}_{\LindbladEff} \circ \,\mathcal C_\Pi
        +
        \mathcal O(\tau^2).
    \end{aligned}
\end{equation*}

The effective generator of this map is presented in \Cref{lem:physical-twirl-qec-reshaping-generator}. This map is used for dissipator structure learning  in Appendix~\ref{sec:appendix_dissipator_structure_learning}.
\begin{lemma}[Effective generator of Physical twirling + QEC reshaping]
\label{lem:physical-twirl-qec-reshaping-generator}
Let $\Lindblad$ be an $n$-qubit GKSL generator as in Eq.~\eqref{eq:lindblad-eq-reshaping}, and let $\Lindblad \otimes I_{n'-n}$ denote its trivial extension to the ancilla register. Let $C$ be a stabilizer code on $n' \ge n$ qubits with code projector $\Pi$ and corner algebra $\mathfrak A_c = \Pi \mathcal B(\mathcal H_{n'}) \Pi$. Assume that the target $\VDetect = \emptyset$ and that $\VCorrect$ is a subset of the dissipator structure, i.e. $\VCorrect \subseteq \DiagDissStruct$. Suppose all Paulis in $\VCorrect \subseteq \PnHerm$ have nonzero mutually distinguishable syndromes. Let $\RMap: \mathcal B(\mathcal H_{n'}) \to \mathfrak A_c$ be the associated recovery map from \Cref{def:recovery-channel}.

Then for any $X \in \mathfrak A_c$, the effective generator of the physical twirling + QEC reshaping map is
\begin{equation}
\label{eq:Leff-physical-twirl-qec-reshaping}
    \begin{aligned}
        \LindbladEff (X)
        \coloneq &\RMap
        \circ \frac{1}{4^n}\sum_{P\in \PnHerm}
            \bigl(\mathcal U_P\otimes I_{n'-n}\bigr)\circ
            \bigl(\Lindblad\otimes I_{n'-n}\bigr)\circ
            \bigl(\mathcal U_P\otimes I_{n'-n})(X) = \\
        = & \sum_{P_k\in \DiagDissStruct \setminus \VCorrect} a_{kk}\Bigl(
                  (Q_{P_k} \otimes I_{n'-n}) X (Q_{P_k}^\dagger \otimes I_{n'-n}) - X\Bigr)
    \end{aligned}  
\end{equation}
where $Q_P \coloneqq \widetilde R_{r(P)}P \in \mathbb P_n$ are Pauli operators composed of $P$ and the corresponding recovery actions $\widetilde R_{r(P)}$ as specified in
Eq.~\eqref{eq:syndrome-correction-unitaries} with $R_{r(P)}=\widetilde R_{r(P)}\otimes I_{n'-n}$. The resulting $\LindbladEff$ is purely dissipative and excludes all terms corresponding to $P_k \in \VCorrect$. 
\end{lemma}
\begin{proof}
    First, the standard Pauli twirling \cite{emerson2007symmetrized} (which we call physical twirling) reshapes the generator by removing its Hamiltonian part as well as off-diagonal Kossakowski elements
    \begin{equation*}
        \frac{1}{4^n}\sum_{P\in \PnHerm}
            \bigl(\mathcal U_P\otimes I_{n'-n}\bigr)\circ
            \bigl(\Lindblad\otimes I_{n'-n}\bigr)\circ
            \bigl(\mathcal U_P\otimes I_{n'-n}) (X)= \sum_{P_k\in \DiagDissStruct} a_{kk}\Bigl(
                (P_k\otimes I_{n'-n}) X (P_k\otimes I_{n'-n})   - X\Bigr)
    \end{equation*}
    
    Acting on $X \in \mathfrak A_c$ and composing with the recovery channel $\RMap$, similarly to Eq.~\eqref{eq:diag-dissipator-qec-reshaped} we get:
    \begin{equation}
        \begin{aligned}
            \LindbladEff (X) &= \sum_{P_k\in \DiagDissStruct} a_{kk}\Bigl(
                      R_{r(P_k)} (P_k\otimes I_{n'-n}) X (P_k\otimes I_{n'-n})  R_{r(P_k)} - X\Bigr) = \\
                      &= \sum_{P_k\in \DiagDissStruct \setminus \VCorrect} a_{kk}\Bigl(
                      (Q_{P_k}\otimes I_{n'-n}) X (Q_{P_k}^\dagger\otimes I_{n'-n}) - X\Bigr) 
        \end{aligned}
    \end{equation}
    where the second equality follows directly from the fact that Paulis from $\VCorrect$ are correctable by design of $\RMap$. 

    Finally, since $r(PQ) = r(P)\oplus r(Q)$, the reshaped operators $Q_{P_k} \coloneq R_{r(P_k)} P_k$ all have $r(Q_{P_k}) = 0$ by construction of $\RMap$ in \Cref{def:recovery-channel}. Since elements of $\VCorrect$ have non-zero syndromes, they are absent from the expansion of $\LindbladEff$ for inputs $X \in \mathfrak A_c$.
\end{proof}

The third relevant map combines QEC reshaping with partial logical twirling (Eq.~\eqref{eq:logical-twirling-QEC-reshaping}). In Appendix~\ref{sec:appendix-hamilt-coeff-learning}, this map is used to approximately isolate evolution under a single logical Pauli, thereby reducing the problem to single-parameter metrology. Assume that the target Pauli $P_a \otimes I_{n'-n} = \overline P_a$ is a nontrivial logical operator for the chosen code, and take any $Y \in \mathcal B(\mathcal H_{n'})$. Using the partial logical twirling set $\overline{\mathcal K}_{\overline P_a}$ defined in Eq.~\eqref{eq:logical-commuting-set}, we have
\begin{equation*}
    \begin{aligned}
        A_{\tau,H}^{II}(Y) \coloneq
        &\frac{1}{|\overline{\mathcal K}_{\overline P_a}|}
        \sum_{\overline Q\in \overline{\mathcal K}_{\overline P_a}}
        \Bigl(\mathcal U_{\overline Q} \circ \RMap \circ (e^{\Lindblad \tau} \otimes I_{n'-n}) \circ \mathcal U_{\overline Q}\circ\mathcal C_\Pi\Bigr)(Y) \\
        =
        &\mathcal C_\Pi(Y)
        +
        \tau \underbrace{\frac{1}{|\overline{\mathcal K}_{\overline P_a}|}
        \sum_{\overline Q\in \overline{\mathcal K}_{\overline P_a}}
        \mathcal U_{\overline Q} \circ \RMap \circ (\Lindblad \otimes I_{n'-n}) \circ \mathcal U_{\overline Q}}_{\to \LindbladEff}\circ\, \mathcal C_\Pi(Y)
        +
        \mathcal O(\tau^2).
    \end{aligned}
\end{equation*}

The effective generator of the ``QEC reshaping combined with partial logical twirling'' map is presented in \Cref{lem:qec-reshape-logical-twirl-generator}.

\begin{lemma}[Effective generator of QEC reshaping + partial logical twirl]
\label{lem:qec-reshape-logical-twirl-generator}
Let $\Lindblad$ be an $n$-qubit GKSL generator as in Eq.~\eqref{eq:lindblad-eq-reshaping}, and let $\Lindblad \otimes I_{n'-n}$ denote its trivial extension to the ancilla register. Let $C$ be a stabilizer code on $n' \ge n$ qubits with code projector $\Pi$ and corner algebra $\mathfrak A_c \coloneq \Pi \mathcal B(\mathcal H_{n'}) \Pi$. Let $\overline P_a \coloneq P_a \otimes I_{n'-n}$ be the nontrivial logical Pauli to be isolated. Suppose $\VCorrect \subseteq \DiagDissStruct \subseteq \PnHerm$ is a specified set and let $\RMap:\mathcal B(\mathcal H_{n'}) \to \mathfrak A_c$ be the associated recovery map from \Cref{def:recovery-channel} based on $\VCorrect$. Assume furthermore that all elements of $\VCorrect$ have pairwise distinct nonzero syndromes. Assume also that $P_a$ is the unique $n$-qubit Pauli whose restriction to the code space contributes to the logical Pauli $\overline P_a$. Equivalently, for every $P_k\in \PnHerm$ with $P_k\neq P_a$, the compressed operator $\Pi (P_k\otimes I_{n'-n})\Pi$ has zero component along $\Pi \overline P_a \Pi$. Assume moreover that no nontrivial product of zero-syndrome Pauli operators appearing in the QEC-reshaped generator of $\Lindblad$ from \Cref{lem:qec-reshaping-generator} is proportional to a stabilizer.

Then for any $X\in \mathfrak A_c$, the effective generator of QEC reshaping + the partial logical twirl over $\overline{\mathcal K}_{\overline P_a}$ is
\begin{equation}
\label{eq:Leff-qec-reshaping-logical-twirl}
    \begin{aligned}
        \LindbladEff(&X)
        \coloneq
        \frac{1}{|\overline{\mathcal K}_{\overline P_a}|}
        \sum_{\overline Q \in \overline{\mathcal K}_{\overline P_a}}
        \mathcal U_{\overline Q} \circ \RMap \circ (\Lindblad \otimes I_{n'-n}) \circ \mathcal U_{\overline Q} (X) =\\
        &= -i h_a [\overline P_a, X]
        \,-\,i \sum_{\substack{
        P_k \in \VCorrect, \\ P_j \in \DiagDissStruct \setminus \VCorrect,\\P_k P_j \propto P_a}} \left({\operatorname{Re}(a_{kj})\left[\frac{i[P_k,P_j]}{2}  \otimes I_{n'-n}\,,\,X\right] + \operatorname{Im}(a_{kj})\left[\frac{\{P_k,P_j\}}{2}  \otimes I_{n'-n}\,,\,X\right]}\right) + \\
        &+\sum_{\substack{P_k\in \DiagDissStruct \setminus \VCorrect,\\\Herm(P_k P_a) \in \DiagDissStruct \setminus \VCorrect;}} \sum_{\substack{P_j \in \DiagDissStruct \setminus \VCorrect,\\ \Herm(P_k P_j) \in \{I_n, P_a\}}} a_{kj}
        \Bigl((Q_{P_k} \otimes I_{n'-n}) X (Q_{P_j}^\dagger \otimes I_{n'-n}) - \frac{1}{2} \{Q_{P_j}^\dagger Q_{P_k}\otimes I_{n'-n},X\}\Bigr) \\
        &+ \sum_{\substack{P_k\in \DiagDissStruct \setminus \VCorrect,\\ \Herm(P_k P_a) \notin \DiagDissStruct \setminus \VCorrect}} a_{kk}
        \Bigl((Q_{P_k} \otimes I_{n'-n}) X (Q_{P_k}^\dagger \otimes I_{n'-n}) - X\Bigr).
    \end{aligned}
\end{equation}
where $Q_P \coloneqq \widetilde R_{r(P)}P \in \mathbb P_n$ are Pauli operators composed of $P$ and the corresponding recovery actions $\widetilde R_{r(P)}$ as specified in
Eq.~\eqref{eq:syndrome-correction-unitaries} with $R_{r(P)}=\widetilde R_{r(P)}\otimes I_{n'-n}$. Here $h_a$ is understood to be zero if $P_a\notin\HamiltStruct$.
\end{lemma}
\begin{proof}
    See Apppendix~\ref{sec:deferred-proofs-effective-QEC-generators}.
\end{proof}

\begin{corollary}[Effective generator of QEC reshaping + partial logical twirl under distinct dissipator syndromes]\label{cor:qec-reshaping-logical-twirl-generator-distinct-syndromes}
    Assuming additionally that all Paulis $P \in \DiagDissStruct$ have pairwise distinct nonzero syndromes, one can significantly simplify the $\LindbladEff$ from Eq.~\eqref{eq:Leff-qec-reshaping-logical-twirl} to:
    \begin{equation}
    \label{eq:Leff-qec-reshaping-logical-twirl-simplified}
            \LindbladEff(X)  =  -i h_a [\overline P_a, X] +\sum_{P_k\in \DiagDissStruct \setminus \VCorrect} a_{kk}
        \Bigl((Q_{P_k} \otimes I_{n'-n}) X (Q_{P_k}^\dagger \otimes I_{n'-n}) - X\Bigr) 
    \end{equation}
\end{corollary}

\begin{remark}[Full-space realization of the effective generators]
\label{rem:full-space-effective-generators}
Although \Cref{lem:qec-reshaping-generator,lem:physical-twirl-qec-reshaping-generator,lem:qec-reshape-logical-twirl-generator} derive $\LindbladEff$ only for inputs $X\in\mathfrak A_c$, in each case we may use the same explicit expression to define a linear map on the full operator algebra $\mathcal B(\mathcal H_{n'})$. In all three cases, the resulting full-space map has the form of a valid GKSL generator on $\mathcal B(\mathcal H_{n'})$, and we continue to denote it by $\LindbladEff$. Therefore, for every $t\ge 0$, the semigroup $e^{t\LindbladEff}$ is CPTP on $\mathcal B(\mathcal H_{n'})$, and thus
\begin{equation*}
    \|e^{t\LindbladEff}\|_\diamond = 1.
\end{equation*}
Moreover, by construction, each $\LindbladEff$ inherits its coefficients from the original Lindbladian Eq.~\eqref{eq:lindblad-eq-reshaping} and contains no more Pauli terms than $\Lindblad$. In particular, if $\Lindblad$ is $M$-sparse, then so is $\LindbladEff$. Hence, by \Cref{lem:diamond-sparse-lindblad},
\begin{equation*}
    \|\LindbladEff\|_\diamond \le 2M.
\end{equation*}
\end{remark}

\subsection{Random stabilizer code construction} \label{sec:qec-construction}
Within our learning framework, the reshaping maps Eq.~\eqref{eq:QEC-reshaping}, Eq.~\eqref{eq:physical-twirling-QEC-reshaping}, and Eq.~\eqref{eq:logical-twirling-QEC-reshaping} are used to selectively suppress previously learned terms. To make this useful algorithmically, we must also ensure that terms not yet learned survive in the effective dynamics with non-negligible probability over the random code draw. A second, independent issue is \emph{injectivity of logical action}: two Pauli strings that commute with all stabilizers, hence are undetectable, but differ by a stabilizer element act as the same logical operator on the code space. If such terms, for example, enter the Hamiltonian with opposite signs, they can cancel in $H_{\mathrm{eff}}$ and hide the signal we wish to learn.

In this section, we present a random stabilizer code construction that addresses these issues while satisfying the detectability and correctability assumptions used in Appendix~\ref{sec:effective-generators}. Recall from Eq.~\eqref{eq:detect-and-correct-sets} that $\VCorrect \subseteq \DiagDissStruct$ keeps track of the learned dissipator terms and $\VDetect \subseteq \HamiltStruct$ keeps track of the learned Hamiltonian terms. We work with the associated target suppression set
\begin{equation} \label{eq:target-suppression-set}
    \mathcal W_{\mathrm{tar}}
    \coloneq
    \VDetect \cup \VCorrect \cup \{\Herm(P'P'') : P',P'' \in \VCorrect,\ P' \neq P''\},
\end{equation}

For a Pauli $P \in \PnHerm \setminus \{I\}$, let $A_P$ denote the event that $P$ survives reshaping, namely that it has zero syndrome under the sampled checks. Using the syndrome map $r()$ from Eq.~\eqref{eq:syndrome-map}, we also define the target suppression event
\begin{equation} \label{eq:Etar}
    E_{\mathrm{tar}}
    \coloneq
    \{r(Q)\neq 0 \text{ for all } Q \in \mathcal W_{\mathrm{tar}}\},
\end{equation}

The random-code construction will be used to guarantee the following:
\begin{itemize}
    \item \textbf{Targeted detectability/correctability.} Every Pauli in $\VDetect$ is detectable and all Paulis in $\VCorrect$ have nonzero mutually distinguishable syndromes. This is equivalent to event $E_{\mathrm{tar}}$ since $r(P' P'') = r(P')\oplus r(P'')$.

    \item \textbf{Survival probability.} Any Pauli operator $P$ outside the target set $\mathcal W_{\mathrm{tar}}$ survives reshaping with non-negligible probability over the random code draw.

    \item \textbf{Injective logical action.} Distinct $n$-qubit probe Pauli terms that survive reshaping induce distinct logical Paulis on the code space, so no two survivors are equal modulo the stabilizer.
\end{itemize}

These guarantees are recorded formally in \Cref{thm:random-qec-for-reshaping}. 

Let us proceed to the code construction. As a starting point in \Cref{alg:qec_for_lindbladian_reshaping}, we will draw a set of random $n$-qubit Pauli strings $G = \{G_1, \dots, G_m\}$ that don't necessarily commute. To abelianize them and hence form a valid stabilizer code, one may append ancillas and use the standard $X/Z$-padding \cite{Brun2006EntanglementQEC, Wilde2008OptimalEntanglementQEC};
this preserves commutation/anticommutation on the data:
\(\operatorname{sgn}(\tilde G_i, P\otimes I_{\rm anc})=\operatorname{sgn}(G_i,P)\), where $\operatorname{sgn}()$ denotes a commutation sign from Eq.~\eqref{eq:pauli_commutation_sign}. Thus, we import the following lemma:

\begin{lemma}[ZX padding to enforce commutation {\cite{Brun2006EntanglementQEC,Wilde2008OptimalEntanglementQEC}}]
\label{lem:random-code-zx-padding}
Let $G=\{G_1,\ldots,G_m\}\subset\PnHerm$ be $m$ Hermitian Pauli checks (not necessarily commuting).
There exists an integer $c\le \lfloor m/2\rfloor$, a register of $c$ ancillary qubits, and Hermitian Pauli
operators $A_1,\ldots,A_m\in \mathcal P_c$ acting on the ancillas such that the padded checks
\[
\widetilde G_k \;:=\; G_k\otimes A_k
\]
pairwise commute: $[\widetilde G_i,\widetilde G_j]=0$ for all $i,j$.
\end{lemma}
\begin{proof}[Proof sketch]
Even though the elements of $G$ do not necessarily commute, we may represent them in the binary symplectic
formalism (Appendix~\ref{app:binary-symplectic}). Let $H\in\{0,1\}^{m\times 2n}$ be the check matrix whose $k$th row is
the binary vector of $G_k$. Then the syndrome map $r(P)=H\Lambda v_P$ (Eq.~\eqref{eq:symp-syndrome}) encodes the
commutation pattern of any $P\in\PnHerm$ with $G$, and the mutual commutation of the checks is encoded by
$\Omega_H\coloneq H\Lambda H^T$.

By symplectic Gram--Schmidt orthogonalization (SGSOP)~\cite{Wilde2008OptimalEntanglementQEC}, there exists an
invertible $m\times m$  matrix $Y$ that encodes symplectic row operations such that $H'=YH$ satisfies
\begin{equation} \label{eq:symplectic-gram-schmidt}
    \Omega_{H'} \;=\; H'\Lambda H'^T \;=\; Y\Omega_HY^T
    \;=\; \bigoplus_{i=1}^{c}
    \begin{pmatrix}0&1\\[2pt]1&0\end{pmatrix}
    \;\oplus\;
    \bigoplus_{j=1}^{\,m-2c}[0],  
\end{equation}

where $[0]$ denotes the zero-matrix. Thus the transformed check set can be written as $\{Q_1,\ldots,Q_m\}$ where $(Q_{2i-1},Q_{2i})$ form $c$
disjoint anticommuting pairs, $\{Q_{2i-1},Q_{2i}\}=0$, and all other pairs commute. By Eq.~\eqref{eq:symplectic-gram-schmidt},
$c=\tfrac12\mathrm{rank}(\Omega_H)\le \lfloor m/2\rfloor$. Introduce $c$ ancillas $a_1,\ldots,a_c$ and pad each
anticommuting pair by
\[
\widetilde Q_{2i-1}\coloneq Q_{2i-1}\otimes Z_{a_i},
\qquad
\widetilde Q_{2i}\coloneq Q_{2i}\otimes X_{a_i},
\]
while padding all remaining $Q_j$ by identity on the ancillas; the resulting $\{\widetilde Q_j\}$ commute.
Finally, applying the inverse row transformation $Y^{-1}$ to $\{\widetilde Q_j\}$ (i.e., multiplying and reordering generators
according to $Y^{-1}$) yields a commuting padded generating set $\widetilde G$ corresponding to the original
list $G$. For details, see \cite{Wilde2008OptimalEntanglementQEC}.
\end{proof}

In Lemmas (\ref{lem:prob-of-detectability-on-target-set})-(\ref{lem:prob-of-survival-non-target}), we analyze the commutation properties (and thus detectability/correctabilitity) of randomly chosen Pauli strings. 

\begin{lemma}[Probability of nonzero syndromes on a finite Pauli set.]
\label{lem:prob-of-detectability-on-target-set}
Let $\mathcal W\subset\PnHerm\setminus\{I\}$ be a target set and let 
$\mathcal T=\{P^{(t)}\}_{t=1}^m$ be a multiset of $m$ Paulis sampled i.i.d. uniformly (with replacement) from $\PnHerm$. Let $E$ denote the event that each $Q \in \mathcal W$ anti-commutes with at least one element of $\mathcal T$. Then
\begin{equation}
    \Pr_{\mathcal T}(E) \ge 1-|\mathcal W|\,2^{-m}.
\end{equation}
\end{lemma}

\begin{proof}

For each $\widetilde Q \in \mathcal W$, let $E_{\widetilde Q}$ denote the event that all $m$ checks in $\mathcal T$ commute with $\widetilde Q$. For a fixed nonidentity Pauli $\widetilde Q$ and a single uniformly random check $P^{(t)}$, we have $\Pr(\{\widetilde Q,P^{(t)}\}=0)=\tfrac12$. Since the $m$ draws are independent,
\begin{equation}
    \Pr(E_{\widetilde Q}) = 2^{-m} \qquad \text{for all } \widetilde Q \in \mathcal W.
\end{equation}

By the union bound,
\begin{equation}
    \Pr_{\mathcal T}(E^c) = \Pr_{\mathcal T}\!\Bigl[\exists\,\widetilde Q \in \mathcal W :  \widetilde Q \text{ commutes with all checks in } \mathcal T \Bigr] \le |\mathcal W|\,2^{-m}.
\end{equation}

Hence
\begin{equation}
    \Pr_{\mathcal T}(E) \ge 1-|\mathcal W|\,2^{-m}.
\end{equation}
proving the claim.
\end{proof}

\begin{lemma}[Probability of survival of non-target set member]
\label{lem:prob-of-survival-non-target}
Let $\mathcal W\subset\PnHerm\setminus\{I\}$ and let 
$\mathcal T=\{P^{(t)}\}_{t=1}^m$ be a multiset of $m$ Paulis sampled i.i.d. uniformly (with replacement) from $\PnHerm$. Let $E$ be the event that each $Q \in \mathcal W$ anti-commutes with at least one element of $\mathcal T$.  Let $A_P$ be the event that a given non-identity Pauli string $P\notin \mathcal W$ commutes with all elements in $\mathcal{T}$. Then conditioned on $E$, the probability that $P$ is undetectable by checks in $\mathcal T$ obeys: 
\begin{equation*}
    \frac{2^{-m}}{(1-\abs{\mathcal W}2^{-m})}\ge \Pr_{\mathcal T}\Bigl[\,A_P\, 
     \Big| \,E\, \Bigr]
    \ \ge\
    2^{-m}(1-|\mathcal W|2^{-m})
\end{equation*}
where the upper bound is only meaningful when $|\mathcal W|2^{-m} <1$.  
\end{lemma}
\begin{proof}
    Using Bayes' theorem, 
    \begin{equation*} 
        \Pr(\,A_P\, | \,E\,) = \frac{\Pr(\,E\, | A_P)\Pr(A_P)}{\Pr(E)}.
    \end{equation*}
    
    Therefore, 
    \begin{equation}\label{eq:bayes-rule-survival-prob}
        \frac{\Pr(A_P)}{\Pr(E)}\ge \Pr(\,A_P\, | \,E\,) \ge \Pr(\,E\, | A_P)\Pr(A_P)
    \end{equation}
    
    For a uniformly random sample $P^{(t)}$, the probability  $\Pr([P, P^{(t)}] = 0) = \frac{1}{2}$. Since the $m$ samples of $\mathcal T$ are independent:
    \begin{equation*}
        \Pr(A_P) = 2^{-m}
    \end{equation*}

    Combining this with $\Pr(E) \ge 1-|\mathcal W|2^{-m}$ (from \Cref{lem:prob-of-detectability-on-target-set}), we obtain:
    \begin{equation*}
        \frac{2^{-m}}{(1-\abs{\mathcal W}2^{-m})}\ge \frac{\Pr(A_P)}{\Pr(E)}\ge \Pr(\,A_P\, | \,E\,)
    \end{equation*}

    It remains to lower bound $\Pr(\,A_P\, | \,E\,)$. To do that, we use the lower bound on $\Pr(\,E\, | A_P)$. Fix $Q \in \mathcal W$. Since $Q \ne P$, using \Cref{lem:gottesman-315},
    \begin{equation*}
        \Pr_{T\sim \PnHerm}([Q, T] = 0 \mid [P,T]=0) = \frac{\text{\# of Paulis that commute w both $Q,P$}}{\text{\# of Paulis that commute w $P$}} = \frac{2^{2n-2}}{2^{2n-1}} = \frac{1}{2}
    \end{equation*}
    
    By independence of the $m$ draws,
    \[
      \Pr\bigl(\,[Q,T]=0\ \forall T\in\mathcal T \,\big|\, [P,T]=0\ \forall T\in\mathcal T\,\bigr)
      \;=\; 2^{-m}.
    \]
   
    A union bound over $|\mathcal W|$ gives
    \[
      \Pr(E^c\mid A_P)\ \le\ |\mathcal W|\,2^{-m}\qquad\Longrightarrow\qquad
      \Pr(E\mid A_P)\ \ge\ 1 - |\mathcal W|\,2^{-m}.
    \]

    Plugging into Eq.~\eqref{eq:bayes-rule-survival-prob} with $\Pr(A_P)=2^{-m}$ concludes the proof:
    \[
      \Pr(\,A_P\,|\,E\,)
      \;\ge\; \Pr(E\mid A_P)\Pr(A_P)
      \;\ge\; \bigl(1 - |\mathcal W|\,2^{-m}\bigr)\,2^{-m},
    \]    
\end{proof}

\begin{algorithm}[H]
    \caption{Random stabilizer code construction for targeted reshaping}
    \label{alg:qec_for_lindbladian_reshaping}
    \DontPrintSemicolon
    \SetKwInOut{Input}{Inputs}
    \SetKwInOut{Output}{Output}
    \SetKwBlock{Schedule}{Schedule}{}

    \Input{%
    \begin{enumerate}[label=(\arabic*), leftmargin=*, nosep]
        \item Target detectable set $\VDetect \subseteq \PnHerm \setminus \{I\}$. We require that each term in $\VDetect$ has a non-zero syndrome.
        \item Target correctable set $\VCorrect \subseteq \PnHerm \setminus \{I\}$. We require that all terms in $\VCorrect$ have mutually distinguishable non-zero syndromes.
        \item Detectability-size bound $M_{\mathrm{det}} \ge |\mathcal W_{\mathrm{tar}}|$ (Eq.~\eqref{eq:target-suppression-set}). This upper bound bounds the number of Pauli terms that we want to have non-zero syndromes under the code with non-negligible probability. For example, for dissipator structure learning one may take $M_{\mathrm{det}} = |\DiagDissStruct|^2$, while for Hamiltonian learning one may take $M_{\mathrm{det}} = M$.
        \item Failure probability $\delta \in (0,1)$.
        \item Arbitrary positive constant $c>1$.
        \item Access to ancilla qubits.
    \end{enumerate}
    }
    \Output{%
    Commuting stabilizer checks $\widetilde{\mathcal G}=\{\widetilde G_\ell\}_{\ell=1}^m$ and a recovery channel $\RMap$. Upon return, the code satisfies $E_{\mathrm{tar}}$ \eqref{eq:Etar}, and distinct zero-syndrome probe Paulis induce distinct logical Paulis on the code space.
    }

    \Schedule{
      $\mathcal W_{\mathrm{tar}} \gets
        \VDetect \cup \VCorrect \cup \{\Herm(P'P'') : P',P'' \in \VCorrect,\ P' \neq P''\}$ \tcp*{target suppression set}
      $m \gets \left\lceil\log_2(c M_{\mathrm{det}})\right\rceil$ \tcp*{number of random checks}
      $N_{\mathrm{attempts}} \gets \left\lceil\log_c\!\left(\frac{1}{\delta}\right)\right\rceil$ \tcp*{repeat-until-success budget}
      $N_{\mathrm{ancillas}} \gets \frac{5}{2}\left\lceil\log_2(cM_{\mathrm{det}})\right\rceil$ \tcp*{$\tfrac{m}{2}$ - abelianity, $m$ - logical injectivity, $m$ - syndrome meas.}
    }

    \For{$r \gets 1$ \KwTo $N_{\mathrm{attempts}}$}{
        \tcc{Each indep. draw of $\mathcal{G}$ satisfies the target event $E_{\mathrm{tar}}$ with probability $> 1-1/c$.}
        $\mathcal{G} \gets \{\, m \text{ random Pauli strings on } n \text{ qubits}\,\}$ \tcp*{Baseline generators}
        $\widetilde{\mathcal G} \gets$ Do ZX padding to ensure abelianity + adjust signs to exclude $-I$ \tcp*{$ \le 1/2\log_2(cM_{\mathrm{det}})$ ancillas}
        $\widetilde{\mathcal G} \gets $ Do Z padding for each check to ensure injective logical action \tcp*{$\log_2(cM_{\mathrm{det}})$ ancillas}

        Compute the syndromes $r(P)$ for all $P \in \mathcal W_{\mathrm{tar}}$\;
        \If{$r(P) \neq 0$ for all $P \in \mathcal W_{\mathrm{tar}}$}{
            Construct recovery channel $\RMap$ \tcp*{based on $\VDetect,\VCorrect$ as in \Cref{def:recovery-channel}}
            \Return $\widetilde{\mathcal G}, \RMap$\;
        }
    }
    \Return FAIL\;
\end{algorithm}

\begin{theorem}[Random stabilizer code for targeted reshaping]
\label{thm:random-qec-for-reshaping}
Let $\VDetect,\VCorrect \subseteq \PnHerm \setminus \{I\}$, and define the corresponding target suppression set
\begin{equation*}
    \mathcal W_{\mathrm{tar}}
    \coloneq
    \VDetect \cup \VCorrect \cup \{\Herm(P'P'') : P',P'' \in \VCorrect,\ P' \neq P''\}.
\end{equation*}
Let $M_{\mathrm{det}} \ge |\mathcal W_{\mathrm{tar}}|$ be the detectability-size bound supplied to \Cref{alg:qec_for_lindbladian_reshaping}, and let $c > 1$ be a positive constant. Use
\begin{equation*}
    m \coloneq \left\lceil \log_2(cM_{\mathrm{det}})\right\rceil
\end{equation*}
to denote the stabilizer generator cardinality. In this setting, each code draw in \Cref{alg:qec_for_lindbladian_reshaping} succeeds with probability 
\begin{equation*}
    p_{\mathrm{succ}} \coloneq \Pr(E_{\mathrm{tar}}) \ge 1-1/c,
\end{equation*}
where success means that the target event $E_{\mathrm{tar}}$ holds, namely that for the constructed code (with stabilizer subgroup $\mathcal S$), every element of $\mathcal W_{\mathrm{tar}}$ has nonzero syndrome. Equivalently, all Paulis in $\VDetect$ have nonzero syndromes and all Paulis in $\VCorrect$ have nonzero mutually distinguishable syndromes. Additionally, the random code satisfies:
\begin{enumerate}[label=(\alph*), leftmargin=*]
    \item \emph{Injective logical action of surviving terms.} If $P_i,P_j \in \PnHerm$ with $P_i\neq P_j$ are such that $\Pi P_i\Pi\ne0$ and $\Pi P_j\Pi\ne0$, then $\Pi P_i\Pi\ne \pm\,\Pi P_j\Pi$.
    \item \emph{Survival probability for non-target Paulis.}
    Any Pauli $P \notin \mathcal W_{\mathrm{tar}}$ satisfies:
    \begin{equation*}
        \Pr\!\Bigl[\,P\ \text{is undetected by the $m$ checks} \,\big|\, E_{\mathrm{tar}}\,\Bigr]
        \ \ge\ 2^{-m}\bigl(1-M_{\mathrm{det}}2^{-m}\bigr)
    \ge
    \frac{1}{2cM_{\mathrm{det}}}\left(1-\frac1c\right).
    \end{equation*}
    \item The ancilla cost of \Cref{alg:qec_for_lindbladian_reshaping} is at most $5/2\,m = 5/2\left\lceil \log_2(cM_{\mathrm{det}})\right\rceil$.
\end{enumerate}
\end{theorem}

\begin{proof}
Consider a single code draw within \Cref{alg:qec_for_lindbladian_reshaping}. Let $G_1,\dots,G_m \in \PnHerm$ be the corresponding i.i.d.\ uniformly sampled checks. To ensure abelianity, we perform standard $ZX$ padding (\Cref{lem:random-code-zx-padding}) using at most $\lfloor m/2\rfloor$ ancillas and obtain commuting checks
\begin{equation*}
    \widehat G_1,\dots,\widehat G_m.
\end{equation*}

We then use additional $Z$ padding to ensure the logical action is injective on data Paulis. In particular, we append $m$ fresh ancillas (one per check) $b_1,\dots,b_m$ and define our final stabilizer checks:
\begin{equation*}
    \widetilde G_\ell \coloneq \widehat G_\ell \otimes Z_{b_\ell},
    \qquad
    \ell=1,\dots,m.
\end{equation*}
Since the resulting checks $\{\widetilde G_\ell\}_{\ell=1}^{m}$ can only generate operators that are non-trivial on the padded ancillas, no two distinct probe Paulis can differ by a stabilizer. This ensures injectivity of logical action upon survival, proving (a).

Importantly, both $ZX$ and $Z$ padding steps preserve the commutativity of any data Pauli $P \in \PnHerm$ with the original random checks $\{G_\ell\}_{\ell = 1}^m$. Concretely, using the commutation sign $\operatorname{sgn}(\cdot,\cdot)$ defined in Eq.~\eqref{eq:pauli_commutation_sign}, we have
\begin{equation*}
    \operatorname{sgn}(\widetilde G_\ell, P \otimes I)
    =
    \operatorname{sgn}(\widehat G_\ell, P \otimes I)
    =
    \operatorname{sgn}(G_\ell,P).
\end{equation*}
Hence the $\{\widetilde G_\ell\}_{\ell=1}^{m}$ checks inherit the syndrome properties from $\{G_\ell\}_{\ell=1}^{m}$.

Recall that $E_{\mathrm{tar}}$ denotes the event that every element of $\mathcal W_{\mathrm{tar}}$ has nonzero syndrome. By \Cref{lem:prob-of-detectability-on-target-set},
\begin{equation}
    \Pr(E_{\mathrm{tar}})
    \ge
    1-|\mathcal W_{\mathrm{tar}}|2^{-m}
    \ge
    1-M_{\mathrm{det}}2^{-m}.
\end{equation}
Since $m=\lceil \log_2(cM_{\mathrm{det}})\rceil$, we get $\Pr(E_{\mathrm{tar}}) \ge 1-1/c$. This proves the claimed success probability.

For the survival bound (b), apply \Cref{lem:prob-of-survival-non-target} with $\mathcal W = \mathcal W_{\mathrm{tar}}$. For every fixed Pauli $P \notin \mathcal W_{\mathrm{tar}}$ we obtain
\begin{equation*}
    \Pr\!\bigl[P \text{ is undetectable by the } m \text{ checks} \mid E_{\mathrm{tar}}\bigr]
    \ge
    2^{-m}\bigl(1-|\mathcal W_{\mathrm{tar}}|2^{-m}\bigr)
    \ge
    2^{-m}\bigl(1-M_{\mathrm{det}}2^{-m}\bigr)
    \ge
    \frac{1}{2cM_{\mathrm{det}}}\left(1-\frac1c\right).
\end{equation*}
where the last inequality follows from $m=\lceil \log_2(cM_{\mathrm{det}})\rceil$.
This concludes the proof of part (b).

Finally, ZX padding uses at most $\lfloor m/2\rfloor$ ancillas, the additional $Z$ padding uses $m$ ancillas, and syndrome measurement can be implemented using $m$ ancillas. Thus the total ancilla budget is at most
\begin{equation}
    \left\lfloor \frac{m}{2}\right\rfloor + m + m \le \frac{5m}{2} \le \frac{5}{2}\left\lceil \log_2(cM_{\mathrm{det}})\right\rceil.
\end{equation}
which proves (c).
\end{proof}

\begin{corollary}[Amplified success probability]
\Cref{thm:random-qec-for-reshaping} proves that a single code draw from \Cref{alg:qec_for_lindbladian_reshaping} succeeds with probability $p_{\mathrm{succ}} \ge 1-1/c$. Simply repeating this step $N_{\mathrm{attempts}} = \left\lceil\log_c\!\left(\frac{1}{\delta}\right)\right\rceil$ times ensures that \Cref{alg:qec_for_lindbladian_reshaping} succeeds with probability at least $1-\delta$.
\end{corollary}

\subsection{Effective evolution under QEC reshaping}
\label{sec:qec-effective-evolution}

In this section, we quantify, how well the repeated QEC–based reshaping maps (Eq.~\eqref{eq:QEC-reshaping}, Eq.~\eqref{eq:physical-twirling-QEC-reshaping}, and Eq.~\eqref{eq:logical-twirling-QEC-reshaping}) approximate the intended effective evolution.
Our analysis has two parts:
(i) a one–step comparison between QEC–reshaped evolution and the target $e^{\tau\LindbladEff}$ with an $O(\tau^2)$ remainder (\Cref{lem:one-step-reshaping-error});
(ii) an $N$–step bound showing $\|A_\tau^{N}-(e^{t\LindbladEff}\circ \,\mathcal C_{\Pi})\|_\diamond=O(t^2/N)$ (\Cref{lem:N-step-reshaping-error}).

\begin{lemma}[One–step reshaping error] \label{lem:one-step-reshaping-error} Let $\Lindblad$ be an $n$-qubit GKSL generator as in Eq.~\eqref{eq:lindblad-eq-reshaping}. Let $C$ be a stabilizer code on $n' \ge n$ qubits with code projector $\Pi$, corner algebra $\mathfrak A_c \coloneq \Pi \mathcal B(\mathcal H_{n'}) \Pi$ and corner projection $\mathcal C_\Pi(X) \coloneq \Pi X\Pi$ for any $X \in \mathcal B(\mathcal H_{n'})$. Let $A_\tau: \mathcal B(\mathcal H_{n'}) \to \mathfrak A_c$ be any of the three Lindbladian ($\Lindblad$) reshaping maps from Eq.~\eqref{eq:QEC-reshaping}, Eq.~\eqref{eq:physical-twirling-QEC-reshaping}, and Eq.~\eqref{eq:logical-twirling-QEC-reshaping}. Let $\LindbladEff$ be the corresponding effective generator of $A_\tau$ (from \Cref{lem:qec-reshaping-generator,lem:physical-twirl-qec-reshaping-generator,lem:qec-reshape-logical-twirl-generator}) defined on the full operator algebra $\mathcal B(\mathcal H_{n'})$ in \Cref{rem:full-space-effective-generators}.

Then the following holds for the diamond norm error:
\begin{equation}
    \|A_\tau - (e^{\LindbladEff \tau} \,\circ \mathcal C_\Pi)\|_\diamond \le  \frac{(\|\Lindblad\|_\diamond^2 + \|\LindbladEff\|_\diamond^2)\tau^2}{2}
\end{equation}

\end{lemma}
\begin{proof}
Assume $A_\tau$ is the physical twirling + QEC reshaping map from Eq.~\eqref{eq:physical-twirling-QEC-reshaping}. The proof for the other two maps (Eq.~\eqref{eq:QEC-reshaping} and Eq.~\eqref{eq:logical-twirling-QEC-reshaping}) is identical. Thus, we analyze the following map:
\begin{equation*}
    A_\tau =
        \RMap \circ \frac{1}{4^n}\sum_{P\in \PnHerm}
        \bigl(\mathcal U_P\otimes I_{n'-n}\bigr)\circ
        \bigl(e^{\Lindblad \tau}\otimes I_{n'-n}\bigr)\circ
        \bigl(\mathcal U_P\otimes I_{n'-n}\bigr) \circ \mathcal C_\Pi.
\end{equation*}
where $\RMap$ is the recovery channel of the stabilizer code $C$ (\Cref{def:recovery-channel}) and $\mathcal U_P(X)\coloneq P(X)P^{\dagger}$ is the unitary conjugation map .

Use \Cref{lem:lindblad-taylor-tail}:
\(
e^{\tau\Lindblad}
= I + \tau\Lindblad + \int_0^\tau (\tau-s)\,\Lindblad^2 e^{s\Lindblad}\,ds
\) and plug it into $A_\tau$:
\begin{equation*}
    \begin{aligned}
    A_\tau  = &\RMap \circ C_\Pi+  
        \frac{\tau}{4^n}\sum_{P\in \PnHerm}\RMap \circ
        \bigl(\mathcal U_P\otimes I_{n'-n}\bigr)\circ
        \bigl(\Lindblad\otimes I_{n'-n}\bigr)\circ
        \bigl(\mathcal U_P\otimes I_{n'-n}\bigr) \circ \mathcal C_\Pi \\
         &+ \frac{1}{4^n}\sum_{P\in \PnHerm}\RMap \circ
        \bigl(\mathcal U_P\otimes I_{n'-n}\bigr)\circ
        \int_{0}^{\tau}\biggl[(\tau-s)\bigl(\Lindblad^2\otimes I_{n'-n}\bigr)\circ
        \bigl(e^{s\Lindblad }\otimes I_{n'-n}\bigr)\,ds\biggr] \circ
        \bigl(\mathcal U_P\otimes I_{n'-n}\bigr) \circ \mathcal C_\Pi
    \end{aligned}
\end{equation*}
Since $\RMap$ acts as an identity on $\mathfrak A_c$ and using $\LindbladEff$ from \Cref{lem:physical-twirl-qec-reshaping-generator}, this simplifies to:
\begin{equation*}
    A_\tau  = C_\Pi+  
        \tau\LindbladEff\circ \mathcal C_\Pi
         + \frac{1}{4^n}\sum_{P\in \PnHerm}\RMap \circ
        \bigl(\mathcal U_P\otimes I_{n'-n}\bigr)\circ
        \int_{0}^{\tau}\biggl[(\tau-s)\bigl(\Lindblad^2\otimes I_{n'-n}\bigr)\circ
        \bigl(e^{s \Lindblad}\otimes I_{n'-n}\bigr)\,ds\biggr] \circ
        \bigl(\mathcal U_P\otimes I_{n'-n}\bigr) \circ \mathcal C_\Pi
\end{equation*}

Now, applying \Cref{lem:lindblad-taylor-tail} for $\LindbladEff$:
\(
e^{\tau\LindbladEff}
= I + \tau\LindbladEff + \int_0^\tau (\tau-s)\,\LindbladEff^2 e^{s\LindbladEff}\,ds
\), we get:
\begin{equation*}
    \begin{aligned}
        \|&A_\tau - (e^{\LindbladEff \tau} \circ \,\mathcal C_\Pi)\|_\diamond  = \\ &=\norm{\frac{1}{4^n}\sum_{P\in \PnHerm}\RMap \circ
        \bigl(\mathcal U_P\otimes I_{n'-n}\bigr)\circ
        \int_{0}^{\tau}\biggl[(\tau-s)\bigl(\Lindblad^2e^{s \Lindblad}\otimes I_{n'-n}\bigr)\circ \,ds\biggr] \circ
        \bigl(\mathcal U_P\otimes I_{n'-n}\bigr) \circ \mathcal C_\Pi - \int_0^\tau (\tau-s)\,\LindbladEff^2 e^{s\LindbladEff}\,ds\circ \mathcal C_\Pi}_\diamond.
    \end{aligned}
\end{equation*}
Using triangle inequality, submultiplicativity of the diamond norm and $\|\RMap\|_\diamond, \|\mathcal C_\Pi\|_\diamond, \|\mathcal U_P\|_\diamond = 1$, we get:
\begin{equation*}
        \|A_\tau - (e^{\LindbladEff \tau} \circ \,\mathcal C_\Pi)\|_\diamond \le 
        \int_{0}^{\tau}(\tau-s) \|\Lindblad\otimes I_{n'-n}\|_\diamond^2 \,\|e^{s\Lindblad}\otimes I_{n'-n}\|_\diamond \,ds + \int_0^\tau (\tau-s)\,\|\LindbladEff\|_\diamond^2 \,\|e^{s\LindbladEff}\|_\diamond\,ds
\end{equation*}

Since $\LindbladEff$, like $\Lindblad$, is a valid Lindbladian generator (\Cref{rem:full-space-effective-generators}), $e^{\LindbladEff s}$ is a CPTP map and hence $\|e^{\LindbladEff s}\|_\diamond = 1$. Therefore, 
\begin{equation*}
        \|A_\tau - (e^{\LindbladEff \tau} \circ \,\mathcal C_\Pi)\|_\diamond \le 
        \int_{0}^{\tau}(\tau-s) \|\Lindblad\|_\diamond^2 \,ds + \int_0^\tau (\tau-s)\,\|\LindbladEff\|_\diamond^2ds = \frac{(\|\Lindblad\|_\diamond^2 + \|\LindbladEff\|_\diamond^2)\tau^2}{2}
\end{equation*}
concluding the proof of the lemma.
\end{proof}

We next obtain the $N$–fold product error bounds:

\begin{lemma}[N-step reshaping error]
\label{lem:N-step-reshaping-error}
Let $t=N\tau$. With the setting from \Cref{lem:one-step-reshaping-error}, the following holds:
\begin{equation}
\label{eq:N-step-reshaping-bound}
\Big\|\,A_\tau^{N} - e^{t\LindbladEff} \,\circ \mathcal C_\Pi\,\Big\|_\diamond
\;\le\; \frac{t^2(\|\Lindblad\|_\diamond^2 + \|\LindbladEff\|_\diamond^2)}{2N}.
\end{equation}
\end{lemma}

\begin{proof}
First, note that effective generator $\LindbladEff: \mathcal B(\mathcal H_{n'}) \to \mathcal B(\mathcal H_{n'})$ satisfies $\LindbladEff |\mathfrak A_c:  \mathfrak A_c \to  \mathfrak A_c$ for each of the three maps (see \Cref{lem:qec-reshaping-generator,lem:physical-twirl-qec-reshaping-generator,lem:qec-reshape-logical-twirl-generator}). Therefore: $e^{\LindbladEff (t_1+t_2)} \circ \mathcal C_\Pi = e^{\LindbladEff t_1}\circ \mathcal C_\Pi \circ e^{\LindbladEff t_2} \circ \mathcal C_\Pi$. Also, by definition $A_\tau \circ \mathcal C_\Pi = A_\tau$.

Hence, we can write the telescoping sum as
\begin{equation*}
A_\tau^{N} - e^{t\LindbladEff} \circ \mathcal C_\Pi
=\sum_{k=0}^{N-1} A_\tau^{N-1-k}\,\Big(A_\tau-e^{\tau\LindbladEff}\circ \mathcal C_\Pi\Big)\,e^{k\tau\LindbladEff} \circ \mathcal C_\Pi.
\end{equation*}

Using triangle inequality, $\|A_\tau\|_\diamond \le 1$ (could be directly checked from the definitions Eq.~\eqref{eq:QEC-reshaping}, Eq.~\eqref{eq:physical-twirling-QEC-reshaping}, and Eq.~\eqref{eq:logical-twirling-QEC-reshaping}), $\|e^{\tau\LindbladEff}\|_\diamond = 1$ and the submultiplicativity of the diamond norm: 
\begin{equation*}
\norm{A_\tau^{N} - e^{t\LindbladEff}\circ \mathcal C_\Pi}_\diamond
\le N\,\norm{A_\tau-e^{\tau\LindbladEff}\circ \mathcal C_\Pi}_\diamond.
\end{equation*}

Finally, using \Cref{lem:one-step-reshaping-error} and $t = \tau N$, we conclude the proof of the lemma: 
\begin{equation*}
    \norm{A_\tau^{N} - e^{t\LindbladEff}\circ \mathcal C_\Pi}_\diamond \le \frac{N\tau^2(\|\Lindblad\|_\diamond^2 + \|\LindbladEff\|_\diamond^2)}{2}=\frac{t^2(\|\Lindblad\|_\diamond^2 + \|\LindbladEff\|_\diamond^2)}{2N}
\end{equation*}
\end{proof}

\begin{corollary}[N-step reshaping error for sparse Lindbladians]
\label{cor:N-step-reshaping-error-sparse}
Given $\Lindblad$ is $M$-sparse and all coefficient are bounded in magnitude by one, \Cref{lem:N-step-reshaping-error} simplifies to 
\begin{equation*}
    \norm{A_\tau^{N} - e^{t\LindbladEff}\circ \mathcal C_\Pi}_\diamond \le \frac{4M^2 t^2}{N}
\end{equation*}
since $\|\Lindblad\|_\diamond \le 2M$ by \Cref{lem:diamond-sparse-lindblad} and $\|\LindbladEff\|_\diamond \le 2M$ by \Cref{rem:full-space-effective-generators}.
\end{corollary}

\begin{corollary}[Comparison with the coherent part of the reshaped generator]
\label{cor:N-step-reshaping-error-hamiltonian-target}
Let $A_\tau$ be either the QEC reshaping map Eq.~\eqref{eq:QEC-reshaping} or the QEC reshaping + partial logical twirl map Eq.~\eqref{eq:logical-twirling-QEC-reshaping}. Let $\LindbladEff$ be the corresponding effective generator of $A_\tau$ (from \Cref{lem:qec-reshaping-generator,lem:qec-reshape-logical-twirl-generator}) defined on the full operator algebra $\mathcal B(\mathcal H_{n'})$ in \Cref{rem:full-space-effective-generators}.

Consider the decomposition $\LindbladEff = \Lindblad_{\mathrm{coh}} + \Delta_D,$ into the coherent part $\Lindblad_{\mathrm{coh}}$ and the dissipator part $\Delta_D$. Then
\begin{equation*}
    \|A_\tau^{N} - e^{t\Lindblad_{\mathrm{coh}}}\circ \mathcal C_\Pi\|_\diamond
    \le
    \frac{t^2(\|\Lindblad\|_\diamond^2 + \|\LindbladEff\|_\diamond^2)}{2N}
    +
    t\|\Delta_D\|_\diamond \le \frac{t^2(\|\Lindblad\|_\diamond^2 + \|\LindbladEff\|_\diamond^2)}{2N}
    +
    2M_D^2\,t\,\max_{P_k\in \DiagDissStruct \setminus \VCorrect} a_{kk}.
\end{equation*}
\end{corollary}

\begin{proof}
By triangle inequality and \Cref{lem:N-step-reshaping-error},
\begin{equation*}
    \|A_\tau^{N} - e^{t\Lindblad_{\mathrm{coh}}}\circ \mathcal C_\Pi\|_\diamond
    \le
    \|A_\tau^{N} - e^{t\LindbladEff}\circ \mathcal C_\Pi\|_\diamond
    +
    \|(e^{t\LindbladEff} - e^{t\Lindblad_{\mathrm{coh}}})\circ \mathcal C_\Pi\|_\diamond.
\end{equation*}
For the second term, Duhamel’s formula gives
\begin{equation*}
    e^{t\LindbladEff} - e^{t\Lindblad_{\mathrm{coh}}}
    =
    \int_0^t e^{(t-s)\LindbladEff}\,\Delta_D\,e^{s\Lindblad_{\mathrm{coh}}}\,ds.
\end{equation*}
Using $\|e^{s\LindbladEff}\|_\diamond = \|e^{s\Lindblad_{\mathrm{coh}}}\|_\diamond = \|\mathcal C_\Pi\|_\diamond = 1$ and Eq.~\eqref{eq:dissipator-diamond-norm} with an explicit bound on $|a_{kj}|\le \sqrt{a_{kk} a_{jj}}$, we get
\begin{equation*}
    \|(e^{t\LindbladEff} - e^{t\Lindblad_{\mathrm{coh}}})\circ \mathcal C_\Pi\|_\diamond
    \le
    t\|\Delta_D\|_\diamond \le 2tM_D^2\max_{P_k\in \DiagDissStruct \setminus \VCorrect} a_{kk}.
\end{equation*}
\end{proof}

\subsection{Deferred proofs for effective generators of QEC reshaping}\label{sec:deferred-proofs-effective-QEC-generators}

\begin{restatelem}{lem:qec-reshaping-generator}[Effective generator under QEC reshaping]
    Let $\Lindblad$ be an $n$-qubit GKSL generator as in Eq.~\eqref{eq:lindblad-eq-reshaping}, and let $\Lindblad \otimes I_{n'-n}$ denote its trivial extension to the ancilla register. Let $C$ be a stabilizer code on $n' \ge n$ qubits with code projector $\Pi$, corner algebra $\mathfrak A_c \coloneq \Pi \mathcal B(\mathcal H_{n'}) \Pi$ and syndrome map $r(\cdot)$ (as in Eq.~\eqref{eq:syndrome-map}). Suppose $\VDetect \subseteq \PnHerm$ and $\VCorrect \subseteq \DiagDissStruct \subseteq \PnHerm$ are specified sets, and that all Paulis in $\VDetect$ have non-zero syndromes. Assume furthermore that all elements of $\VCorrect$, have pairwise distinct nonzero syndromes. Let $\RMap: \mathcal B(\mathcal H_{n'}) \to \mathfrak A_c$ be the associated recovery map from \Cref{def:recovery-channel} based on the specified sets $\VDetect$ and $\VCorrect$. 

    Then, for any $X\in\mathfrak A_c$, the effective generator of the QEC reshaping map from Eq.~\eqref{eq:QEC-reshaping} is
    \begin{equation}
    \label{eq:Leff-qec-reshaping-deferred}
        \begin{aligned}
            &\LindbladEff(X) \coloneq \RMap \circ (\Lindblad \otimes I_{n'-n})(X)  =  \sum_{\substack{P_k,P_j\in \DiagDissStruct \setminus \VCorrect,\\ r(P_k) = r(P_j)}} a_{kj}
        \Bigl((Q_{P_k} \otimes I_{n'-n}) X (Q_{P_j}^\dagger \otimes I_{n'-n}) - \frac{1}{2}\{Q_{P_j}^\dagger Q_{P_k}\otimes I_{n'-n} ,X\}\Bigr) \\
        &-i\underbrace{\sum_{\substack{
        P_k \in \HamiltStruct, \\ r(P_k) = 0}} h_k[P_k  \otimes I_{n'-n},X]}_{[H_\mathrm{tar}, X]}
        \,-\,i \underbrace{\sum_{\substack{
        P_k \in \VCorrect, \\ P_j \in \DiagDissStruct \setminus \VCorrect, \\r(P_k) = r(P_j)}} \left({\operatorname{Re}(a_{kj})\left[\frac{i[P_k,P_j]}{2}  \otimes I_{n'-n}\,,\,X\right] + \operatorname{Im}(a_{kj})\left[\frac{\{P_k,P_j\}}{2}  \otimes I_{n'-n}\,,\,X\right]}\right)}_{[H_\mathrm{bias},X]}
        \end{aligned} 
    \end{equation}
    where $Q_P \coloneqq \widetilde R_{r(P)}P \in \mathbb P_n$ are Pauli operators composed of $P$ and the corresponding recovery actions $\widetilde R_{r(P)}$ as specified in Eq.~\eqref{eq:syndrome-correction-unitaries} with $R_{r(P)}=\widetilde R_{r(P)}\otimes I_{n'-n}$. Notably, the resulting Hamiltonian part of $\LindbladEff$ contains only zero-syndrome Paulis. In particular, no Pauli term from
    \begin{equation*}
    \VDetect \cup \VCorrect \cup \{\Herm(PP') : P,P' \in \VCorrect,\ P\neq P'\}
    \end{equation*}
    is present in the Pauli decompositions of $H_{\mathrm{tar}}$ and $H_{\mathrm{bias}}$.
\end{restatelem}

\begin{proof}    
    Throughout the proof, for any probe-system Pauli $P \in \mathcal P_n$, we write again $P$ for its canonical embedding $P \otimes I_{n'-n} \in \mathcal P_{n'}$. Likewise, we write $H$ and $\Lindblad$ for their trivial extensions to the ancilla register. Expand the Lindbladian $\Lindblad$ acting on $X \in \mathfrak A_c$ in the Pauli basis:
    \begin{equation*}
    \Lindblad(X)
    = -i \sum_{P_k\in \HamiltStruct} h_k \,[P_k,X]
    \;+\;
    \sum_{P_k,P_j\in \DiagDissStruct} a_{kj}\!\left(P_k X P_j - \tfrac12\{P_j P_k,X\}\right) = -i[H, X] \,+\, \Diss(X)
    \end{equation*}   
    We split the dissipator into four sums:
    \begin{equation*}
        \begin{aligned}
            \Diss(X) = \left(\sum_{P_k,P_j\in \VCorrect} +\sum_{\substack{P_k\in \VCorrect,\\ P_j \in \DiagDissStruct\setminus\VCorrect}}+\sum_{\substack{P_k\in\DiagDissStruct\setminus\VCorrect,\\P_j \in \VCorrect}}+\sum_{P_k,P_j\in \DiagDissStruct\setminus \VCorrect}\right)a_{kj}\!\left(P_k X P_j - \tfrac12\{P_j P_k,X\}\right) = \\
            = \Diss_{1}(X) + \Diss_{2}(X) + \Diss_{3}(X) + \Diss_{4}(X).
        \end{aligned}
    \end{equation*}
    where $\Diss_{j}(X)$ denotes the $j$-th sum in the parentheses from the left. Therefore, since $X = \Pi X \Pi$ 
    \begin{equation*}
            \LindbladEff(X)
            = \RMap\circ\Lindblad\,(X)  = \RMap(
               -i \, [H,\Pi X \Pi])+
               \RMap(\Diss_{1}(\Pi X \Pi)) + \RMap(\Diss_{2}(\Pi X \Pi)) + \RMap(\Diss_{3}(\Pi X \Pi)) + \RMap(\Diss_{4}(\Pi X \Pi))
    \end{equation*}

    We first analyze the $\Diss_{1}$ term by explicitly expanding the recovery channel $\RMap$ from \Cref{def:recovery-channel}:
    \begin{equation*}
         \RMap(\Diss_{1}(\Pi X \Pi)) = \sum_{s\in\{0,1\}^m} \Biggl[\sum_{P_k,P_j\in \VCorrect}a_{kj}\!\left(R_s\Pi_s P_k \Pi X \Pi P_j \Pi_s R_s - \tfrac12 R_s\Pi_s\{P_j P_k,\Pi X \Pi\} \Pi_s R_s \right)\Biggr]
    \end{equation*}

    Exchanging the summation order and using the projector properties $\Pi_s\Pi_{s'}=\delta_{s,s'}\Pi_s$ and $P_k\Pi_s = \Pi_{s\oplus r(P_k)}P_k$, $\Pi_s P_k= P_k\Pi_{s\oplus r(P_k)}$ from Eq.~\eqref{eq:syndrome-shift-rule}, rewrite:
     \begin{equation}
        \label{eq:diag-dissipator-qec-reshaped}
        \begin{aligned}
            \RMap(\Diss_{1}(\Pi X \Pi)) 
            &= \sum_{P_k,P_j\in \VCorrect}a_{kj}\!\left( \sum_{s\in\{0,1\}^m} \Bigl[R_s P_k \Pi_{s\oplus r(P_k)}\,\Pi X \Pi\,\Pi _{s\oplus r(P_j)} P_j R_s\Bigr] - \tfrac12 \Pi\{P_j P_k,\Pi X \Pi\} \Pi \right) = \\
            &= \sum_{\substack{P_k,P_j\in \VCorrect,\\ r(P_k) =r(P_j)}}a_{kj}\!\left( R_{r(P_k)} P_k \Pi X \Pi P_j R_{r(P_k)} - \tfrac12 \{\Pi P_j P_k \Pi,\Pi X \Pi\} \right)
        \end{aligned}
    \end{equation}
    By assumption of the lemma all Paulis $P\in \VCorrect$ have mutually distinguishable nonzero syndromes. Then, by construction of $\RMap$ (see \Cref{def:recovery-channel}), all Paulis $P_k\in \VCorrect$ have a corresponding $R_{r(P_k)} = P_k$. Therefore:
    \begin{equation} \label{eq:Diss1_term}
        \RMap(\Diss_{1}(\Pi X \Pi))  = \sum_{P_k\in \VCorrect}a_{kk}\!\left( P_k^2 \Pi X \Pi P_k^2) - \Pi X\Pi \right) = 0
    \end{equation}

    Similarly to Eq.~\eqref{eq:diag-dissipator-qec-reshaped} with $X = \Pi X\Pi$:
    \begin{equation*}
        \RMap(\Diss_{2}(X))  =\sum_{\substack{P_k\in \VCorrect,\\ P_j \in \DiagDissStruct\setminus\VCorrect ,\\ r(P_k)=r(P_j)}}a_{kj}\!\left( R_{r(P_k)} P_k  X  P_j R_{r(P_k)} - \tfrac12 \{\Pi P_j P_k \Pi,X \} \right) = \sum_{\substack{P_k\in \VCorrect,\\ P_j \in \DiagDissStruct\setminus\VCorrect ,\\ r(P_k)=r(P_j)}}\frac{a_{kj}}{2}\!\left( X P_j P_k - P_j P_k X \right)
    \end{equation*}
    where in the last equality we used $R_{r(P_k)} = P_k $ for all $ P_k\in \VCorrect$ combined with $r(P_kP_j) = r(P_k)\oplus r(P_j) =0$ implying commutativity of $\Pi$ with $P_kP_j$. Same calculation for $\RMap(\Diss_{3}(X))$ yields:
    \begin{equation*}
        \RMap(\Diss_{3}(X))  =\sum_{\substack{P_k\in \DiagDissStruct\setminus\VCorrect,\\ P_j \in  \VCorrect,\\ r(P_k)=r(P_j)}}a_{kj}\!\left( R_{r(P_j)} P_k  X  P_j R_{r(P_j)} - \tfrac12 \{\Pi P_j P_k \Pi,X \} \right) = \sum_{\substack{P_k\in \DiagDissStruct\setminus\VCorrect,\\ P_j \in  \VCorrect,\\ r(P_k)=r(P_j)}}\frac{a_{kj}}{2}\!\left( P_jP_k X - X P_j P_k\right)
    \end{equation*}

    Combining $\RMap(\Diss_{2}(X))$ and $\RMap(\Diss_{3}(X))$ and using the positivity of the Kossakowski matrix $a$, we get:
    \begin{equation}\label{eq:Diss2and3_term}
        \RMap((\Diss_{2}+\Diss_{3})(X)) = \,-\,i\sum_{\substack{
    P_k \in \VCorrect, \\ P_j \in \DiagDissStruct \setminus \VCorrect, \\r(P_k) = r(P_j)}} \left({\operatorname{Re}(a_{kj})\left[\frac{i[P_k,P_j]}{2}  \otimes I_{n'-n}\,,\,X\right] + \operatorname{Im}(a_{kj})\left[\frac{\{P_k,P_j\}}{2}  \otimes I_{n'-n}\,,\,X\right]}\right)
    \end{equation} 
    Repeating this calculation for $\RMap(\Diss_{4}(X))$, we get
    \begin{equation}\label{eq:Diss4_term}
        \begin{aligned}
            \RMap(\Diss_{4}(X))  &=\sum_{\substack{P_k,P_j\in \DiagDissStruct\setminus\VCorrect,\\ r(P_k)=r(P_j)}}a_{kj}\!\left( R_{r(P_k)} P_k  X  P_j R_{r(P_j)} - \tfrac12 \{ P_jR_{r(P_j)}R_{r(P_k)} P_k ,X \} \right) \\
            &= \sum_{\substack{P_k,P_j\in \DiagDissStruct \setminus \VCorrect,\\ r(P_k) = r(P_j)}} a_{kj}
            \Bigl((Q_{P_k} \otimes I_{n'-n}) X (Q_{P_j}^\dagger \otimes I_{n'-n}) - \frac{1}{2}\{Q_{P_j}^\dagger Q_{P_k}\otimes I_{n'-n} ,X\}\Bigr)
        \end{aligned}
    \end{equation}
    where $Q_{P_k} \coloneq R_{r(P_k)}P_k \in \mathbb P_n$ are Pauli operators composed of $P_k$ and the corresponding recovery actions $R_{r(P_k)}$ from \Cref{def:recovery-channel}.

    Now, expand the Hamiltonian part of $\LindbladEff(X)$ using the recovery channel $\RMap$ from \Cref{def:recovery-channel}:
     \begin{equation*}
         \begin{aligned}
                \RMap(
                   -i \, [H,\Pi X \Pi]) =  
                \sum_{s\in\{0,1\}^m}\Biggl[
                   -i \sum_{P_i\in \HamiltStruct} h_i\, R_s\Pi_s [P_i,\Pi X \Pi] \Pi_s R_s \Biggr] = \\
                 =\sum_{s\in\{0,1\}^m}\Biggl[
                   -i \sum_{P_i\in \HamiltStruct} h_i\, R_s\Pi_s P_i\, \Pi X \Pi\, \Pi_s R_s  +i \sum_{P_i\in \HamiltStruct} h_i\, R_s\Pi_s \Pi X \Pi\, P_i\, \Pi_s R_s\Biggr]     
        \end{aligned}
    \end{equation*}
    
    Since $\Pi_s\Pi_{s'}=\delta_{s,s'}\Pi_s$, $\Pi_0 = \Pi$ and $R_0 = I$,  the Hamiltonian part of $\LindbladEff(X)$ simplifies to:
    \begin{equation*}
                \RMap(
                   -i \, [H,\Pi X \Pi])
                 = 
                   -i \sum_{P_i\in \HamiltStruct} h_i\, \Pi P_i \Pi X \Pi\,  +i \sum_{P_i\in \HamiltStruct} h_i\, \Pi X \Pi P_i \Pi.
    \end{equation*}
     Since $\Pi P \Pi = 0$ for every Pauli $P$ with $r(P)\neq 0$, and $[\Pi,Q]=0$ for every Pauli $Q$ with $r(Q)=0$, the above expression further simplifies to 
     \begin{equation*}
                \RMap(
                   -i \, [H,\Pi X \Pi])
                 = 
                   -i \sum_{P_i\in \HamiltStruct\,:\, r(P_i)=0} h_i \left(\ P_i \Pi^2 X \Pi\,  - \, \Pi X \Pi^2 P_i\right) = -i \sum_{P_i\in \HamiltStruct\,:\, r(P_i)=0} h_i [P_i,X] .
    \end{equation*}
    which combined with the reshaped dissipator terms (Eq.~\eqref{eq:Diss1_term}, Eq.~\eqref{eq:Diss2and3_term}, and Eq.~\eqref{eq:Diss4_term}) yields Eq.~\eqref{eq:Leff-qec-reshaping-deferred} for inputs in $\mathfrak A_c$.

    Finally, by detectability on $\VDetect$ and syndrome distinguishability on $\VCorrect$, all Paulis in the following set have non-zero syndromes:
    \begin{equation*}
        \VDetect \cup \VCorrect \cup \{\Herm(PP') : P,P' \in \VCorrect, \ P \neq P'\},
    \end{equation*}
    and are thus excluded from the Pauli decompositions of the resulting Hamiltonians $H_{\mathrm{tar}}$ and $H_{\mathrm{bias}}$.
\end{proof}

\begin{restatelem}{lem:qec-reshape-logical-twirl-generator}[Effective generator of QEC reshaping + partial logical twirl]
Let $\Lindblad$ be an $n$-qubit GKSL generator as in Eq.~\eqref{eq:lindblad-eq-reshaping}, and let $\Lindblad \otimes I_{n'-n}$ denote its trivial extension to the ancilla register. Let $C$ be a stabilizer code on $n' \ge n$ qubits with code projector $\Pi$ and corner algebra $\mathfrak A_c \coloneq \Pi \mathcal B(\mathcal H_{n'}) \Pi$. Let $\overline P_a \coloneq P_a \otimes I_{n'-n}$ be the nontrivial logical Pauli to be isolated. Suppose $\VCorrect \subseteq \DiagDissStruct \subseteq \PnHerm$ is a specified set and let $\RMap:\mathcal B(\mathcal H_{n'}) \to \mathfrak A_c$ be the associated recovery map from \Cref{def:recovery-channel} based on $\VCorrect$. Assume furthermore that all elements of $\VCorrect$ have pairwise distinct nonzero syndromes. Assume also that $P_a$ is the unique $n$-qubit Pauli whose restriction to the code space contributes to the logical Pauli $\overline P_a$. Equivalently, for every $P_k\in \PnHerm$ with $P_k\neq P_a$, the compressed operator $\Pi (P_k\otimes I_{n'-n})\Pi$ has zero component along $\Pi \overline P_a \Pi$. Assume moreover that no nontrivial product of zero-syndrome Pauli operators appearing in the QEC-reshaped generator of $\Lindblad$ from \Cref{lem:qec-reshaping-generator} is proportional to a stabilizer.

Then for any $X\in \mathfrak A_c$, the effective generator of QEC reshaping + the partial logical twirl over $\overline{\mathcal K}_{\overline P_a}$ is
\begin{equation}
\label{eq:Leff-qec-reshaping-logical-twirl-deferred}
    \begin{aligned}
        \LindbladEff(&X)
        \coloneq
        \frac{1}{|\overline{\mathcal K}_{\overline P_a}|}
        \sum_{\overline Q \in \overline{\mathcal K}_{\overline P_a}}
        \mathcal U_{\overline Q} \circ \RMap \circ (\Lindblad \otimes I_{n'-n}) \circ \mathcal U_{\overline Q} (X) =\\
        &= -i h_a [\overline P_a, X]
        \,-\,i \sum_{\substack{
        P_k \in \VCorrect, \\ P_j \in \DiagDissStruct \setminus \VCorrect,\\P_k P_j \propto P_a}} \left({\operatorname{Re}(a_{kj})\left[\frac{i[P_k,P_j]}{2}  \otimes I_{n'-n}\,,\,X\right] + \operatorname{Im}(a_{kj})\left[\frac{\{P_k,P_j\}}{2}  \otimes I_{n'-n}\,,\,X\right]}\right) + \\
        &+\sum_{\substack{P_k\in \DiagDissStruct \setminus \VCorrect,\\\Herm(P_k P_a) \in \DiagDissStruct \setminus \VCorrect;}} \sum_{\substack{P_j \in \DiagDissStruct \setminus \VCorrect,\\ \Herm(P_k P_j) \in \{I_n, P_a\}}} a_{kj}
        \Bigl((Q_{P_k} \otimes I_{n'-n}) X (Q_{P_j}^\dagger \otimes I_{n'-n}) - \frac{1}{2} \{Q_{P_j}^\dagger Q_{P_k}\otimes I_{n'-n},X\}\Bigr) \\
        &+ \sum_{\substack{P_k\in \DiagDissStruct \setminus \VCorrect,\\ \Herm(P_k P_a) \notin \DiagDissStruct \setminus \VCorrect}} a_{kk}
        \Bigl((Q_{P_k} \otimes I_{n'-n}) X (Q_{P_k}^\dagger \otimes I_{n'-n}) - X\Bigr).
    \end{aligned}
\end{equation}
where $Q_P \coloneqq \widetilde R_{r(P)}P \in \mathbb P_n$ are Pauli operators composed of $P$ and the corresponding recovery actions $\widetilde R_{r(P)}$ as specified in
Eq.~\eqref{eq:syndrome-correction-unitaries} with $R_{r(P)}=\widetilde R_{r(P)}\otimes I_{n'-n}$
\end{restatelem}

\begin{proof}
    First, apply \Cref{lem:qec-reshaping-generator} with $\VDetect = \emptyset$ for the chosen code and recovery map. Since all elements of $\VCorrect$ have pairwise distinct nonzero syndromes, the QEC-reshaped generator on $X \in \mathfrak A_c$ is
    \begin{equation*}
        \begin{aligned}
            &\RMap \circ (\Lindblad \otimes I_{n'-n})(X)  =  \underbrace{\sum_{\substack{P_k,P_j\in \DiagDissStruct \setminus \VCorrect,\\ r(P_k) = r(P_j)}} a_{kj}
        \Bigl((Q_{P_k} \otimes I_{n'-n}) X (Q_{P_j}^\dagger \otimes I_{n'-n}) - \frac{1}{2}\{Q_{P_j}^\dagger Q_{P_k}\otimes I_{n'-n} ,X\}\Bigr)}_{A} \\
        &-i\underbrace{\sum_{\substack{
        P_k \in \HamiltStruct, \\ r(P_k) = 0}} h_k[P_k  \otimes I_{n'-n},X]}_{B}
        \,-\,i \underbrace{\sum_{\substack{
        P_k \in \VCorrect, \\ P_j \in \DiagDissStruct \setminus \VCorrect, \\r(P_k) = r(P_j)}} \left({\operatorname{Re}(a_{kj})\left[\frac{i[P_k,P_j]}{2}  \otimes I_{n'-n}\,,\,X\right] + \operatorname{Im}(a_{kj})\left[\frac{\{P_k,P_j\}}{2}  \otimes I_{n'-n}\,,\,X\right]}\right)}_{C}
        \end{aligned} 
    \end{equation*}
    where $r(\cdot)$ is the syndrome map of code $C$.

    We first analyze the contribution of term $B$ to $\LindbladEff$. Applying the logical twirl to $B$, we get that for any $X \in \mathfrak A_c$:
    \begin{equation*}
        \frac{-i}{|\overline{\mathcal K}_{\overline P_a}|}
        \sum_{\overline Q \in \overline{\mathcal K}_{\overline P_a}}
        \mathcal U_{\overline Q}  \sum_{\substack{P_k \in \HamiltStruct,\\ r(P_k)=0}}h_k[P_k\otimes I_{n'-n,\,}\mathcal U_{\overline Q} (X)] = -i\left[
        \frac{1}{|\overline{\mathcal K}_{\overline P_a}|}
        \sum_{\overline Q \in \overline{\mathcal K}_{\overline P_a}}
        \mathcal U_{\overline Q}\bigl(\Pi (H \otimes I_{n'-n}) \Pi\bigr),
        X
        \right]
    \end{equation*}
    where the right-hand side follows from the fact that $\mathcal U_{\overline Q}$ is a unitary conjugation.
    
    By the assumptions of the lemma, the projected Hamiltonian $\Pi (H \otimes I_{n'-n}) \Pi$ has exactly one contribution proportional to $\overline P_a$, namely $h_a \overline P_a$. Applying \Cref{lem:logical-pauli-twirling}, the average over $\overline{\mathcal K}_{\overline P_a}$ removes all other non-identity logical Pauli terms, while any multiple of $\Pi$ can be absorbed into $X$. Therefore the contribution of term $B$ to $\LindbladEff$ is
    \begin{equation}\label{eq:term_B_QEC_logical_twirl}
        -i h_a [\overline P_a, X].
    \end{equation}

    The same exact argument applies to term $C$. Since each $P_kP_j$ in term $C$ has zero syndrome, the only surviving Pauli terms are those whose restriction to the code space contributes to $\overline P_a$, namely the terms with $P_kP_j \propto P_a$. Thus the contribution of term $C$ to $\LindbladEff$ is Eq.~\eqref{eq:term_C_QEC_logical_twirl}.
    \begin{equation}\label{eq:term_C_QEC_logical_twirl}
        -\,i \sum_{\substack{
        P_k \in \VCorrect, \\ P_j \in \DiagDissStruct \setminus \VCorrect,\\P_k P_j \propto P_a}} \left({\operatorname{Re}(a_{kj})\left[\frac{i[P_k,P_j]}{2}  \otimes I_{n'-n}\,,\,X\right] + \operatorname{Im}(a_{kj})\left[\frac{\{P_k,P_j\}}{2}  \otimes I_{n'-n}\,,\,X\right]}\right)
    \end{equation}

    It remains to quantify the contribution of term $A$. For that, consider logically twirling its single summand:
    \begin{equation*}
        \frac{a_{kj}}{|\overline{\mathcal K}_{\overline P_a}|}
        \sum_{\overline Q \in \overline{\mathcal K}_{\overline P_a}} \mathcal U_{\overline Q}\Bigl((Q_{P_k} \otimes I_{n'-n}) \mathcal \,U_{\overline Q}(X)\, (Q_{P_j}^\dagger \otimes I_{n'-n}) - \frac{1}{2}\{Q_{P_j}^\dagger Q_{P_k}\otimes I_{n'-n} ,\mathcal U_{\overline Q}(X)\}\Bigr)
    \end{equation*}
    
    If $P_k = P_j$ and thus $Q_{P_k} = Q_{P_j}$ for $r(P_k) = r(P_j)$, a single summand contributes
    \begin{equation}\label{eq:diagonal_summand_logical_twirl_qec}
        \frac{a_{kk}}{|\overline{\mathcal K}_{\overline P_a}|}
        \sum_{\overline Q \in \overline{\mathcal K}_{\overline P_a}} \mathcal U_{\overline Q}\Bigl((Q_{P_k} \otimes I_{n'-n}) \mathcal \,U_{\overline Q}(X)\, (Q_{P_k}^\dagger \otimes I_{n'-n}) - \mathcal U_{\overline Q}(X)\}\Bigr) = a_{kk}
        \Bigl((Q_{P_k} \otimes I_{n'-n})X (Q_{P_k}^\dagger \otimes I_{n'-n}) - X\Bigr)
    \end{equation}
    where the right-hand side follows from each $Q_{P_k}$ being a Pauli. Concretely, conjugation of $Q_{P_k}$ by any $\overline Q \in \overline{\mathcal K}_{\overline P_a}$ maps $Q_{P_k}$ to $\pm Q_{P_k}$, and since $Q_{P_k}$ is conjugated twice, the final sign cancels.

    Now consider a summand where $\Herm(P_k P_j) \notin \{I_n, \pm P_a\}$ and $r(P_k)=r(P_j)$. Then $Q_{P_j}^\dagger Q_{P_k}$ is a nonidentity zero-syndrome probe Pauli obtained as a product of Pauli operators appearing in the QEC-reshaped generator. By the assumption of the lemma, it is not proportional to a stabilizer, and since $\Herm(P_k P_j)\neq \pm P_a$, its logical action is neither equal to the identity nor to $\overline P_a$. Therefore the anti-commutator part is zero directly by \Cref{lem:logical-pauli-twirling}. The sandwiched term $\overline Q(Q_{P_k} \otimes I_{n'-n}) \overline Q\, X \,\overline Q (Q_{P_j}^\dagger \otimes I_{n'-n})\overline Q$ also sums to zero, since half of $\overline{\mathcal K}_{\overline P_a}$ commute with the $Q_{P_j}^\dagger Q_{P_k}$ and half anti-commute:
    \begin{equation}\label{eq:non-Pa_summand_logical_twirl_qec}
        \frac{a_{kj}}{|\overline{\mathcal K}_{\overline P_a}|}
        \sum_{\overline Q \in \overline{\mathcal K}_{\overline P_a}} \overline Q(Q_{P_k} \otimes I_{n'-n}) \overline QX\overline Q\, (Q_{P_j}^\dagger \otimes I_{n'-n})\overline Q - \frac{1}{2}\{\overline Q (Q_{P_j}^\dagger Q_{P_k}\otimes I_{n'-n})\overline Q ,X\}\Bigr) = 0
    \end{equation}
    
    Finally, if $P_k P_j \propto P_a$ and thus $Q_{P_k} Q_{P_j}^\dagger \propto P_a$ for $r(P_k) = r(P_j)$, a single summand contributes:
    \begin{equation}\label{eq:Pa_summand_logical_twirl_qec}
    \begin{aligned}
        \frac{a_{kj}}{|\overline{\mathcal K}_{\overline P_a}|}
        \sum_{\overline Q \in \overline{\mathcal K}_{\overline P_a}} \overline Q(Q_{P_k} \otimes I_{n'-n}) \overline QX\overline Q\, (Q_{P_j}^\dagger \otimes I_{n'-n})\overline Q - \frac{1}{2}\{\overline Q (Q_{P_j}^\dagger Q_{P_k}\otimes I_{n'-n})\overline Q ,X\}\Bigr) =\\
        =a_{kj}
         (Q_{P_k} \otimes I_{n'-n}) X (Q_{P_j}^\dagger \otimes I_{n'-n}) - \frac{1}{2}\{Q_{P_j}^\dagger Q_{P_k}\otimes I_{n'-n},X\}
        \end{aligned}
    \end{equation}
    where again the anti-commutator simplifies directly by \Cref{lem:logical-pauli-twirling}. The $(Q_{P_k} \otimes I_{n'-n}) X (Q_{P_j}^\dagger \otimes I_{n'-n})$ term follows from the fact that $Q_{P_k} Q_{P_j}^\dagger \propto  P_a$ and thus all $\overline Q \in {\mathcal K}_{\overline P_a}$ commute with $Q_{P_k} Q_{P_j}^\dagger$. This implies that the commutation sign (defined in Eq.~\eqref{eq:pauli_commutation_sign}) of these Paulis satisfies $\operatorname{sgn}(\overline Q, Q_{P_k}) = \operatorname{sgn}(\overline Q, Q_{P_j})$ and hence:
    \begin{equation*}
        \overline Q(Q_{P_k} \otimes I_{n'-n}) \overline QX\overline Q\, (Q_{P_j}^\dagger \otimes I_{n'-n})\overline Q  = \overline Q^2(Q_{P_k} \otimes I_{n'-n}) X (Q_{P_j}^\dagger \otimes I_{n'-n})\overline Q^2 = (Q_{P_k} \otimes I_{n'-n}) X (Q_{P_j}^\dagger \otimes I_{n'-n})
    \end{equation*}

    Adding the summands together and simplifying according to Eq.~\eqref{eq:Pa_summand_logical_twirl_qec}, Eq.~\eqref{eq:diagonal_summand_logical_twirl_qec}, and Eq.~\eqref{eq:non-Pa_summand_logical_twirl_qec}, we obtain that the contribution of term $A$ to $\LindbladEff$ is:
    \begin{equation}\label{eq:term_A_QEC_logical_twirl}
    \begin{aligned}
        \sum_{\substack{P_k,P_j\in \DiagDissStruct \setminus \VCorrect,\\ r(P_k) = r(P_j)}}
            \frac{a_{kj}}{|\overline{\mathcal K}_{\overline P_a}|}
            \sum_{\overline Q \in \overline{\mathcal K}_{\overline P_a}} \mathcal U_{\overline Q}
            \Bigl((Q_{P_k} \otimes I_{n'-n}) \,\mathcal U_{\overline Q}(X)\, (Q_{P_j}^\dagger \otimes I_{n'-n}) - \frac{1}{2}\{Q_{P_j}^\dagger Q_{P_k}\otimes I_{n'-n},\,\mathcal U_{\overline Q}(X)\}\Bigr) = \\
            = \sum_{\substack{P_k\in \DiagDissStruct \setminus \VCorrect,\\\Herm(P_k P_a) \in \DiagDissStruct \setminus \VCorrect;}} \sum_{\substack{P_j \in \DiagDissStruct \setminus \VCorrect,\\\Herm(P_k P_j) \in \{I_n, P_a\}}} a_{kj}
            \Bigl((Q_{P_k} \otimes I_{n'-n}) X (Q_{P_j}^\dagger \otimes I_{n'-n}) - \frac{1}{2} \{Q_{P_j}^\dagger Q_{P_k}\otimes I_{n'-n},X\}\Bigr) +\\
            + \sum_{\substack{P_k\in \DiagDissStruct \setminus \VCorrect,\\ \Herm(P_k P_a) \notin \DiagDissStruct \setminus \VCorrect}} a_{kk}
            \Bigl((Q_{P_k} \otimes I_{n'-n}) X (Q_{P_k}^\dagger \otimes I_{n'-n}) - X\Bigr).
    \end{aligned}
    \end{equation}

    Combining the contributions of terms $A,B,C$ from Eq.~\eqref{eq:term_A_QEC_logical_twirl}, Eq.~\eqref{eq:term_B_QEC_logical_twirl}, and Eq.~\eqref{eq:term_C_QEC_logical_twirl} proves the $\LindbladEff$ in Eq.~\eqref{eq:Leff-qec-reshaping-logical-twirl-deferred}.
\end{proof}

%% file: appendix/dissipator_structure_learning.tex
\section{Dissipator Structure Learning}
\label{sec:appendix_dissipator_structure_learning}

Recall the $n$-qubit Lindbladian we wish to learn:
\begin{equation}\label{eq:lindblad-eq-diss-struct}
    \Lindblad(\rho)
    =
    -i\sum_{P_k\in \HamiltStruct} h_k \comm{P_k}{\rho}
    +
    \sum_{P_k,P_j \in \DiagDissStruct} a_{kj}
    \left(P_k \rho P_j - \tfrac12\acomm{P_j P_k}{\rho}\right).
\end{equation}
As in the rest of the work, we assume without loss of generality that the Lindbladian has been rescaled so that all coefficients have magnitude at most one: $|h_k|\le 1, |a_{kj}|\le 1.$ In this section, however, we do not assume sparsity.

In this section, we present our results on learning the $\eta$-heavy subset of the dissipator structure. In particular, we present \Cref{alg:appendix-diss-struct-learning} that identifies:
\begin{equation}\label{eq:def-eta-heavy-diss-struct}
    \EtaDiagDissStruct \coloneq \{P_k \in \DiagDissStruct: a_{kk} \ge \eta\}
\end{equation}
given an upper bound on $\|\Lindblad\|_\diamond$. To attain the $\SoftBigO{1/\eta}$ evolution time, our algorithm extends the hierarchical learning strategy of \cite{hu2025ansatz} to open-system dynamics using the QEC-based Lindbladian reshaping introduced in Eq.~\eqref{eq:physical-twirling-QEC-reshaping}. More specifically, the algorithm first samples the strongest dissipative terms and then applies a testing step to eliminate false positives. It subsequently updates the error correcting code adaptively to suppress the identified terms, and then iterates the same procedure on the remaining weaker terms.

The main result of this section is the following theorem:
\begin{theorem}[Hierarchical dissipator structure learning]
\label{thm:appendix-hier-diss-struct-learning}
Let $\Lindblad$ be an $n$-qubit Lindbladian as in Eq.~\eqref{eq:lindblad-eq-diss-struct} with $\|\Lindblad\|_\diamond \le \Lambda_L$. Fix $\eta, \delta_{d} \in (0,1)$ and assume black-box access to $e^{t\Lindblad}$. Let $\EtaDiagDissStruct \subseteq \DiagDissStruct$ denote the $\eta$-heavy part of the dissipator structure defined in Eq.~\eqref{eq:def-eta-heavy-diss-struct}. Let $\Gamma_{\mathrm{tot}}$ denote the total Lindbladian decay rate defined in Eq.~\eqref{eq:total-decay-rate-diss-testing}. Then running \Cref{alg:appendix-diss-struct-learning} returns an estimator $\EstEtaDiagDissStruct$ of $\EtaDiagDissStruct$, satisfying: 

\begin{enumerate}[leftmargin=*, label=\arabic*.]
\item \textit{(Accuracy)} With probability $\ge1-\delta_d$, the output estimator obeys
$\EtaDiagDissStruct \subseteq \EstEtaDiagDissStruct \subseteq \HalfEtaDiagDissStruct$.

\item \textit{(Total evolution time)} It applies $e^{\Lindblad t}$ for a total evolution time
\begin{equation*}
t_{\mathrm{tot}}=\SoftBigO{\frac{1}{\eta}}.
\end{equation*}

\item \textit{(Minimum evolution time)} It only ever applies $e^{\Lindblad t}$ for $t \ge t_{\mathrm{min}}$, where
\begin{equation*}
t_{\mathrm{min}}=\Omega\Bigl(\frac{\eta}{\Lambda_L^2}\Bigr)  .
\end{equation*}

\item \textit{(Number of experiments)} The total number of experiments is 
\begin{equation*}
N_{\exp} \;=\; \SoftBigO{\frac{\Gamma_{\mathrm{tot}}}{\eta}+1}.
\end{equation*}
with the total number of interleaved QEC rounds bounded by $\SoftBigO{\Lambda_L^2/\eta^2}$.

\item \textit{(Ancilla cost)}
The number of ancillary qubits used by the procedure (dominated by Bell sampling + QEC) is at most
\begin{equation*}
 n + \BigO{\log(\frac{\Gamma_{\mathrm{tot}}}{\eta}+1)}
\end{equation*}

\item \textit{(Classical overhead)} The classical cost of the algorithm (dominated by repeated QEC in structure testing) is:
    \begin{equation*}
    \SoftBigO{\bigl(n + (\Gamma_{\mathrm{tot}}/\eta)^4\bigr)\Lambda_L^2/\eta^2}
    \end{equation*}
\end{enumerate}
\end{theorem}
\begin{proof}
    See Appendix~\ref{sec:diss-struct-hierarchical-learning}. 
\end{proof}

\begin{remark}[Ancilla-reduced dissipator structure learner]
    In Appendix~\ref{sec:appendix-reducing-the-ancilla-cost}, following \cite{hu2025ansatz}, we also briefly describe an ancilla-reduced variant of \Cref{thm:appendix-hier-diss-struct-learning} that improves the total ancilla count from  $n + \BigO{\log(\frac{\Gamma_{\mathrm{tot}}}{\eta}+1)}$ to $ \BigO{\log(\frac{\Gamma_{\mathrm{tot}}}{\eta}+1)}$ at the cost of increased black-box evolution time $t_{\mathrm{tot}} = \SoftBigO{\frac{M_{D,\mathrm{eff}}(\eta)\log(n)}{\eta}}$, where $M_{D,\mathrm{eff}}(\eta)$ is the $\eta$-effective sparsity:
    \begin{equation*}
        M_{D,\mathrm{eff}}(\eta)\coloneq \sum_{P_k\in \DiagDissStruct}\min\left\{1, \frac{a_{kk}}{\eta}\right\} = |\{P_k:a_{kk}\ge\eta\}|+\frac{1}{\eta}\sum_{\substack{P_k\in \DiagDissStruct,\\ a_{kk}\le \eta}}a_{kk}
    \end{equation*}
\end{remark}

This section is organized as follows:
\begin{itemize}
    \item \textbf{Dissipator structure hierarchy \& notation.} In Appendix~\ref{sec:dissipator-hier-learning-notation}, we describe the hierarchical learning procedure and introduce the necessary notation. 
    \item \textbf{Candidate identification via Bell sampling.} In Appendix~\ref{sec:diss-structure-sampling}, we present \Cref{alg:appendix-round-j-diss-struct-sampling} for sampling the strongest unknown dissipator terms via Bell sampling and prove its guarantees in \Cref{thm:appendix-round-j-diss-struct-sampling}. 
    \item \textbf{Candidate testing via syndrome measurements.} In Appendix~\ref{sec:diss-struct-testing}, we present \Cref{alg:appendix-diss-rate-testing} that enables removal of false positives of the dissipator structure identification by testing the decay rates of identified terms. The testing is done via syndrome measurements of random stabilizer codes. 
    \item \textbf{Dissipator structure learning algorithm.} In Appendix~\ref{sec:diss-struct-hierarchical-learning}, we combine the dissipator candidate identification (\Cref{alg:appendix-round-j-diss-struct-sampling}) and dissipator structure testing (\Cref{alg:appendix-diss-rate-testing}) in a hierarchical fashion into \Cref{alg:appendix-diss-struct-learning}, which learns the $\eta$-heavy part of the dissipator structure $\EtaDiagDissStruct$ in $\SoftBigO{1/\eta}$ interaction time. 
    \item \textbf{Deferred proofs.}
    In Appendix~\ref{sec:deferred-diss-struct-sampling-proofs} and Appendix~\ref{sec:deferred-diss-struct-testing-proofs}, we provide the deferred proofs related to dissipator structure identification and testing.
\end{itemize}

\subsection{Hierarchy and notation} \label{sec:dissipator-hier-learning-notation}

The hierarchy is formed by splitting the learning problem according to the following definition:
\begin{definition}[Round-$j$ dissipator set]\label{def:round-j-dissipator-set}
    Given a Lindbladian from Eq.~\eqref{eq:lindblad-eq-diss-struct} and target accuracy $\eta >0$, define the round-$j$ dissipator Pauli set as:
    \begin{equation}
        \DHierSet_j\;\coloneq\; \bigl\{\,P_k\in \DiagDissStruct : 2^{-(j+1)} < a_{kk} \le 2^{-j} \,\bigr\},
  \qquad
  j=0,1,2,\ldots, \lceil \log_2(1/\eta)\rceil-1 .
    \end{equation}
\end{definition}

By construction, learning all $\DHierSet_j$ is sufficient for learning the $\eta$-heavy dissipator structure $\EtaDiagDissStruct$:
\begin{equation*}
    \EtaDiagDissStruct
    \subseteq
    \bigcup_{j=0}^{\lceil \log_2(1/\eta)\rceil-1} \DHierSet_j
    \subseteq
    \HalfEtaDiagDissStruct.
\end{equation*}

Hierarchical learning proceeds in rounds indexed by $j$, where round $j$ targets the terms in $\DHierSet_j$. The algorithm begins by identifying the largest diagonal dissipator terms and then updates the QEC target-removal set $\VCorrect$ (\Cref{lem:physical-twirl-qec-reshaping-generator}) to suppress the terms that have been learned. It subsequently increases the interrogation time and repeats the same procedure on progressively weaker bins. Within each round, the algorithm first samples the strongest remaining unknown dissipator terms at the current scale and then tests the resulting candidates to eliminate false positives.

Across the hierarchical rounds, we keep track of all terms identified so far. Let $\widehat{\mathcal S}_{D,j-1}$ denote the set of Hermitian Paulis known at the beginning of round $j$. For every $j$, we maintain the following invariant:
\begin{equation}
\label{eq:diss-hierarchical-invariant}
   \DHierSet_k \subseteq \widehat{\mathcal{S}}_{D,j-1}
   \subseteq \HalfEtaDiagDissStruct
   \qquad
   \text{for all } k \in \{0,\dots,j-1\}.
\end{equation}

Another relevant quantity is the deterministic residual tail-rate $\Gamma_j$:
\begin{equation}\label{eq:def-gamma-j}
    \Gamma_j \coloneq \sum_{\substack{P_k \in \DiagDissStruct,\\ a_{kk}\le 2^{-j}}}a_{kk}
\end{equation}

\subsection{Candidate identification via Bell sampling}
\label{sec:diss-structure-sampling}

We now describe \Cref{alg:appendix-round-j-diss-struct-sampling}, which identifies a superset of $\DHierSet_j \setminus \widehat{\mathcal{S}}_{D,j-1}$ via Bell sampling. In round $j$, we set $\VCorrect=\widehat{\mathcal S}_{D,j-1}$ and draw a random stabilizer code via \Cref{alg:qec_for_lindbladian_reshaping}, together with its recovery map. For the resulting round-$j$ code, recovery map, and code projector, define the corresponding physical-twirling plus QEC-reshaping map
\begin{equation}
\label{eq:round-j-physical-twirling-QEC-reshaping}
    A_{\tau,D}^{(j)}
    \coloneq
    \RMap_j \circ \frac{1}{4^n}\sum_{P\in \PnHerm}
    \bigl(\mathcal U_P\otimes I_{n'-n}\bigr)\circ
    \bigl(e^{\Lindblad \tau}\otimes I_{n'-n}\bigr)\circ
    \bigl(\mathcal U_P\otimes I_{n'-n}\bigr) \circ \mathcal C_{\Pi_j}.
\end{equation}
This is the round-specific instance of Eq.~\eqref{eq:physical-twirling-QEC-reshaping}. For an intended interrogation time $t$ and an integer $N$, we write
\begin{equation}
\label{eq:round-j-reshaped-dynamics}
   A_D^{(j)}(t,N)
    \coloneq
    \left(A_{t/N,D}^{(j)}\right)^N .
\end{equation}

Conditioned on the success of the random QEC construction, \Cref{lem:physical-twirl-qec-reshaping-generator} and \Cref{rem:full-space-effective-generators} imply that the effective generator of $A_{\tau,D}^{(j)}$ on $\mathcal B(\mathcal H_{n'})$ is purely dissipative and takes the form
\begin{equation}
\label{eq:round-j-Leff-diss-structure}
    \LindbladEff^{(j)}(\rho)
    =
    \sum_{P_k\in \DiagDissStruct\setminus \widehat{\mathcal S}_{D,j-1}}
    a_{kk}
    \Bigl(
        (Q_{P_k}\otimes I_{n'-n})\rho(Q_{P_k}^{\dagger}\otimes I_{n'-n})
        -\rho
    \Bigr),
\end{equation}
where
$Q_P\coloneqq \widetilde R_{r(P)}P\in\mathbb P_n$ is the reshaped Pauli obtained by composing $P$ with the recovery action
specified in Eq.~\eqref{eq:syndrome-correction-unitaries}, with
$R_{r(P)}=\widetilde R_{r(P)}\otimes I_{n'-n}$. Therefore, every unknown term $P_k\in\DHierSet_j\setminus\widehat{\mathcal S}_{D,j-1}$ contributes to the reshaped dynamics through the logical Pauli $Q_{P_k}$. We then use Bell sampling to identify these recovered logical Paulis modulo stabilizers. To convert the sampled logical Paulis back into physical Pauli candidates, the algorithm composes each Bell-sampling outcome with the recovery actions associated with all possible syndromes and retains the resulting probe-supported Paulis. This produces a candidate set $\widehat{\DHierSet}_j$ satisfying $\widehat{\DHierSet}_j \supseteq \DHierSet_j \setminus \widehat{\mathcal S}_{D,j-1}$ with probability at least $1-\delta_j$.

Given $\widehat{\mathcal S}_{D,j-1}$ and a particular drawn code, let
\begin{equation}
\label{eq:gamma-eff-j-dissipator-struct}
    \Gamma_j^{\mathrm{eff}}
    \coloneq
    \sum_{\substack{P_k \in \DiagDissStruct \setminus \widehat{\mathcal S}_{D,j-1}\\ Q_{P_k}\not\propto I_n}}
    a_{kk}
\end{equation}
denote the total effective decay rate of the reshaped dissipator. By Eq.~\eqref{eq:diss-hierarchical-invariant}, all bins above the current scale have already been included in $\widehat{\mathcal S}_{D,j-1}$. Hence every term in $\DiagDissStruct\setminus\widehat{\mathcal S}_{D,j-1}$ has rate at most $2^{-j}$, and therefore $\Gamma_j^{\mathrm{eff}}\le\Gamma_j$.

The interrogation time is chosen as $t_j\coloneq 1/\widehat{\Gamma}_j$, where the estimator $\widehat{\Gamma}_j$ is obtained by applying the pre-calibration routine \Cref{lem:precalibrate-diag-diss-rate} to the reshaped dynamics $A_D^{(j)}$ down to the current bin scale $2^{-(j+1)}$. With probability at least $1-\delta_j/4$, it satisfies
\begin{equation}
\label{eq:gammahat-j-guarantee-dissipator-struct}
    \max\{\Gamma_j^{\mathrm{eff}},2^{-(j+1)}\}
    \le
    \widehat{\Gamma}_j
    \le
    4\max\{\Gamma_j^{\mathrm{eff}},2^{-(j+1)}\}.
\end{equation}
and admits a trivial upper bound $\widehat{\Gamma}_j = \BigO{\Gamma_j^{\mathrm{eff}}+2^{-(j+1)}}\le \BigO{\Gamma_j+2^{-(j+1)}}.$

Notably, because of QEC reshaping, the sampled Paulis might extend beyond the probe qubits via stabilizer products. Therefore, for any Pauli $A\in\mathcal P_{n'}$, we define its probe-supported stabilizer reduction with respect to the round-$j$ code by
\begin{equation}
\label{eq:probe-supported-stabilizer-reduction}
    \operatorname{Probe}_j(A)
    \coloneq
    \left\{P\in\PnHerm:
    \exists \sigma\in S_j \text{ such that }
    \sigma A \propto P\otimes I_{n'-n}
    \right\},
\end{equation}
where $S_j$ is the stabilizer subgroup of the $j$-th round code. If no stabilizer representative of $A$ is supported only on the probe subsystem, then $\operatorname{Probe}_j(A)=\emptyset$. By the $Z$-padding step in \Cref{alg:qec_for_lindbladian_reshaping}, no two distinct probe Paulis are equivalent modulo $S_j$. Hence for every $A\in\mathcal P_{n'}$, we have $\abs{\operatorname{Probe}_j(A)} \le 1.$

When $\widehat{\mathcal S}_{D,j-1}=\emptyset$, the QEC step is trivial: we take
$\mathcal G_j=\emptyset$, $S_j=\{I_n\}$, $\Pi_j=I_n$, $\RMap_j=I$, $m_j=0$, and $n'=n$. Equivalently, the reshaping map is just physical Pauli twirling, without error correction. All bounds below are stated with $(|\widehat{\mathcal S}_{D,j-1}|+1)^2$ so that this initial case is included uniformly.

\begin{algorithm}[H]
    \caption{Dissipator candidate identification subroutine, $j$th round}
    \label{alg:appendix-round-j-diss-struct-sampling}
    \DontPrintSemicolon
    \SetKwInOut{Input}{Inputs}
    \SetKwInOut{Output}{Output}
    \SetKwBlock{Schedule}{Schedule}{end}

    \Input{
      \begin{enumerate}[label=(\arabic*), leftmargin=*, nosep]
         \item Hierarchical round index $j$.
        \item Set $\widehat{\mathcal{S}}_{D,j-1}$ of Hermitian Paulis identified in prior hierarchical rounds satisfying the invariant in eq.\eqref{eq:diss-hierarchical-invariant}.
        \item Upper bound $\Lambda_L \ge \|\Lindblad\|_\diamond$.
        \item Confidence parameter $\delta_j\in(0,1)$.
        \item Access to $n$-qubit evolution $e^{t\Lindblad}$.
        \item Access to $n+\BigO{\log(|\widehat{\mathcal{S}}_{D,j-1}|+1)}$ noiseless ancilla qubits.
      \end{enumerate}
    }

    \Output{
      A set $\widehat{\DHierSet}_j \subseteq \PnHerm$ with
      $|\widehat{\DHierSet}_j| =\SoftBigO{(2^j \Gamma_j+1)(|\widehat{\mathcal{S}}_{D,j-1}|+1)^2}$ such that $\widehat{\DHierSet}_j \supseteq \DHierSet_j \setminus \widehat{\mathcal{S}}_{D,j-1}$ with probability $\ge1-\delta_j$.
    }

    \BlankLine

    \If{$\widehat{\mathcal S}_{D,j-1}=\emptyset$}{
        $\mathcal G_j\gets \emptyset,\quad S_j\gets\{I_n\},\quad \Pi_j\gets I_n,\quad \RMap_j\gets I,\quad n'\gets n$ \tcp*{no error correction needed}
    }
    \Else{
        $\VCorrect \gets \widehat{\mathcal{S}}_{D,j-1}$ \tcp*{QEC target correctable set}
        $(\mathcal G_j,\RMap_j) \gets \texttt{RandomQEC}(\emptyset,\VCorrect,|\widehat{\mathcal{S}}_{D,j-1}|^2,\delta_j/4,2)$ \tcp*{Alg~\ref{alg:qec_for_lindbladian_reshaping} stab. generator \& recovery ch.}
        $S_j\gets \langle \mathcal G_j\rangle$ \tcp*{corresponding stabilizer group}
        $\Pi_j \gets \prod_{G\in\mathcal G_j}(1+G)/2$\;
        $n' \gets n + \frac{5}{2}\left\lceil \log_2(2|\widehat{\mathcal{S}}_{D,j-1}|^2)\right\rceil$ \tcp*{left register size accounting for QEC ancillas}
    }
    $m_j\gets |\mathcal G_j|$ \tcp*{cardinality of the stabilizer generator  $m_j=\BigO{\log( |\widehat{\mathcal{S}}_{D,j-1}|^2+1)}$}
    Define access to $A_D^{(j)}(t,N)$ using Eq.~\eqref{eq:round-j-reshaped-dynamics}\;
    $\widehat{\Gamma}_j \gets \texttt{EstimateGamma}(\max\{\Lambda_L,2^{-(j+1)}\},2^{-(j+1)},\delta_j/4,A_D^{(j)})$ \tcp*{estimate from \Cref{lem:precalibrate-diag-diss-rate}}
    $N_{\mathrm{reshape}} \gets 10e\cdot 2^j\Lambda_L^2/\widehat{\Gamma}_j$ \tcp*{\# of reshape cycles per sample}
    $t_j \gets 1/\widehat{\Gamma}_j, \quad  \tau \gets t_j/N_{\mathrm{reshape}}$ \tcp*{interrogation time and reshape step}
    $N_{\mathrm{meas}} \gets 4e\cdot\widehat{\Gamma}_j\, 2^j\log(8\widehat{\Gamma}_j2^j/\delta_j)$ \tcp*{\# of Bell-basis measurements}
      
    $\widehat{\DHierSet}_j \gets \bigcup_{s\in\{0,1\}^{m_j}}\operatorname{Probe}_j(R_s)$ \tcp*{identity-coset candidates, $R_s$ -- recovery operators of $\RMap_j$}

    \For{$i \gets 1$ \KwTo $N_{\mathrm{meas}}$}{
        Prepare $\ket{\Phi_0} \gets \ket{\Phi_+}^{\otimes n'}$ on $L\otimes R$ \tcp*{split into left and right registers}
        Project to code corner algebra: $\rho_{\mathrm{prep}} \propto (\Pi_j\otimes I_{n'})\ket{\Phi_0}\bra{\Phi_0}(\Pi_j\otimes I_{n'})$ \tcp*{$\SoftBigO{|\widehat{\mathcal{S}}_{D,j-1}|^2}$ attempts}
        \For{$k \gets 1$ \KwTo $N_{\mathrm{reshape}}$}{
            Randomly sample $P \sim \PnHerm$ \tcp*{for Physical Pauli twirling}
            Apply $P\otimes I_{n'-n}$ to the left register \;
            Evolve the left register qubits by $e^{\tau \Lindblad} \otimes I_{n'-n}$ \;
            Apply $P\otimes I_{n'-n}$ to the left register \;
            Apply $\RMap_j$ to the left register \tcp*{error detection + correction,  trivial if $\widehat{\mathcal S}_{D,j-1}=\emptyset$}
        }
        Measure in the Bell basis and let $Q\in\mathcal P_{n'}$ be the Pauli outcome\;
        $\widehat{\DHierSet}_j \gets \widehat{\DHierSet}_j \cup \bigcup_{s\in\{0,1\}^{m_j}}\operatorname{Probe}_j(R_s Q)$ \tcp*{compose sampled logical w all recovery operators of $\RMap_j$}
    }
    \Return $\widehat{\DHierSet}_j\setminus\{I_n\}$
\end{algorithm}

The key ingredient of this algorithm is the following lemma, which we invoke on the reshaped dynamics:

\begin{lemma}[Bell sampling of effective diagonal dissipator]
\label{lem:bell-sampling-of-eff-diss-struct}
Let $\LindbladEff$ be an $n'$-qubit Lindbladian with purely diagonal dissipator part
\begin{equation*}
    \LindbladEff(\rho)
    =
    \sum_{Q_k \in \mathcal S_\mathrm{eff} \,\subseteq\, \mathcal P_{n'} \setminus \{I\}}
    \alpha_{kk}
    \left(Q_k \rho Q_k - \rho\right)
\end{equation*}
where $\mathcal S_\mathrm{eff}$ is a set of distinct Hermitian Paulis and $\alpha_{kk}\ge 0$ for all $Q_k\in\mathcal S_{\mathrm{eff}}$. Define the total effective dissipative rate $\Gamma_{\mathrm{eff}} \coloneq \sum_{Q_k\in\mathcal S_{\mathrm{eff}}} \alpha_{kk}.$ Fix a target Pauli $Q_r\in\mathcal S_\mathrm{eff}$. Let $C$ be a stabilizer code on $n'$ qubits with code projector $\Pi$ and stabilizer subgroup $S$. Let $G_{\mathrm{eff}} \coloneq \langle \mathcal S_{\mathrm{eff}} \rangle \subseteq \mathbb P_{n'}$ denote the subgroup generated by $\mathcal S_{\mathrm{eff}}$. Assume that all Paulis
$Q \in \mathcal S_{\mathrm{eff}}$ have zero syndrome $r(Q)=0$ with respect to $C$, and that for every $R \in G_{\mathrm{eff}}$, $R\propto s\in S \implies s = I_{n'}$. Equivalently, no nontrivial product of elements of $\mathcal S_{\mathrm{eff}}$ is proportional to a stabilizer. Let $\rho_0$ denote a $2n'$-qubit maximally entangled state $\rho_0 \coloneq \ket{\Phi_0}\bra{\Phi_0}$ where $\ket{\Phi_0} \coloneq \ket{\Phi_+}^{\otimes n'}$ and $\ket{\Phi_+} = (\ket{00} + \ket{11})/\sqrt{2}.$
Let $\widetilde \rho$ denote the state immediately before the Bell basis measurement
\begin{equation*}
    \widetilde \rho \coloneq (e^{\LindbladEff t}\otimes I_{n'}) \left(\frac{\Pi\otimes I_{n'} \,\rho_0 \,\Pi\otimes I_{n'}}{\tr(\Pi\otimes I_{n'} \,\rho_0)}\right).
\end{equation*}
For any Pauli $Q\in \mathbb P_{n'}$, let $O_Q \coloneq (Q\otimes I_{n'})\ket{\Phi_0}\bra{\Phi_0}(Q^\dagger\otimes I_{n'})$ denote the corresponding Bell-measurement projector. Then, for any $t\ge 0$, a single run of Bell sampling on $\widetilde \rho$ satisfies
\begin{align}
    \Pr(\text{sample $Q_r$ modulo stabilizers})
    &\coloneq
    \tr\left(\sum_{\sigma\in S} O_{\sigma Q_r}\,\widetilde{\rho}\right)
    \ge
    e^{-\Gamma_{\mathrm{eff}}t}\alpha_{rr}t.
\end{align}
\end{lemma}
\begin{proof}
    See Appendix~\ref{sec:deferred-diss-struct-sampling-proofs}.
\end{proof}

\begin{theorem}[Round-$j$ dissipator structure sampling]
\label{thm:appendix-round-j-diss-struct-sampling}
Let $\Lindblad$ be an $n$-qubit Lindbladian as in Eq.~\eqref{eq:lindblad-eq-diss-struct} with $\|\Lindblad\|_\diamond \le \Lambda_L$. Assume black-box access to $e^{t\Lindblad}$. Fix a hierarchical round $j$ and a failure probability $\delta_j\in(0,1)$. Let $\EtaDiagDissStruct \subseteq \DiagDissStruct$ denote the $\eta$-heavy part of the dissipator structure defined in Eq.~\eqref{eq:def-eta-heavy-diss-struct}. Let $\widehat{\mathcal{S}}_{D,j-1}$ denote the set of Hermitian Paulis known at the beginning of the $j$-th hierarchical round, and assume it satisfies the invariant in Eq.~\eqref{eq:diss-hierarchical-invariant}. Let $\Gamma_j$ denote the round-$j$ residual rate from Eq.~\eqref{eq:def-gamma-j} and let $\widehat{\Gamma}_j$ denote the output of the pre-calibration subroutine from \Cref{lem:precalibrate-diag-diss-rate}.

Then running \Cref{alg:appendix-round-j-diss-struct-sampling} outputs a candidate set $\widehat{\DHierSet}_j$ and satisfies:

\begin{enumerate}[leftmargin=*, label=\arabic*, nosep]
\item \textit{(Coverage and size)} With probability $\ge 1-\delta_j$,
\begin{equation*}
\widehat{\DHierSet}_j \supseteq \DHierSet_j\setminus\widehat{\mathcal{S}}_{D,j-1},
\qquad
|\widehat{\DHierSet}_j|
=
\BigO{ 2^j\widehat{\Gamma}_j \bigl(|\widehat{\mathcal{S}}_{D,j-1}|+1\bigr)^2 \log(2^j\widehat{\Gamma}_j/\delta_j)}
=
\SoftBigO{(2^j\Gamma_j+1)\bigl(|\widehat{\mathcal{S}}_{D,j-1}|+1\bigr)^2}.
\end{equation*}

\item \textit{(Total evolution time)}
The total evolution time under $e^{t\Lindblad}$, including pre-calibration, is
\begin{equation*}
t_{\mathrm{tot}} = \BigO{2^j\log\left(\frac{2^j\Gamma_j+1}{\delta_j}\right) + 2^j\log\left(\frac{\log(2^j\Lambda_L+4)}{\delta_j}\right)}.
\end{equation*}

\item \textit{(Minimum evolution time)} It only ever applies $e^{\Lindblad t}$ for $t \ge t_{\mathrm{min}}$, where
\begin{equation*}
t_{\mathrm{min}} = \Omega\Bigl(\frac{1}{\Lambda_L^2 2^j}\Bigr).
\end{equation*}

\item \textit{(Number of experiments)}
The number of Bell-sampling experiments, including pre-calibration, is
\begin{equation*}
N_{\exp} = \BigO{(2^j\Gamma_j+1)\log\left(\frac{2^j\Gamma_j+1}{\delta_j}\right) + \log(2^j\Lambda_L+4) \log\left(\frac{\log(2^j\Lambda_L+4)}{\delta_j}\right) }.
\end{equation*}
The number of raw state-preparation attempts is larger by a factor
$\SoftBigO{|\widehat{\mathcal{S}}_{D,j-1}|^2}$ due to the probabilistic projection to the code space before reshaping. Including pre-calibration, the total number of interleaved QEC rounds is $\SoftBigO{\Lambda_L^24^j}$.

\item \textit{(Ancilla cost)}
The procedure uses $n + 5\left\lceil\log_2\bigl(2(|\widehat{\mathcal{S}}_{D,j-1}|+1)^2\bigr)\right\rceil$ ancillary qubits in addition to the $n$ probe qubits.

\item \textit{(Classical overhead)} The classical cost of the algorithm is dominated by repeated QEC and the syndrome-lift post-processing of Bell outcomes, and is bounded by
\begin{equation*}
\SoftBigO{\bigl(n+(|\widehat{\mathcal{S}}_{D,j-1}|+1)^2\bigr)\Lambda_L^24^j}.
\end{equation*}
\end{enumerate}
\end{theorem}

\begin{proof}
    Most of the proof is for item $(1)$, the coverage of the algorithm. We then verify the resource bounds in items $(2)$--$(4)$, including the pre-calibration overhead. Items $(5)$--$(6)$ are proved at the end. 

    Let us begin with item $(1)$. To suppress the already known Lindbladian terms we use the physical twirling + partial QEC reshaping map from
    Eq.~\eqref{eq:physical-twirling-QEC-reshaping}:
    \begin{equation*}
        A_{\tau, D}
        =
        \RMap_j \circ \frac{1}{4^n}\sum_{P\in \PnHerm}
        \bigl(\mathcal U_P\otimes I_{n'-n}\bigr)\circ
        \bigl(e^{\Lindblad \tau}\otimes I_{n'-n}\bigr)\circ
        \bigl(\mathcal U_P\otimes I_{n'-n}\bigr) \circ \mathcal C_{\Pi_j} .
    \end{equation*}
    If $\widehat{\mathcal S}_{D,j-1}=\emptyset$, we use the trivial code:
    $\mathcal G_j=\emptyset$, $S_j=\{I_n\}$, $\Pi_j=I_n$, $\RMap_j=I$, and $n'=n$; in this case let
    $E_{\mathrm{tar}}$ denote the sure event. Otherwise, $\RMap_j$ is the recovery map of the random
    stabilizer code draw from \Cref{alg:qec_for_lindbladian_reshaping} with inputs
    $\VCorrect=\widehat{\mathcal S}_{D,j-1}$, $c = 2$ and
    $M_{\mathrm{det}}=|\widehat{\mathcal S}_{D,j-1}|^2$, and $E_{\mathrm{tar}}$ denotes the event that this
    code corrects all Paulis in $\VCorrect$. In both cases,
    \begin{equation}\label{eq:diss-struct-sampling-stab-size}
        |S_j| \le 4\bigl(|\widehat{\mathcal S}_{D,j-1}|+1\bigr)^2 .
    \end{equation}
    
    Conditioned on $E_{\mathrm{tar}}$, by
    \Cref{lem:physical-twirl-qec-reshaping-generator} and
    \Cref{rem:full-space-effective-generators}, the effective generator
    of this reshaping on $\rho \in \mathcal B(\mathcal H_{n'})$ is
    \begin{equation} \label{eq:effective-generator-diss-struct-identification}
        \LindbladEff^{(j)}(\rho)
        =
        \sum_{P_k\in \DiagDissStruct\setminus \widehat{\mathcal S}_{D,j-1}}
        a_{kk}
        \Bigl(
            (Q_{P_k}\otimes I_{n'-n})\rho(Q_{P_k}^{\dagger}\otimes I_{n'-n})
            -\rho
        \Bigr),
    \end{equation}
    where
    $Q_P\coloneqq \widetilde R_{r(P)}P\in\mathbb P_n$
    is the Pauli obtained by composing $P$ with the recovery action
    $\widetilde R_{r(P)}$ specified in
    Eq.~\eqref{eq:syndrome-correction-unitaries}, with
    $R_{r(P)}=\widetilde R_{r(P)}\otimes I_{n'-n}$. Thus, every unknown term $P_k\in\DHierSet_j\setminus\widehat{\mathcal S}_{D,j-1}$ contributes to the reshaped dynamics through the logical Pauli $Q_{P_k}$.     When several physical Paulis induce the same nonidentity recovered Pauli, we aggregate their rates in the effective diagonal dissipator. Thus, the coefficient of an effective Pauli $Q$ is the sum of all $a_{kk}$ with $Q_{P_k}\propto Q$. In particular, for any target $P_r$ with $Q_{P_r}\not\propto I_n$, the aggregate rate of $Q_{P_r}$ is at least $a_{rr}$. Terms for which $Q_{P_k}\propto I_n$ do not contribute to the effective rate $\Gamma_j^{\mathrm{eff}}$ and are not detected through nonidentity Bell outcomes. These terms are nevertheless accounted for by the initialization
    \begin{equation*}
        \widehat{\DHierSet}_j
        \gets
        \bigcup_{s\in\{0,1\}^{m_j}}\operatorname{Probe}_j(R_s).
    \end{equation*}
    Indeed, if $Q_{P_k}\propto I_n$, then up to stabilizers, $P_k\otimes I_{n'-n}$ is equivalent to $R_{r(P_k)}$, and therefore $P_k\in \operatorname{Probe}_j(R_{r(P_k)})$. Thus every target Pauli reshaped into the identity coset is included in the candidate set before the Bell-sampling loop begins.
    
    Before proceeding to candidate identification via Bell sampling, we first record the Lindbladian reshaping error. From \Cref{lem:N-step-reshaping-error} with
    $\tau\coloneq t/N_{\mathrm{reshape}}$ and using $\Gamma_j$ from Eq.~\eqref{eq:def-gamma-j},
    \begin{equation}\label{eq:diss-struct-sampling-reshaping-error}
        \|A_{\tau, D}^{N_{\mathrm{reshape}}}
        -
        e^{\LindbladEff^{(j)} t}\circ \mathcal C_{\Pi_j}\|_\diamond
        \le
        \frac{t^2(\|\Lindblad\|_\diamond^2+\|\LindbladEff^{(j)}\|_\diamond^2)}
        {2N_{\mathrm{reshape}}}
        \le
        \frac{t^2(\Lambda_L^2 + 4\Gamma_j^2)}{2N_{\mathrm{reshape}}} \le \frac{5t^2\Lambda_L^2}{2N_{\mathrm{reshape}}}.
    \end{equation}
    where the last inequality follows from $\Gamma_j \le \|\Lindblad_{\mathrm{diag}}\|_\diamond \le \|\Lindblad\|_\diamond$, with $\Lindblad_{\mathrm{diag}}$ denoting the Pauli-twirled version of $\Lindblad$. Thus, going forward in the analysis, we calculate the measurement statistics as
    if we had access to $e^{\LindbladEff^{(j)} t}$ and then account for the reshaping
    error separately.

    Now, suppose that both the random QEC construction (event $E_\mathrm{tar}$) and the $\widehat{\Gamma}_j$ pre-calibration from \Cref{lem:precalibrate-diag-diss-rate} have succeeded. Let $\Pi_j$ denote the corner algebra projection of the code. We are interested in the probability to sample the logical $\{Q_{P_k}: a_{kk}\ge 2^{-(j+1)}\}$ of $\LindbladEff^{(j)}$ using Bell sampling. For that, we use \Cref{lem:bell-sampling-of-eff-diss-struct}, but first we argue why the assumptions of this lemma are satisfied in our setting. Let $\rho_0$ denote a $2n'$-qubit maximally entangled state $\rho_0 \coloneq \ket{\Phi_0}\bra{\Phi_0}$ where $\ket{\Phi_0} \coloneq \ket{\Phi_+}^{\otimes n' }$ and $\ket{\Phi_+} = (\ket{00} + \ket{11})/\sqrt{2}.$
    To prepare the state:
    \begin{equation*}
        \widetilde \rho \coloneq (e^{\LindbladEff^{(j)} t}\otimes I_{n'}) \left(\frac{\Pi_j\otimes I_{n'} \,\rho_0 \,\Pi_j\otimes I_{n'}}{\tr(\Pi_j \otimes I_{n'} \,\rho_0)}\right).
    \end{equation*}
    we first do a projection of the $2n'$-qubit maximally entangled state into the corner algebra (Eq.~\eqref{eq:corner-algebra-def}). Operationally, this could be done via measuring the stabilizers and discarding if the measured syndrome is not zero. The probability of successful projection is 
    \begin{equation}\label{eq:diss-struct-sampling-prob-projection}
        \tr(\Pi_j \otimes I_{n'} \,\rho_0) = \frac{1}{|S_j|} \ge 1/(4(|\widehat{\mathcal S}_{D,j-1}|+1)^2)
    \end{equation} where we used Eq.~\eqref{eq:diss-struct-sampling-stab-size}. 
    After projection, we evolve the state under $e^{\LindbladEff^{(j)} t}$ (using Lindbladian reshaping). The remaining assumption of \Cref{lem:bell-sampling-of-eff-diss-struct} is that the group generated by the $\LindbladEff^{(j)}$ Pauli terms has no non-identity stabilizer elements. In the trivial-code case this is immediate since $S_j=\{I_n\}$. In the nontrivial-code case, this is guaranteed by the $Z$-padding of \Cref{alg:qec_for_lindbladian_reshaping}: each non-identity stabilizer $\sigma \in S_j\setminus\{I\}$ acts non-trivially on the ancilla qubits, while the $n$-qubit Lindbladian terms act trivially on ancillas. Therefore, we can use \Cref{lem:bell-sampling-of-eff-diss-struct}. Let $P_r \in \DHierSet_j\setminus \widehat{\mathcal{S}}_{D,j-1}$ with $Q_{P_r}\not\propto I_n$ denote a particular Pauli we wish to identify, where $Q_{P_r}$ is the corresponding logical Pauli from Eq.~\eqref{eq:effective-generator-diss-struct-identification}. By \Cref{lem:bell-sampling-of-eff-diss-struct} and using $\widehat{\Gamma}_j \ge \Gamma_j^{\mathrm{eff}}$, we have
    \begin{equation*}
        \Pr_{\text{ideal}}(\text{sample $Q_{P_r}$ modulo stabilizers}) \ge e^{-\Gamma_j^{\mathrm{eff}}t}\alpha_{rr}t \ge e^{-\widehat{\Gamma}_j t}2^{-(j+1)}t 
    \end{equation*}
    accounting for the reshaping diamond norm error from Eq.~\eqref{eq:diss-struct-sampling-reshaping-error} and using $t = 1/\widehat{\Gamma}_j$, 
    \begin{equation} \label{eq:diss-struct-sample_pauli-modulo-stabilizers}
        \Pr(\text{sample $Q_{P_r}$ modulo stabilizers})  \ge e^{-\widehat{\Gamma}_j t}2^{-(j+1)}t - \frac{5t^2\Lambda_L^2}{2N_{\mathrm{reshape}}} \ge \frac{1}{2e\,2^{j}\,\widehat{\Gamma}_j} - \frac{5\Lambda_L^2}{2\widehat{\Gamma}_j^2N_{\mathrm{reshape}}}
    \end{equation}
    By setting $N_{\mathrm{reshape}} \ge 10e\cdot 2^j\Lambda_L^2/\widehat{\Gamma}_j$, we get
    \begin{equation}\label{eq:diss-struct-sampling-prob-bell-sampling}
       \Pr(\text{sample $Q_{P_r}$ modulo stabilizers}) \ge \frac{1}{4e\,2^{j}\,\widehat{\Gamma}_j}.
    \end{equation}

    We now have the necessary ingredients to guarantee the success of the algorithm by enforcing the following four events:
    \begin{enumerate}[label=\Alph*]
        \item – the random code suppresses all Paulis from $\VCorrect$ (event $E_{\mathrm{tar}}$) $\rightarrow$ this is sure in the trivial-code case ($\VCorrect = \emptyset$) and succeeds with probability $\ge 1-\delta_j/4$ otherwise by \Cref{alg:qec_for_lindbladian_reshaping};
        \item – the precalibration $\widehat{\Gamma}_j$ estimate satisfies Eq.~\eqref{eq:precalibrate-diag-rate-guarantee}  $\rightarrow$ succeeds w. prob $\ge 1-\delta_j/4$ by \Cref{lem:precalibrate-diag-diss-rate};
        \item – all $N_\mathrm{meas}$ experiments succeed at preparing the code-projected Bell states $\rightarrow$ Eq.~\eqref{eq:diss-struct-sampling-event-C};
        \item – each Pauli $P \in \DHierSet_j\setminus\widehat{\mathcal{S}}_{D,j-1}$ is covered at least once (via initialization or coupon collector) $\rightarrow$ Eq.~\eqref{eq:diss-struct-sampling-event-D}.
    \end{enumerate}
    
    First, let us lower bound the probability that each Pauli $P \in \DHierSet_j\setminus\widehat{\mathcal{S}}_{D,j-1}$ is covered at least once (event $D$). Fix a single $P_r\in \DHierSet_j\setminus\widehat{\mathcal{S}}_{D,j-1}$. By Eq.~\eqref{eq:diss-struct-sampling-prob-bell-sampling}, repeating Bell sampling $N_{\mathrm{meas}} \gets 4e\,2^j\,\widehat{\Gamma}_j\, \log(8\widehat{\Gamma}_j2^j/\delta_j)$ times guarantees:
    \begin{equation*}
        \Pr(\text{$Q_{P_r}$ is not sampled once}) =  (1-\Pr(\text{sample $Q_{P_r}$}))^{N_{\mathrm{meas}}} \le  \frac{\delta_j}{8\widehat{\Gamma}_j2^j}
    \end{equation*}
    To recover the probe-only $P_r \in \PnHerm$ from the sampled $Q_{P_r}$, for every sampled logical $Q\in \mathbb{P}_n$, we compose $Q$ with all possible recovery operators $R_s$ of $\RMap_j$ and add their probe representations $\bigcup_{s\in\{0,1\}^{m_j}}\operatorname{Probe}_j(R_s Q)$ to the candidate set. For every sampled logical, this introduces a candidate-size overhead of $|S_j| \le 4(|\widehat{\mathcal S}_{D,j-1}|+1)^2$. Using a union bound over elements of $\DHierSet_j\setminus\widehat{\mathcal{S}}_{D,j-1}$ that induce nonidentity recovered Paulis, of which there are at most $\widehat{\Gamma}_j/2^{-(j+1)}$, we get
    \begin{equation*} 
        \Pr[\text{$\ge1$  Pauli from $\DHierSet_j\setminus\widehat{\mathcal{S}}_{D,j-1}$ not covered}] \le \frac{\widehat{\Gamma}_j}{2^{-(j+1)}}\Pr(\text{given Pauli not sampled once}) \le \frac{\delta_j}{4}
    \end{equation*}
    Hence,
    \begin{equation}\label{eq:diss-struct-sampling-event-D}
        \Pr(D|A\wedge B \wedge C) = 1 - \Pr[\text{$\ge1$  Pauli from $\DHierSet_j\setminus\widehat{\mathcal{S}}_{D,j-1}$ not covered}] \ge 1-\delta_j/4
    \end{equation}

    Combining the identity-coset initialization with the Bell sampling outcomes, the total candidate size $|\widehat{\DHierSet}_j|$ is upper bounded by $(N_\mathrm{meas}+1) |S_j|$ and hence:
    \begin{equation*}
        |\widehat{\DHierSet}_j| \le (N_\mathrm{meas}+1) |S_j| =   \BigO{2^j\widehat{\Gamma}_j  (|\widehat{\mathcal{S}}_{D,j-1}|+1)^2\log(2^j\widehat{\Gamma}_j/\delta_j)}
    \end{equation*}
    
    The event $A$ is guaranteed to succeed with probability at least $1-\delta_j/4$ by running the random code construction \Cref{alg:qec_for_lindbladian_reshaping} with probability of failure $\delta_j/4$. Similarly, the event $B$ is guaranteed to succeed with probability at least $1-\delta_j/4$ by running the pre-calibration from \Cref{lem:precalibrate-diag-diss-rate} and \Cref{rem:precalibration-reshaped-implementation} with probability of failure $\delta_j/4$.
    
    To guarantee that the code-space-projected state preparation succeeds for each of the $N_{\mathrm{meas}}$ experiments (event $C$), we distinguish the trivial and nontrivial code cases. If $\widehat{\mathcal S}_{D,j-1}=\emptyset$, then $\Pi_j=I_n$ and event $C$ is sure. Otherwise, we attempt the preparation + projection $N_{\mathrm{prep}} = 4|\widehat{\mathcal{S}}_{D,j-1}|^2\ln(4N_{\mathrm{meas}}/\delta_j)$ times per experiment. Then, using a union bound over experiments and Eq.~\eqref{eq:diss-struct-sampling-prob-projection}, we get
    \begin{equation}\label{eq:diss-struct-sampling-event-C}
        \Pr[\neg C]
        \le
        N_\mathrm{meas}
        \left(1- \frac{1}{4|\widehat{\mathcal{S}}_{D,j-1}|^2}\right)^{N_{\mathrm{prep}}}
        \le \frac{\delta_j}{4}.
    \end{equation}
    
    Combining the $\delta_j/4$ failure probabilities of events $A$ and $B$ with Eq.~\eqref{eq:diss-struct-sampling-event-C} and Eq.~\eqref{eq:diss-struct-sampling-event-D}, we get:
    \begin{equation*}
        \Pr[\text{success}] \ge \Pr(A \wedge B \wedge C\wedge D)  \ge 1- \Pr(\neg A) -\Pr(\neg B) - \Pr(\neg C) - \Pr(\neg D|A \wedge B\wedge C) \ge 1-4\,\frac{\delta_j}{4} = 1-\delta_j
    \end{equation*}
    where we used the conditional probability and union bound. This concludes the proof of item $(1)$.

    We next prove item $(2)$. In the Bell-sampling stage, each experiment uses interrogation time $t_j=1/\widehat{\Gamma}_j$, and the number of Bell-sampling experiments is $N_{\mathrm{meas}} = \BigO{2^j\widehat{\Gamma}_j\log\left(2^j\widehat{\Gamma}_j/\delta_j\right)}$. On the pre-calibration success event, $\widehat{\Gamma}_j=\BigO{\Gamma_j+2^{-(j+1)}}$, and hence $2^j\widehat{\Gamma}_j=\BigO{2^j\Gamma_j+1}$. Therefore the Bell-sampling stage uses total evolution time
    \begin{equation*}
        N_{\mathrm{meas}}t_j
        =
        \BigO{
        2^j\log\left(\frac{2^j\Gamma_j+1}{\delta_j}\right)
        }.
    \end{equation*}
    The pre-calibration is invoked with $\gamma_{\min}=2^{-(j+1)}$ and $\Lambda_D\le \max\{\Lambda_L,2^{-(j+1)}\}$, so by \Cref{lem:precalibrate-diag-diss-rate} it contributes
    \begin{equation*}
        \BigO{
        2^j\log\left(\frac{\log(2^j\Lambda_L+4)}{\delta_j}\right)
        }
    \end{equation*}
    additional black-box evolution time. This proves item $(2)$.
    
    For item $(3)$, the Bell-sampling stage applies $e^{\tau\Lindblad}$ with $\tau = t_j/N_{\mathrm{reshape}} = \Omega\left(1/(\Lambda_L^2 2^j)\right)$.
    By \Cref{rem:precalibration-reshaped-implementation}, the pre-calibration stage has the same minimum evolution time scaling. Thus item $(3)$ follows.
    
    For item $(4)$, the Bell-sampling stage uses
    \begin{equation*}
        N_{\mathrm{meas}}
        =
        \BigO{
        (2^j\Gamma_j+1)
        \log\left(\frac{2^j\Gamma_j+1}{\delta_j}\right)
        }
    \end{equation*}
    experiments on the pre-calibration success event. By \Cref{lem:precalibrate-diag-diss-rate}, pre-calibration contributes
    \begin{equation*}
        \BigO{
        \log(2^j\Lambda_L+4)
        \log\left(\frac{\log(2^j\Lambda_L+4)}{\delta_j}\right)
        }
    \end{equation*}
    additional Bell-sampling experiments. The raw state-preparation attempts are larger by the stated projection overhead. Finally, the Bell-sampling stage uses $N_{\mathrm{meas}}N_{\mathrm{reshape}} = \SoftBigO{\Lambda_L^2 4^j}$
    interleaved QEC rounds, with the pre-calibration stage contributing the same order. This proves item $(4)$.

    We now prove item $(5)$. In the trivial-code case, no QEC ancillas are needed. In the nontrivial-code case, by \Cref{thm:random-qec-for-reshaping} with $c=2$ and $M_{\mathrm{det}}=|\widehat{\mathcal{S}}_{D,j-1}|^2$, the QEC construction uses at most $5/2\lceil\log_2(2|\widehat{\mathcal{S}}_{D,j-1}|^2)\rceil$ ancillas for both stabilizers and syndrome extraction. Forming Bell pairs uses at most extra $n + 5/2\lceil\log_2(2(|\widehat{\mathcal{S}}_{D,j-1}|+1)^2)\rceil$ ancillas. Thus, uniformly, the total number is at most
    \begin{equation*}
        n + 5\left\lceil\log_2\bigl(2(|\widehat{\mathcal{S}}_{D,j-1}|+1)^2\bigr)\right\rceil .
    \end{equation*}

    The classical overhead (item $(6)$) is dominated by repeated syndrome extraction + correction and is equal to:
    \begin{equation*}
    \text{Classical cost}
    =
    \BigO{(|S_j| + n)N_{\mathrm{meas}} N_{\mathrm{reshape}}}
    =
    \SoftBigO{\bigl(n + (|\widehat{\mathcal{S}}_{D,j-1}|+1)^2\bigr)\Lambda_L^24^j}.
\end{equation*}
    
    This concludes the proof of the theorem. 
\end{proof}

\subsection{Candidate testing via syndrome measurements}
\label{sec:diss-struct-testing}

After identifying a superset $\widehat{\DHierSet}_j \supseteq \DHierSet_j \setminus \widehat{\mathcal{S}}_{D,j-1}$, the next step is to eliminate false positives in order to construct $\widehat{\mathcal{S}}_{D,j}\subseteq \HalfEtaDiagDissStruct$. In this section, we present \Cref{alg:appendix-diss-rate-testing}, which performs batched tolerant thresholding over all candidates in $\widehat{\DHierSet}_j$, given an upper bound on the total diagonal decay rate
\begin{equation}\label{eq:total-decay-rate-diss-testing}
    \Gamma_{\mathrm{tot}}
    \coloneq
    \sum_{P_k \in \DiagDissStruct} a_{kk}.
\end{equation}

\begin{problem}[Batched tolerant testing of diagonal dissipator rates]
\label{prob:batched-diag-rate-tolerant-test}
Fix a candidate set $\mathcal Z\subseteq\PnHerm\setminus\{I\}$ and parameters $\eta>0$ and $\delta\in(0,1)$. Given black-box access to short-time channels $e^{t\Lindblad}$, design a procedure that outputs a subset $\mathcal Z_{\mathrm{acc}}\subseteq\mathcal Z$ such that, with probability at least $1-\delta$,
\begin{equation*}
    \{P_r\in\mathcal Z:a_{rr}\ge\eta\}
    \subseteq
    \mathcal Z_{\mathrm{acc}}
    \subseteq
    \{P_r\in\mathcal Z:a_{rr}>\eta/2\}.
\end{equation*}
For candidates with $a_{rr}\in(\eta/2,\eta)$, the procedure may either accept or reject.
\end{problem}

At a high level, our approach uses stabilizer QEC to convert Pauli jumps into syndrome events. For a candidate Pauli $P_r$, the syndrome frequency of $r(P_r)$ estimates the total diagonal dissipative rate of all Paulis that share this syndrome. Instead of requiring $P_r$ to have a unique syndrome among all nonzero dissipator terms, we draw random stabilizer codes with enough checks so that, with high probability, every candidate in $\widehat{\DHierSet}_j$ has colliding syndrome mass small compared to $\eta$. For each code, we run repeated QEC cycles and test all candidates in parallel from the resulting syndrome histogram. This yields a batched tolerant test that accepts all candidates with $a_{rr}\ge\eta$ and rejects all candidates with $a_{rr}\le\eta/2$ with high probability.

Since our goal is to learn from the syndrome measurements, we first simplify the dynamics by resorting to the physical twirling  map \cite{emerson2007symmetrized}:
\begin{equation} \label{eq:physical-twirling-one-step-map}
    \widetilde{A}_{\tau}
    \coloneq \frac{1}{4^n}\sum_{P\in \PnHerm}
    \bigl(\mathcal U_P\otimes I_{n'-n}\bigr)\circ
    \bigl(e^{\Lindblad \tau}\otimes I_{n'-n}\bigr)\circ
    \bigl(\mathcal U_P\otimes I_{n'-n}\bigr).
\end{equation}

Crucially, the effective generator of $\widetilde{A}_{\tau}$ only retains the diagonal dissipator terms: 
    \begin{equation}\label{eq:lindblad-diag}
        \LindbladDiag(X) \coloneq \frac{1}{4^n}\sum_{P\in \PnHerm}
            \bigl(\mathcal U_P\otimes I_{n'-n}\bigr)\circ
            \bigl(\Lindblad\otimes I_{n'-n}\bigr)\circ
            \bigl(\mathcal U_P\otimes I_{n'-n}\bigr) (X)= \sum_{P_k\in \DiagDissStruct} a_{kk}\Bigl(
                P_k\otimes I_{n'-n} X P_k\otimes I_{n'-n}   - X\Bigr)
    \end{equation}

\begin{lemma}[One-step twirling error]
\label{lem:one-step-twirling-error}
Let $\Lindblad$ be an $n$-qubit GKSL generator as in Eq.~\eqref{eq:lindblad-eq-diss-struct} with $\|\Lindblad\|_\diamond\le \Lambda_L$ and $\Gamma_{\mathrm{tot}} \coloneq \sum_{P_k \in \DiagDissStruct} a_{kk}.$ Let $\widetilde{A}_{\tau}: \mathcal B(\mathcal H_{n'}) \to \mathcal B(\mathcal H_{n'})$ be the physical twirling map from Eq.~\eqref{eq:physical-twirling-one-step-map}, and let $\LindbladDiag$ be the corresponding effective generator from Eq.~\eqref{eq:lindblad-diag}. Then
\begin{equation}
\label{eq:one-step-twirling-error}
    \|\widetilde{A}_{\tau} - e^{\LindbladDiag \tau} \|_\diamond
    \le
    \frac{(\|\Lindblad\|_\diamond^2 + \|\LindbladDiag\|_\diamond^2)\tau^2}{2}
    \le
    \frac{(\Lambda_L^2 + 4\Gamma_{\mathrm{tot}}^2)\tau^2}{2}
    \le
    \frac{5\Lambda_L^2\tau^2}{2}.
\end{equation}
\end{lemma}

\begin{proof}
The proof of the first inequality is identical to \Cref{lem:one-step-reshaping-error}. For the second inequality, note that $\|\Lindblad\|_\diamond\le \Lambda_L$ by assumption. Moreover, $\LindbladDiag$ is a diagonal Pauli dissipator with total decay rate $\Gamma_{\mathrm{tot}}$, and hence
\begin{equation*}
    \|\LindbladDiag\|_\diamond
    \le
    \sum_{P_k\in\DiagDissStruct} a_{kk}\,\|\mathcal U_{P_k}-I\|_\diamond
    \le
    2\Gamma_{\mathrm{tot}}.
\end{equation*}
Finally, $\Gamma_{\mathrm{tot}}\le \|\LindbladDiag\|_\diamond\le \|\Lindblad\|_\diamond\le \Lambda_L$, where the first inequality is the standard Pauli-diagonal lower bound and the second follows from contractivity of physical Pauli twirling in diamond norm. This gives the last inequality in Eq.~\eqref{eq:one-step-twirling-error}.
\end{proof}

Next, we relate the \emph{diagonal rate} $a_{rr}$ of the simplified dynamics to an observable syndrome statistic. \Cref{lem:syndrome-hit-prob} relates the target syndrome probability to the sum of all diagonal rates sharing that syndrome.

\begin{lemma}[Syndrome hit probability for a diagonal Pauli jump]
\label{lem:syndrome-hit-prob}
Assume the Lindbladian $\LindbladDiag$ has the diagonal Pauli-dissipator form
as in \eqref{eq:lindblad-diag} with dissipator structure $\DiagDissStruct \subseteq \PnHerm$. Let $\{G_\ell\}_{\ell=1}^m \subseteq \mathcal P_{n'}$ be stabilizer generators with syndrome projectors $\{\Pi_s\}_{s\in\{0,1\}^m}$ as in \eqref{eq:syndrome-projectors}. Fix a non-identity
target Pauli $P_r$ and assume that its syndrome is nonzero, i.e. $r(P_r)\neq \mathbf 0$. Let $\rho=\Pi_0\rho\Pi_0 \in \mathfrak A_c \subseteq \mathcal{B}(\mathcal H_{n'})$ be any code state. Then for any $\tau\ge 0$,
\begin{equation*}
 p_r(\tau) \coloneq \Pr\bigl[\text{measuring syndrome }r(P_r)\bigr]
\;=\;\tr \bigl(\Pi_{r(P_r)}\,e^{\tau\LindbladDiag}(\rho)\bigr)
\;=\; \sum_{\substack{r(P_k)=r(P_r) \\ P_k \in \DiagDissStruct}}a_{kk}\,\tau \;+\; \Delta_r(\tau),
\end{equation*}
where the remainder satisfies $\abs{\Delta_r(\tau)}\ \le\  \frac12\,\|\LindbladDiag\|_\diamond^2\,\tau^2.$
\end{lemma}
\begin{proof}
    Use \Cref{lem:lindblad-taylor-tail} to expand $e^{\tau\LindbladDiag} = I + \tau \LindbladDiag + \int_0^\tau (\tau-s)\,\LindbladDiag^2 e^{s\LindbladDiag}\,ds$ and study the resulting terms. See Appendix~\ref{sec:deferred-diss-struct-testing-proofs} for the full proof. 
\end{proof}

\begin{corollary}[Syndrome hit probability for twirled evolution]
\label{cor:twirled-syndrome-hit-prob}
Combining \Cref{lem:one-step-twirling-error} and \Cref{lem:syndrome-hit-prob} and using Hölder's inequality, we get that for any $\rho \in \mathfrak A_c$ and $\tau \ge 0$:
\begin{equation*}
    \Pr\bigl[\text{measuring syndrome }r(P_r)\bigr]
    =
    \tr(\Pi_{r(P_r)} \widetilde{A}_\tau(\rho))
    =
    \sum_{\substack{r(P_k)=r(P_r) \\ P_k \in \DiagDissStruct}}a_{kk}\,\tau
    +
    \widetilde{\Delta}(\tau),
\end{equation*}
where $|\widetilde{\Delta}(\tau)| \le \frac12\|\LindbladDiag\|_\diamond^2\tau^2 +\frac12(\Lambda_L^2+4\Gamma_{\mathrm{tot}}^2)\tau^2 \le 5\Lambda_L^2\tau^2$. Here we used $\|\LindbladDiag\|_\diamond\le2\Gamma_{\mathrm{tot}}$ and $\Gamma_{\mathrm{tot}}\le\Lambda_L$.
\end{corollary}

To control the contribution of non-target dissipator terms to the syndrome statistic in \Cref{cor:twirled-syndrome-hit-prob}, we bound the total diagonal rate assigned to terms with the same syndrome as a candidate Pauli. The following lemma shows that a random stabilizer code makes this colliding rate small simultaneously for all candidates in a finite set.

\begin{lemma}[Small syndrome-collision mass for random checks]
\label{lem:random-code-small-collision-mass}
Let $\mathcal Z\subseteq\PnHerm\setminus\{I\}$ be a finite candidate set, and let
$\Gamma_{\mathrm{tot}} \coloneq \sum_{P_k\in\DiagDissStruct}a_{kk}$. Let $\mathcal T=\{P^{(t)}\}_{t=1}^m$ be sampled i.i.d.\ uniformly from $\PnHerm$, and let $r(\cdot)\in\{0,1\}^m$ be the syndrome map with $[r(Q)]_t=1$ iff $\{Q,P^{(t)}\}=0$. For each candidate $P_r\in\mathcal Z$, define its colliding syndrome mass
\begin{equation*}
    W_r
    \coloneq
    \sum_{\substack{P_k\in\DiagDissStruct\setminus\{P_r\}\\ r(P_k)=r(P_r)}}a_{kk}.
\end{equation*}
Then, for any $\xi>0$,
\begin{equation*}
    \Pr_{\mathcal T}\!\left[
    \exists P_r\in\mathcal Z:
    r(P_r)=0 \text{ or } W_r>\xi
    \right]
    \le
    |\mathcal Z|2^{-m}
    +
    \frac{|\mathcal Z|\Gamma_{\mathrm{tot}}}{\xi\,2^m}.
\end{equation*}
In particular, if $m =\left\lceil \log_2\left(\frac{32|\mathcal Z|(\Gamma_{\mathrm{tot}}/\eta+1)}{\delta}\right)\right\rceil$, then with probability at least $1-\delta$, every $P_r\in\mathcal Z$ has nonzero syndrome and colliding syndrome mass $W_r\le \eta/16$.
\end{lemma}

\begin{proof}
Fix $P_r\in\mathcal Z$. Since $P_r\neq I$, a uniformly random check commutes with $P_r$ with probability $1/2$. Hence
\begin{equation*}
    \Pr_{\mathcal T}[r(P_r)=0]=2^{-m}.
\end{equation*}
Next fix $P_k\in\DiagDissStruct\setminus\{P_r\}$. Since $P_kP_r$ is nonidentity, a uniformly random check commutes with $P_kP_r$ with probability $1/2$. Equivalently, for a single random check, $P_k$ and $P_r$ have the same syndrome bit with probability $1/2$. Since the $m$ checks are independent,
\begin{equation*}
    \Pr_{\mathcal T}\bigl[r(P_k)=r(P_r)\bigr]=2^{-m}.
\end{equation*}
Therefore,
\begin{equation*}
    \mathbb E_{\mathcal T} W_r
    =
    \sum_{P_k\in\DiagDissStruct\setminus\{P_r\}}
    a_{kk}\Pr_{\mathcal T}\bigl[r(P_k)=r(P_r)\bigr]
    \le
    \Gamma_{\mathrm{tot}}2^{-m}.
\end{equation*}
By Markov's inequality,
\begin{equation*}
    \Pr_{\mathcal T}[W_r>\xi]
    \le
    \frac{\Gamma_{\mathrm{tot}}}{\xi\,2^m}.
\end{equation*}
Taking a union bound over $P_r \in\mathcal Z$ proves the first claim. The second claim follows directly from setting $\xi=\eta/16$ and the chosen value of $m$.
\end{proof}

For our final ingredient, we use a standard result from classical hypothesis-testing:
\begin{lemma}[Threshold test for adapted Bernoulli trials]
\label{lem:martingale-threshold-test}
Let $X_1,\dots,X_N\in\{0,1\}$ be Bernoulli random variables, not necessarily independent, and let
$\widehat p \coloneq \frac{1}{N}\sum_{\ell=1}^N X_\ell$.
Let $\epsilon\in(0,1]$ and $\delta\in(0,1)$. Then there exists a universal constant
$C_{\mathrm{mg}}>0$ such that, if
$N \ge C_{\mathrm{mg}}\,\epsilon^{-1}\log(1/\delta)$, then:
\begin{enumerate}[label=(\roman*)]
    \item if $\mathbb E[X_\ell \mid X_1,\dots,X_{\ell-1}] \le \frac{5}{8}\epsilon$ for every $\ell\in[N]$, then
    $\Pr\left(\widehat p > \frac{3}{4}\epsilon\right) \le \delta$;
    \item if $\mathbb E[X_\ell \mid X_1,\dots,X_{\ell-1}] \ge \frac{7}{8}\epsilon$ for every $\ell\in[N]$, then
    $\Pr\left(\widehat p \le \frac{3}{4}\epsilon\right) \le \delta$.
\end{enumerate}
\end{lemma}
\begin{proof}
    The proof uses the multiplicative Azuma's inequality \cite{freedman1975martingales,kuszmaul2025multiplicative} (martingale version of the multiplicative Chernoff bound). An admissible value for the universal constant is $C_{\mathrm{mg}} = 112$. See Appendix~\ref{sec:deferred-diss-struct-testing-proofs} for details.
\end{proof}

\begin{algorithm}[htbp]
    \caption{Batched dissipator candidates testing subroutine, $j$th round}
    \label{alg:appendix-diss-rate-testing}
    \DontPrintSemicolon
    \SetKwInOut{Input}{Inputs}
    \SetKwInOut{Output}{Output}
    \SetKwBlock{Schedule}{Schedule}{end}

    \Input{
      \begin{enumerate}[label=(\arabic*), leftmargin=*, nosep]
        \item Nonempty set of candidate terms $\widehat{\DHierSet}_j \subseteq \PnHerm\setminus\{I\}$.
        \item Resolution parameter $\eta \in(0,1)$.
        \item Confidence parameter $\delta_j\in(0,1)$.
        \item Upper bound $\Lambda_L \ge \|\Lindblad\|_\diamond$.
        \item Upper bound $\widehat{\Gamma}_{\mathrm{tot}} \ge \Gamma_{\mathrm{tot}}$; one may take $\widehat{\Gamma}_{\mathrm{tot}} = \Lambda_L$.
        \item Access to $n$-qubit evolution $e^{t\Lindblad}$.
        \item Access to $\BigO{\log\left(| \widehat{\DHierSet}_j|(\widehat{\Gamma}_{\mathrm{tot}}/\eta+1)\right)}$ noiseless ancilla qubits.
      \end{enumerate}
    }

    \Output{Set $\widehat{\DHierSet}_{j,\mathrm{acc}}$ that, with probability at least $1-\delta_j$, solves \Cref{prob:batched-diag-rate-tolerant-test}, i.e.
    \begin{equation*}
    \{P_r\in\widehat{\DHierSet}_j:a_{rr}\ge\eta\} \subseteq
    \widehat{\DHierSet}_{j,\mathrm{acc}}
    \subseteq
    \{P_r\in\widehat{\DHierSet}_j:a_{rr}>\eta/2\}
    \end{equation*}
    }

    \Schedule{
      $m \gets \left\lceil {\log_2\bigl(64| \widehat{\DHierSet}_j|(\widehat{\Gamma}_{\mathrm{tot}}/\eta+1)\bigr)}\right\rceil$ \tcp*{\# of random checks}
      $N_{\mathrm{codes}} \gets \lceil\log_2(2/\delta_j)\rceil$ \tcp*{\# of random QEC draws}
      $\tau \gets \eta/(80\Lambda_L^2)$ \tcp*{single step evolution time}
      $N_{\mathrm{QEC}} \gets  \left\lceil\frac{112}{\eta \tau} \ln\left(\frac{2N_{\mathrm{codes}}|\widehat{\DHierSet}_j|}{\delta_j}\right)\right\rceil$ \tcp*{\# of syndrome measurements per code}
      $N_{\mathrm{ancillas}} \gets 3m/2$ \tcp*{$m/2$ - abelianity, $m$ - syndrome meas.}
    }

    \BlankLine
    \ForEach{$P_k\in\widehat{\DHierSet}_j$}{
        $\widehat p_k \gets 1$ \tcp*{to track minimum empirical syndrome rate across codes}
    }

    \For{$i \gets 1$ \KwTo $N_{\mathrm{codes}}$}{
        $\mathcal{G}_i \gets \{\, m \text{ random $n$-qubit Pauli strings}\,\}$ \tcp*{baseline generators}
        $\widetilde{\mathcal G}_i \gets$ Do ZX padding to ensure abelianity \tcp*{$n' \gets n+\lceil m/2\rceil$}
        Construct an arbitrary recovery channel $\RMap_i$\;
        Apply $\RMap_i$ to prepare some logical state\;

        \ForEach{$P_k\in\widehat{\DHierSet}_j$}{
            Compute the syndrome $r_i(P_k)$\;
            $\mathrm{cnt}_i(P_k)\gets 0$ \tcp*{initialize empirical counter to zero for each Pauli}
        }

        \For{$\ell \gets 1$ \KwTo $N_{\mathrm{QEC}}$}{
            Randomly sample $P \sim \PnHerm$ \tcp*{for physical Pauli twirling}
            Apply $P\otimes I_{n'-n}$\;
            Evolve the system by $e^{\tau \Lindblad} \otimes I_{n'-n}$\;
            Apply $P\otimes I_{n'-n}$\;
            Measure the syndrome $s_\ell$ and apply the recovery channel $\RMap_i$\;
            \ForEach{$P_k\in\widehat{\DHierSet}_j$ with $r_i(P_k)=s_\ell\neq 0$}{
                $\mathrm{cnt}_i(P_k)\gets \mathrm{cnt}_i(P_k)+1$ \tcp*{record syndrome outcomes}
            }
        }

        \ForEach{$P_k\in\widehat{\DHierSet}_j$}{
            \If{$r_i(P_k)\neq 0$}{
                $\widehat p_k \gets \min\left(\widehat p_k,\mathrm{cnt}_i(P_k)/N_{\mathrm{QEC}}\right)$\;
            }
        }
    }

    $\widehat{\DHierSet}_{j,\mathrm{acc}}\gets\emptyset$\;
    \ForEach{$P_k\in\widehat{\DHierSet}_j$}{
        \If{$\widehat p_k > 3\eta\tau/4$}{
            $\widehat{\DHierSet}_{j,\mathrm{acc}}\gets\widehat{\DHierSet}_{j,\mathrm{acc}}\cup\{P_k\}$\;
        }
    }
    \Return $\widehat{\DHierSet}_{j,\mathrm{acc}}$
\end{algorithm}

Crucially, the sequential syndrome outcomes can be statistically dependent, because the recovery channel may alter the code state between rounds. For this reason, we use the martingale-based concentration result in \Cref{lem:martingale-threshold-test} instead of an independence-based tail bound. 

\begin{theorem}[Batched dissipator rate testing]
\label{thm:appendix-diss-rate-testing}
Let $\Lindblad$ be an $n$-qubit Lindbladian as in Eq.~\eqref{eq:lindblad-eq-diss-struct} with $\|\Lindblad\|_\diamond \le \Lambda_L$. Let $\widehat{\DHierSet}_j\subseteq\PnHerm\setminus\{I\}$ be a nonempty finite candidate set, and let $\eta,\delta_j\in(0,1)$. Let $\widehat{\Gamma}_{\mathrm{tot}}\ge \Gamma_{\mathrm{tot}}$, where $\Gamma_{\mathrm{tot}}$ is the total decay rate defined in Eq.~\eqref{eq:total-decay-rate-diss-testing}. Assume black-box access to $e^{t\Lindblad}$.

Then running \Cref{alg:appendix-diss-rate-testing} with schedule
\begin{equation*}
    m=\BigO{\log\left(|\widehat{\DHierSet}_j|(\widehat{\Gamma}_{\mathrm{tot}}/\eta+1)\right)},\;\;
    \tau=\Theta(\eta/\Lambda_L^2),\;\;N_{\mathrm{codes}}=\BigO{\log(1/\delta_j)},\;\;
    N_{\mathrm{QEC}}
    =
    \BigO{\frac{\Lambda_L^2}{\eta^2}
    \log\left(\frac{N_{\mathrm{codes}}|\widehat{\DHierSet}_j|}{\delta_j}\right)}
\end{equation*}
outputs $\widehat{\DHierSet}_{j,\mathrm{acc}}$ and satisfies:
\begin{enumerate}[leftmargin=*, label=\arabic*, nosep]
\item \textit{(Correctness)} With probability $\ge 1- \delta_j$, the algorithm solves \Cref{prob:batched-diag-rate-tolerant-test}. Concretely, the output $\widehat{\DHierSet}_{j,\mathrm{acc}}$ satisfies:
\begin{equation*}
    \{P_r\in\widehat{\DHierSet}_j:a_{rr}\ge\eta\}
    \subseteq
    \widehat{\DHierSet}_{j,\mathrm{acc}}
    \subseteq
    \{P_r\in\widehat{\DHierSet}_j:a_{rr}>\eta/2\}.
\end{equation*}
\item \textit{(Total evolution time)}
The total evolution time under $e^{t\Lindblad}$ is
\begin{equation*}
t_{\mathrm{tot}}
=
N_{\mathrm{codes}}N_{\mathrm{QEC}}\tau
=
\BigO{
\frac{1}{\eta}
\log\left(\frac{1}{\delta_j}\right)
\log\left(\frac{|\widehat{\DHierSet}_j|\log(1/\delta_j)}{\delta_j}\right)
}.
\end{equation*}

\item \textit{(Minimum evolution time)}
It only ever applies $e^{\Lindblad t}$ for $t\ge t_{\mathrm{min}}$, where
\begin{equation*}
t_{\mathrm{min}}
\coloneq
\tau
=
\Theta(\eta/\Lambda_L^2).
\end{equation*}

\item \textit{(Number of experiments)}
The number of random-code experiments is
\begin{equation*}
N_{\exp}
\coloneq
N_{\mathrm{codes}}
=
\BigO{\log(1/\delta_j)}.
\end{equation*}
The number of interleaved QEC rounds, or syndrome extractions, is larger by a factor of $N_{\mathrm{QEC}} = \SoftBigO{\Lambda_L^2/\eta^2}$

\item \textit{(Ancilla cost)}
The procedure uses at most $ 3/2 \left\lceil \log_2\bigl(64|\widehat{\DHierSet}_j|(\widehat{\Gamma}_{\mathrm{tot}}/\eta+1)\bigr)\right\rceil$
ancillary qubits for QEC.

\item \textit{(Classical overhead)}
The classical cost of the algorithm is dominated by repeated QEC rounds and is bounded by
\begin{equation*}
\SoftBigO{
\bigl(n+|\widehat{\DHierSet}_j|(\widehat{\Gamma}_{\mathrm{tot}}/\eta+1)\bigr)
\frac{\Lambda_L^2}{\eta^2}}.
\end{equation*}
\end{enumerate}
\end{theorem}

\begin{proof}
    We first prove the correctness claim in item $(1)$ under the stated scheduling. Items $(2)$--$(4)$ then follow directly from the schedule, and items $(5)$--$(6)$ are briefly proved at the end.
    
    Let $\mathcal Z\coloneq \widehat{\DHierSet}_j$. We use the convention that $a_{kk}=0$ for $P_k\notin\DiagDissStruct$.

    \Cref{alg:appendix-diss-rate-testing} draws random checks to form random stabilizer codes. The abelianity is ensured by the $ZX$-padding from \Cref{lem:random-code-zx-padding}, which lifts the random checks to a valid stabilizer code without changing the commutation pattern with Paulis on the $n$ probe qubits. Hence the syndrome-collision guarantees from \Cref{lem:random-code-small-collision-mass} apply directly to the generated code.

    Call a code good if every $P_k\in\mathcal Z$ has nonzero syndrome and colliding syndrome mass at most $\eta/16$. By \Cref{lem:random-code-small-collision-mass} with $\xi=\eta/16$ and the chosen value of $m$, a single random code is good with probability at least $1/2$. Therefore, with
    $N_{\mathrm{codes}}=\lceil\log_2(2/\delta_j)\rceil$, the probability that none of the sampled codes is good is at most
    \begin{equation}
    \label{eq:batched-rate-testing-good-code}
        \Pr(\text{no sampled code is good})
        \le
        (1/2)^{N_{\mathrm{codes}}}
        \le
        \delta_j/2.
    \end{equation}

    We next analyze syndrome-hit probabilities for a fixed code. By \Cref{cor:twirled-syndrome-hit-prob}, for any candidate $P_k$ with nonzero syndrome,
    \begin{equation*}
        \Pr(\text{measure }r(P_k))
        =
        \sum_{\substack{r(P_\ell)=r(P_k)\\ P_\ell\in\DiagDissStruct}}a_{\ell\ell}\tau
        +
        \widetilde{\Delta}(\tau),
        \qquad
        |\widetilde{\Delta}(\tau)|\le 5\Lambda_L^2\tau^2.
    \end{equation*}
    Since $\tau=\eta/(80\Lambda_L^2)$, the error term obeys $|\widetilde{\Delta}(\tau)|\le \eta\tau/16.$

    If $a_{kk}\le\eta/2$ and the code is good, then the colliding syndrome mass for $P_k$ is at most $\eta/16$, so
    \begin{equation*}
        \Pr(\text{measure }r(P_k))
        \le
        \left(\frac{\eta}{2}+\frac{\eta}{16}\right)\tau+\frac{\eta\tau}{16}
        =
        \frac{5}{8}\eta\tau.
    \end{equation*}
    On the other hand, if $a_{kk}\ge\eta$, then for every code with $r(P_k)\neq0$,
    \begin{equation*}
        \Pr(\text{measure }r(P_k))
        \ge
        \eta\tau-\frac{\eta\tau}{16}
        >
        \frac{7}{8}\eta\tau.
    \end{equation*}

    We now apply the martingale threshold test from \Cref{lem:martingale-threshold-test} with $\epsilon\coloneq\eta\tau$. For each code $i$ and candidate $P_k$ with $r_i(P_k)\neq0$, let $X_1,\dots,X_{N_{\mathrm{QEC}}}$ be the indicators of observing syndrome $r_i(P_k)$ in the sequential syndrome measurements. The conditional expectations of these adapted Bernoulli variables satisfy the upper or lower bounds above, because \Cref{cor:twirled-syndrome-hit-prob} holds for any code state. With
    \begin{equation*}
        N_{\mathrm{QEC}}
        =
        \left\lceil
        \frac{112}{\eta\tau}
        \ln\left(\frac{2N_{\mathrm{codes}}|\mathcal Z|}{\delta_j}\right)
        \right\rceil,
    \end{equation*}
    each such empirical syndrome frequency fails its threshold test with probability at most $\delta_j/(2N_{\mathrm{codes}}|\mathcal Z|)$. Taking a union bound over all codes and all candidates, all empirical tests used by the algorithm are simultaneously correct with probability at least $1-\delta_j/2$.

    Condition on the event that at least one sampled code is good and all empirical threshold tests are correct. If $a_{kk}\le\eta/2$, then for a good code the empirical syndrome frequency of $P_k$ is at most $3\eta\tau/4$, and therefore the minimum over codes satisfies $\widehat p_k\le 3\eta\tau/4$. Hence $P_k$ is rejected. If $a_{kk}\ge\eta$, then every code with $r_i(P_k)\neq0$ has empirical syndrome frequency greater than $3\eta\tau/4$; codes with zero syndrome are ignored by the algorithm. Thus $\widehat p_k>3\eta\tau/4$, and $P_k$ is accepted. Combining the two failure probabilities gives total failure probability at most $\delta_j$, proving item $(1)$.

    We next prove the ancilla cost in item $(5)$. Since the algorithm draws $m = \left\lceil \log_2\bigl(64|\widehat{\DHierSet}_j| (\widehat{\Gamma}_{\mathrm{tot}}/\eta+1)\bigr)\right\rceil$
    random checks per code, \Cref{lem:random-code-zx-padding} uses at most $\lceil m/2\rceil$ ancillas for abelianity, and we reserve $m$ ancillas for syndrome extraction. Thus the total number of ancillas is at most $3m/2$, as claimed.

    Finally, the classical overhead in item $(6)$ is dominated by repeated QEC rounds. Since the stabilizer group has size $2^m=\SoftBigO{|\widehat{\DHierSet}_j|(\widehat{\Gamma}_{\mathrm{tot}}/\eta+1)}$, the total classical cost is bounded by
    \begin{equation*}
        \SoftBigO{
        \bigl(n+|\widehat{\DHierSet}_j|(\widehat{\Gamma}_{\mathrm{tot}}/\eta+1)\bigr)N_{\mathrm{codes}}
        N_{\mathrm{QEC}}
        } = \SoftBigO{
        \bigl(n+|\widehat{\DHierSet}_j|(\widehat{\Gamma}_{\mathrm{tot}}/\eta+1)\bigr)
        \frac{\Lambda_L^2}{\eta^2}
        }.
    \end{equation*}
    This concludes the proof of the theorem.
\end{proof}

\subsection{Full hierarchical learning algorithm}\label{sec:diss-struct-hierarchical-learning}

We now combine candidate sampling \Cref{alg:appendix-round-j-diss-struct-sampling} and candidate testing \Cref{alg:appendix-diss-rate-testing} into the full dissipator structure learning algorithm:

\begin{algorithm}[H]
    \caption{Dissipator structure learning algorithm}
    \label{alg:appendix-diss-struct-learning}
    \DontPrintSemicolon
    \SetKwInOut{Input}{Inputs}
    \SetKwInOut{Output}{Output}
    \SetKwBlock{Schedule}{Schedule}{end}

    \Input{
      \begin{enumerate}[label=(\arabic*), leftmargin=*, nosep]
        \item Resolution parameter $\eta \in(0,1)$.
        \item Confidence parameter $\delta_{d}\in(0,1)$.
        \item Upper bound $\Lambda_L \ge \|\Lindblad\|_\diamond$.
        \item Access to $n$-qubit evolution $e^{t\Lindblad}$, where the total decay rate of $\Lindblad$ is $\Gamma_{\mathrm{tot}}$ from Eq.~\eqref{eq:total-decay-rate-diss-testing}.
        \item Access to $n+\BigO{\log (\Gamma_\mathrm{tot}/\eta+1)}$ noiseless ancilla qubits.
      \end{enumerate}
    }

    \Output{Estimator $\EstEtaDiagDissStruct$ of the $\eta$-heavy part of the dissipator structure $\EtaDiagDissStruct$ (Eq.~\eqref{eq:def-eta-heavy-diss-struct}) such that $\EtaDiagDissStruct \subseteq \EstEtaDiagDissStruct \subseteq \HalfEtaDiagDissStruct$ with probability at least $1-\delta_{d}$.}

    \Schedule{
      $N_{\mathrm{rounds}} \gets \lceil\log_2(1/\eta)\rceil$ \tcp*{\# of hierarchical rounds}
      $\delta_{\mathrm{round}} \gets \delta_{d}/(N_{\mathrm{rounds}}+1)$  \tcp*{failure probability per hierarchical round}
    }

    \BlankLine
    $\widehat{\Gamma}_\mathrm{tot} \gets \texttt{EstimateGamma}(\max\{\Lambda_L,\eta/2\},\eta/2,\delta_{\mathrm{round}}, \text{ Pauli twirled }e^{\Lindblad t})$ \tcp*{estimate from \Cref{lem:precalibrate-diag-diss-rate}}
    $\EstEtaDiagDissStruct \gets  \emptyset$ \tcp*{Initialize the running dissipator structure estimator.}
    
    \For{$j \gets 0$ \KwTo $N_{\mathrm{rounds}} - 1$}{
        $\widehat{\DHierSet}_{j} \gets  \texttt{Diss\_Struct\_Sampling}(j, \EstEtaDiagDissStruct, \Lambda_L, \delta_{\mathrm{round}}/2, e^{\Lindblad t})$ \tcp*{superset estimator of $\DHierSet_{j}\setminus \EstEtaDiagDissStruct$}
        \If{$\widehat{\DHierSet}_{j}\neq\emptyset$}{
            $\EstEtaDiagDissStruct \gets  \EstEtaDiagDissStruct \cup \texttt{Diss\_Struct\_Testing}( \widehat{\DHierSet}_{j}, \eta, \delta_{\mathrm{round}}/2, \Lambda_L, \widehat{\Gamma}_\mathrm{tot}, e^{\Lindblad t})$ \tcp*{update structure estimator}}
    }
    \Return $\EstEtaDiagDissStruct$
\end{algorithm}

\begin{restatethm}{thm:appendix-hier-diss-struct-learning}[Hierarchical dissipator structure learning]
Let $\Lindblad$ be an $n$-qubit Lindbladian as in Eq.~\eqref{eq:lindblad-eq-diss-struct} with $\|\Lindblad\|_\diamond \le \Lambda_L$. Fix $\eta, \delta_{d} \in (0,1)$ and assume black-box access to $e^{t\Lindblad}$. Let $\EtaDiagDissStruct \subseteq \DiagDissStruct$ denote the $\eta$-heavy part of the dissipator structure defined in Eq.~\eqref{eq:def-eta-heavy-diss-struct}. Let $\Gamma_{\mathrm{tot}}$ denote the total Lindbladian decay rate defined in Eq.~\eqref{eq:total-decay-rate-diss-testing}. Then running \Cref{alg:appendix-diss-struct-learning} returns an estimator $\EstEtaDiagDissStruct$ of $\EtaDiagDissStruct$, satisfying: 

\begin{enumerate}[leftmargin=*, label=\arabic*.]
\item \textit{(Accuracy)} With probability $\ge1-\delta_d$, the output estimator obeys
$\EtaDiagDissStruct \subseteq \EstEtaDiagDissStruct \subseteq \HalfEtaDiagDissStruct$.

\item \textit{(Total evolution time)} It applies $e^{\Lindblad t}$ for a total evolution time
\begin{equation*}
t_{\mathrm{tot}}=\SoftBigO{\frac{1}{\eta}}.
\end{equation*}

\item \textit{(Minimum evolution time)} It only ever applies $e^{\Lindblad t}$ for $t \ge t_{\mathrm{min}}$, where
\begin{equation*}
t_{\mathrm{min}}=\Omega\Bigl(\frac{\eta}{\Lambda_L^2}\Bigr)  .
\end{equation*}

\item \textit{(Number of experiments)} The total number of experiments is 
\begin{equation*}
N_{\exp} \;=\; \SoftBigO{\frac{\Gamma_{\mathrm{tot}}}{\eta}+1}.
\end{equation*}
with the total number of interleaved QEC rounds bounded by $\SoftBigO{\Lambda_L^2/\eta^2}$.

\item \textit{(Ancilla cost)}
The number of ancillary qubits used by the procedure (dominated by Bell sampling + QEC) is at most
\begin{equation*}
 n + \BigO{\log(\frac{\Gamma_{\mathrm{tot}}}{\eta}+1)}
\end{equation*}

\item \textit{(Classical overhead)} The classical cost of the algorithm (dominated by repeated QEC in structure testing) is:
    \begin{equation*}
    \SoftBigO{\bigl(n + (\Gamma_{\mathrm{tot}}/\eta)^4\bigr)\Lambda_L^2/\eta^2}
    \end{equation*}
\end{enumerate}
\end{restatethm}

\begin{proof}
Let $N_{\mathrm{rounds}}\coloneq \lceil \log_2(1/\eta)\rceil$ and
$\delta_{\mathrm{round}}\coloneq \delta_d/(N_{\mathrm{rounds}}+1)$. Let
$\widehat{\mathcal S}_{D,j}$ denote the value of $\EstEtaDiagDissStruct$ after round $j$, with $\widehat{\mathcal S}_{D,-1}\coloneq\emptyset$. First condition on the success of the initial call to \texttt{EstimateGamma}, which occurs with
probability at least $1-\delta_{\mathrm{round}}$ and gives
$\widehat{\Gamma}_{\mathrm{tot}}\ge \Gamma_{\mathrm{tot}}$ and
$\widehat{\Gamma}_{\mathrm{tot}}/\eta+1=\BigO{\Gamma_{\mathrm{tot}}/\eta+1}$ by
\Cref{lem:precalibrate-diag-diss-rate}.

We prove correctness by induction. At the beginning of round $j$, assume
\begin{equation}\label{eq:hier-diss-learning-proof-invariant}
    \DHierSet_k\subseteq \widehat{\mathcal S}_{D,j-1}\subseteq \HalfEtaDiagDissStruct \qquad\text{for all }k<j .
\end{equation}
For $j=0$ this is vacuous. In round $j$, \Cref{thm:appendix-round-j-diss-struct-sampling} gives
$\widehat{\DHierSet}_j\supseteq \DHierSet_j\setminus \widehat{\mathcal S}_{D,j-1}$ with probability
at least $1-\delta_{\mathrm{round}}/2$. If $\widehat{\DHierSet}_j=\emptyset$, the testing step is skipped and
returns the empty set. Otherwise, \Cref{thm:appendix-diss-rate-testing} applies with failure probability
at most $\delta_{\mathrm{round}}/2$ and accepts every candidate with $a_{rr}\ge\eta$ while accepting only
candidates with $a_{rr}>\eta/2$. Therefore the update preserves
$\widehat{\mathcal S}_{D,j}\subseteq \HalfEtaDiagDissStruct$. Moreover, for every non-final round
$j<N_{\mathrm{rounds}}-1$, every element of $\DHierSet_j$ has rate strictly larger than $\eta$, and hence is
accepted. In the final round, all elements of $\DHierSet_j$ with rate at least $\eta$ are accepted. Thus,
conditioned on the success of all rounds,
\begin{equation*}
    \EtaDiagDissStruct\subseteq \EstEtaDiagDissStruct\subseteq \HalfEtaDiagDissStruct .
\end{equation*}
The initial calibration and the $N_{\mathrm{rounds}}$ rounds fail with total probability $\le (N_{\mathrm{rounds}}+1)\delta_{\mathrm{round}}=\delta_d$, proving item~$(1)$.

We now collect resources. Since structure identification internally also invokes \texttt{EstimateGamma}, we absorb the resources used by \texttt{EstimateGamma} into the structure sampling subroutine. On the success event, $\EstEtaDiagDissStruct\subseteq \HalfEtaDiagDissStruct$, so $|\EstEtaDiagDissStruct|\le 2\Gamma_{\mathrm{tot}}/\eta$ throughout. Hence, by \Cref{thm:appendix-round-j-diss-struct-sampling}, $|\widehat{\DHierSet}_j|=\SoftBigO{(2^j\Gamma_j+1)(\Gamma_{\mathrm{tot}}/\eta+1)^2}.$ Combining \Cref{thm:appendix-round-j-diss-struct-sampling} with \Cref{thm:appendix-diss-rate-testing}, the round-$j$ evolution time is bounded by
\begin{equation*}
    T_j=\BigO{2^j\log\!\left(\frac{2^j\Gamma_j+1}{\delta_{\mathrm{round}}}\right)+2^j\log\!\left(\frac{\log(2^j\Lambda_L+4)}{\delta_{\mathrm{round}}}\right)}+\SoftBigO{\frac{1}{\eta}\log\!\left(\frac{1}{\delta_{\mathrm{round}}}\right)\log\!\left(\frac{(2^j\Gamma_j+1)(\Gamma_{\mathrm{tot}}/\eta+1)^2\log(1/\delta_{\mathrm{round}})}{\delta_{\mathrm{round}}}\right)}.
\end{equation*}
Uusing $\sum_{j=0}^{N_{\mathrm{rounds}}-1}2^j=\BigO{1/\eta}$, this proves item~$(2)$:
\begin{equation*}
    t_{\mathrm{tot}}=\sum_{j=0}^{N_{\mathrm{rounds}}-1}T_j+\SoftBigO{1/\eta}=\SoftBigO{1/\eta}.
\end{equation*}

The minimum round-$j$ evolution time is $\Omega(1/(\Lambda_L^2 2^j))$ during sampling and $\Theta(\eta/\Lambda_L^2)$ during testing. Since $2^j\le 1/\eta$,
\begin{equation*}
    t_{\mathrm{min}}=\Omega\Bigl(\frac{\eta}{\Lambda_L^2}\Bigr),
\end{equation*}
proving item~$(3)$.

For the number of experiments (item~$(4)$), \Cref{thm:appendix-round-j-diss-struct-sampling} gives
\begin{equation*}
    N_{\exp}^{\mathrm{samp}}=\sum_{j=0}^{N_{\mathrm{rounds}}-1}\BigO{(2^j\Gamma_j+1)\log\!\left(\frac{2^j\Gamma_j+1}{\delta_{\mathrm{round}}}\right)+\log(2^j\Lambda_L+4)\log\!\left(\frac{\log(2^j\Lambda_L+4)}{\delta_{\mathrm{round}}}\right)}=\SoftBigO{\frac{\Gamma_{\mathrm{tot}}}{\eta}+1}.
\end{equation*}
The testing step contributes only $\SoftBigO{1}$ random-code experiments per round, so $N_{\exp}=\SoftBigO{\Gamma_{\mathrm{tot}}/\eta+1}$. The number of interleaved QEC rounds is
\begin{equation*}
    \sum_{j=0}^{N_{\mathrm{rounds}}-1}\SoftBigO{\Lambda_L^2 4^j}+\SoftBigO{N_{\mathrm{rounds}}\Lambda_L^2/\eta^2}=\SoftBigO{\Lambda_L^2/\eta^2}.
\end{equation*}

For ancillas (item~$(5)$), the sampling stage uses $n+\BigO{\log(|\EstEtaDiagDissStruct|+1)}=n+\BigO{\log(\Gamma_{\mathrm{tot}}/\eta+1)}$ ancillas. The testing stage uses
\begin{equation*}
    \BigO{\log\!\left(|\widehat{\DHierSet}_j|(\widehat{\Gamma}_{\mathrm{tot}}/\eta+1)\right)}=\BigO{\log\!\left((2^j\Gamma_j+1)(\Gamma_{\mathrm{tot}}/\eta+1)^3\right)}
\end{equation*}
ancillas, which is absorbed into $n+\BigO{\log(\Gamma_{\mathrm{tot}}/\eta+1)}$ up to the logarithmic factors.

Finally, the total classical cost (item~$(6)$) is dominated by the testing stage. Using the candidate-size bound from \Cref{thm:appendix-round-j-diss-struct-sampling}, the calibration guarantee
$\widehat{\Gamma}_{\mathrm{tot}}/\eta+1=\BigO{\Gamma_{\mathrm{tot}}/\eta+1}$, and
\Cref{thm:appendix-diss-rate-testing}, the total classical overhead is
\begin{equation*}
    C_{\mathrm{tot}}=\SoftBigO{\left(n+\left(\frac{\Gamma_{\mathrm{tot}}}{\eta}+1\right)^4\right)\frac{\Lambda_L^2}{\eta^2}}.
\end{equation*}
\end{proof}

\subsection{Deferred proofs for Bell-sampling identification}\label{sec:deferred-diss-struct-sampling-proofs}
Here, we present the deferred proofs related to dissipator structure identification via Bell sampling (\Cref{alg:appendix-round-j-diss-struct-sampling}). We restate the lemmas for convenience. 

\begin{restatelem}{lem:bell-sampling-of-eff-diss-struct}[Bell sampling of effective diagonal dissipator]
Let $\LindbladEff$ be $n'$-qubit Lindbladian with purely diagonal dissipator part
\begin{equation*}
    \LindbladEff(\rho)
    =
    \sum_{Q_k \in \mathcal S_\mathrm{eff} \,\subseteq\, \mathcal P_{n'} \setminus \{I\}}
    \alpha_{kk}
    \left(Q_k \rho Q_k - \rho\right).
\end{equation*}
where $\mathcal S_\mathrm{eff}$ is a set of distinct Hermitian Paulis and $\alpha_{kk}\ge 0$ for all $Q_k\in\mathcal S_{\mathrm{eff}}$. Define the total effective dissipative rate $\Gamma_{\mathrm{eff}} \coloneq \sum_{Q_k\in\mathcal S_{\mathrm{eff}}} \alpha_{kk}.$ Fix a target Pauli $Q_r\in\mathcal S_\mathrm{eff}$. Let $C$ be a stabilizer code on $n'$ qubits with code projector $\Pi$ and stabilizer subgroup $S$. Let $G_{\mathrm{eff}} \coloneq \langle \mathcal S_{\mathrm{eff}} \rangle \subseteq \mathbb P_{n'}$ denote the subgroup generated by $\mathcal S_{\mathrm{eff}}$. Assume that all Paulis
$Q \in \mathcal S_{\mathrm{eff}}$ have zero syndrome $r(Q)=0$ with respect to $C$, and that for every $R \in G_{\mathrm{eff}}$, $R\propto s\in S \implies s = I_{n'}$. Equivalently, no nontrivial product of elements of $\mathcal S_{\mathrm{eff}}$ is proportional to a stabilizer. Let $\rho_0$ denote a $2n'$-qubit maximally entangled state $\rho_0 \coloneq \ket{\Phi_0}\bra{\Phi_0}$ where $\ket{\Phi_0} \coloneq \ket{\Phi_+}^{\otimes n' }$ and $\ket{\Phi_+} = (\ket{00} + \ket{11})/\sqrt{2}.$
Let $\widetilde \rho$ denote the state immediately before the Bell basis measurement
\begin{equation*}
    \widetilde \rho \coloneq (e^{\LindbladEff t}\otimes I_{n'}) \left(\frac{\Pi\otimes I_{n'} \,\rho_0 \,\Pi\otimes I_{n'}}{\tr(\Pi\otimes I_{n'} \,\rho_0)}\right).
\end{equation*}
For any Pauli $Q\in \mathbb P_{n'}$, let $O_Q \coloneq (Q\otimes I_{n'})\ket{\Phi_0}\bra{\Phi_0}(Q^\dagger\otimes I_{n'})$ denote the corresponding Bell-measurement projector. Then, for any $t\ge 0$, a single run of Bell sampling on $\widetilde \rho$ satisfies
\begin{align}
    \Pr(\text{sample $Q_r$ modulo stabilizers})
    &\coloneq
    \tr\left(\sum_{\sigma\in S} O_{\sigma Q_r}\,\widetilde{\rho}\right)
    \ge
    e^{-\Gamma_{\mathrm{eff}}t}\alpha_{rr}t.
\end{align}
\end{restatelem}

\begin{proof}
    Note that since the code projector can be expressed as $\Pi = \frac{1}{|S|}\sum_{s\in S} s$ (Eq.~\eqref{eq:stabilizer-projector}), we have $\tr(\Pi\otimes I_{n'} \,\rho_0) = 1/|S|$. To further simplify the $\widetilde{\rho}$, we 
    commute the corner projection $\mathcal C_\Pi \coloneq \Pi(\cdot)\Pi$ with the action of $e^{\LindbladEff t}$ since all $Q_k \in \mathcal S_{\mathrm{eff}}$ have zero syndromes and thus commute with $\Pi$:
    \begin{equation*}
        \widetilde{\rho} = |S|  (\mathcal C_\Pi \otimes I_{n'}) \circ (e^{\LindbladEff t}\otimes I_{n'}) (\rho_0)
    \end{equation*}
    Since $\LindbladEff$ is just a diagonal dissipator, $e^{\LindbladEff t}$ admits the following chi-matrix representation for any $\rho$:
    \begin{equation*}
        e^{\LindbladEff t}(\rho)
        =
        \sum_{Q_m\in G_{\mathrm{eff}}}
        \chi_{mm}(t) Q_m \rho Q_m .
    \end{equation*}
    In this chi representation, the sum over $G_{\mathrm{eff}}$ is understood modulo global phases, with one Hermitian representative chosen for each phase class; we choose the representative of the target phase class to be $Q_r$.

    Fix $\sigma\in S$ and write
    $Z_r\coloneq \sigma Q_r$ for the corresponding representative of the coset $SQ_r$.
    Using the chi-matrix representation of $e^{\LindbladEff t}$ together with $\widetilde{\rho}$ and the fact that $\sigma \Pi = \Pi\sigma = \Pi$, we get
    \begin{equation*}
        \begin{aligned}
            &\Pr(\text{sample $Z_r$})
            \coloneq \tr(O_{Z_r}\widetilde{\rho}) = |S| \sum_{Q_m\in G_{\mathrm{eff}}} \chi_{mm}(t)
            \bigg|\bra{\Phi_0} (Z_r^\dagger \Pi Q_m\otimes I_{n'}) \ket{\Phi_0}\bigg|^2  = \\
            &= |S| \sum_{Q_m\in G_{\mathrm{eff}}} \chi_{mm}(t)
            \bigg|\bra{\Phi_0} (Q_r \Pi Q_m\otimes I_{n'}) \ket{\Phi_0}\bigg|^2 
            = \frac{1}{|S|}\sum_{Q_m\in G_{\mathrm{eff}}} \chi_{mm}(t)
            \bigg|\bra{\Phi_0} (Q_r Q_m\otimes I_{n'}) \ket{\Phi_0}\bigg|^2 .
        \end{aligned}
    \end{equation*}
    where the last equality follows from $Q_rQ_m \in G_{\mathrm{eff}}$, $\Pi = \frac{1}{|S|}\sum_{s\in S} s$ and the assumption that no nontrivial product of elements of $\mathcal S_{\mathrm{eff}}$ is proportional to a stabilizer.
    By the property of Bell pairs, this further simplifies to
    \begin{equation*}
            \Pr(\text{sample $Z_r$}) 
            = \frac{1}{|S|}\chi_{rr}(t) \,\bigg|\bra{\Phi_0} (Q_r Q_r\otimes I_{n'}) \ket{\Phi_0}\bigg|^2 = \frac{\chi_{rr}(t)}{|S|}.
    \end{equation*}
    Therefore, combining the probabilities across all representatives of the coset $SQ_r$,
    \begin{equation} \label{eq:prob-sample-Qr-modulo-stabilizers}
        \Pr(\text{sample $Q_r$ modulo stabilizers})
        =
        \tr\left(\sum_{\sigma\in S} O_{\sigma Q_r}\,\widetilde{\rho}\right)
        =
        \sum_{\sigma\in S}\Pr(\text{sample $\sigma Q_r$})
        =
        \chi_{rr}(t).
    \end{equation}

    We now lower bound $\chi_{rr}(t)$. For each $Q_k\in\mathcal S_{\mathrm{eff}}$, let
    $\mathcal U_k(\rho)\coloneq Q_k\rho Q_k$. The superoperators $\mathcal U_k$ commute with each other, since Pauli products commute up to phase and the global phase disappears under conjugation. Hence
    \begin{equation*}
        e^{\LindbladEff t}
        =
        \prod_{Q_k\in\mathcal S_{\mathrm{eff}}}
        \exp\bigl(\alpha_{kk}t(\mathcal U_k-I)\bigr) = e^{-\Gamma_{\mathrm{eff}}t} \prod_{Q_k\in\mathcal S_{\mathrm{eff}}}
        \exp\bigl(\alpha_{kk}t \,\mathcal U_k\bigr).
    \end{equation*}
    For each $k$,
    \begin{equation*}
        \exp\bigl(\alpha_{kk}t \,\mathcal U_k\bigr)
        =
        \sum_{\ell=0}^{\infty}
        \frac{(\alpha_{kk}t)^\ell}{\ell!}\mathcal U_k^\ell .
    \end{equation*}
    Multiplying these expansions gives a Pauli-channel representation in which every term has nonnegative weight. In particular, the event that exactly one jump of type $Q_r$ occurs and no other jump occurs contributes the channel
    $\rho\mapsto Q_r\rho Q_r$ with weight
    \begin{equation*}
        e^{-\Gamma_{\mathrm{eff}}t}\alpha_{rr}t .
    \end{equation*}
    Other jump patterns may also multiply to $Q_r$ up to phase and hence can only increase $\chi_{rr}(t)$. Therefore
    \begin{equation*}
        \chi_{rr}(t)
        \ge
        e^{-\Gamma_{\mathrm{eff}}t}\alpha_{rr}t .
    \end{equation*}
    Combining this lower bound with
    $\Pr(\text{sample $Q_r$ modulo stabilizers})=\chi_{rr}(t)$ gives
    \begin{equation*}
        \Pr(\text{sample $Q_r$ modulo stabilizers}) = \chi_{rr}(t)
        \ge
        e^{-\Gamma_{\mathrm{eff}}t}\alpha_{rr}t.
    \end{equation*}
    concluding the proof.
\end{proof}

The following two lemmas are necessary steps for pre-calibrating the total remaining decay rate in \Cref{alg:appendix-round-j-diss-struct-sampling}. 

\begin{lemma}[Non-identity Bell-sampling probability for a diagonal dissipator]
\label{lem:nonidentity-bell-sampling-diag-diss}
Let $\LindbladEff$ be an $n'$-qubit Lindbladian with purely diagonal dissipator part
\begin{equation*}
    \LindbladEff(\rho)
    =
    \sum_{Q_k \in \mathcal S_\mathrm{eff} \subseteq \mathcal P_{n'} \setminus \{I\}}
    \alpha_{kk}(Q_k\rho Q_k-\rho),
\end{equation*}
where $\mathcal S_\mathrm{eff}$ is a set of distinct Hermitian Paulis and $\alpha_{kk}\ge0$. Define $\Gamma_{\mathrm{eff}} \coloneq \sum_{Q_k\in\mathcal S_{\mathrm{eff}}}\alpha_{kk}.$ Let $p(t)$ denote the probability that Bell sampling $(e^{t\LindbladEff}\otimes I)(\rho_0)$ returns a non-identity Pauli outcome, where $\rho_0=\ket{\Phi_0}\bra{\Phi_0}$ and $\ket{\Phi_0}=\ket{\Phi_+}^{\otimes n'}$. Then, for every $t\ge0$,
\begin{equation*}
    \frac12\left(1-e^{-\Gamma_{\mathrm{eff}}t}\right)
    \le
    p(t)
    \le
    1-e^{-\Gamma_{\mathrm{eff}}t}.
\end{equation*}
Moreover, $p(t)$ is nondecreasing in $t$.
\end{lemma}

\begin{proof}
The channel $e^{t\LindbladEff}$ is a Pauli channel. Thus, writing
\begin{equation*}
    e^{t\LindbladEff}(\rho)
    =
    \sum_{R\in G_{\mathrm{eff}}}\chi_{RR}(t)R\rho R,
\end{equation*}
where $G_{\mathrm{eff}}\coloneq\langle\mathcal S_{\mathrm{eff}}\rangle$ is the phase-stripped Pauli subgroup generated by $\mathcal S_{\mathrm{eff}}$, Bell sampling returns $R$ with probability $\chi_{RR}(t)$. Hence
\begin{equation*}
    p(t)=\sum_{R\in G_{\mathrm{eff}}\setminus\{I\}}\chi_{RR}(t).
\end{equation*}

We now rewrite this Pauli channel as a continuous-time random walk in the binary symplectic representation from Appendix~\ref{app:binary-symplectic}. Let $v_k\coloneq v_{Q_k}\in\mathbb F_2^{2n'}$ and let
\begin{equation*}
    V_{\mathrm{eff}}
    \coloneq
    \operatorname{span}_{\mathbb F_2}\{v_k:Q_k\in\mathcal S_{\mathrm{eff}}\}.
\end{equation*}
Let $N_k(t)$ be independent Poisson random variables with rates $\alpha_{kk}t$. Since multiplication of phase-stripped Paulis corresponds to addition of their binary symplectic vectors modulo $2$, the Bell-sampling outcome is distributed as
\begin{equation*}
    Y_t
    \coloneq
    \sum_{Q_k\in\mathcal S_{\mathrm{eff}}}
    \bigl(N_k(t)\bmod 2\bigr)v_k
    \in V_{\mathrm{eff}}.
\end{equation*}
In this notation, $p(t)=\Pr[Y_t\neq0]$.

The upper bound follows immediately. If $Y_t\neq0$, then at least one Poisson jump must have occurred. Since the total number of jumps is Poisson with rate $\Gamma_{\mathrm{eff}}t$,
\begin{equation*}
    p(t)
    \le
    \Pr[\text{at least one jump by time }t]
    =
    1-e^{-\Gamma_{\mathrm{eff}}t}.
\end{equation*}

For the lower bound, we use the Fourier analysis of Boolean functions \cite{o2014boolean}. Consider parity functions on $V_{\mathrm{eff}}$. We write them as
\begin{equation*}
    \pi_a(y)
    \coloneq
    (-1)^{\langle a,y\rangle_s},
\end{equation*}
where $a\in\mathbb F_2^{2n'}$ and the function is restricted to $V_{\mathrm{eff}}$. For such a parity function, define
\begin{equation*}
    \lambda_a
    \coloneq
    \sum_{\substack{Q_k\in\mathcal S_{\mathrm{eff}},\\\pi_a(v_k)=-1}}\alpha_{kk}.
\end{equation*}
Averaging over all parity functions on $V_{\mathrm{eff}}$, every nonzero $v_k$ has parity $-1$ for exactly half of them. Hence
\begin{equation*}
    \mathbb E_a[\lambda_a]
    =
    \frac12\sum_k\alpha_{kk}
    =
    \frac{\Gamma_{\mathrm{eff}}}{2}.
\end{equation*}
Therefore there exists a parity function $\pi_a$ such that $\lambda_a\ge \Gamma_{\mathrm{eff}}/2$.

For this parity function, define the two-state process $X_t^{(a)}\coloneq \pi_a(Y_t)$. It starts at $+1$ and flips precisely when a jump $v_k$ with $\pi_a(v_k)=-1$ occurs. The cumulative flip process is therefore Poisson with rate $\lambda_a$, so
\begin{equation*}
    \Pr[X_t^{(a)}=-1]
    =
    \frac{1-e^{-2\lambda_a t}}{2}.
\end{equation*}
Since $\pi_a(0)=1$, the event $\{X_t^{(a)}=-1\}$ implies $Y_t\neq0$. Thus
\begin{equation*}
    p(t)
    =
    \Pr[Y_t\neq0]
    \ge
    \Pr[X_t^{(a)}=-1]
    =
    \frac{1-e^{-2\lambda_a t}}{2}
    \ge
    \frac12\left(1-e^{-\Gamma_{\mathrm{eff}}t}\right),
\end{equation*}
where the last inequality uses $\lambda_a\ge\Gamma_{\mathrm{eff}}/2$.

It remains to prove monotonicity. By Fourier inversion of the indicator function $\mathbf 1_{Y_t =0}$ on $V_{\mathrm{eff}}$,
\begin{equation*}
    \Pr[Y_t=0]
    =
    \frac{1}{|V_{\mathrm{eff}}|}
    \sum_{\pi}
    \mathbb E[\pi(Y_t)]
    =
    \frac{1}{|V_{\mathrm{eff}}|}
    \sum_{\pi}
    e^{-2\lambda_\pi t},
\end{equation*}
where the sum is over all parity functions $\pi:V_{\mathrm{eff}}\to\{\pm1\}$ and $\lambda_\pi\coloneq\sum_{k:\pi(v_k)=-1}\alpha_{kk}\ge0$. Hence $\Pr[Y_t=0]$ is nonincreasing in $t$, and therefore $p(t)=1-\Pr[Y_t=0]$ is nondecreasing.
\end{proof}

\begin{lemma}[Pre-calibration of total effective dissipative rate]
\label{lem:precalibrate-diag-diss-rate}
Let $\LindbladEff$ be an $n'$-qubit Lindbladian with purely diagonal dissipator part
\begin{equation*}
    \LindbladEff(\rho)
    =
    \sum_{Q_k \in \mathcal S_\mathrm{eff} \,\subseteq\, \mathcal P_{n'} \setminus \{I\}} \alpha_{kk} \left(Q_k \rho Q_k - \rho\right),
\end{equation*}
where $\mathcal S_\mathrm{eff}$ is a set of distinct Hermitian Paulis and $\alpha_{kk}\ge 0$ for all $Q_k\in\mathcal S_{\mathrm{eff}}$. Define the total effective dissipative rate $\Gamma_{\mathrm{eff}} \coloneq \sum_{Q_k\in\mathcal S_{\mathrm{eff}}} \alpha_{kk}$. Fix an upper bound $\Lambda_D \ge \Gamma_{\mathrm{eff}}$, a relevance scale $\gamma_{\min}>0$ satisfying $\gamma_{\min}\le \Lambda_D$, and a failure probability $\delta\in(0,1)$. Let $\widetilde{\Gamma}_{\mathrm{eff}} \coloneq \max\{\Gamma_{\mathrm{eff}},\gamma_{\min}\}$.

Assume Bell-sampling access to $e^{t\LindbladEff}$: namely, for any chosen $t\ge0$, one may prepare $\rho_0\coloneq\ket{\Phi_0}\bra{\Phi_0}$ with $\ket{\Phi_0}\coloneq\ket{\Phi_+}^{\otimes n'}$, apply $e^{t\LindbladEff}$ to the left register, and measure in the Bell basis. Then there is a procedure $\texttt{EstimateGamma}$ which returns an estimate $\widehat{\Gamma}$ satisfying, with probability at least $1-\delta$,
\begin{equation}
\label{eq:precalibrate-diag-rate-guarantee}
    \widetilde{\Gamma}_{\mathrm{eff}}
    \le
    \widehat{\Gamma}
    \le
    4\widetilde{\Gamma}_{\mathrm{eff}}.
\end{equation}
The procedure uses Bell samples at a geometric grid of interrogation times from $1/\gamma_{\min}$ down to $1/(4\Lambda_D)$. Its total evolution time is
\begin{equation*}
    \BigO{
    \frac{1}{\gamma_{\min}}
    \log\left(
    \frac{\log(4\Lambda_D/\gamma_{\min})}{\delta}
    \right)
    },
\end{equation*}
and the number of Bell-sampling experiments is
\begin{equation*}
    \BigO{
    \log\left(\frac{4\Lambda_D}{\gamma_{\min}}\right)
    \log\left(
    \frac{\log(4\Lambda_D/\gamma_{\min})}{\delta}
    \right)
    }.
\end{equation*}
\end{lemma}

\begin{proof}
Let $p(t)$ denote the probability to Bell sample a non-identity Pauli outcome after time $t$. By \Cref{lem:nonidentity-bell-sampling-diag-diss}, for every $t\ge0$,
\begin{equation}
\label{eq:precalibration-p-bounds}
    \frac12\left(1-e^{-\Gamma_{\mathrm{eff}}t}\right)
    \le
    p(t)
    \le
    1-e^{-\Gamma_{\mathrm{eff}}t},
\end{equation}
and $p(t)$ is nondecreasing in $t$. Set $\epsilon\coloneq 1/100$ and define
\begin{equation*}
    p_{\mathrm{th}}
    \coloneq
    \frac12\left(1-e^{-1/4}+\frac12(1-e^{-3/4})\right).
\end{equation*}
One readily verifies that
\begin{equation}
\label{eq:pth-with-errors}
    1-e^{-1/4}+\epsilon
    <
    p_{\mathrm{th}}
    <
    \frac12(1-e^{-3/4})-\epsilon .
\end{equation}

Consider the geometric schedule
$t_\ell \coloneq \gamma_{\min}^{-1}(3/4)^\ell$ for $\ell=0,1,\dots,L,$
where $L$ is the smallest integer such that $t_L\le1/(4\Lambda_D)$. At each time $t_\ell$, run Bell sampling $N_0\coloneq C_{\mathrm{pc}}\log\left(\frac{L+1}{\delta}\right)$
times, and let $\widehat p_\ell$ denote the empirical frequency of non-identity outcomes. The procedure scans the times in decreasing order. If $\widehat p_0<p_{\mathrm{th}}$, it returns $\widehat{\Gamma}\coloneq\gamma_{\min}$. Otherwise, let $\ell_\star$ be the largest index such that $\widehat p_{\ell_\star}\ge p_{\mathrm{th}}$, and return
\begin{equation*}
    \widehat{\Gamma}
    \coloneq
    \frac{1}{t_{\ell_\star}}.
\end{equation*}

By Hoeffding's inequality and the union bound, choosing $C_{\mathrm{pc}}$ sufficiently large ensures that, with probability at least $1-\delta$, $|\widehat p_\ell-p(t_\ell)|\le \epsilon$ simultaneously for all $\ell=0,\dots,L$. We condition on this event for the remainder of the proof.

If $\Gamma_{\mathrm{eff}}t_\ell\le1/4$, then Eq.~\eqref{eq:precalibration-p-bounds} and Eq.~\eqref{eq:pth-with-errors} give
\begin{equation*}
    p(t_\ell)
    \le
    1-e^{-1/4}
    <
    p_{\mathrm{th}}-\epsilon,
\end{equation*}
and therefore $\widehat p_\ell<p_{\mathrm{th}}$. Hence the empirical test rejects at $\ell$. Conversely, if $\Gamma_{\mathrm{eff}}t_\ell\ge3/4$, then Eq.~\eqref{eq:precalibration-p-bounds} and Eq.~\eqref{eq:pth-with-errors} give
\begin{equation*}
    p(t_\ell)
    \ge
    \frac12(1-e^{-3/4})
    >
    p_{\mathrm{th}}+\epsilon,
\end{equation*}
and therefore $\widehat p_\ell\ge p_{\mathrm{th}}$. Hence the empirical test accepts at $\ell$.

If the procedure returns $\widehat{\Gamma}=\gamma_{\min}$, then the test rejected at $\ell=0$. Hence we cannot have $\Gamma_{\mathrm{eff}}t_0\ge3/4$, so $\Gamma_{\mathrm{eff}}<\gamma_{\min}$. Thus $\widetilde{\Gamma}_{\mathrm{eff}}=\gamma_{\min}$, and the returned estimate satisfies Eq.~\eqref{eq:precalibrate-diag-rate-guarantee}.

It remains to consider the case where the procedure stops at $\ell_\star$. Since the test accepts at $\ell_\star$, the rejection implication gives $\Gamma_{\mathrm{eff}}t_{\ell_\star}>1/4$, and therefore
\begin{equation*}
    \widehat{\Gamma}
    =
    \frac{1}{t_{\ell_\star}}
    <
    4\Gamma_{\mathrm{eff}}.
\end{equation*}
Moreover, $\ell_\star<L$, since $t_L\le1/(4\Lambda_D)$ and $\Gamma_{\mathrm{eff}}\le\Lambda_D$ imply $\Gamma_{\mathrm{eff}}t_L\le1/4$, so the final grid point rejects. By maximality of $\ell_\star$, the test rejects at $\ell_\star+1$. Hence we cannot have $\Gamma_{\mathrm{eff}}t_{\ell_\star+1}\ge3/4$. Since $t_{\ell_\star+1}=(3/4)t_{\ell_\star}$, this implies $\Gamma_{\mathrm{eff}}t_{\ell_\star}<1$, and therefore
\begin{equation*}
    \widehat{\Gamma}
    =
    \frac{1}{t_{\ell_\star}}
    >
    \Gamma_{\mathrm{eff}}.
\end{equation*}
Also, since $t_{\ell_\star}\le t_0=1/\gamma_{\min}$, we have $\widehat{\Gamma}\ge\gamma_{\min}$. Combining these bounds gives
\begin{equation*}
    \widetilde{\Gamma}_{\mathrm{eff}}
    \le
    \widehat{\Gamma}
    \le
    4\widetilde{\Gamma}_{\mathrm{eff}},
\end{equation*}
as claimed.

Finally, $L+1=\BigO{\log(4\Lambda_D/\gamma_{\min})}$, and each grid point uses $N_0=\BigO{\log((L+1)/\delta)}$ Bell samples. This gives
\begin{equation*}
    \BigO{
    \log\left(\frac{4\Lambda_D}{\gamma_{\min}}\right)
    \log\left(
    \frac{\log(4\Lambda_D/\gamma_{\min})}{\delta}
    \right)
    }
\end{equation*}
Bell-sampling experiments. Since the times $t_\ell$ decrease geometrically,
\begin{equation*}
    \sum_{\ell=0}^{L} N_0t_\ell
    \le
    4N_0\gamma_{\min}^{-1}
    =
    \BigO{
    \frac{1}{\gamma_{\min}}
    \log\left(
    \frac{\log(4\Lambda_D/\gamma_{\min})}{\delta}
    \right)
    },
\end{equation*}
which proves the claimed total evolution time.
\end{proof}

\begin{remark}[Implementation of pre-calibration using reshaped dynamics]
\label{rem:precalibration-reshaped-implementation}
In \Cref{alg:appendix-round-j-diss-struct-sampling}, the pre-calibration routine from \Cref{lem:precalibrate-diag-diss-rate} is not given direct access to the ideal semigroup $e^{t\LindbladEff^{(j)}}$. Instead, it is implemented using the round-$j$ reshaped dynamics $A_D^{(j)}(t,N)$ from Eq.~\eqref{eq:round-j-reshaped-dynamics}, which consists of physical Pauli twirling interleaved with repeated QEC recovery.

There are two minor implementation points. First, the use of $A_D^{(j)}(t,N)$ introduces a reshaping error. By \Cref{lem:N-step-reshaping-error},
\begin{equation*}
    \left\|
    A_D^{(j)}(t,N)
    -
    e^{t\LindbladEff^{(j)}}\circ \mathcal C_{\Pi_j}
    \right\|_\diamond
    \le
    \frac{t^2(\|\Lindblad\|_\diamond^2+\|\LindbladEff^{(j)}\|_\diamond^2)}{2N} \le \frac{5\Lambda_L^2t^2}{2N}.
\end{equation*}
where the second inequality follow from $\Lambda_L\ge\|\Lindblad\|_\diamond$, $\Gamma_{\mathrm{eff},j}\le\Lambda_L$, and
$\|\LindbladEff^{(j)}\|_\diamond\le2\Gamma_{j}^{\mathrm{eff}}\le2\Lambda_L$ due to the diagonal form of $\LindbladEff^{(j)}$.  Thus we choose the number of QEC-reshaping cycles as a function of the interrogation time:
\begin{equation*}
    N_{\mathrm{pc}}(t)
    \coloneq
    \left\lceil 500\,\Lambda_L^2 t^2 \right\rceil \le \BigO{\frac{\Lambda_L^2}{\gamma_{\min}^2}}.
\end{equation*}
where the second inequality is due to the longest time $t\le 1/\gamma_{\min}$. This achieves the reshaping error is at most $1/200$ for every interrogation time in the geometric pre-calibration grid. Since the grid is geometric, the total number of pre-calibration QEC-reshaping cycles is dominated by the largest interrogation time and is also $\BigO{\frac{\Lambda_L^2}{\gamma_{\min}^2}}$.

Since the diamond-norm distance upper bounds the additive error in every Bell-measurement probability, the proof of \Cref{lem:precalibrate-diag-diss-rate} goes through by estimating the empirical non-identity probability to accuracy $1/200$ and allocating the remaining $1/200$ to reshaping error.

The corresponding minimum evolution time is:
\begin{equation*}
    \frac{t}{N_\mathrm{pc}(t)} = \Omega\left(\frac{1}{t\Lambda_L^2}\right) \ge  \Omega\left(\frac{\gamma_{\min}}{\Lambda_L^2}\right)
\end{equation*}

Second, the implemented experiment uses the Bell state projected into the code corner algebra. This does not change the effective Bell-sampling probabilities once outcomes are grouped modulo stabilizers. Indeed, for a Pauli-channel coefficient $\chi_{rr}(t)$ corresponding to a logical Pauli $Q_r$, Eq.~\eqref{eq:prob-sample-Qr-modulo-stabilizers} gives
\begin{equation*}
    \Pr(\text{sample $Q_r$ modulo stabilizers})=\chi_{rr}(t).
\end{equation*}
Therefore the abstract pre-calibration lemma applies to the code-projected implementation with ``non-identity'' interpreted as a nontrivial stabilizer coset. Terms whose reshaped Pauli lies in the identity coset are excluded from $\Gamma_j^{\mathrm{eff}}$ and are handled separately by the identity-coset initialization in \Cref{alg:appendix-round-j-diss-struct-sampling}.
\end{remark}

\subsection{Deferred proofs for syndrome-based testing }\label{sec:deferred-diss-struct-testing-proofs}
Here, we present the deferred proofs related to dissipator structure testing via syndrome measurements (\Cref{thm:appendix-diss-rate-testing}). We restate the lemmas for convenience. 

\begin{restatelem}{lem:syndrome-hit-prob}[Syndrome hit probability for a diagonal Pauli jump]

Assume the Lindbladian $\LindbladDiag$ has the diagonal Pauli-dissipator form
as in \eqref{eq:lindblad-diag} with dissipator structure $\DiagDissStruct \subseteq \PnHerm$. Let $\{G_\ell\}_{\ell=1}^m \subseteq \mathcal P_{n'}$ be stabilizer generators with syndrome projectors $\{\Pi_s\}_{s\in\{0,1\}^m}$ as in \eqref{eq:syndrome-projectors}. Fix a non-identity
target Pauli $P_r$ and assume that its syndrome is nonzero, i.e. $r(P_r)\neq \mathbf 0$. Let $\rho=\Pi_0\rho\Pi_0 \in \mathfrak A_c \subseteq \mathcal{B}(\mathcal H_{n'})$ be any code state. Then for any $\tau\ge 0$,
\begin{equation}
\label{eq:syndrome-hit-linear-plus-remainder}
 p_r(\tau) \coloneq \Pr\bigl[\text{measuring syndrome }r(P_r)\bigr]
\;=\;\tr \bigl(\Pi_{r(P_r)}\,e^{\tau\LindbladDiag}(\rho)\bigr)
\;=\; \sum_{\substack{r(P_k)=r(P_r) \\ P_k \in \DiagDissStruct}}a_{kk}\,\tau \;+\; \Delta_r(\tau),
\end{equation}
where the remainder satisfies $\abs{\Delta_r(\tau)}\ \le\ \frac12\,\|\LindbladDiag\|_\diamond^2\,\tau^2.$
\end{restatelem}

\begin{proof}
By \Cref{lem:lindblad-taylor-tail}, $e^{\tau\LindbladDiag} = I + \tau \LindbladDiag + \int_0^\tau (\tau-s)\,\LindbladDiag^2 e^{s\LindbladDiag}\,ds,$ and therefore
\begin{equation}
\label{eq:pj-expansion}
    p_r(\tau)
    =
    \tr\bigl(\Pi_{r(P_r)} \rho\bigr)
    +
    \tau\,\tr\bigl(\Pi_{r(P_r)} \LindbladDiag(\rho)\bigr)
    +
    \tr\left(
        \Pi_{r(P_r)}
        \int_0^\tau (\tau-s)\,\LindbladDiag^2 e^{s\LindbladDiag}(\rho)\,ds
    \right).
\end{equation}
Since $\rho = \Pi_0 \rho \Pi_0$ is a code state and $r(P_r) \neq \mathbf 0$, we have $\Pi_{r(P_r)} \rho = 0$, so the first term vanishes.

We now evaluate the linear term. Using the diagonal Pauli-dissipator form,
\begin{equation*}
    \LindbladDiag(\rho)
    =
    \sum_{P_k \in \DiagDissStruct}
    a_{kk}
    \bigl(
        (P_k \otimes I_{n'-n}) \rho (P_k \otimes I_{n'-n})
        -
        \rho
    \bigr),
\end{equation*}
and again, since $\tr(\Pi_{r(P_r)}\rho)=0$, it follows that
\begin{equation*}
    \tr\bigl(\Pi_{r(P_r)} \LindbladDiag(\rho)\bigr)
    =
    \sum_{P_k \in \DiagDissStruct}
    a_{kk}\,
    \tr\bigl(\Pi_{r(P_r)} (P_k \otimes I_{n'-n}) \rho (P_k \otimes I_{n'-n})\bigr).
\end{equation*}
By syndrome projector shift property Eq.~\eqref{eq:syndrome-shift-rule} and trace cyclic property:
\begin{equation*}
    \tr\bigl(\Pi_{r(P_r)} \LindbladDiag(\rho)\bigr)= \sum_{P_k \in \DiagDissStruct}
    a_{kk}
    \tr\bigl((P_k \otimes I_{n'-n}) \Pi_{r(P_r)\oplus r(P_k)}\rho (P_k \otimes I_{n'-n})\bigr) = \sum_{P_k \in \DiagDissStruct}
    a_{kk}
    \tr\bigl(\Pi_{r(P_r)\oplus r(P_k)}\rho\bigr)
\end{equation*}
Since $\rho = \Pi_0 \rho \Pi_0$, the $k$th summand can be nonzero only if $r(P_k)=r(P_r)$. Together with $\tr(\rho)=1$ this yields
\begin{equation*}
    \tr\bigl(\Pi_{r(P_r)} \LindbladDiag(\rho)\bigr)
    =
    \sum_{\substack{r(P_k)=r(P_r) \\ P_k \in \DiagDissStruct}}
    a_{kk}
\end{equation*}

Define
\begin{equation*}
    \Delta_r(\tau)
    \coloneq
    \tr\left(
        \Pi_{r(P_r)}
        \int_0^\tau (\tau-s)\,\LindbladDiag^2 e^{s\LindbladDiag}(\rho)\,ds
    \right).
\end{equation*}
Then Eq.~\eqref{eq:pj-expansion} becomes
\begin{equation*}
    p_r(\tau)
    =
    \sum_{\substack{r(P_k)=r(P_r) \\ P_k \in \DiagDissStruct}}
    a_{kk}\tau + \Delta_r(\tau),
\end{equation*}
which proves Eq.~\eqref{eq:syndrome-hit-linear-plus-remainder}.

It remains to bound the remainder. Using Hölder's inequality and $\|\Pi_{r(P_r)}\|_\infty \le 1$,
\begin{equation*}
    |\Delta_r(\tau)|
    \le
    \left\|
        \int_0^\tau (\tau-s)\,\LindbladDiag^2 e^{s\LindbladDiag}(\rho)\,ds
    \right\|_1 \|\Pi_{r(P_r)}\|_\infty \le \left\|
        \int_0^\tau (\tau-s)\,\LindbladDiag^2 e^{s\LindbladDiag}(\rho)\,ds
    \right\|_1
\end{equation*}
Since $\|\rho\|_1=1$ and $\|e^{s\LindbladDiag}\|_\diamond=1$, we further obtain
\begin{equation*}
    |\Delta_r(\tau)|
    \le
    \left\|
        \int_0^\tau (\tau-s)\,\LindbladDiag^2 e^{s\LindbladDiag}\,ds
    \right\|_\diamond
    \le
    \int_0^\tau
    (\tau-s)\,
    \|\LindbladDiag\|_\diamond^2
    \|e^{s\LindbladDiag}\|_\diamond
    \,ds \le \frac{1}{2}\|\LindbladDiag\|_\diamond^2 \tau^2,
\end{equation*}
concluding the proof.
\end{proof}

\begin{restatelem}{lem:martingale-threshold-test}[Threshold test for adapted Bernoulli trials]
Let $X_1,\dots,X_N\in\{0,1\}$ be Bernoulli random variables, not necessarily independent, and let
$\widehat p \coloneq \frac{1}{N}\sum_{\ell=1}^N X_\ell$.
Let $\epsilon\in(0,1]$ and $\delta\in(0,1)$. Then there exists a universal constant
$C_{\mathrm{mg}}>0$ such that, if
$N \ge C_{\mathrm{mg}}\,\epsilon^{-1}\log(1/\delta)$, then:
\begin{enumerate}[label=(\roman*)]
    \item if $\mathbb E[X_\ell \mid X_1,\dots,X_{\ell-1}] \le \frac{5}{8}\epsilon$ for every $\ell\in[N]$, then
    $\Pr\left(\widehat p > \frac{3}{4}\epsilon\right) \le \delta$;
    \item if $\mathbb E[X_\ell \mid X_1,\dots,X_{\ell-1}] \ge \frac{7}{8}\epsilon$ for every $\ell\in[N]$, then
    $\Pr\left(\widehat p \le \frac{3}{4}\epsilon\right) \le \delta$.
\end{enumerate}
\end{restatelem}

\begin{proof}
    Let $S_N \coloneq \sum_{\ell=1}^N X_\ell.$ Since $X_\ell\in\{0,1\}$ for every $\ell$, both Corollary~6 and Corollary~14 of \cite{kuszmaul2025multiplicative} apply with $c=1$.
    
    For part (i), set $a_\ell \coloneq \frac{5}{8}\epsilon$ and $\mu \coloneq \sum_{\ell=1}^N a_\ell = \frac{5}{8}\epsilon N$. Then
    \begin{equation*}
        \frac{3}{4}\epsilon N = \frac{6}{5}\mu = \left(1+\frac{1}{5}\right)\mu.
    \end{equation*}
    Since $\mathbb E[X_\ell\mid X_1,\dots,X_{\ell-1}] \le a_\ell$ for all $\ell$, Corollary~6 of \cite{kuszmaul2025multiplicative} with deviation parameter $\frac{1}{5}$ gives
    \begin{equation*}
        \Pr\left(\widehat p > \frac{3}{4}\epsilon\right)
        \le
        \Pr\left(S_N \ge \left(1+\frac{1}{5}\right)\mu\right)
        \le
        \exp\left(-\frac{(1/5)^2}{2+1/5}\mu\right)
        =
        \exp\left(-\frac{\epsilon N}{88}\right).
    \end{equation*}
    Thus part (i) holds whenever $N \ge 88\,\epsilon^{-1}\log(1/\delta)$.
    
    For part (ii), set $a_\ell \coloneq \frac{7}{8}\epsilon$ and $\mu \coloneq \sum_{\ell=1}^N a_\ell = \frac{7}{8}\epsilon N$. Then
    \begin{equation*}
        \frac{3}{4}\epsilon N = \frac{6}{7}\mu = \left(1-\frac{1}{7}\right)\mu.
    \end{equation*}
    Since $\mathbb E[X_\ell\mid X_1,\dots,X_{\ell-1}] \ge a_\ell$ for all $\ell$, Corollary~14 of \cite{kuszmaul2025multiplicative} with deviation parameter $\frac{1}{7}$ gives
    \begin{equation*}
        \Pr\left(\widehat p \le \frac{3}{4}\epsilon\right)
        \le
        \Pr\left(S_N \le \left(1-\frac{1}{7}\right)\mu\right)
        \le
        \exp\left(-\frac{(1/7)^2}{2}\mu\right)
        =
        \exp\left(-\frac{\epsilon N}{112}\right).
    \end{equation*}
    Thus part (ii) holds whenever $N \ge 112\,\epsilon^{-1}\log(1/\delta)$.
    
    Therefore both conclusions hold with the universal constant $C_{\mathrm{mg}} = 112$.
\end{proof}

%% file: appendix/hamiltonian-HDD-learning.tex
\section{\texorpdfstring{\HND{} (\HNDAbbrev{})}{Hamiltonian-Disjoint-from-Dissipator (HDD)} Learning}
\label{sec:appendix-hamilt-learning}

Let us write down the unknown $n$-qubit Lindbladian yet again:
\begin{equation}
\label{eq:lindblad-eq-hamilt}
    \Lindblad(\rho)
    =
    -i\sum_{P_k\in \HamiltStruct} h_k \comm{P_k}{\rho}
    +
    \sum_{P_k,P_j \in \DiagDissStruct}
    a_{kj}
    \left(
        P_k \rho P_j
        -
        \tfrac12\acomm{P_j P_k}{\rho}
    \right).
\end{equation}

In general, Markovian dissipation precludes Heisenberg-limited learning of arbitrary Hamiltonian coefficients~\cite{zhou2018achieving,demkowicz2017qecmetrology}. If the dissipator is known exactly, one can in principle attempt to learn the Hamiltonian-not-in-the-Lindblad-span component (HNLS~\cite{zhou2018achieving,demkowicz2017qecmetrology,gorecki2020optimal,abbasgholinejad2026error}) at Heisenberg-limited scaling. In the ansatz-free setting considered here, however, the dissipator is itself unknown, and learning it is standard-quantum-limited~\cite{brady2026precision}, restricting the Heisenberg-learnable Hamiltonian target even further. We refer to this more restricted set of Hamiltonian targets as the \HND{} (\HNDAbbrev{}) Pauli set, defined below.

Here, we develop a protocol (\Cref{alg:hamilt-hier-learning}) for learning the Hamiltonian coefficients associated with the \HNDAbbrev{} Pauli set. Later in Appendix~\ref{sec:hdd-boundary}, we show that the \HNDAbbrev{} set is in fact maximal: as $\varepsilon\to0$, no strictly larger set of Hamiltonian Pauli targets admits a uniform Heisenberg-limited learning guarantee in the ansatz-free black-box setting.

\begin{definition}[Dissipator footprint]
\label{def:diss-footprint}
The dissipator structure $\DiagDissStruct$ induces the \emph{full dissipator footprint}
\begin{equation*}
    \FullDissStruct
    \coloneq
    \DiagDissStruct
    \cup
    \bigl\{
        \Herm(P_\ell P_m):
        P_\ell,P_m\in\DiagDissStruct,\,
        P_\ell\neq P_m
    \bigr\}.
\end{equation*}
Thus, $\FullDissStruct$ contains both the Paulis appearing directly in the dissipator and all Hermitian Paulis generated, up to phase, by pairwise products of dissipator Paulis.
\end{definition}

\begin{definition}[\HND{} (\HNDAbbrev{})]
\label{def:hamilt-not-in-diss}
Given the full dissipator footprint from \cref{def:diss-footprint}, we define the \emph{\HND{} (\HNDAbbrev{})} Pauli set as
\begin{equation*}
    \HNDStruct
    \coloneq
    \HamiltStruct\setminus\FullDissStruct.
\end{equation*}
\end{definition}

Our algorithm learns the coefficients of every Pauli in $\HNDStruct$ at the Heisenberg limit. At finite target accuracy, however, one can generally learn a larger Hamiltonian sector if the obstructing dissipative directions are sufficiently weak. This motivates finite-resolution versions of the above structures.

\begin{definition}[Dissipator footprint, $\eta$-heavy]
\label{def:diss-footprint-eta}
For a resolution scale $\eta>0$, define the $\eta$-heavy dissipator structure by
\begin{equation*}
    \EtaDiagDissStruct
    \coloneq
    \{P_k\in\DiagDissStruct:a_{kk}\ge\eta\}.
\end{equation*}
The corresponding $\eta$-heavy dissipator footprint is
\begin{equation*}
    \EtaFullDissStruct
    \coloneq
    \EtaDiagDissStruct
    \cup
    \bigl\{
        \Herm(P_\ell P_m):
        P_\ell,P_m\in\EtaDiagDissStruct,\,
        P_\ell\neq P_m
    \bigr\}.
\end{equation*}
\end{definition}

\begin{definition}[\HND{} (\HNDAbbrev{}) at scale $\eta$]
\label{def:hamilt-not-in-diss-eta}
Given the $\eta$-heavy dissipator footprint from \cref{def:diss-footprint-eta}, define
\begin{equation*}
    \EtaHNDStruct
    \coloneq
    \HamiltStruct\setminus\EtaFullDissStruct.
\end{equation*}
Since $\EtaFullDissStruct\subseteq\FullDissStruct$, one has
\begin{equation*}
    \HNDStruct\subseteq\EtaHNDStruct.
\end{equation*}
The additional targets are precisely the Hamiltonian Paulis lying in the weak dissipator footprint,
\begin{equation*}
    \EtaHNDStruct\setminus\HNDStruct
    =
    \HamiltStruct
    \cap
    \bigl(
        \FullDissStruct\setminus\EtaFullDissStruct
    \bigr).
\end{equation*}
\end{definition}

In this section, we first prove a conditional Heisenberg-limited learning
result. Assuming the $\eta$-heavy dissipator structure
$\EtaDiagDissStruct$ is known for $\eta$ sufficiently small,
\Cref{alg:hamilt-hier-learning} learns the support and coefficients of
the corresponding \HND{} Hamiltonian terms to additive accuracy
$\varepsilon$, without prior knowledge of the Hamiltonian structure.
We then remove the known-dissipator assumption in
Appendix~\ref{sec:appendix-removing-known-dissipator-assumption} by
composing this routine with the dissipator structure learner from
Appendix~\ref{sec:appendix_dissipator_structure_learning}. This yields
a Heisenberg-limited learner of all coefficients associated with
$\HNDStruct$ (\cref{cor:ansatz-free-hnd}). At finite resolution, the
algorithm may additionally learn Hamiltonian coefficients in the weak
dissipator footprint. In general this requires
$\eta\propto\varepsilon^2$
(\Cref{cor:ansatz-free-hnd-worst-case}); under an additional regularity
assumption, one may instead take $\eta\propto\varepsilon$, yielding
Heisenberg-limited learning outside the learned dissipator footprint
with no prior knowledge of either the Hamiltonian or dissipator
structures (\Cref{cor:ansatz-free-hnd-balanced-tail}).

At a high level, \Cref{alg:hamilt-hier-learning} parallels the dissipator structure learner from Appendix~\ref{sec:diss-struct-hierarchical-learning}. It samples the strongest remaining \HNDAbbrev{} terms, estimates their coefficients to remove false positives, updates the error-correcting code to suppress the identified terms, and iterates on the weaker residual Hamiltonian. The reshaping scheme uses only Pauli check measurements and applications of Pauli operators, eliminating any need for quantum controls that simulate inverse-time evolution.

The main result of this section is summarized in the following theorem.
The exact role of the bias and residual parameters
$B_\infty,K_\infty,R_\diamond$ used in
\Cref{thm:appendix-hier-hamilt_learning} is described in
\Cref{thm:appendix-round-j-hamilt-coeff-learn,thm:appendix-round-j-hamilt-struct-learn},
but intuitively they control the errors caused by unknown dissipator
terms. In Appendix~\ref{sec:appendix-removing-known-dissipator-assumption},
we bound these quantities for the finite-resolution guarantees and give
a separate favorable-code argument for the full \HNDAbbrev{} (from \cref{def:hamilt-not-in-diss}) guarantee.

\begin{theorem}[Hierarchical \HNDAbbrev{} Hamiltonian learning]
\label{thm:appendix-hier-hamilt_learning}
    Let $\Lindblad$ be an $n$-qubit $M$-sparse Lindbladian as in Eq.~\eqref{eq:lindblad-eq-hamilt}. Fix $\varepsilon,\delta\in(0,1)$ and assume black-box access to $e^{t\Lindblad}$. Assume the $\eta$-heavy dissipator structure $\EtaDiagDissStruct$ is given, and that it satisfies the following uniform small-bias/small-residual condition. For every successful QEC code draw used by \Cref{alg:appendix-round-j-hamilt-struct-learning} (and reused by \Cref{alg:appendix-round-j-hamilt-coeff-learning}), the corresponding effective generator $\LindbladEff$ written in the form of Eq.~\eqref{eq:generic-effective-generator-hnd} has bias and residual parameters from Eq.~\eqref{eq:generic-bias-residual-def} satisfying
    \begin{equation*}
        B_\infty\le \varepsilon/50,
        \qquad
        K_\infty\le \varepsilon/50,
        \qquad
        R_\diamond\le \frac{\varepsilon}{8 \sqrt2 \pi}.
    \end{equation*}
    
    Then running \Cref{alg:hamilt-hier-learning} with $\VCorrect=\EtaDiagDissStruct$ returns coefficient estimates $\widehat h_{\mathrm{\HNDAbbrev},\eta}$ for the \HND{} Hamiltonian at scale $\eta$ satisfying the following guarantees.
    
    \begin{enumerate}[leftmargin=*, label=\arabic*.]
    \item \textit{(Accuracy)}
    With probability at least $1-\delta$,
    \begin{equation*}
        \|\widehat h_{\mathrm{\HNDAbbrev},\eta}
        -
        h_{\mathrm{\HNDAbbrev},\eta}\|_\infty
        \le \varepsilon .
    \end{equation*}
    
    \item \textit{(Total evolution time)}
    The total evolution time under $e^{t\Lindblad}$ is
    \begin{equation*}
        t_{\mathrm{tot}}
        =
        \SoftBigO{\frac{M^2}{\varepsilon}} .
    \end{equation*}
    
    \item \textit{(Minimum evolution time)}
    The procedure only applies $e^{t\Lindblad}$ for times at least
    \begin{equation*}
        t_{\mathrm{min}}
        =
        \Omega\left(\frac{\varepsilon}{M^3}\right).
    \end{equation*}
    
    \item \textit{(Number of experiments)}
    The total number of experiments is
    \begin{equation*}
        \SoftBigO{M^3},
    \end{equation*}
    dominated by Bell sampling during the candidate-identification stage.
    
    \item \textit{(Ancilla cost)}
    The procedure uses at most
    \begin{equation*}
        n+\BigO{\log M}
    \end{equation*}
    noiseless ancilla qubits in addition to the $n$ probe qubits.
    
    \item \textit{(Classical overhead)}
    The classical processing cost is
    \begin{equation*}
        \SoftBigO{\frac{M^6+nM^5+n^2M^4}{\varepsilon^2}} .
    \end{equation*}
    The first and second terms come from real-time QEC processing in the candidate-identification stage, while the last term comes from logical-Pauli sampling in coefficient learning.
    \end{enumerate}
\end{theorem}

\begin{proof} 
    See Appendix~\ref{sec:hamilt-hier-learning-algorithm}.
\end{proof}

\begin{remark}[Ancilla-reduced \HNDAbbrev{} Hamiltonian learner]
    In Appendix~\ref{sec:appendix-reducing-the-ancilla-cost}, we also briefly describe an ancilla-reduced variant of \Cref{thm:appendix-hier-hamilt_learning} that improves the total ancilla count from  $n+\BigO{\log M}$ to $\BigO{\log M}$ at the cost of increased black-box evolution time $t_{\mathrm{tot}} = \SoftBigO{\frac{M^4\log(n)}{\varepsilon}}$ following \cite{hu2025ansatz}.
\end{remark}

This section is organized as follows:
\begin{itemize}
    \item \textbf{\HND{} hierarchy \& notation.} 
    In Appendix~\ref{sec:hamilt-hier-learning-notation}, we describe the hierarchical learning procedure and introduce the notation used throughout the conditional construction. 

    \item \textbf{Candidate identification via Bell sampling.} 
    In Appendix~\ref{sec:appendix-hamilt-struct-sampling}, we present \Cref{alg:appendix-round-j-hamilt-struct-learning}, which samples the strongest unknown \HNDAbbrev{} terms via Bell sampling, and prove its guarantees in \Cref{thm:appendix-round-j-hamilt-struct-learn}. 

    \item \textbf{\HNDAbbrev{} coefficient learning via robust frequency estimation.} 
    In Appendix~\ref{sec:appendix-hamilt-coeff-learning}, we present \Cref{alg:appendix-round-j-hamilt-coeff-learning}, which estimates the coefficients of the candidate Hamiltonian terms and removes the false positives produced by the \HNDAbbrev{} candidate-identification step. 

    \item \textbf{\HND{} hierarchical learning algorithm.} 
    In Appendix~\ref{sec:hamilt-hier-learning-algorithm}, we combine \HND{} candidate identification (\Cref{alg:appendix-round-j-hamilt-struct-learning}) and coefficient learning (\Cref{alg:appendix-round-j-hamilt-coeff-learning}) into the full hierarchical algorithm \Cref{alg:hamilt-hier-learning}, which learns the \HNDAbbrev{} Hamiltonian with Heisenberg-limited interaction time.

    \item \textbf{Removing the known dissipator structure assumption.}
    In Appendix~\ref{sec:appendix-removing-known-dissipator-assumption}, we lift the known dissipator-structure assumption by invoking the learner from Appendix~\ref{sec:appendix_dissipator_structure_learning}.

    \item \textbf{Deferred proofs.}
    In Appendix~\ref{sec:deferred-hamilt-struct-sampling-proofs}, we provide the deferred proofs for the \HNDAbbrev{} candidate-identification guarantees.
\end{itemize}

\subsection{Hierarchy and notation} 
\label{sec:hamilt-hier-learning-notation}

Throughout the conditional \HNDAbbrev{} learning analysis, we take the $\eta$-heavy dissipator structure $\EtaDiagDissStruct$ as input. Similar to dissipator structure learning (Appendix~\ref{sec:appendix_dissipator_structure_learning}), the \HNDAbbrev{} learner uses QEC-based Lindbladian reshaping (Eq.~\eqref{eq:QEC-reshaping} and Eq.~\eqref{eq:logical-twirling-QEC-reshaping}) together with a hierarchical learning protocol adapted from \cite{hu2025ansatz}. The role of $\eta$ is to control the light dissipative terms not corrected by QEC: after reshaping, these terms appear as a residual Lindbladian and as a small Hamiltonian bias.

The hierarchy is formed by splitting the learning problem according to the following definition:
\begin{definition}[Round-$j$ \HND{} set]
\label{def:round-j-hamilt-set}
Given the \HND{} set $\EtaHNDStruct$ from \Cref{def:hamilt-not-in-diss-eta} and target accuracy $\varepsilon>0$, define the round-$j$ \HNDAbbrev{} Pauli set as
\begin{equation}
    \HHierSet_j
    \coloneq
    \bigl\{P_k\in \EtaHNDStruct :
    2^{-(j+1)} < |h_k| \le 2^{-j}\bigr\},
    \qquad
    j=0,1,\ldots,\lceil \log_2(1/\varepsilon)\rceil-1 .
\end{equation}
\end{definition}

By construction, learning all sets $\HHierSet_j$ and the corresponding Hamiltonian coefficients to accuracy $\varepsilon$ is sufficient for learning the entire \HND{} at scale $\eta$.

Hierarchical learning proceeds in rounds indexed by $j$, where round $j$ uses interrogation time $T_j = 2^jT$ and targets the bin $\HHierSet_j$. The algorithm starts from the largest \HND{} terms and then moves to progressively weaker bins. At the beginning of each round, we choose a QEC code that corrects the known $\eta$-heavy dissipator structure $\EtaDiagDissStruct$ and suppresses the \HND{} Hamiltonian terms learned in previous rounds. Within the round, we first use Bell sampling to identify a candidate set containing the strongest remaining terms, and then use coefficient learning to estimate the candidate coefficients and remove false positives. The newly confirmed terms are then added to the suppressed set, the interrogation time is doubled, and the procedure is repeated. Since the QEC code is drawn randomly, the Bell-sampling step is repeated with multiple independent code draws, ensuring that with high probability every $P_k\in\HHierSet_j$ is probed by at least one QEC instance.

More concretely, for each round $j$, let 
$\widehat{\mathcal S}_{\mathrm{\HNDAbbrev{}},\eta}^{(j-1)} \subseteq \EtaHNDStruct$
denote the set of \HND{} Paulis identified in the previous rounds. Throughout the hierarchical learning algorithm, we maintain the invariant
\begin{equation}
\label{eq:hnd-round-invariant}
    \{P_k\in\EtaHNDStruct:\ |h_k|> 2^{-j}\}
    \subseteq
    \widehat{\mathcal S}_{\mathrm{\HNDAbbrev{}},\eta}^{(j-1)} .
\end{equation}

The \HNDAbbrev{} candidate-identification and coefficient-learning steps use different reshaping primitives, but both are analyzed through the same effective-generator interface. In each case, the reshaped dynamics on $\mathcal B(\mathcal H_{n'})$ is described, using the full-space realization from \Cref{rem:full-space-effective-generators}, by an effective Lindbladian of the form
\begin{equation}
\label{eq:generic-effective-generator-hnd}
    \LindbladEff(\rho)
    =
    -i[H_{\mathrm{tar}} + H_{\mathrm{bias}},\rho]
    +
    \mathcal L_{\mathrm{resid}}(\rho).
\end{equation}
For candidate identification, $\LindbladEff$ is obtained from QEC reshaping alone (\Cref{lem:qec-reshaping-generator}); here $H_{\mathrm{tar}}$ is the Hamiltonian signal that survives the QEC reshaping. For coefficient learning of a candidate Pauli $P_a$, $\LindbladEff$ is obtained from QEC reshaping followed by a partial logical twirl (\Cref{lem:qec-reshape-logical-twirl-generator}); in this case, with $\overline P_a \coloneq P_a\otimes I_{n'-n}$, the target Hamiltonian is $H_{\mathrm{tar}}=h_a\overline P_a$. In both cases, $H_{\mathrm{bias}}$ denotes the Hamiltonian bias induced by corrected--uncorrected dissipator cross terms, and $\mathcal L_{\mathrm{resid}}$ denotes the residual dissipator supported on uncorrected light dissipator Paulis. We write the residual dissipator in the aggregate Pauli-Kossakowski form:
\begin{equation}
\label{eq:generic-residual-pauli-kossakowski}
    \mathcal L_{\mathrm{resid}}(\rho)
    =
    \sum_{Q_k,Q_j\in\mathcal S_{\mathrm{D,res}}}
    \alpha_{kj}
    \left(
        Q_k \rho Q_j
        -
        \frac12\{Q_jQ_k,\rho\}
    \right),
\end{equation}
where $\mathcal S_{\mathrm{D,res}}\subseteq\mathcal P_{n'}$ is the residual Pauli support and
$\alpha\succeq 0$ is the corresponding Kossakowski matrix. Explicit expressions for the $\LindbladEff$ generators are given in Eq.~\eqref{eq:Leff-qec-reshaping} and Eq.~\eqref{eq:Leff-qec-reshaping-logical-twirl}. Since the coefficients $\alpha_{kj}$ are aggregate, they need not coincide with the original Kossakowski entries $a_{kj}$ appearing in Eq.~\eqref{eq:Leff-qec-reshaping} and Eq.~\eqref{eq:Leff-qec-reshaping-logical-twirl}.

For the Bell sampling analysis, we also use the quantum jump unraveling. In particular, define
\begin{equation}
\label{eq:generic-residual-no-jump-operator}
    K_{\mathrm{resid}}
    \coloneq
    \sum_{Q_k,Q_j\in\mathcal S_{\mathrm{D,res}}}
    \alpha_{kj} Q_jQ_k .
\end{equation}
The no-jump branch of the effective Lindblad evolution is generated by the non-Hermitian operator
$H_{\mathrm{tar}}+H_{\mathrm{bias}}-\frac{i}{2}K_{\mathrm{resid}}$.

To bound the bias and residual errors in Eq.~\eqref{eq:generic-effective-generator-hnd}, we define the following parameters:
\begin{equation}
\label{eq:generic-bias-residual-def}
    B_\infty \coloneq \|H_{\mathrm{bias}}\|_{\infty,\mathrm{coeff}},
    \qquad
    R_\diamond \coloneq \|\mathcal L_{\mathrm{resid}}\|_{\diamond},
    \qquad
    K_\infty \coloneq \|K_{\mathrm{resid}}\|_{\infty,\mathrm{coeff}} .
\end{equation}
where $\|\cdot\|_{\infty,\mathrm{coeff}}$ denotes the maximum absolute Pauli coefficient (including the identity in the case of $K_{\mathrm{resid}}$). The candidate-identification and coefficient-learning subsections state their admissible upper bounds on $B_\infty$, $R_\diamond$ and $K_\infty$. In Appendix~\ref{sec:appendix-removing-known-dissipator-assumption}, we choose $\eta$ so that the corresponding bounds hold uniformly over all rounds and code draws for the finite-resolution
guarantees and use a separate favorable-code argument for the full
\HNDAbbrev{} guarantee.

\subsection{Candidate identification via Bell sampling} 
\label{sec:appendix-hamilt-struct-sampling}

We now describe \Cref{alg:appendix-round-j-hamilt-struct-learning},
the candidate-identification subroutine for round $j$. Under the round
invariant Eq.~\eqref{eq:hnd-round-invariant}, all \HND{} terms above
scale $2^{-j}$ have already been identified, and the goal is to output
a candidate set containing the current bin $\HHierSet_j$. For the
conditional analysis in this subsection, we instantiate the QEC
reshaping routine from \Cref{lem:qec-reshaping-generator} with

\begin{equation}
\label{eq:round-j-bell-qec-targets}
    \VCorrect = \EtaDiagDissStruct,
    \qquad
    \VDetect =
    \widehat{\mathcal S}_{\mathrm{\HNDAbbrev{}},\eta}^{(j-1)} .
\end{equation}
Thus, the $\eta$-heavy dissipator terms are corrected, while previously learned \HND{} Hamiltonian terms are detected and suppressed.

The procedure has a \emph{two-level coupon-collector} structure. First, we draw random stabilizer codes via \Cref{alg:qec_for_lindbladian_reshaping}. For each $P_k\in\HHierSet_j$, a fresh code makes $P_k$ survive, i.e. $r(P_k)=0$, as a nontrivial \emph{logical} operator with non-negligible probability. Repeating the code draw ensures, with high probability, that every target term survives the QEC construction at least once. This differs from dissipator structure identification in Appendix~\ref{sec:diss-structure-sampling}, where the unknown terms of interest need not have zero syndrome: detectability is sufficient to suppress Hamiltonian terms to leading order, whereas a dissipative Pauli jump with nonzero syndrome can still contribute after recovery as a logical Pauli. Second, conditioned on survival, Bell sampling on the projected EPR state must hit one of the stabilizer-coset representatives of the survived logical Pauli. Repeating the Bell measurements for each code ensures that we see at least one representative for every survived term. Together, these two coverage steps yield a candidate set $\widehat{\HHierSet}_j$ that contains $\HHierSet_j$ with probability at least $1-\delta_j$.

\begin{algorithm}[H]
    \caption{\HNDAbbrev{} Hamiltonian candidate identification subroutine, $j$th round}
    \label{alg:appendix-round-j-hamilt-struct-learning}
    \DontPrintSemicolon
    \SetKwInOut{Input}{Inputs}
    \SetKwInOut{Output}{Output}
    \SetKwBlock{Schedule}{Schedule}{end}
    \SetKwFunction{RandomQEC}{RandomQECLindbladianReshaping}

    \Input{
      \begin{enumerate}[label=(\arabic*), leftmargin=*, nosep]
        \item Hierarchical round index $j$.
        \item Access to $n$-qubit evolution $e^{t\Lindblad}$
      (Eq.~\eqref{eq:lindblad-eq-hamilt}) and a known sparsity bound
      $M\ge M_H+M_D^2$.
        \item Target correction set   $\VCorrect\subseteq\DiagDissStruct$.
        \item Target detection set $\VDetect\subseteq\PnHerm\setminus\{I\}$.
        \item Confidence parameter $\delta_j\in(0,1)$.
        \item Access to $n+5\lceil \log_2(2M)\rceil$ noiseless ancilla qubits in addition to the $n$ probe qubits.
      \end{enumerate}
    }

    \Output{
    A candidate set $\widehat{\HHierSet}_j\subseteq\mathcal P_n$,
    together with, for each candidate, a QEC code under which it was
    observed as a nontrivial logical Pauli.
    }

    \Schedule{
      $N_{\mathrm{codes}} \gets \left\lceil32M\ln\left(\frac{3M}{\delta_j}\right)\right\rceil$ \tcp*{\# of random QEC draws}
      $N_{\mathrm{Bell}} \gets 210M^2$ \tcp*{\# of Bell-basis measurements per code}
      $N_{\mathrm{cand}}\gets840M$
      \tcp*{maximum number of distinct candidates retained per code}
      $N_{\mathrm{reshape}} \gets 600M^2 4^j$ \tcp*{\# of error-correction cycles per sample}
      $n' \gets n + \frac{5}{2}\left\lceil \log_2(2M)\right\rceil$ \tcp*{left-register size including QEC ancillas}
      $t \gets 2^j/(2M)$,\quad $\tau \gets t/N_{\mathrm{reshape}}$ \tcp*{interrogation time and reshape step}
    }

    \BlankLine
    $\widehat{\HHierSet}_j \gets \emptyset$ \tcp*{candidate Pauli set}

    \For{$i \gets 1$ \KwTo $N_{\mathrm{codes}}$}{
        $(\mathcal G_i, \RMap_i) \gets \texttt{RandomQEC}(\VDetect,\VCorrect, M,\,\frac{\delta_j}{3N_{\mathrm{codes}}}, 2)$ \tcp*{Alg.~\ref{alg:qec_for_lindbladian_reshaping} stab. generator \& recovery ch.}
        $\Pi_i \gets \prod_{G\in\mathcal G_i} (1+G)/2$ \tcp*{code-space projector}
        $\mathcal C_i\gets\emptyset$ \tcp*{candidates retained from code $i$}
        \For{$r \gets 1$ \KwTo $N_{\mathrm{Bell}}$}{
            Prepare $\ket{\Phi_0} \gets \ket{\Phi_+}^{\otimes n'}$ on $L\otimes R$ \tcp*{left and right registers}
            Project to the code corner algebra:
            $\rho_{\mathrm{prep}} \propto (\Pi_i\otimes I_{n'})\ket{\Phi_0}\bra{\Phi_0}(\Pi_i\otimes I_{n'})$ \tcp*{$\SoftBigO{M}$ attempts}
            
            \For{$k \gets 1$ \KwTo $N_{\mathrm{reshape}}$}{
                Evolve the left register by $e^{\tau \Lindblad}\otimes I_{n'-n}$\;
                Apply $\RMap_i$ to the left register \tcp*{error detection and correction}
            }
            
            Measure in the Bell basis and determine the
            nontrivial logical probe-Pauli representative $P$\;
            \If{$P$ exists, $P\notin\mathcal C_i$, and
                $|\mathcal C_i|<N_{\mathrm{cand}}$}{
                $\mathcal C_i\gets\mathcal C_i\cup\{P\}$\;
            }
        }
        For every $P\in\mathcal C_i\setminus\widehat{\HHierSet}_j$,
        set $\mathrm{QEC}_P\gets(\mathcal G_i,\RMap_i)$\;
        $\widehat{\HHierSet}_j \gets \widehat{\HHierSet}_j \cup \mathcal C_i$\;
    }

    \Return
    $(\widehat{\HHierSet}_j,
    \{\mathrm{QEC}_P\}_{P\in\widehat{\HHierSet}_j})$
\end{algorithm}

The key ingredient of this algorithm is the following lemma, which we invoke on the reshaped dynamics. 

\begin{lemma}[Hamiltonian Bell sampling under bias and residual dissipation]
\label{lem:bell-sampling-of-eff-hamilt-struct}
Let $\LindbladEff$ be an $n'$-qubit effective generator of the form in
Eq.~\eqref{eq:generic-effective-generator-hnd}, with bias and residual parameters $B_\infty,K_\infty$ as in
Eq.~\eqref{eq:generic-bias-residual-def}. Write it in the jump unraveling picture using Eq.~\eqref{eq:generic-bias-residual-def} and Eq.~\eqref{eq:generic-residual-no-jump-operator}:

\begin{equation*}
    \LindbladEff(\rho)
    =
    -iH_{\rm eff}\rho +i\rho H_{\rm eff}^\dagger + \sum_{w,m}\alpha_{wm}P_w\rho P_m, \qquad H_{\rm eff}\coloneq H_{\rm tar}+H_{\rm bias}-\frac i2K_{\rm resid}.
\end{equation*}
Expand the target Hamiltonian in the Pauli basis as $H_{\mathrm{tar}}=\sum_{P_m\in\mathcal S_{\mathrm{H,tar}}} h_mP_m$, the bias as $H_{\mathrm{bias}}=\sum_{P_m\in\mathcal S_{\mathrm{H,bias}}} b_mP_m$ and the residual non-Hermitian contribution as $K_{\mathrm{resid}}= \sum_{P_m\in\mathcal S_{K}} k_mP_m$, where
$\mathcal S_{\mathrm{H,tar}},\mathcal S_{\mathrm{H,bias}}\subseteq\mathcal P_{n'}\setminus\{I\}$, $\mathcal S_K\subseteq\mathcal P_{n'}$ and $\{h_m\}, \{b_m\}, \{k_m\}$ are real coefficients.
Let $\mathcal S_{\mathrm{H,eff}}\coloneq\mathcal S_{\mathrm{H,tar}}\cup\mathcal S_{\mathrm{H,bias}}\cup \mathcal S_K$ and assume $|\mathcal S_{\mathrm{H,eff}}|\le M$.

Fix an integer $j\ge 0$. Assume that $|h_m|\le 2^{-j}$ for every
$P_m\in\mathcal S_{\mathrm{H,tar}}$ and that a target Pauli
$P_r\in\mathcal S_{\mathrm{H,tar}}$ satisfies $|h_r|\ge 2^{-(j+1)}$. Assume that the bias and residual parameters satisfy $B_\infty\le 2^{-(j+1)}/25$ and $K_\infty\le 2^{-(j+1)}/25.$

Let $C$ be a stabilizer code on $n'$ qubits with code projector $\Pi$ and stabilizer subgroup
$S\subseteq\mathcal P_{n'}$. Let $\mathcal S_{\mathrm{eff}} \coloneq \mathcal S_{\mathrm{H,eff}} \cup \mathcal S_{\mathrm{D,res}}$ and let $G_{\mathrm{eff}}\coloneq
\langle\mathcal S_{\mathrm{eff}}\rangle\subseteq\mathbb P_{n'}$ denote the subgroup generated by $\mathcal S_{\mathrm{eff}}$. Assume that all Paulis
$P\in\mathcal S_{\mathrm{eff}}$ have zero syndrome $r(P)=0$ with respect to $C$, and that no
nontrivial product of elements of $\mathcal S_{\mathrm{eff}}$ is proportional to a stabilizer; equivalently,
for every $Q\in G_{\mathrm{eff}}$, $Q\propto s\in S$ implies $s=I_{n'}$.

Let $\rho_0\coloneq\ket{\Phi_0}\bra{\Phi_0}$, where
$\ket{\Phi_0}\coloneq\ket{\Phi_+}^{\otimes n'}$ and
$\ket{\Phi_+}=(\ket{00}+\ket{11})/\sqrt{2}$. Let $\widetilde\rho$ denote the state immediately before Bell-basis measurement,
\begin{equation*}
    \widetilde \rho
    \coloneq
    (e^{t\LindbladEff}\otimes I_{n'})
    \left(
        \frac{(\Pi\otimes I_{n'})\rho_0(\Pi\otimes I_{n'})}
        {\tr((\Pi\otimes I_{n'})\rho_0)}
    \right).
\end{equation*}
For any Pauli $P\in\mathbb P_{n'}$, let
$O_P\coloneq(P\otimes I_{n'})\ket{\Phi_0}\bra{\Phi_0}(P^\dagger\otimes I_{n'})$
denote the corresponding Bell-measurement projector. Then, for
$t=2^j/(2M)$, a single Bell-sampling experiment on $\widetilde\rho$ satisfies
\begin{equation*}
    \begin{aligned}
        \Pr(\text{sample $P_r$ modulo stabilizers})
        &\coloneq
        \tr\left(\sum_{s\in S} O_{sP_r}\,\widetilde\rho\right)
        \ge
        \frac{1}{200M^2},\\
        \Pr(\text{sample a stabilizer})
        &\coloneq
        \tr\left(\sum_{s\in S} O_s\,\widetilde\rho\right)
        \ge
        1-\frac{1}{2M}.
    \end{aligned}
\end{equation*}
\end{lemma}

\begin{proof}
    See Appendix~\ref{sec:deferred-hamilt-struct-sampling-proofs}.
\end{proof}

\begin{theorem}[Round-$j$ \HNDAbbrev{} candidate identification via Bell sampling]
\label{thm:appendix-round-j-hamilt-struct-learn}
Let $\Lindblad$ be an $n$-qubit $M$-sparse Lindbladian as in Eq.~\eqref{eq:lindblad-eq-hamilt}. Assume black-box access to $e^{t\Lindblad}$. Fix a hierarchical round $j$ and a failure probability $\delta_j\in(0,1)$. Let $\widehat{\mathcal S}_{\mathrm{\HNDAbbrev{}},\eta}^{(j-1)}$ denote the set of \HNDAbbrev{} terms identified in previous rounds and assume that $\widehat{\mathcal S}_{\mathrm{\HNDAbbrev{}},\eta}^{(j-1)}$ satisfies the round invariant Eq.~\eqref{eq:hnd-round-invariant}. 

In \Cref{alg:appendix-round-j-hamilt-struct-learning}, instantiate
$\VCorrect=\EtaDiagDissStruct$ and
$\VDetect=\widehat{\mathcal S}_{\mathrm{\HNDAbbrev{}},\eta}^{(j-1)}$ for QEC reshaping. For every successful QEC code draw (\Cref{alg:qec_for_lindbladian_reshaping}), let $\LindbladEff$ denote the resulting $n'$-qubit effective generator, written in the form of Eq.~\eqref{eq:generic-effective-generator-hnd}. Suppose the corresponding bias and no-jump residual parameters $B_\infty,K_\infty$ from Eq.~\eqref{eq:generic-bias-residual-def} satisfy $B_\infty\le 2^{-(j+1)}/25$ and $K_\infty\le 2^{-(j+1)}/25$. Then running \Cref{alg:appendix-round-j-hamilt-struct-learning} with schedule
\begin{equation*}
    t=\frac{2^j}{2M},
    \qquad
    N_{\mathrm{codes}}=\BigO{M\log(M/\delta_j)},
    \qquad
    N_{\mathrm{Bell}}=\BigO{M^2},
    \qquad
    N_{\mathrm{reshape}}=\BigO{M^2 4^j}
\end{equation*}
outputs a candidate set $\widehat{\HHierSet}_j\subseteq\mathcal P_n$ satisfying the following guarantees:
\begin{enumerate}[leftmargin=*, label=\arabic*., nosep]
\item \textit{(Coverage and size)} With probability at least $1-\delta_j$,
\begin{equation*}
    \widehat{\HHierSet}_j
    \supseteq
    \HHierSet_j\setminus
    \widehat{\mathcal S}_{\mathrm{\HNDAbbrev{}},\eta}^{(j-1)},
    \qquad
    |\widehat{\HHierSet}_j|
    =
    \BigO{M^2\log(M/\delta_j)} .
\end{equation*}

\item \textit{(Total evolution time)}
The total evolution time under $e^{t\Lindblad}$ is
\begin{equation*}
    t_{\mathrm{tot}}
    =
    N_{\mathrm{codes}}N_{\mathrm{Bell}}t
    =
    \BigO{2^j M^2\log(M/\delta_j)} .
\end{equation*}

\item \textit{(Minimum evolution time)}
The procedure only applies $e^{\Lindblad t}$ for times at least
\begin{equation*}
    t_{\mathrm{min}}
    \coloneq
    \tau
    =
    \frac{t}{N_{\mathrm{reshape}}}
    \ge
    \Omega\left(\frac{1}{M^3 2^j}\right).
\end{equation*}

\item \textit{(Number of experiments)}
The number of Bell-sampling experiments is
\begin{equation*}
    N_{\exp}
    =
    N_{\mathrm{codes}}N_{\mathrm{Bell}}
    =
    \BigO{M^3\log(M/\delta_j)} .
\end{equation*}
The number of interleaved QEC rounds is larger by a factor $\BigO{M^24^j}$, while the number of raw state-preparation attempts is larger by a factor $\SoftBigO{M}$ due to the probabilistic projection to the code space before reshaping.

\item \textit{(Ancilla cost)}
The procedure uses at most
\(
    n+5\lceil\log_2(2M)\rceil
\)
ancillary qubits in addition to the $n$ probe qubits.

\item \textit{(Classical overhead)}
The classical cost is dominated by repeated QEC and is
\begin{equation*}
    \SoftBigO{(n+M)M^5 4^j}.
\end{equation*}
\end{enumerate}
\end{theorem}

\begin{proof}
    Most of the proof concerns item $(1)$, namely coverage of the candidate set. 
    Items $(2)$--$(4)$ follow directly from the schedule, and items $(5)$--$(6)$ are verified at the end.
    
    We first prove coverage. The candidate-identification subroutine has a \emph{two-level coupon-collector} structure. The first level draws random error-correcting codes until every term $P\in \HHierSet_j\setminus \widehat{\mathcal S}_{\mathrm{\HNDAbbrev{}},\eta}^{(j-1)}$
    survives the reshaping, i.e. has zero syndrome $r(P)=0$ for at least one code. The second level, conditioned on a code in which a target term survives, repeats Bell sampling until one stabilizer-coset representative of that survived logical Pauli is observed.
    
    We now analyze the first level. In \Cref{alg:appendix-round-j-hamilt-struct-learning}, the QEC reshaping map is
    \begin{equation}
    \label{eq:Hamilt-struc-learning-QEC-map}
        A_{\tau,H}^{I}
        \coloneq
        \RMap \circ \bigl(e^{\Lindblad \tau}\otimes I_{n'-n}\bigr) \circ \mathcal C_\Pi ,
    \end{equation}
    where the original Lindbladian acts on the $n$ system qubits and is trivially extended to the $n'-n$ QEC ancillas. The recovery map $\RMap$ is obtained from a random stabilizer code drawn by \Cref{alg:qec_for_lindbladian_reshaping} with $c=2$ and inputs
    \begin{equation*}
        \VCorrect = \EtaDiagDissStruct,
        \qquad
        \VDetect = \widehat{\mathcal S}_{\mathrm{\HNDAbbrev{}},\eta}^{(j-1)},
        \qquad
        M_{\mathrm{det}}=M .
    \end{equation*}
    Let $E_{\mathrm{tar}}$ denote the success event from \Cref{thm:random-qec-for-reshaping}: the sampled code corrects all Paulis in $\VCorrect$ and detects all Paulis in $\VDetect$. By the same theorem, the number of stabilizer generators is $\lceil \log_2(2M)\rceil$, and therefore the stabilizer group satisfies
    \begin{equation}
    \label{eq:hamilt-struct-sampling-stab-size}
        |S|\le 4M .
    \end{equation}
    
    Conditioned on $E_{\mathrm{tar}}$, \Cref{lem:qec-reshaping-generator} and the effective-generator notation in Eq.~\eqref{eq:generic-effective-generator-hnd} imply that the effective generator of the reshaping map in Eq.~\eqref{eq:Hamilt-struc-learning-QEC-map}, realized on $\mathcal B(\mathcal H_{n'})$ as in \Cref{rem:full-space-effective-generators}, can be written as
    \begin{equation}
    \label{eq:hamilt-struct-learning-Leff}
        \LindbladEff(\rho)
        =
        -i[H_{\mathrm{tar}}+H_{\mathrm{bias}},\rho]
        +\mathcal L_{\mathrm{resid}}(\rho) 
        =
        -i\sum_{\substack{P_k\in\HamiltStruct\\ r(P_k)=0}}
        h_k[P_k\otimes I_{n'-n},\rho]
        -i[H_{\mathrm{bias}},\rho]
        +\mathcal L_{\mathrm{resid}}(\rho).
    \end{equation}
    Here $H_{\mathrm{tar}}$ is the Hamiltonian signal surviving QEC reshaping; $H_{\mathrm{bias}}$ and $\mathcal L_{\mathrm{resid}}$ are controlled by the bias and residual parameters $B_\infty$ and $K_\infty$ from Eq.~\eqref{eq:generic-bias-residual-def}. The explicit expression for $\LindbladEff$ is given in Eq.~\eqref{eq:Leff-qec-reshaping}.
    
    For a nonidentity Hermitian Pauli $P$, let $A_P$ denote the event that $P$ has zero syndrome under the sampled checks. Recall the target-suppression set from Eq.~\eqref{eq:target-suppression-set}. With the present choices of $\VCorrect$ and $\VDetect$, it becomes
    \begin{equation*}
        \mathcal W_{\mathrm{tar}}
        \coloneq
        \VDetect \cup \VCorrect \cup \{\Herm(P'P'') : P',P''\in \VCorrect,\ P'\neq P''\}
        =
        \widehat{\mathcal S}_{\mathrm{\HNDAbbrev{}},\eta}^{(j-1)}
        \cup \EtaFullDissStruct .
    \end{equation*}
   By \Cref{thm:random-qec-for-reshaping} with $c=2$ and $M_{\mathrm{det}}=M$, any given $P_k \in 
    \HamiltStruct \setminus \mathcal W_{\mathrm{tar}} = \EtaHNDStruct \setminus \widehat{\mathcal S}_{\mathrm{\HNDAbbrev{}},\eta}^{(j-1)} \supseteq \HHierSet_j\setminus \widehat{\mathcal S}_{\mathrm{\HNDAbbrev{}},\eta}^{(j-1)}$ survives with conditional probability
    \begin{equation}
    \label{eq:hamilt-struct-sampling-prob-of-survival}
        p_{\mathrm{surv}}
        \coloneq
        \Pr[A_{P_k}\mid E_{\mathrm{tar}}]
        \ge \frac{1}{8M}.
    \end{equation}
    This completes the first coupon-collector layer. 
    
    To quantify the reshaping error, we use  \Cref{cor:N-step-reshaping-error-sparse} with $\tau \coloneq t/N_{\mathrm{reshape}}$:
    \begin{equation}\label{eq:hamilt-struct-sampling-reshaping-error}
    \|(A_{\tau, H}^{I})^{N_{\mathrm{reshape}}} - e^{t\LindbladEff}\circ \mathcal C_\Pi\|_\diamond
    \le \frac{t^2(\|\Lindblad\|_\diamond^2 + \|\LindbladEff\|_\diamond^2)}{2N_{\mathrm{reshape}}} \le \frac{4t^2M^2}{N_{\mathrm{reshape}}}.
    \end{equation}
    thus, going forward in our analysis, we calculate the measurement statistics as if we had access to $e^{t\LindbladEff}$ and then account for the reshaping error separately.

    For the second level of the coupon collector, suppose that the $i$-th QEC code draw satisfies $E_{\mathrm{tar}}$, and fix a particular $P_r \in \HHierSet_j\setminus \widehat{\mathcal S}_{\mathrm{\HNDAbbrev{}},\eta}^{(j-1)}$ that survives this draw, i.e. $A_{P_r}$ holds. We are interested in the probability of measuring $P_r$ using Bell sampling. For that, we use \Cref{lem:bell-sampling-of-eff-hamilt-struct}, but first we argue why the assumptions of this lemma are satisfied in our setting.
    
    Conditioned on $E_{\mathrm{tar}}$, the QEC-reshaped generator is $\LindbladEff$ from Eq.~\eqref{eq:hamilt-struct-learning-Leff}, which has the form required by \Cref{lem:bell-sampling-of-eff-hamilt-struct}. The effective Hamiltonian support has size at most $M$: every Hamiltonian term in $H_{\mathrm{tar}}$ is inherited from a nonzero Hamiltonian coefficient, and the terms in $H_{\mathrm{bias}}$ and $K_\mathrm{resid}$ are inherited from dissipator cross term in the original $M$-sparse Lindbladian (see Eq.~\eqref{eq:Leff-qec-reshaping}).
    
    The coefficient assumptions also hold. Since $\VDetect=\widehat{\mathcal S}_{\mathrm{\HNDAbbrev{}},\eta}^{(j-1)}$, \Cref{lem:qec-reshaping-generator} removes all previously learned \HND{} terms from $H_{\mathrm{tar}}$. Combining this with the round invariant Eq.~\eqref{eq:hnd-round-invariant}, every Pauli term appearing in $H_{\mathrm{tar}}$ has coefficient at most $2^{-j}$. On the other hand, because $P_r\in\HHierSet_j$, its coefficient satisfies $|h_r|>2^{-(j+1)}$. Finally, the theorem assumes the Bell-admissible bias and residual bounds $B_\infty\le 2^{-(j+1)}/25$ and $K_\infty\le 2^{-(j+1)}/25$.
    
    We also need to verify that no nontrivial product of effective Hamiltonian Pauli terms is proportional to a stabilizer. This is guaranteed by the $Z$-padding in \Cref{alg:qec_for_lindbladian_reshaping}: every nonidentity stabilizer $\sigma\in S\setminus\{I\}$ acts nontrivially on the padding ancillas, while all Pauli terms inherited from the original Lindbladian act trivially on these ancillas. Hence, for every $Q\in G_{\mathrm{eff}}$, the implication $Q\propto s\in S \Rightarrow s=I_{n'}$ holds, as required by \Cref{lem:bell-sampling-of-eff-hamilt-struct}.
    
    We now instantiate the Bell-sampling experiment. Let $\rho_0\coloneq\ket{\Phi_0}\bra{\Phi_0}$, where $\ket{\Phi_0}\coloneq\ket{\Phi_+}^{\otimes n'}$ and $\ket{\Phi_+}=(\ket{00}+\ket{11})/\sqrt{2}$. The left half of $\rho_0$ is projected into the code corner algebra:
    \begin{equation*}
        \rho_{\mathrm{code}}
        \coloneq
        \frac{(\Pi_i\otimes I_{n'})\rho_0(\Pi_i\otimes I_{n'})}{\tr((\Pi_i\otimes I_{n'})\rho_0)} .
    \end{equation*}
    Operationally, this is implemented by measuring the stabilizers on the left register and postselecting on the trivial syndrome. Using Eq.~\eqref{eq:hamilt-struct-sampling-stab-size}, the success probability of this post-selection is
    \begin{equation}
    \label{eq:hamilt-struct-sampling-prob-projection}
        \tr((\Pi_i\otimes I_{n'})\rho_0)
        =
        \frac{1}{|S|}
        \ge
        \frac{1}{4M}.
    \end{equation}
    
    For the ideal effective evolution under
    $(e^{t\LindbladEff}\otimes I_{n'})$ with $t=2^j/(2M)$, \Cref{lem:bell-sampling-of-eff-hamilt-struct} gives
    \begin{equation}
    \label{eq:hamilt-struct-sampling-prob-bell-sampling-ideal}
        \Pr_{\mathrm{ideal}}(\text{sample $P_r$ modulo stabilizers})
        \ge
        \frac{1}{200M^2}.
    \end{equation}

     Accounting for the reshaping diamond norm error from Eq.~\eqref{eq:hamilt-struct-sampling-reshaping-error} and setting $N_{\mathrm{reshape}}\ge 600M^2\,4^j$, we obtain
    \begin{equation}
    \label{eq:hamilt-struct-sampling-prob-bell-sampling}
        \Pr(\text{sample $P_r$ modulo stabilizers}\mid A_{P_r}\wedge E_{\mathrm{tar}})
        \ge
        \frac{1}{200M^2}-\frac{1}{600M^2}
        =
        \frac{1}{300M^2}.
    \end{equation} 
    This completes the second coupon-collector layer.
    
    We can guarantee the success of the algorithm by enforcing the following
three events:
\begin{enumerate}[label=\(\mathsf{\Alph*}\), leftmargin=*]
    \item all $N_{\mathrm{codes}}$ random code constructions succeed,
    i.e.\ satisfy the target event $E_{\mathrm{tar}}$
    $\rightarrow$ Eq.~\eqref{eq:hamilt-struct-sampling-event-A};
    
    \item all $N_{\mathrm{codes}}N_{\mathrm{Bell}}$ experiments succeed
    at preparing the code-projected Bell states
    $\rightarrow$ Eq.~\eqref{eq:hamilt-struct-sampling-event-B};
    
    \item every Pauli
    $P\in\HHierSet_j\setminus
    \widehat{\mathcal S}_{\mathrm{\HNDAbbrev{}},j-1}$
    is retained in the candidate set at least once
    $\rightarrow$ Eq.~\eqref{eq:hamilt-struct-sampling-event-C}.
\end{enumerate}

We first show that the per-code candidate cap preserves the coverage
guarantee. Fix
$P_r\in\HHierSet_j\setminus
\widehat{\mathcal S}_{\mathrm{\HNDAbbrev{}},j-1}$
and consider a returned code under which $P_r$ survives.
By Eq.~\eqref{eq:hamilt-struct-sampling-prob-bell-sampling}, each Bell
experiment samples $P_r$ modulo stabilizers with probability at least
$1/(300M^2)$. Hence, for
$N_{\mathrm{Bell}}=210M^2$,
\begin{equation*}
    \Pr(\text{$P_r$ is sampled at least once})
    \ge
    1-\left(1-\frac{1}{300M^2}\right)^{210M^2}
    \ge
    1-e^{-7/10}
    >
    \frac12 .
\end{equation*}

On the other hand, the stabilizer-probability bound in
\Cref{lem:bell-sampling-of-eff-hamilt-struct}, together with
Eq.~\eqref{eq:hamilt-struct-sampling-reshaping-error}, implies that
each Bell experiment has nontrivial-outcome probability at most $1/(2M)+1/(600M^2)
    \le
    1/M.$
Let $X$ denote the total number of nontrivial Bell outcomes among the
$N_{\mathrm{Bell}}$ experiments for this code. Then
$\mathbb E[X]\le210M$, so Markov's inequality gives
\begin{equation*}
    \Pr(X>840M)
    \le
    \frac14.
\end{equation*}
Whenever $X\le840M$, the cap
$N_{\mathrm{cand}}=840M$ cannot discard any distinct nontrivial
candidate that was observed. Therefore, without requiring independence
between these events,
\begin{equation}
\label{eq:hamilt-struct-sampling-retained-per-surviving-code}
    \Pr(
        \text{$P_r$ is retained}
        \mid
        A_{P_r}\wedge E_{\mathrm{tar}}
    )
    \ge
    1-\frac12-\frac14
    =
    \frac14 .
\end{equation}

Combining
Eq.~\eqref{eq:hamilt-struct-sampling-prob-of-survival} and
Eq.~\eqref{eq:hamilt-struct-sampling-retained-per-surviving-code},
a fixed target is retained from a single returned code with probability
at least
\begin{equation*}
    \frac{1}{8M}\cdot\frac14
    =
    \frac{1}{32M}.
\end{equation*}
Using a union bound over at most $M_H\le M$ target terms and
$N_{\mathrm{codes}}
\ge32M\ln(3M/\delta_j)$, we obtain
\begin{equation}
\label{eq:hamilt-struct-sampling-event-C}
    \Pr(\neg\mathsf C\mid\mathsf A\wedge\mathsf B)
    \le
    M\left(1-\frac{1}{32M}\right)^{N_{\mathrm{codes}}}
    \le
    \frac{\delta_j}{3}.
\end{equation}

The event $\mathsf A$ succeeds with probability at least
$1-\delta_j/3$ by running each random code construction
\Cref{alg:qec_for_lindbladian_reshaping} with failure probability
$\delta_j/(3N_{\mathrm{codes}})$. By a union bound,
\begin{equation}
\label{eq:hamilt-struct-sampling-event-A}
    \Pr(\mathsf A)
    \ge
    1-\frac{\delta_j}{3}.
\end{equation}

Let
$N_{\mathrm{exp}}
\coloneq
N_{\mathrm{codes}}N_{\mathrm{Bell}}
=
\BigO{M^3\log(M/\delta_j)}$.
To guarantee that the code-space-projected state preparation succeeds
for every experiment, we attempt preparation and projection $N_{\mathrm{prep}} = 4M\ln\left(3N_{\mathrm{exp}}/\delta_j\right)$
times per experiment. Using a union bound over experiments and
Eq.~\eqref{eq:hamilt-struct-sampling-prob-projection},
\begin{equation}
\label{eq:hamilt-struct-sampling-event-B}
    \Pr(\neg\mathsf B)
    \le
    N_{\mathrm{codes}}N_{\mathrm{Bell}}
    \left(1-\frac{1}{4M}\right)^{N_{\mathrm{prep}}}
    \le
    \frac{\delta_j}{3}.
\end{equation}

Combining
Eq.~\eqref{eq:hamilt-struct-sampling-event-A},
Eq.~\eqref{eq:hamilt-struct-sampling-event-B}, and
Eq.~\eqref{eq:hamilt-struct-sampling-event-C}, we obtain
\begin{equation*}
    \Pr(\text{success})
    \ge
    1
    -
    \Pr(\neg\mathsf A)
    -
    \Pr(\neg\mathsf B)
    -
    \Pr(\neg\mathsf C\mid\mathsf A\wedge\mathsf B)
    \ge
    1-\frac{3\delta_j}{3}
    \ge
    1-\delta_j .
\end{equation*}

Finally, the per-code cap gives the deterministic candidate-set bound
\begin{equation*}
    |\widehat{\HHierSet}_j|
    \le
    N_{\mathrm{cand}}N_{\mathrm{codes}}
    =
    \BigO{M^2\log(M/\delta_j)}.
\end{equation*}
This completes the proof of item $(1)$.

We now prove item $(5)$. By
\Cref{thm:random-qec-for-reshaping} with $c=2$ and
$M_{\mathrm{det}}=M$, our QEC construction uses at most
$5/2\lceil\log_2(2M)\rceil$ ancillas for stabilizers and syndrome
extraction. Forming Bell pairs uses at most an additional
$n+5/2\lceil\log_2(2M)\rceil$ ancillas. Thus the total number is
$n+5\lceil\log_2(2M)\rceil$.

The classical overhead (item $(6)$) is dominated by repeated syndrome
extraction and correction and is
\begin{equation*}
    \text{Classical cost}
    =
    \BigO{
        (|S|+n)
        N_{\mathrm{exp}}
        N_{\mathrm{reshape}}
    }
    =
    \SoftBigO{(n+M)M^5 4^j}.
\end{equation*}
This concludes the proof of the theorem.
\end{proof}

\subsection{Coefficient learning via robust frequency estimation} \label{sec:appendix-hamilt-coeff-learning}

In this section, we describe \HND{} coefficient learning at the Heisenberg limit. Given the candidate set $\widehat{\HHierSet}_j$ produced by Bell sampling, the goal is twofold: estimate the coefficients of the true round-$j$ terms and remove the false positives. This produces the updated set $\widehat{\mathcal S}_{\mathrm{\HNDAbbrev{}},\eta}^{(j)}\subseteq \EtaHNDStruct$ needed to maintain the round invariant in Eq.~\eqref{eq:hnd-round-invariant}.

For each candidate Pauli $P_a$, the coefficient-learning routine uses QEC reshaping followed by a partial logical twirl, as in Eq.~\eqref{eq:logical-twirling-QEC-reshaping}, to isolate the corresponding logical Hamiltonian signal. In the notation of Eq.~\eqref{eq:generic-effective-generator-hnd}, the resulting effective generator has target Hamiltonian $H_{\mathrm{tar}}=h_a\overline P_a$, where $\overline P_a=P_a\otimes I_{n'-n}$, together with a bias Hamiltonian and residual dissipator of sizes $B_\infty$ and $R_\diamond$ from Eq.~\eqref{eq:generic-bias-residual-def}. When $B_\infty$ and $R_\diamond$ are below the error budget specified below, the residual dynamics gives only a controlled perturbation of the ideal single-frequency signal. We can then apply Robust Frequency Estimation (RFE) within the code to estimate $h_a$ with Heisenberg-limited interaction time. This extends the Hamiltonian coefficient-learning methods of \cite{huang2023learning,hu2025ansatz,mirani2024learning,ma2024learning} to the present dissipative setting.

 At a high level, RFE is a binary search algorithm that starts with an initial coefficient guess $\hat h_a \in [-A, A]$ and iteratively narrows down this range using a test statistic \(F(t)\) built from expectation values; after $\BigO{\log(1/\varepsilon)}$ steps it returns \(\hat h_a\) with \(|\hat h_a- h_a|\le \varepsilon\). The main requirement on \(F(t)\) is the separation property in Lemma~\ref{lem:rfe-lemma} (copied from \cite{ma2024learning,hu2025ansatz}), which our algorithm satisfies for encoded (logical) dynamics.

\begin{lemma}[Lemma 10 of \cite{ma2024learning}]\label{lem:rfe-lemma}
Let \( \mu \in [a,b] \). Let \( F(t) \) be a random variable such that
\[
|F(t) - e^{i\mu t}| \le \tfrac{1}{2}.
\]
Then we can correctly conclude whether \( \mu \in [a, (a+2b)/3] \) or \( \mu \in [(2a+b)/3, b] \) using a single sample of \( F(t) \) at \( t = \pi / (b - a) \).
\end{lemma}

\begin{proof}
Set \( t = \pi / (b - a) \) and define the real-valued function
\[
f(\mu) := \sin\left( \frac{\pi}{b - a} \left( \mu - \frac{a + b}{2} \right) \right).
\]

Apply phase shift $e^{-i \frac{(a+b)\pi}{2(b - a)}}$ to $(F(t) - e^{i\mu t})$ and take the imaginary part:
\[
\left| \operatorname{Im}\left( e^{-i \frac{(a+b)\pi}{2(b - a)}} F\left( \frac{\pi}{b - a} \right) \right) - f(\mu) \right| \le \frac{1}{2}.
\]
This centers the signal $F(t)$ such that the sign of its imaginary part distinguishes between the two cases:
\[
\begin{aligned}
&\text{If }\operatorname{Im}\left( e^{-i \frac{(a+b)\pi}{2(b - a)}} F\left( \frac{\pi}{b - a} \right) \right) \le 0,\ \text{then } f(\mu) \le \frac{1}{2},\ \text{which implies } \mu \in \left[a, \frac{a+2b}{3} \right].\\[4pt]
&\text{If }\operatorname{Im}\left( e^{-i \frac{(a+b)\pi}{2(b - a)}} F\left( \frac{\pi}{b - a} \right) \right) > 0,\ \text{then } f(\mu) \ge -\frac{1}{2},\ \text{which implies } \mu \in \left[\frac{2a+b}{3}, b \right].
\end{aligned}
\]
\end{proof}

With QEC-based Lindbladian reshaping (\Cref{lem:qec-reshape-logical-twirl-generator} and Eq.~\eqref{eq:generic-effective-generator-hnd}) and the RFE (\Cref{lem:rfe-lemma}), we now have all the necessary ingredients for learning \HNDAbbrev{} coefficients.

\begin{algorithm}[H]
    \caption{\HNDAbbrev{} Hamiltonian coefficient learning subroutine}
    \label{alg:appendix-round-j-hamilt-coeff-learning}
    \DontPrintSemicolon
    \SetKwInOut{Input}{Inputs}
    \SetKwInOut{Output}{Output}
    \SetKwBlock{Schedule}{Schedule}{end}

    \Input{
          \begin{enumerate}[label=(\arabic*), leftmargin=*, nosep]
            \item Candidate nonidentity Pauli $P_a\in\PnHerm\setminus\{I\}$
            whose Hamiltonian coefficient $h_a$ we wish to learn.
            \item Accuracy parameter $\varepsilon\in(0,1)$.
            \item Confidence parameter $\delta_\mathrm{coeff}\in(0,1)$.
            \item QEC code with stabilizer $S$ and recovery channel $\RMap$
            under which
            $\overline P_a\coloneq P_a\otimes I_{n'-n}$ is a nontrivial
            logical Pauli.
            \item Access to $n$-qubit evolution $e^{t\Lindblad}$
            (Eq.~\eqref{eq:lindblad-eq-hamilt}), where the total
            Lindbladian sparsity is bounded by $M$.
            \item Access to $\BigO{\log M}$ noiseless ancilla qubits used for
            QEC padding and check measurements.
          \end{enumerate}
        }

    \Output{
    An estimator $\widehat h_a$ of the Hamiltonian coefficient $h_a$.
    }

    \Schedule{
        $(x,y)\gets (-2,2)$ \tcp*{initial range ansatz}
        $L\gets \left\lceil \log_{3/2}(4/\varepsilon)\right\rceil$ \tcp*{\# of RFE iterations}
        $N_{\mathrm{exp}}
        \gets 2\left\lceil64\ln(4L/\delta_{\mathrm{coeff}})\right\rceil$ \tcp*{\# of samples per iteration}
        $\overline T \gets \text{any logical Pauli s.t. } \acomm{\overline T}{\overline P_a}=0$ \tcp*{used for state preparation}
    }

    \For{$i \gets 0$ \KwTo $L-1$}{
        $t_i\gets (\pi/8)(3/2)^i$ \tcp*{physical evolution time}
        $N_{\mathrm{reshape}} \gets \left\lceil 32\sqrt2\,M^2t_i^2\right\rceil$,\quad $\tau_i \gets t_i/N_{\mathrm{reshape}}$ \tcp*{\# reshape steps \& step length}
        $\hat{o}_s^{+},\hat{o}_s^{-} \gets 0,0$ \;
    
        \For{$r \gets 1$ \KwTo $N_{\mathrm{exp}}$}{
            Prepare $\rho_c \gets \RMap(\ket{0}\bra{0}^{\otimes n'})$ \tcp*{state-prep to the code}
            $v \gets$ record logical measurement of $\overline T;$ the new state is $\rho_v$ \tcp*{$+1$ or $-1$}
            $\rho(t_i) \gets \prod_{k=1}^{N_{\mathrm{reshape}}}(\mathcal U_{\overline Q_k}\circ \RMap \circ e^{\Lindblad \tau_i} \circ \mathcal U_{\overline Q_k})(\rho_v)$ \tcp*{$\overline Q_k\in\overline{\mathcal K}_{\overline P_a}$ uniformly}
            $\hat{o}_s^{+} \gets \hat{o}_s^{+} + v\,\cdot$ logical measurement of $\overline T$ on $\rho(t_i)$ \tcp*{for half of $N_{\mathrm{exp}}$ sample "cosine"}
            $\hat{o}_s^{-} \gets \hat{o}_s^{-} + v\,\cdot$ logical measurement of $(i\overline T\,\overline P_a)$ on $\rho(t_i)$ \tcp*{for half of $N_{\mathrm{exp}}$ sample "sine"}
        }
    
        $\widehat F \gets (\hat{o}_s^{+}+i\,\hat{o}_s^{-})/(N_{\mathrm{exp}}/2)$ \tcp*{estimator of $e^{2i\omega_a t_i}$}
    
        \If{\(\operatorname{Im}\left(e^{-i (x+y)t_i}\,\widehat F\right) \le 0\)}{
          \(y \gets (x+2y)/3\)\;
        }
        \Else{
          \(x \gets (2x+y)/3\)\;
        }   
    }
    \Return $\widehat h_a \gets (x+y)/2$
\end{algorithm}

\begin{theorem}[\HNDAbbrev{} coefficient learning via QEC reshaping and RFE]
\label{thm:appendix-round-j-hamilt-coeff-learn}
Fix a target accuracy $\varepsilon\in(0,1)$ and failure probability $\delta_\mathrm{coeff}\in(0,1)$. Fix a candidate Pauli $P_a\in \PnHerm\setminus\EtaFullDissStruct$. Assume that the QEC reshaping plus partial logical twirl used in \Cref{alg:appendix-round-j-hamilt-coeff-learning} produces an $n'$-qubit effective generator $\LindbladEff$ of the form in Eq.~\eqref{eq:generic-effective-generator-hnd}, with target Hamiltonian
$H_{\mathrm{tar}}=h_a\overline P_a$ and bias $H_{\mathrm{bias}}=b_a\overline P_a$, where $\overline P_a\coloneq P_a\otimes I_{n'-n}\in\mathcal P_{n'}$ is a nontrivial logical Pauli. Assume additionally that the corresponding bias and residual parameters $B_\infty,R_\diamond$ from Eq.~\eqref{eq:generic-bias-residual-def} satisfy $B_\infty\le \varepsilon/2$ and $R_\diamond\le \varepsilon/(4\sqrt{2}\pi)$. 

Then running \Cref{alg:appendix-round-j-hamilt-coeff-learning} outputs an estimator $\widehat h_a$ satisfying the following guarantees:
\begin{enumerate}
\item \textit{(Accuracy)} With probability $\ge 1- \delta_{\mathrm{coeff}}$,
\begin{equation*}
    |\widehat h_a - h_a| \le \varepsilon
\end{equation*}

\item \textit{(Total evolution time)} The total evolution time under
$e^{t\Lindblad}$ is
\[
t_{\mathrm{tot}}
\;=\; \mathcal O\biggl({\frac{1}{\varepsilon}\left[\log\biggl(\frac{1}{\delta_{\mathrm{coeff}}}\biggr) + \log\log\biggl(\frac{1}{\varepsilon}\biggr)\right]}\biggr),
\]

\item \textit{(Minimum evolution time)} It only ever applies $e^{\Lindblad t}$ for $t \ge t_{\mathrm{min}}$, where
\[
t_{\mathrm{min}}\;=\;\min_i \frac{t_i}{N_{\mathrm{reshape}}(t_i)}
\;=\;\Omega\left(\frac{\varepsilon}{M^2}\right),
\]

\item \textit{(Number of experiments)} The total number of state preparations / evolutions is
\[
N_{\exp}^{\mathrm{(tot)}} \;=\; L\,N_{\exp}
\;=\;\mathcal O\left(\log\biggl(\frac{1}{\varepsilon}\biggr)\left[\log\biggl(\frac{1}{\delta_{\mathrm{coeff}}}\biggr) + \log\log\biggl(\frac{1}{\varepsilon}\biggr)\right]\right).
\]

\item \textit{(Ancilla cost)} The procedure uses $\mathcal O(\log M)$ ancillary qubits for code padding and check measurements (the same code reused from the structure-learning round in which $P_a$ was sampled).

\item \textit{(Classical overhead)} The classical cost is:
\[
    \SoftBigO{n^2M^2/\varepsilon^2}
\]
\end{enumerate}
\end{theorem}

\begin{proof}
    Matching the QEC reshaping plus logical twirl generator Eq.~\eqref{eq:Leff-qec-reshaping-logical-twirl} with the effective-generator form from Eq.~\eqref{eq:generic-effective-generator-hnd}, we can write
    \begin{equation*}
        \LindbladEff(\rho)
        =
        -i[(h_a+b_a)\overline P_a,\rho]
        +
        \mathcal L_{\mathrm{resid}}(\rho),
        \qquad
        |b_a|\le B_\infty,\quad
        \|\mathcal L_{\mathrm{resid}}\|_\diamond\le R_\diamond .
    \end{equation*}
    Let $\omega_a\coloneq h_a+b_a$. We first analyze RFE under the ideal unitary evolution $e^{-i\omega_a\overline P_a t}$, and then account for finite-step Lindbladian reshaping and the residual Lindbladian. Since $B_\infty\le\varepsilon/2$, estimating $\omega_a$ to accuracy $\varepsilon/2$ suffices to estimate $h_a$ to accuracy $\varepsilon$.

    To perform iterative improvements via \Cref{lem:rfe-lemma}, we want to prepare a measurement statistic $\widehat F(s_i)$ that approximates $e^{i\omega_a s_i}$. To get a better sense of how $F(s_i)$ is generated, consider a toy example of learning a single-qubit reshaped Hamiltonian $H=\omega Z$. By preparing an equal superposition of the $+1$ and $-1$ eigenstates of $H$, namely $|\Phi_0\rangle=|+\rangle$, evolving by $e^{-i\omega Zs/2}$, and measuring the $X$ and $Y$ observables, one obtains $\langle X\rangle=\cos(\omega s)$ and $\langle Y\rangle=\sin(\omega s)$. Thus $F(s)=\langle X\rangle+i\langle Y\rangle=e^{i\omega s}$.

    Within \Cref{alg:appendix-round-j-hamilt-coeff-learning}, for $L\coloneq \left\lceil\log_{3/2}(4/\varepsilon)\right\rceil$, we set
    \begin{equation*}
        s_i
        \coloneq
        \frac{\pi}{4}
        \left(\frac32\right)^i,
        \qquad
        t_i
        \coloneq
        \frac{s_i}{2}
        =
        \frac{\pi}{8}
        \left(\frac32\right)^i,
        \qquad
        i=0,\ldots,L-1.
    \end{equation*}
    With this choice, RFE narrows the initial interval to width at most $\varepsilon$, and hence estimates $\omega_a$ to accuracy $\varepsilon/2$.

    For a general $n$-qubit target Pauli $P_a$, choose a Hermitian logical Pauli $\overline T$ such that $\{\overline T,\overline P_a\}=0$, which exists by \Cref{cor:anticommute-partner}. It can be obtained in the symplectic picture (Appendix~\ref{app:binary-symplectic}) by solving an under-specified system of linear equations encoding the commutation relation of $\overline T$ with the stabilizer matrix and with the symplectic representation of $\overline P_a$.

    Let $\RMap$ denote the recovery map of the QEC used by \Cref{alg:appendix-round-j-hamilt-coeff-learning}. We initialize an arbitrary state in the code space,
    \begin{equation*}
        \rho_0=\RMap(\ket{0}\bra{0}^{\otimes n'}).
    \end{equation*}
    We then measure the logical Pauli $\overline T$ using the projectors
$\Pi_v^T=(I+v\overline T)/2$, record the outcome $v\in\{\pm1\}$, and
obtain
\begin{equation}
\label{eq:rho_outcome_v}
    \rho_v
    =
    \frac{\Pi_v^T\rho_0\Pi_v^T}{\tr(\Pi_v^T\rho_0)} .
\end{equation}
Under ideal unitary evolution by $e^{-i\omega_a\overline P_a t_i}$,
anticommutation gives
\begin{equation}
\label{eq:rfe-ideal-observables}
    \expval{\overline T}_{\rho_v(t_i)}
    =
    v\cos(\omega_a s_i),
    \qquad
    \expval{i\overline T\,\overline P_a}_{\rho_v(t_i)}
    =
    v\sin(\omega_a s_i),
\end{equation}
where $s_i=2t_i$. Hence the preparation sign $v$ can be used to remove
the random sign of the initial logical eigenstate.

We now account for shot noise. For each RFE iteration, use
$N_{\mathrm{exp}}/2$ independent shots for each of the two observables.
For the $\ell$th cosine shot, let $v_\ell^{(+)}$ be the preparation
outcome and $x_\ell$ the subsequent measurement outcome of
$\overline T$, and define $X_\ell:=v_\ell^{(+)}x_\ell$. Similarly,
for the sine shots define $Y_\ell:=v_\ell^{(-)}y_\ell$, where $y_\ell$
is the measurement outcome of $i\overline T\,\overline P_a$.
Thus $X_\ell,Y_\ell\in\{\pm1\}$ and, under ideal evolution,
$\mathbb E[X_\ell]=\cos(\omega_a s_i)$ and
$\mathbb E[Y_\ell]=\sin(\omega_a s_i)$.

Let $\mu_{X,i}$ and $\mu_{Y,i}$ denote their expectations under the
implemented reshaped evolution. Hoeffding's inequality gives
\begin{equation}
\label{eq:hoeffding-coeff-learning}
    \Pr\left(
        \left|
        \frac{2}{N_{\mathrm{exp}}}
        \sum_{\ell=1}^{N_{\mathrm{exp}}/2}X_\ell
        -
        \mu_{X,i}
        \right|
        \ge
        \frac{1}{4\sqrt2}
    \right)
    \le
    2e^{-N_{\mathrm{exp}}/128},
\end{equation}
and identically for the $Y_\ell$ samples. Thus, choosing
\begin{equation}
\label{eq:rfe-n-exp}
    N_{\mathrm{exp}}
    =
    2\left\lceil
        64\ln\left(\frac{4L}{\delta_{\mathrm{coeff}}}\right)
    \right\rceil
    =
    \mathcal O\left(
        \log\frac{1}{\delta_{\mathrm{coeff}}}
        +
        \log\log\frac{1}{\varepsilon}
    \right)
\end{equation}
and union bounding over the $2L$ estimates ensures, with probability
at least $1-\delta_{\mathrm{coeff}}$, a complex shot-noise error
\begin{equation}
\label{eq:rfe-shot-noise-total}
    E_{\mathrm{shot}}
    \le
    \frac14
\end{equation}
at every RFE iteration.

We next bound the error from finite-step reshaping and residual
dissipation. Let $\widetilde B_{t_i}$ denote the implemented channel and
$B_{t_i}(\cdot):=e^{-i[\omega_a\overline P_a,\cdot]t_i}(\cdot)$ the
ideal unitary channel. By \Cref{lem:N-step-reshaping-error}, the bounds
$\|\Lindblad\|_\diamond,\|\LindbladEff\|_\diamond\le2M$, and Duhamel's
formula,
\begin{equation*}
    \|\widetilde B_{t_i}-B_{t_i}\|_\diamond
    \le
    \frac{4M^2t_i^2}{N_{\mathrm{reshape}}(t_i)}
    +
    t_iR_\diamond .
\end{equation*}
Since $t_i<\pi/(2\varepsilon)$ and
$R_\diamond\le\varepsilon/(4\sqrt2\pi)$, the residual contribution is
at most $1/(8\sqrt2)$. Choosing
$N_{\mathrm{reshape}}(t_i)\ge32\sqrt2\,M^2t_i^2$ bounds the finite-step
contribution by the same amount. Hence, uniformly over both preparation
outcomes and both measured observables,
\begin{equation}
\label{eq:rfe-reshape-error}
    \left|
        \tr\left(
            P_{\mathrm{obs}}
            (\widetilde B_{t_i}-B_{t_i})(\rho_v)
        \right)
    \right|
    \le
    \frac{1}{4\sqrt2},
    \qquad
    P_{\mathrm{obs}}\in
    \{\overline T,i\overline T\,\overline P_a\}.
\end{equation}
Multiplication by the preparation sign does not change this bound, so
$|\mu_{X,i}-\cos(\omega_a s_i)|\le1/(4\sqrt2)$ and
$|\mu_{Y,i}-\sin(\omega_a s_i)|\le1/(4\sqrt2)$. Therefore the total
complex reshaping error satisfies
\begin{equation}
\label{eq:rfe-reshape-error-total}
    E_{\mathrm{reshape}}
    \le
    \frac14.
\end{equation}

Define
\begin{equation*}
    \widehat F(s_i)
    \coloneq
    \frac{2}{N_{\mathrm{exp}}}
    \sum_{\ell=1}^{N_{\mathrm{exp}}/2}X_\ell
    +
    i\frac{2}{N_{\mathrm{exp}}}
    \sum_{\ell=1}^{N_{\mathrm{exp}}/2}Y_\ell .
\end{equation*}
Combining the shot-noise and reshaping bounds, for every RFE iteration
simultaneously,
\begin{equation*}
    |\widehat F(s_i)-e^{i\omega_a s_i}|
    \le
    E_{\mathrm{shot}}+E_{\mathrm{reshape}}
    \le
    \frac12
\end{equation*}
with probability at least $1-\delta_{\mathrm{coeff}}$. Hence
\Cref{lem:rfe-lemma} applies. After
$L=\lceil\log_{3/2}(4/\varepsilon)\rceil$ iterations, it returns
$\widehat\omega_a$ satisfying
$|\widehat\omega_a-\omega_a|\le\varepsilon/2$. Since
$\omega_a=h_a+b_a$ and $|b_a|\le B_\infty\le\varepsilon/2$, the output
$\widehat h_a:=\widehat\omega_a$ obeys
\begin{equation*}
    |\widehat h_a-h_a|
    \le
    |\widehat\omega_a-\omega_a|+|b_a|
    \le
    \varepsilon.
\end{equation*}
This proves the accuracy guarantee item $(1)$.

    We now prove item $(2)$, the total evolution time. The total evolution time under $e^{\Lindblad t}$ is
    \begin{equation*}
        t_{\mathrm{tot}}
        =
        N_{\mathrm{exp}}\sum_{i=0}^{L-1}t_i
        =
        \mathcal O\left(
        \frac{1}{\varepsilon}
        \left[
        \log\left(\frac{1}{\delta_{\mathrm{coeff}}}\right)
        +
        \log\log\left(\frac{1}{\varepsilon}\right)
        \right]
        \right),
    \end{equation*}
    where we used Eq.~\eqref{eq:rfe-n-exp} and $\sum_{i=0}^{L-1}t_i=\mathcal O(1/\varepsilon)$.

    The minimum evolution time (item $(3)$) is dictated by the smallest reshaping step:
    \begin{equation*}
        t_{\mathrm{min}}
        \coloneq
        \min_i \frac{t_i}{N_{\mathrm{reshape}}(t_i)}
        =
        \Omega\left(\frac{\varepsilon}{M^2}\right),
    \end{equation*}
    using $t_i<\pi/(2\varepsilon)$ and $N_{\mathrm{reshape}}(t_i) = \BigO{M^2t_i^2}$.

    For the total number of experiments (item $(4)$), each of the $L$ RFE iterations uses $N_{\mathrm{exp}}$ state preparations/evolutions, so
    \begin{equation*}
        N_{\exp}^{\mathrm{(tot)}}
        =
        L\,N_{\mathrm{exp}}
        =
        \mathcal O\left(
        \log\left(\frac{1}{\varepsilon}\right)
        \left[
        \log\left(\frac{1}{\delta_{\mathrm{coeff}}}\right)
        +
        \log\log\left(\frac{1}{\varepsilon}\right)
        \right]
        \right).
    \end{equation*}

    The ancilla cost (item $(5)$) is due to the QEC ancilla padding and check measurement overhead. \Cref{alg:appendix-round-j-hamilt-coeff-learning} uses the same QEC code as the structure-learning round in which $P_a$ was sampled. Hence it inherits the QEC ancilla cost of $\mathcal O(\log M)$.

    We now prove item $(6)$, the classical overhead, which consists of two parts. First, we need to pre-compute $\sum_i N_{\mathrm{reshape}}(t_i)$ different logical Paulis that we use for logical twirling. This corresponds to randomly sampling vectors in the symplectic binary picture from the null space of the stabilizer matrix (with an appended row corresponding to $P_a$) times $\Lambda$ (see Appendix~\ref{app:binary-symplectic}). Let $s=\log(2M)$ be the number of generators of the QEC from \Cref{thm:random-qec-for-reshaping}. Before sampling, there is a one-time pre-processing consisting of taking the matrix product with $\Lambda$ ($\BigO{n^2s}$) and then performing Gaussian elimination ($\BigO{ns^2}$). After that, sampling in the null space is done by drawing $(2n-(s+1))$ iid fair bits, then doing matrix-vector multiplication $\BigO{n^2}$. Therefore, the total classical computation needed to prepare logical Paulis from $\overline{\mathcal K}_{\overline P_a}$ is
    \begin{equation*}
        \BigO{n^2s+ns^2}
        +
        \sum_{i=0}^{L-1}N_{\mathrm{reshape}}(t_i)\cdot \BigO{n^2}
        =
        \BigO{n^2}\sum_{i=0}^{L-1}N_{\mathrm{reshape}}(t_i)
        =
        \BigO{n^2M^2}\sum_{i=0}^{L-1}\left(\frac{3}{2}\right)^{2i}
        =
        \BigO{\frac{n^2M^2}{\varepsilon^2}} .
    \end{equation*}
    Storing these logical Paulis takes $\BigO{n\cdot\sum_{i=0}^{L-1}N_{\mathrm{reshape}}(t_i)}=\BigO{nM^2/\varepsilon^2}$ classical space.

    The second part is the real-time syndrome extraction, lookup-table decoding, and merging of the recovery unitary with the logical Pauli from $\overline{\mathcal K}_{\overline P_a}$. The real-time classical processing cost is
    \begin{equation*}
        \SoftBigO{n}\cdot\sum_{i=0}^{L-1}N_{\mathrm{reshape}}(t_i)
        =
        \SoftBigO{nM^2/\varepsilon^2}.
    \end{equation*}
    Hence the classical processing is dominated by preparing the logical Paulis from $\overline{\mathcal K}_{\overline P_a}$ and equals
    \begin{equation*}
       \mathrm{Classical\,overhead}
       =
       \SoftBigO{n^2M^2/\varepsilon^2}.
    \end{equation*}
\end{proof}

\subsection{Full \HNDAbbrev{} hierarchical learning algorithm}\label{sec:hamilt-hier-learning-algorithm}

In this section, we combine candidate identification (\Cref{alg:appendix-round-j-hamilt-struct-learning}) and coefficient learning (\Cref{alg:appendix-round-j-hamilt-coeff-learning}) into the full \HND{} learning \Cref{alg:hamilt-hier-learning} according to Appendix~\ref{sec:hamilt-hier-learning-notation}. Even though the algorithm closely follows \cite{hu2025ansatz}, the complexity is different due to the underlying QEC-based Lindbladian reshaping. We then summarize its guarantees in \Cref{thm:appendix-hier-hamilt_learning}. 

\begin{algorithm}[H]
  \caption{Hamiltonian (\HNDAbbrev{}) learning}
  \label{alg:hamilt-hier-learning}
  \DontPrintSemicolon
  \SetKwInOut{Input}{Inputs}
  \SetKwInOut{Output}{Output}
  \SetKwBlock{Schedule}{Schedule}{end}

  \Input{
      \begin{enumerate}[label=(\arabic*), leftmargin=*, nosep]
        \item Access to $n$-qubit evolution $e^{t\Lindblad}$, where
        $\Lindblad$ is $M$-sparse in the Pauli basis.
        \item Target accuracy $\varepsilon\in(0,1)$ and total failure
        probability $\delta\in(0,1)$.
        \item A supplied correction set
        $\VCorrect\subseteq\DiagDissStruct$.
        \item Access to $n+\BigO{\log M}$ noiseless ancillas
        (for Bell sampling, QEC code and checks).
      \end{enumerate}
    }
    
    \Output{
    An estimator $\widehat h_{\mathrm{\HNDAbbrev},\eta}$ of the
    Hamiltonian coefficients.
    }

  \Schedule{
  $\mathcal V_{\mathrm{corr}}
    \gets
    \VCorrect
    \cup
    \{\Herm(PP'):P,P'\in\VCorrect,\ P\neq P'\}$
    \tcp*{footprint of the supplied correction set}
    $L_h  \gets \left\lceil \log_2(1/\varepsilon) \right\rceil$ \tcp*{\# hierarchical rounds}
    $ \delta_{\mathrm{struct}},\delta_{\mathrm{coeff}} \gets \frac{\delta}{2L_h}$ \tcp*{per-round failure budgets}
  }

  \BlankLine
  $\widehat h_{\mathrm{\HNDAbbrev},\eta} \gets \mathbf 0$ \tcp*{coefficient estimates (indexed by Pauli)}
  $\widehat{\mathcal S}_{\mathrm{\HNDAbbrev{}},\eta} \gets \emptyset$ \tcp*{identified support}

  \For{$j \gets 0$ \KwTo $L_h-1$}{
    $(\widehat{\HHierSet}_j,\{\mathrm{QEC}_P\}_{P\in\widehat{\HHierSet}_j})
    \gets
    \texttt{Hamilt\_Struct\_Sampling}
    \left(
        j,\,
        e^{t\Lindblad},\,
        \VCorrect,\,
        \widehat{\mathcal S}_{\mathrm{\HNDAbbrev{}},\eta},\,
        \delta_{\mathrm{struct}},\,
        \mathrm{ancillas}
    \right)$
    \tcp*{candidate terms and codes for which they were sampled from alg ~\ref{alg:appendix-round-j-hamilt-struct-learning}}
        
    \For{$P\in\widehat{\HHierSet}_j\setminus\mathcal V_{\mathrm{corr}}$}{
  $(S_P,\RMap_P)\gets\mathrm{QEC}_P$\;
      $\delta_{P} \gets \delta_{\mathrm{coeff}}/\max\{1,|\widehat{\HHierSet}_j|\}$ \tcp*{per-Pauli failure budget}
      
      $\widehat\mu_P \gets \texttt{Hamilt\_Coeff\_Learning}(P,\ \varepsilon/2,\ \delta_P,\ (S_P,\RMap_P),\ e^{\Lindblad t},\ \mathrm{ancillas})$\tcp*{learn coeff. alg.~\ref{alg:appendix-round-j-hamilt-coeff-learning}}

        \If{$\abs{\widehat\mu_P} > \varepsilon/2$}{
            $\widehat h_{\mathrm{\HNDAbbrev},\eta}[P] \gets \widehat\mu_P$ \tcp*{record estimate}
            $\widehat{\mathcal S}_{\mathrm{\HNDAbbrev{}},\eta} \gets \widehat{\mathcal S}_{\mathrm{\HNDAbbrev{}},\eta} \cup \{P\}$ \tcp*{update learned support}     
        }
    }
  }
  \Return $\widehat h_{\mathrm{\HNDAbbrev},\eta}$
\end{algorithm}

We now prove \Cref{thm:appendix-hier-hamilt_learning}. We restate it here for convenience. 

\begin{restatethm}{thm:appendix-hier-hamilt_learning}[Conditional hierarchical \HNDAbbrev{} Hamiltonian learning]
Let $\Lindblad$ be an $n$-qubit $M$-sparse Lindbladian as in Eq.~\eqref{eq:lindblad-eq-hamilt}. Fix $\varepsilon,\delta\in(0,1)$ and assume black-box access to $e^{t\Lindblad}$. Assume the $\eta$-heavy dissipator structure $\EtaDiagDissStruct$ is given, and that $\EtaDiagDissStruct$ satisfies the following uniform small-bias/small-residual condition. For every successful QEC code draw
 used by \Cref{alg:appendix-round-j-hamilt-struct-learning} (and reused by \Cref{alg:appendix-round-j-hamilt-coeff-learning}), the corresponding effective generator $\LindbladEff$ written in the form of Eq.~\eqref{eq:generic-effective-generator-hnd} has bias and residual parameters from Eq.~\eqref{eq:generic-bias-residual-def} satisfying
\begin{equation*}
    B_\infty\le \varepsilon/50,
    \qquad
    K_\infty\le \varepsilon/50,
    \qquad
    R_\diamond\le \frac{\varepsilon}{8 \sqrt2 \pi}.
\end{equation*}

Then running \Cref{alg:hamilt-hier-learning} with
$\VCorrect=\EtaDiagDissStruct$ returns coefficient estimates
$\widehat h_{\mathrm{\HNDAbbrev},\eta}$ for the \HND{} Hamiltonian at
scale $\eta$ satisfying the following guarantees.

\begin{enumerate}[leftmargin=*, label=\arabic*.]
\item \textit{(Accuracy)}
With probability at least $1-\delta$,
\begin{equation*}
    \|\widehat h_{\mathrm{\HNDAbbrev},\eta}
    -
    h_{\mathrm{\HNDAbbrev},\eta}\|_\infty
    \le \varepsilon .
\end{equation*}

\item \textit{(Total evolution time)}
The total evolution time under $e^{t\Lindblad}$ is
\begin{equation*}
    t_{\mathrm{tot}}
    =
    \SoftBigO{\frac{M^2}{\varepsilon}} .
\end{equation*}

\item \textit{(Minimum evolution time)}
The procedure only applies $e^{t\Lindblad}$ for times at least
\begin{equation*}
    t_{\mathrm{min}}
    =
    \Omega\left(\frac{\varepsilon}{M^3}\right).
\end{equation*}

\item \textit{(Number of experiments)}
The total number of experiments is
\begin{equation*}
    \SoftBigO{M^3},
\end{equation*}
dominated by Bell sampling during the candidate-identification stage.

\item \textit{(Ancilla cost)}
The procedure uses at most
\begin{equation*}
    n+\BigO{\log M}
\end{equation*}
noiseless ancilla qubits in addition to the $n$ probe qubits.

\item \textit{(Classical overhead)}
The classical processing cost is
\begin{equation*}
    \SoftBigO{\frac{M^6+nM^5+n^2M^4}{\varepsilon^2}} .
\end{equation*}
The first and second terms come from real-time QEC processing in the candidate-identification stage, while the last term comes from logical-Pauli sampling in coefficient learning.
\end{enumerate}
\end{restatethm}

\begin{proof}
    We first verify correctness. Since $2^{-j}\ge \varepsilon$ for all rounds $j<L_h$, the uniform assumptions $B_\infty\le \varepsilon/50$ and $K_\infty\le \varepsilon/50$ imply the bias/residual assumptions required by \Cref{thm:appendix-round-j-hamilt-struct-learn}. $B_\infty\le \varepsilon/50$ together with $R_\diamond\le \frac{\varepsilon}{8 \sqrt2 \pi}$ imply the assumptions of \Cref{thm:appendix-round-j-hamilt-coeff-learn} when coefficient learning is called with accuracy $\varepsilon/2$. Thus, in each round, candidate identification contains all terms in $\HHierSet_j$ not already learned, and coefficient learning estimates all candidates to accuracy $\varepsilon/2$. The strict threshold $|\widehat\mu_P|>\varepsilon/2$ removes false positives with $h_P=0$, while any true candidate with $|h_P|>\varepsilon$ is retained. Hence the round invariant Eq.~\eqref{eq:hnd-round-invariant} is maintained, and after $L_h=\lceil\log_2(1/\varepsilon)\rceil$ rounds all \HND{} coefficients are estimated to accuracy at most $\varepsilon$. The union bound over the structure and coefficient-learning calls, with $\delta_{\mathrm{struct}}=\delta_{\mathrm{coeff}}=\delta/(2L_h)$, gives success probability at least $1-\delta$.

    We now prove the resource bounds. In round $j$, the evolution time is the sum of candidate-identification time and coefficient-learning time:
    \begin{equation*}
        T_j
        =
        \BigO{2^jM^2\log(M/\delta_{\mathrm{struct}})}
        +
        \BigO{\frac{|\widehat{\HHierSet}_j|}{\varepsilon}
        \left[
        \log\left(\frac{|\widehat{\HHierSet}_j|}{\delta_{\mathrm{coeff}}}\right)
        +
        \log\log\left(\frac{1}{\varepsilon}\right)
        \right]} .
    \end{equation*}
    By the per-code cap in
    \Cref{alg:appendix-round-j-hamilt-struct-learning},
    $|\widehat{\HHierSet}_j|
    =\BigO{M^2\log(M/\delta_{\mathrm{struct}})}$ deterministically. Therefore,
    \begin{equation*}
        T_j
        =
        \BigO{2^jM^2\log(M/\delta_{\mathrm{struct}})}
        +
        \SoftBigO{\frac{M^2}{\varepsilon}} .
    \end{equation*}
    Summing over $j=0,\ldots,L_h-1$ and using $\sum_j2^j=\BigO{1/\varepsilon}$, we obtain
    \begin{equation*}
        t_{\mathrm{tot}}
        =
        \sum_{j=0}^{L_h-1}T_j
        =
        \SoftBigO{\frac{M^2}{\varepsilon}} .
    \end{equation*}

    The total number of experiments in round $j$ is
    \begin{equation*}
        N_{\mathrm{exp}}^{(j)}
        =
        \BigO{M^3\log(M/\delta_{\mathrm{struct}})}
        +
        |\widehat{\HHierSet}_j|\cdot
        \mathcal O\left(
        \log\left(\frac{1}{\varepsilon}\right)
        \left[
        \log\left(\frac{|\widehat{\HHierSet}_j|}{\delta_{\mathrm{coeff}}}\right)
        +
        \log\log\left(\frac{1}{\varepsilon}\right)
        \right]\right).
    \end{equation*}
    Since $|\widehat{\HHierSet}_j|=\SoftBigO{M^2}$, it is dominated by the candidate identification step and summing over all rounds gives
    \begin{equation*}
        N_{\mathrm{exp}}
        =
        \SoftBigO{M^3},
    \end{equation*}

    The minimum evolution time is bottle-necked by candidate identification. By \Cref{thm:appendix-round-j-hamilt-struct-learn}, round $j$ has resolution $\Omega\left(\frac{1}{M^3 2^j}\right).$
    It is smallest in the final round, where $2^j=\Theta(1/\varepsilon)$, so
    \begin{equation*}
        t_{\mathrm{min}}
        =
        \Omega\left(\frac{\varepsilon}{M^3}\right).
    \end{equation*}

    The ancilla bound follows directly from
    \Cref{thm:appendix-round-j-hamilt-struct-learn}: candidate
    identification uses at most
    $n+5\lceil\log_2(2M)\rceil$ noiseless ancillas in addition to the
    $n$ probe qubits, while coefficient learning reuses the source QEC
    code stored when the candidate was sampled. Hence the total ancilla
    cost is $n+\BigO{\log M}$.

    The classical overhead is the sum of the candidate-identification and coefficient-learning contributions. From \Cref{thm:appendix-round-j-hamilt-struct-learn,thm:appendix-round-j-hamilt-coeff-learn},
    \begin{equation*}
        \mathrm{Classical\,overhead}
        =\sum_{j=0}^{L_h-1}\left(\SoftBigO{(n+M)M^5 4^j} +\SoftBigO{n^2 M^4/\varepsilon^2}\right)
        =
        \SoftBigO{\frac{M^6+nM^5+n^2M^4}{\varepsilon^2}}.
    \end{equation*}
    This completes the proof.
\end{proof}

\subsection{Removing the known dissipator structure assumption}
\label{sec:appendix-removing-known-dissipator-assumption}

The conditional \HNDAbbrev{} Hamiltonian learner in
\Cref{thm:appendix-hier-hamilt_learning} takes the $\eta$-heavy
dissipator structure as input. We now remove this known-structure
assumption by composing it with the dissipator-structure learner from
Appendix~\ref{sec:appendix_dissipator_structure_learning}. We first control the residual dissipation and establish the full
\HNDAbbrev{} guarantee using favorable random codes. We then bound the
Hamiltonian bias for the finite-resolution extensions. This yields
three guarantees:
Heisenberg-limited learning of the full \HNDAbbrev{} set $\HNDStruct$
with $\eta\propto\varepsilon$; learning of the enlarged finite-resolution
target with $\eta\propto\varepsilon^2$ in the worst case; and
Heisenberg-limited learning of the same enlarged target with
$\eta\propto\varepsilon$ under an additional regularity condition on
the Kossakowski matrix.

The dissipator-structure learner returns a slightly enlarged estimate.
When run at scale $\eta$, \Cref{alg:appendix-diss-struct-learning}
outputs, with high probability, a set $\EstEtaDiagDissStruct$ satisfying
\begin{equation*}
    \EtaDiagDissStruct
    \subseteq
    \EstEtaDiagDissStruct
    \subseteq
    \HalfEtaDiagDissStruct .
\end{equation*}
Throughout the remainder of this subsection, we condition on this
success event and set $\VCorrect=\EstEtaDiagDissStruct$. Define the
corresponding learned dissipator footprint
\begin{equation*}
    \EstEtaFullDissStruct
    \coloneq
    \EstEtaDiagDissStruct
    \cup
    \{\Herm(PP'):\,
      P,P'\in\EstEtaDiagDissStruct,\,
      P\neq P'\}.
\end{equation*}

The conditional learner in \Cref{thm:appendix-hier-hamilt_learning}
is stated using the exact $\eta$-heavy dissipator structure for
simplicity. In the full algorithm, we instead use
$\VCorrect=\EstEtaDiagDissStruct$, so that the Hamiltonian learner acts
on
\begin{equation*}
    \HamiltStruct\setminus\EstEtaFullDissStruct .
\end{equation*}
At a fixed threshold this may be slightly smaller than the ideal
finite-resolution target $\EtaHNDStruct$. This difference amounts only
to a constant-factor ambiguity in the threshold: running the
dissipator-structure learner at threshold $2\eta$ gives
$\EstEtaDiagDissStruct\subseteq\EtaDiagDissStruct$ at scale $\eta$,
and hence the corresponding learned footprint is contained in
$\EtaFullDissStruct$. Thus all Hamiltonian terms in $\EtaHNDStruct$
remain targets of the Hamiltonian learner, while potentially some
additional terms are learned as well. The constant-factor shift only
changes the residual and bias bounds below by constant factors and does
not affect any asymptotic scaling.

Since $\EstEtaFullDissStruct\subseteq\FullDissStruct$, the learned target
always contains the full \HNDAbbrev{} set
\begin{equation*}
    \HNDStruct
    =
    \HamiltStruct\setminus\FullDissStruct,
\end{equation*}
independently of this finite-resolution threshold bookkeeping.
We first establish the guarantee for $\HNDStruct$ itself. The following
lemma provides the required residual bounds.

\begin{lemma}[Residual bounds from a learned dissipator structure]
\label{lem:hnd-learned-dissipator-BR}
Condition on the success event of the dissipator structure learner. Then for both the candidate-identification
effective generator from Eq.~\eqref{eq:Leff-qec-reshaping} and the
coefficient-learning effective generator from
Eq.~\eqref{eq:Leff-qec-reshaping-logical-twirl}, the residual parameters
from Eq.~\eqref{eq:generic-bias-residual-def} satisfy
\begin{equation*}
    R_\diamond \le 2M_D^2\eta,
    \qquad
    K_\infty \le M_D\eta.
\end{equation*}
\end{lemma}

\begin{proof}
Since the dissipator-structure learner succeeds at scale $\eta$, every
uncorrected dissipator Pauli
$P_k\in\DiagDissStruct\setminus\VCorrect$ satisfies $a_{kk}<\eta$.
By positivity of the Kossakowski matrix,
$|a_{kj}|^2\le a_{kk}a_{jj}$. Hence, for any two uncorrected dissipator
Paulis $P_k,P_j\in\DiagDissStruct\setminus\VCorrect$,
\begin{equation*}
    |a_{kj}|
    \le
    \sqrt{a_{kk}a_{jj}}
    \le
    \eta.
\end{equation*}

The residual Lindbladian is a sum of at most $M_D^2$ original
uncorrected--uncorrected contributions, each with coefficient magnitude
at most $\eta$. Each associated dissipative map has diamond norm at most
$2$, and logical twirling cannot increase this norm. Thus
\begin{equation*}
    R_\diamond
    =
    \|\mathcal L_{\mathrm{resid}}\|_\diamond
    \le
    2M_D^2\eta.
\end{equation*}

For each surviving same-syndrome pair,
$Q_{P_j}^\dagger Q_{P_k}=P_jP_k$.
For a fixed $P_k$, at most one $P_j$ produces any specified Pauli label.
Thus every coefficient of $K_{\mathrm{resid}}$ contains at most $M_D$
original contributions, each of magnitude at most $\eta$, giving
$K_\infty\le M_D\eta$.
The same argument applies after logical twirling, which only removes
such contributions.
\end{proof}

\begin{theorem}[Full \HNDAbbrev{} learning from a learned dissipator structure]
\label{thm:full-hdd-learned-dissipator}
Condition on success of the dissipator-structure learner and set
$\VCorrect=\EstEtaDiagDissStruct$. Fix $\varepsilon,\delta\in(0,1)$
and suppose
\begin{equation*}
    M_D\eta\le\frac{\varepsilon}{50},
    \qquad
    2M_D^2\eta\le\frac{\varepsilon}{8\sqrt2\pi}.
\end{equation*}
Then running \Cref{alg:hamilt-hier-learning} learns all Hamiltonian coefficients
associated with $\HNDStruct$ to accuracy $\varepsilon$ with probability
at least $1-\delta$. Its Hamiltonian-learning resource bounds are those
of \Cref{thm:appendix-hier-hamilt_learning}.
\end{theorem}

\begin{proof}
We follow the hierarchical proof of
\Cref{thm:appendix-hier-hamilt_learning}, but use favorable code draws
for candidate identification. At the beginning of each executed round
$j$, we maintain
\begin{equation*}
    \{P\in\HNDStruct:|h_P|>2^{-j}\}
    \subseteq
    \VDetect
    \subseteq
    \HamiltStruct\cup\FullDissStruct.
\end{equation*}
This holds initially because $\VDetect=\emptyset$ and $|h_P|\le1$.

Fix the preceding successful history and an unlearned target
$P_r\in\HNDStruct$ with $|h_r|>2^{-(j+1)}$.
Define, only for the analysis,
$\mathcal W:=\VDetect\cup\FullDissStruct$.
Then $P_r\notin\mathcal W$ and
$|\mathcal W|\le M_H+M_D^2\le M$.
For $m=\lceil\log_2(2M)\rceil$ random checks, the conditional union-bound
argument from \Cref{thm:random-qec-for-reshaping} gives
\begin{equation*}
    \Pr\bigl[
        r(P_r)=0,\;
        r(Q)\neq0\ \text{for every }Q\in\mathcal W
    \bigr]
    \ge
    2^{-m}\bigl(1-|\mathcal W|2^{-m}\bigr)
    \ge
    \frac{1}{8M}.
\end{equation*}
Indeed, conditional on $r(P_r)=0$, every distinct nonidentity
$Q\in\mathcal W$ still has zero syndrome with probability $2^{-m}$.
The favorable event implies the algorithm's code-acceptance event,
so it has probability at least $1/(8M)$ among returned codes as well.

On a favorable code, all dissipator Paulis have distinct nonzero
syndromes. Hence
\Cref{cor:qec-reshaping-generator-distinct-syndromes} gives
$H_{\mathrm{bias}}=0$ and suppresses every original Hamiltonian term
inside $\FullDissStruct$.
The remaining Hamiltonian coefficients have magnitude at most $2^{-j}$,
while the target retains its coefficient $h_r$.
Moreover, \Cref{lem:hnd-learned-dissipator-BR} gives
$K_\infty\le M_D\eta\le\varepsilon/50\le2^{-j}/50$.
The remaining Bell-sampling assumptions follow from the same padded
code construction as before.

Thus \Cref{lem:bell-sampling-of-eff-hamilt-struct} applies with
$B_\infty=0$.
The finite-step and capping argument in
Eq.~\eqref{eq:hamilt-struct-sampling-retained-per-surviving-code}
then retains the target with probability at least $1/4$ conditional
on a favorable code, and hence at least $1/(32M)$ per returned code.
Amplification and the finite code-construction and preparation budgets
proceed as in \Cref{thm:appendix-round-j-hamilt-struct-learn}.
Neither $\mathcal W$ nor the favorable event needs to be known to
the learner.

For coefficient learning, consider any tested candidate
$P\notin\FullDissStruct$.
No corrected--uncorrected dissipator product is proportional to $P$,
so \Cref{lem:qec-reshape-logical-twirl-generator} gives zero
Hamiltonian bias after the partial logical twirl.
This holds on any stored source code, not only on favorable codes.
The residual bound
$R_\diamond\le2M_D^2\eta\le\varepsilon/(8\sqrt2\pi)$ therefore permits
\Cref{thm:appendix-round-j-hamilt-coeff-learn} to be applied with
requested accuracy $\varepsilon/2$ and $B_\infty=0$.

On the coefficient-estimation success event, the threshold
$|\widehat\mu_P|>\varepsilon/2$ rejects every candidate outside
$\HamiltStruct\cup\FullDissStruct$ and retains every discovered
full-HDD term with $|h_P|>\varepsilon$.
Candidates inside $\FullDissStruct$ may have biased estimates, but
accepting them preserves
$\VDetect\subseteq\HamiltStruct\cup\FullDissStruct$.
Since the threshold $2^{-(j+1)}$ is at least $\varepsilon$ whenever
the next round is executed, the round invariant is maintained.
After the final round, every full-HDD coefficient larger than
$\varepsilon$ in magnitude has been learned; unreported coefficients
are estimated as zero.

The existing per-round and per-candidate failure budgets give total
failure probability at most $\delta$, by conditioning on the preceding
successful history at each call.
Finally, the per-code cap bounds the candidate-set size regardless
of the bias on other codes. The experimental schedules and resource
counting are therefore unchanged from
\Cref{thm:appendix-hier-hamilt_learning}.
\end{proof}

\begin{corollary}[Heisenberg-limited ansatz-free \HNDAbbrev{} learning]
\label{cor:ansatz-free-hnd}
Run the dissipator-structure learner from
\Cref{alg:appendix-diss-struct-learning} at scale
\begin{equation*}
    \eta=\Theta\left(\frac{\varepsilon}{M_D^2}\right),
\end{equation*}
with the constant chosen small enough to satisfy the conditions of
\Cref{thm:full-hdd-learned-dissipator}. Then run
\Cref{alg:hamilt-hier-learning} with
$\VCorrect=\EstEtaDiagDissStruct$.
Conditioned on the success of the dissipator-structure learner, the
algorithm learns all Hamiltonian coefficients associated with
\begin{equation*}
    \HNDStruct
    =
    \HamiltStruct\setminus\FullDissStruct
\end{equation*}
to accuracy $\varepsilon$ with probability at least $1-\delta$, when
the Hamiltonian routine is run with failure parameter $\delta$.
This follows from \Cref{thm:full-hdd-learned-dissipator}.

The total resource bounds are obtained by composing the
dissipator-structure learning resources from
\Cref{thm:appendix-hier-diss-struct-learning} with the conditional
\HNDAbbrev{} learning resources from
\Cref{thm:appendix-hier-hamilt_learning}. In particular, the black-box
evolution time is
\begin{equation*}
    t_{\mathrm{tot}}
    =
    \SoftBigO{\frac{1}{\eta}}
    +
    \SoftBigO{\frac{M^2}{\varepsilon}}
    =
    \SoftBigO{\frac{M^2}{\varepsilon}},
\end{equation*}
which is Heisenberg-limited in the target Hamiltonian accuracy. The minimum evolution time satisfies
\begin{equation*}
    t_{\mathrm{min}}
    =
    \Omega\left(
        \min\left\{
            \frac{\eta}{M^2},
            \frac{\varepsilon}{M^3}
        \right\}
    \right)
    =
    \Omega\left(\frac{\varepsilon}{M^3}\right),
\end{equation*}
where we used $M_D^2\le M$ and
$\|\Lindblad\|_\diamond\le 2M$ from
\Cref{lem:diamond-sparse-lindblad}.
\end{corollary}

At finite resolution, in addition to $\HNDStruct$, the candidate sector $\HamiltStruct\setminus\EstEtaFullDissStruct$ may contain
Hamiltonian terms in the weak dissipator footprint
\begin{equation*}
    \HamiltStruct
    \cap
    \bigl(
        \FullDissStruct\setminus\EstEtaFullDissStruct
    \bigr).
\end{equation*}
These intermediate-scale targets can be biased by corrected--uncorrected
dissipator couplings ($H_{\mathrm{bias}}$ in Eq.~\eqref{eq:generic-effective-generator-hnd}). We first derive the worst-case bias implied by
positivity of the Kossakowski matrix alone.

\begin{lemma}[Worst-case bias bound for finite-resolution targets]
\label{lem:worst-case-learned-dissipator-bias}
Condition on the success event of the dissipator structure learner. Then for both the candidate-identification
effective generator from Eq.~\eqref{eq:Leff-qec-reshaping} and the
coefficient-learning effective generator from
Eq.~\eqref{eq:Leff-qec-reshaping-logical-twirl}, the Hamiltonian bias
parameter from Eq.~\eqref{eq:generic-bias-residual-def} satisfies
\begin{equation*}
    B_\infty
    \le
    \frac{M_D}{2}\sqrt{\eta}.
\end{equation*}
\end{lemma}

\begin{proof}
From Eq.~\eqref{eq:Leff-qec-reshaping}, each coefficient of
$H_{\mathrm{bias}}$ is a sum of corrected--uncorrected dissipator
cross terms with
\begin{equation*}
    P_k\in\VCorrect,
    \qquad
    P_j\in\DiagDissStruct\setminus\VCorrect .
\end{equation*}
For any fixed resulting Pauli, the number of contributing pairs is at
most
\begin{equation*}
    \min\{|\VCorrect|,M_D-|\VCorrect|\}
    \le
    \frac{M_D}{2}.
\end{equation*}
Since $P_j$ is uncorrected, $a_{jj}<\eta$, while $a_{kk}\le1$ by
normalization. Positivity of the Kossakowski matrix therefore gives
\begin{equation}\label{eq:worst-case-bias-psd}
    |a_{kj}|
    \le
    \sqrt{a_{kk}a_{jj}}
    \le
    \sqrt{\eta}.
\end{equation}
Each corrected--uncorrected cross term contributes to a Hermitian
Pauli bias coefficient with magnitude at most $|a_{kj}|$, and hence
\begin{equation*}
    B_\infty
    =
    \|H_{\mathrm{bias}}\|_{\infty,\mathrm{coeff}}
    \le
    \frac{M_D}{2}\sqrt{\eta}.
\end{equation*}
The same bound applies after partial logical twirling, since
Eq.~\eqref{eq:Leff-qec-reshaping-logical-twirl} only retains a subset
of these corrected--uncorrected terms.
\end{proof}

Together with \Cref{lem:hnd-learned-dissipator-BR}, this implies that
the small-bias/small-residual assumptions of
\Cref{thm:appendix-hier-hamilt_learning} hold whenever
\begin{equation*}
    \eta
    \le
    c\,\frac{\varepsilon^2}{M_D^2}
\end{equation*}
for a sufficiently small universal constant $c>0$.

\begin{corollary}[Ansatz-free finite-resolution Hamiltonian learning, worst case]
\label{cor:ansatz-free-hnd-worst-case}
Run the dissipator-structure learner from
\Cref{alg:appendix-diss-struct-learning} at scale
\begin{equation*}
    \eta
    =
    \Theta\left(\frac{\varepsilon^2}{M_D^2}\right),
\end{equation*}
with the constant chosen small enough so that the combined bounds of
\Cref{lem:hnd-learned-dissipator-BR,lem:worst-case-learned-dissipator-bias}
satisfy the assumptions of \Cref{thm:appendix-hier-hamilt_learning}.
Then run \Cref{alg:hamilt-hier-learning} with
$\VCorrect=\EstEtaDiagDissStruct$.
Conditioned on the success of the dissipator-structure learner, the
algorithm learns all Hamiltonian coefficients in
\begin{equation*}
    \HamiltStruct\setminus\EstEtaFullDissStruct
\end{equation*}
to accuracy $\varepsilon$ with the guarantees of
\Cref{thm:appendix-hier-hamilt_learning}.

The total resource bounds are obtained by composing the
dissipator-structure learning resources from
\Cref{thm:appendix-hier-diss-struct-learning} with the conditional
Hamiltonian-learning resources from
\Cref{thm:appendix-hier-hamilt_learning}. In particular, because the
worst-case choice of $\eta$ scales quadratically in $\varepsilon$ up
to sparsity-dependent factors, the dissipator-structure preprocessing
is standard-quantum-limited in the target Hamiltonian accuracy.
\end{corollary}

The worst-case SQL scaling for the additional finite-resolution targets
$\HamiltStruct \cap
\bigl(\FullDissStruct\setminus\EstEtaFullDissStruct\bigr)$
arises from PSD-allowed corrected--uncorrected cross terms: a corrected
heavy dissipator direction can couple to an uncorrected light direction
with $|a_{kj}|=O(\sqrt{\eta})$, inducing an $O(\sqrt{\eta})$
Hamiltonian bias after QEC reshaping. Consequently, under positivity
alone, controlling this bias to accuracy $\varepsilon$ requires
$\eta\propto\varepsilon^2$, up to sparsity-dependent factors. In
\cref{rem:necessity-tail-regularity}, we show that this mechanism
fundamentally restricts uniform finite-resolution learning to SQL scaling
unless additional regularity is imposed on the Kossakowski tail. This
motivates the following balanced-tail condition, which rules out precisely
the imbalanced corrected--uncorrected couplings responsible for this
obstruction.

\begin{assumption}[Balanced Kossakowski tail at scale $\eta$]
\label{ass:balanced-kossakowski-tails}
For every pair $P_k,P_j\in\DiagDissStruct$ with $j\neq k$, if
\begin{equation*}
    \min\{a_{kk},a_{jj}\}\le \eta
    \qquad\text{and}\qquad
    a_{kk}a_{jj}\ge \eta^2,
\end{equation*}
then
\begin{equation*}
    |a_{kj}|
    \le
    \kappa \min\{a_{kk},a_{jj}\}
\end{equation*}
for some parameter $\kappa \ge 1$.
\end{assumption}

\begin{lemma}[Bias bound under balanced Kossakowski tails]
\label{lem:balanced-tail-learned-dissipator-BR}
Condition on the dissipator structure learner success event above
and assume \Cref{ass:balanced-kossakowski-tails}. Then, for both the
candidate-identification effective generator from
Eq.~\eqref{eq:Leff-qec-reshaping} and the coefficient-learning
effective generator from
Eq.~\eqref{eq:Leff-qec-reshaping-logical-twirl}, the Hamiltonian bias
parameter from Eq.~\eqref{eq:generic-bias-residual-def} satisfies
\begin{equation*}
    B_\infty
    \le
    \frac{\kappa M_D}{2}\eta.
\end{equation*}
\end{lemma}

\begin{proof}
Consider any corrected--uncorrected pair contributing to
$H_{\mathrm{bias}}$. The uncorrected Pauli has diagonal weight
$a_{jj}<\eta$. If $a_{kk}a_{jj}<\eta^2$, positivity gives
\begin{equation*}
    |a_{kj}|
    \le
    \sqrt{a_{kk}a_{jj}}
    <
    \eta
    \le
    \kappa\eta.
\end{equation*}
If instead $a_{kk}a_{jj}\ge\eta^2$, then
\Cref{ass:balanced-kossakowski-tails} gives
\begin{equation*}
    |a_{kj}|
    \le
    \kappa\min\{a_{kk},a_{jj}\}
    \le
    \kappa\eta.
\end{equation*}
As in the proof of
\Cref{lem:worst-case-learned-dissipator-bias}, any fixed resulting
Pauli receives contributions from at most $M_D/2$ such pairs.
Therefore
\begin{equation*}
    B_\infty
    \le
    \frac{\kappa M_D}{2}\eta.
\end{equation*}
\end{proof}

Combining \Cref{lem:balanced-tail-learned-dissipator-BR} with
\Cref{lem:hnd-learned-dissipator-BR}, there exists a universal constant
$c>0$ such that the small-bias/small-residual conditions of
\Cref{thm:appendix-hier-hamilt_learning}, namely
\begin{equation*}
    B_\infty\le\frac{\varepsilon}{50},
    \qquad
    K_\infty\le\frac{\varepsilon}{50},
    \qquad
    R_\diamond\le\frac{\varepsilon}{8\sqrt{2}\pi},
\end{equation*}
hold whenever
\begin{equation*}
    \eta
    \le
    c\min\left\{
        \frac{\varepsilon}{M_D^2},
        \frac{\varepsilon}{\kappa M_D}
    \right\}.
\end{equation*}

\begin{corollary}[Heisenberg-limited ansatz-free finite-resolution Hamiltonian learning under balanced tails]
\label{cor:ansatz-free-hnd-balanced-tail}
Assume \Cref{ass:balanced-kossakowski-tails} holds with
$\kappa\le M_D$. Run the dissipator-structure learner from
\Cref{alg:appendix-diss-struct-learning} at scale
\begin{equation*}
    \eta=\Theta\left(\frac{\varepsilon}{M_D^2}\right),
\end{equation*}
with the constant chosen small enough so that the combined bounds of
\Cref{lem:hnd-learned-dissipator-BR,lem:balanced-tail-learned-dissipator-BR}
satisfy the assumptions of
\Cref{thm:appendix-hier-hamilt_learning}. Then run
\Cref{alg:hamilt-hier-learning} with
$\VCorrect=\EstEtaDiagDissStruct$.
Conditioned on the success of the dissipator-structure learner, the
algorithm learns all Hamiltonian coefficients in
\begin{equation*}
    \HamiltStruct\setminus\EstEtaFullDissStruct
\end{equation*}
to accuracy $\varepsilon$ with the guarantees of
\Cref{thm:appendix-hier-hamilt_learning}. The total resource bounds are obtained by composing the dissipator-structure learning resources from \Cref{thm:appendix-hier-diss-struct-learning} with the conditional \HNDAbbrev{} learning resources from \Cref{thm:appendix-hier-hamilt_learning}. In particular, the black-box evolution time is
\begin{equation*}
    t_{\mathrm{tot}}
    = \SoftBigO{\frac{1}{\eta}}+ 
    \SoftBigO{\frac{M^2}{\varepsilon}}
    =
    \SoftBigO{\frac{M^2}{\varepsilon}},
\end{equation*}
which is Heisenberg-limited in the target Hamiltonian accuracy. The minimum evolution time is:
\begin{equation*}
    t_{\mathrm{min}}
    =
    \Omega\left({\min\left\{\frac{\eta}{M^2},\frac{\varepsilon}{M^3}\right\}}\right)
    =
    \Omega\left(\frac{\varepsilon}{M^3}\right),
\end{equation*}
where we used $M_D^2\le M$ and $\|\Lindblad\|_\diamond \le 2M$ from \Cref{lem:diamond-sparse-lindblad}.
\end{corollary}

Let us unpack why \Cref{ass:balanced-kossakowski-tails} is a natural
condition for the finite-resolution extension. It does not constrain strong--strong couplings: if both diagonal entries are above the learned dissipator threshold, both Pauli directions are included in the correctable set, so their off-diagonal coupling does not produce a corrected--uncorrected bias. This covers standard coherent Pauli blocks such as amplitude damping when the block is above threshold: a jump operator proportional to $X+iY$ produces an $X/Y$ Kossakowski block whose diagonal and off-diagonal entries are all of the same order, and the two Pauli directions are learned together. The condition also does not constrain weak--weak couplings: if both directions are below threshold, then $a_{kk}a_{jj}<\eta^2$, so positivity already gives $|a_{kj}|<\eta$. Pauli channels satisfy the condition trivially because the Kossakowski matrix is diagonal.

The only excluded regime is an imbalanced weak--strong tail: one diagonal entry lies below the dissipator-learning threshold, the other is much larger, and the off-diagonal entry is large enough to behave like the PSD worst-case geometric mean. The balanced-tail condition rules this out by requiring that, in this intermediate regime, the off-diagonal entry is controlled by the smaller diagonal weight. Thus the assumption does not forbid off-diagonal Kossakowski structure; it only prevents a weak unlearned dissipator component from being coherently tied to a strong corrected component strongly enough to create the bias obstruction.

The condition is still permissive. If the two coupled diagonal weights obey
\begin{equation*}
    \frac{\max\{a_{kk},a_{jj}\}}{\min\{a_{kk},a_{jj}\}}\le \kappa^2,
\end{equation*}
then positivity alone implies
\begin{equation*}
    |a_{kj}|
    \le
    \sqrt{a_{kk}a_{jj}}
    \le
    \kappa\min\{a_{kk},a_{jj}\},
\end{equation*}
so \Cref{ass:balanced-kossakowski-tails} is automatically satisfied for that pair. Moreover, by \Cref{lem:balanced-tail-learned-dissipator-BR}, it suffices to choose
\begin{equation*}
    \eta
    \le
    c\min\left\{
        \frac{\varepsilon}{M_D^2},
        \frac{\varepsilon}{\kappa M_D}
    \right\}.
\end{equation*}
Thus, whenever $\kappa\le M_D$, the residual term is the bottleneck and one may take $\eta=\Theta(\varepsilon/M_D^2)$. In particular, diagonal-weight ratios as large as $\kappa^2\le M_D^2$ are compatible with the Heisenberg-limited composition. For a single jump operator proportional to $X+\alpha Y$, for example, this allows $|\alpha|\gtrsim 1/M_D$ up to constants.

\subsection{Deferred proofs for \HNDAbbrev{} Bell-sampling identification} \label{sec:deferred-hamilt-struct-sampling-proofs}

\begin{restatelem}{lem:bell-sampling-of-eff-hamilt-struct}[Hamiltonian Bell sampling under bias and residual dissipation]
Let $\LindbladEff$ be an $n'$-qubit effective generator of the form in
Eq.~\eqref{eq:generic-effective-generator-hnd}, with bias and residual parameters $B_\infty,K_\infty$ as in
Eq.~\eqref{eq:generic-bias-residual-def}. Write it in the jump unraveling picture using Eq.~\eqref{eq:generic-bias-residual-def} and Eq.~\eqref{eq:generic-residual-no-jump-operator}:

\begin{equation*}
    \LindbladEff(\rho)
    =
    -iH_{\rm eff}\rho +i\rho H_{\rm eff}^\dagger + \sum_{w,m}\alpha_{wm}P_w\rho P_m, \qquad H_{\rm eff}\coloneq H_{\rm tar}+H_{\rm bias}-\frac i2K_{\rm resid}.
\end{equation*}
Expand the target Hamiltonian in the Pauli basis as $H_{\mathrm{tar}}=\sum_{P_m\in\mathcal S_{\mathrm{H,tar}}} h_mP_m$, the bias as $H_{\mathrm{bias}}=\sum_{P_m\in\mathcal S_{\mathrm{H,bias}}} b_mP_m$ and the residual non-Hermitian contribution as $K_{\mathrm{resid}}= \sum_{P_m\in\mathcal S_{K}} k_mP_m$, where
$\mathcal S_{\mathrm{H,tar}},\mathcal S_{\mathrm{H,bias}}\subseteq\mathcal P_{n'}\setminus\{I\}$, $\mathcal S_K\subseteq\mathcal P_{n'}$ and $\{h_m\}, \{b_m\}, \{k_m\}$ are real coefficients.
Let $\mathcal S_{\mathrm{H,eff}}\coloneq\mathcal S_{\mathrm{H,tar}}\cup\mathcal S_{\mathrm{H,bias}}\cup \mathcal S_K$ and assume $|\mathcal S_{\mathrm{H,eff}}|\le M$.

Fix an integer $j\ge 0$. Assume that $|h_m|\le 2^{-j}$ for every
$P_m\in\mathcal S_{\mathrm{H,tar}}$ and that a target Pauli
$P_r\in\mathcal S_{\mathrm{H,tar}}$ satisfies $|h_r|\ge 2^{-(j+1)}$. Assume that the bias and residual parameters satisfy $B_\infty\le 2^{-(j+1)}/25$ and $K_\infty\le 2^{-(j+1)}/25.$

Let $C$ be a stabilizer code on $n'$ qubits with code projector $\Pi$ and stabilizer subgroup
$S\subseteq\mathcal P_{n'}$. Let $\mathcal S_{\mathrm{eff}} \coloneq \mathcal S_{\mathrm{H,eff}} \cup \mathcal S_{\mathrm{D,res}}$ and let $G_{\mathrm{eff}}\coloneq
\langle\mathcal S_{\mathrm{eff}}\rangle\subseteq\mathbb P_{n'}$ denote the subgroup generated by $\mathcal S_{\mathrm{eff}}$. Assume that all Paulis
$P\in\mathcal S_{\mathrm{eff}}$ have zero syndrome $r(P)=0$ with respect to $C$, and that no
nontrivial product of elements of $\mathcal S_{\mathrm{eff}}$ is proportional to a stabilizer; equivalently,
for every $Q\in G_{\mathrm{eff}}$, $Q\propto s\in S$ implies $s=I_{n'}$.

Let $\rho_0\coloneq\ket{\Phi_0}\bra{\Phi_0}$, where
$\ket{\Phi_0}\coloneq\ket{\Phi_+}^{\otimes n'}$ and
$\ket{\Phi_+}=(\ket{00}+\ket{11})/\sqrt{2}$. Let $\widetilde\rho$ denote the state immediately before Bell-basis measurement,
\begin{equation*}
    \widetilde \rho
    \coloneq
    (e^{t\LindbladEff}\otimes I_{n'})
    \left(
        \frac{(\Pi\otimes I_{n'})\rho_0(\Pi\otimes I_{n'})}
        {\tr((\Pi\otimes I_{n'})\rho_0)}
    \right).
\end{equation*}
For any Pauli $P\in\mathbb P_{n'}$, let
$O_P\coloneq(P\otimes I_{n'})\ket{\Phi_0}\bra{\Phi_0}(P^\dagger\otimes I_{n'})$
denote the corresponding Bell-measurement projector. Then, for
$t=2^j/(2M)$, a single Bell-sampling experiment on $\widetilde\rho$ satisfies
\begin{equation*}
    \begin{aligned}
        \Pr(\text{sample $P_r$ modulo stabilizers})
        &\coloneq
        \tr\left(\sum_{s\in S} O_{sP_r}\,\widetilde\rho\right)
        \ge
        \frac{1}{200M^2},\\
        \Pr(\text{sample a stabilizer})
        &\coloneq
        \tr\left(\sum_{s\in S} O_s\,\widetilde\rho\right)
        \ge
        1-\frac{1}{2M}.
    \end{aligned}
\end{equation*}
\end{restatelem}

\begin{proof}
    Since the code projector can be expressed as $\Pi = \frac{1}{|S|}\sum_{s\in S} s$ (Eq.~\eqref{eq:stabilizer-projector}), we have $\tr((\Pi\otimes I_{n'})\rho_0) = 1/|S|$. Define
    \begin{equation*}
        \rho_\mathrm{code}
        \coloneq
        \frac{(\Pi\otimes I_{n'})\rho_0(\Pi\otimes I_{n'})}{\tr((\Pi\otimes I_{n'})\rho_0)} = |S|(\Pi\otimes I_{n'})\rho_0(\Pi\otimes I_{n'}).
    \end{equation*}

    To obtain the lower bounds on the projection probabilities of this lemma, we use the quantum jump unraveling formalism. In particular, we compare the actual evolution of $\rho_\mathrm{code}$ under $\LindbladEff$ to the non-hermitian Hamiltonian evolution generated by
    \begin{equation}\label{eq:hnd_H_eff}
        H_{\mathrm{eff}}
        \coloneq H_{\mathrm{tar}}+H_{\mathrm{bias}} -\frac{i}{2}K_{\mathrm{resid}}
        = \sum_{P_m\in\mathcal S_{\mathrm{H,eff}}}\widetilde h_m P_m,
        \qquad
        \widetilde h_m \coloneq h_m+b_m-\frac{i}2{k_m},
    \end{equation}
    where $h_m$, $b_m$ and $k_m$ are extended by zero outside $\mathcal S_{\mathrm{H,tar}}$, $\mathcal S_{\mathrm{H,bias}}$ and $\mathcal S_{K}$, respectively. Let $\mathcal E_t(\rho) \coloneq e^{-iH_{\mathrm{eff}}t}\rho e^{iH_{\mathrm{eff}}^\dagger t}$ and $\mathcal J(\rho) \coloneq \sum_{P_w,P_m\in \mathcal S_{\mathrm{D,res}}} \alpha_{wm} P_w\rho P_m$.
    By decomposing $e^{\LindbladEff t}$ and using Duhamel's formula:
    \begin{equation*}
        e^{\LindbladEff t} = \mathcal E_t + \underbrace{( e^{\LindbladEff t} - \mathcal E_t)}_{\Phi_t}= \mathcal E_t + \int_{s=0}^te^{\LindbladEff (t-s)} \circ\mathcal J \circ\mathcal E_s\,ds
    \end{equation*}
    Since $\mathcal E_t$ is a conjugation map, it is completely positive (CP). Similarly, $\sum_{P_w,P_m\in \mathcal S_{\mathrm{D,res}}} \alpha_{wm}\,P_w \rho P_m$ is a CP map since one can diagonalize it by positivity of the Kossakowski matrix $\alpha$. Concretely, set $\alpha_{wm} \coloneq \sum_x \gamma_{wx}\gamma^*_{mx}$:
    \begin{equation*}
        \sum_{P_w,P_m\in \mathcal S_{\mathrm{D,res}}} \alpha_{wm}\,P_w \rho P_m = \sum_x \sum_{P_w\in \mathcal S_{\mathrm{D,res}}}\gamma_{wx}P_w \rho \sum_{P_m\in \mathcal S_{\mathrm{D,res}}}\gamma^*_{mx}P_m = \sum_xL_x \rho L_x^\dagger
    \end{equation*}
    where $L_x \coloneq \sum_{P_w\in \mathcal S_{\mathrm{D,res}}}\gamma_{wx}P_w$. Since the integrand is a composition of CP maps for every $s$, and the integral of CP maps is CP, $\Phi_t$ is completely positive. Therefore,
    \begin{equation}\label{eq:hnd-prob-sample-Pr-mod-stabilizers-unraveling}
            \begin{aligned}
                \Pr(\text{sample $P_r$ modulo stabilizers})= \tr\left(\sum_{s\in S} O_{sP_r}\,\mathcal E_t(\rho_{\mathrm{code}})\right) + \tr\left(\sum_{s\in S} O_{sP_r}\,\Phi_t(\rho_{\mathrm{code}})\right)\\
                \ge \tr\left(\sum_{s\in S} O_{sP_r}\,\mathcal E_t(\rho_{\mathrm{code}})\right)
            \end{aligned}
    \end{equation}
    
    The same holds for $\Pr(\text{sample a stabilizer})$ by setting $P_r = I_{n'}$. Thus moving forward, we focus on $\mathcal E_t(\rho) \coloneq e^{-iH_{\mathrm{eff}}t}\rho e^{iH_{\mathrm{eff}}^\dagger t}$. By the assumptions $|h_m|\le 2^{-j}$, $B_\infty\le \frac{2^{-j}}{50}$ and $K_\infty\le \frac{2^{-j}}{50}$ and using Eq.~\eqref{eq:hnd_H_eff}, we have
    \begin{equation}\label{eq:hnd-bell-sampling-beta}
        \beta \coloneq \max_{P_m\in\mathcal S_{\mathrm{H,eff}}}|\widetilde h_m| \le  |h_m|+|b_m|+ |k_m|\le1.04\cdot 2^{-j}.
    \end{equation}
    Moreover, for the target Pauli $P_r$, since $|h_r| \ge 2^{-(j+1)}$,
    \begin{equation*}
        |\operatorname{Re}(\widetilde h_r)| \ge |h_r|-|b_r| \ge 0.96\cdot 2^{-(j+1)}.
    \end{equation*}
    Since every $P_m\in\mathcal S_{\mathrm{eff}}$ has zero syndrome, every $P_m$ commutes with $\Pi$, and hence $H_{\mathrm{eff}}$ and $e^{-iH_{\mathrm{eff}}t}$ commute with $\Pi$.

    Let $\sigma\in S$ be an arbitrary stabilizer. We first lower bound the probability to sample $\sigma P_r$ and then combine the probability across all stabilizers. Using commutativity of $G_{\mathrm{eff}}$ with $\Pi$, $\sigma\Pi=\Pi$ and the hermiticity of $P_r$ and $\sigma$, we get
    \begin{equation}
    \label{eq:hamilt-structure-sampling-hamilt-only-prob}
        \begin{aligned}
        p^{\mathrm H}_{(\sigma P_r)}
        \coloneq \tr\left(O_{(\sigma P_r)}(e^{-iH_{\mathrm{eff}}t}\otimes I_{n'})\,\rho_\mathrm{code} \,(e^{iH_{\mathrm{eff}}^\dagger t}\otimes I_{n'})\right) &= |S|\,\big|\bra{\Phi_0}(P_r^\dagger \sigma^\dagger\otimes I_{n'})(e^{-iH_{\mathrm{eff}}t}\Pi \otimes I_{n'})\ket{\Phi_0}\big|^2 =\\
        &= |S|^{-1}\big|\bra{\Phi_0}(P_r \otimes I_{n'})(\sum_{s\in S} s\, e^{-iH_{\mathrm{eff}}t} \otimes I_{n'})\ket{\Phi_0}\big|^2 .
        \end{aligned}
    \end{equation}

    Since no nontrivial product of elements of $\mathcal S_{\mathrm{eff}}$ is proportional to a stabilizer, by orthogonality of Bell-basis states, the only $s\in S$ that contributes nonzero probability is $s=I_{n'}$. Thus,
    \begin{equation}
    \label{eq:prob-sigma-Pr}
        p^{\mathrm H}_{(\sigma P_r)}
        =
        |S|^{-1}\big|\bra{\Phi_0}(P_r \otimes I_{n'})(e^{-iH_{\mathrm{eff}}t} \otimes I_{n'})\ket{\Phi_0}\big|^2 .
    \end{equation}

    Taylor-expanding $e^{-iH_{\mathrm{eff}}t}=I_{n'}-iH_{\mathrm{eff}}t+\sum_{q\ge2}\frac{(-iH_{\mathrm{eff}}t)^q}{q!}$, and using orthogonality of Bell-basis states, we obtain
    \begin{equation*}
        \begin{aligned}
            p^{\mathrm H}_{(\sigma P_r)}
            &=
            \frac{1}{|S|}\left|-it\widetilde h_r+\sum_{q\ge2}\frac{(-it)^q}{q!}\bra{\Phi_0}(P_rH_{\mathrm{eff}}^q \otimes I_{n'})\ket{\Phi_0}\right|^2 \\
            &\ge
            \frac{1}{|S|}\left(|\operatorname{Re}(\widetilde h_r)| t-\left|\operatorname{Im}\left(\sum_{q\ge2}\frac{(-it)^q}{q!}\bra{\Phi_0}(P_rH_{\mathrm{eff}}^q \otimes I_{n'})\ket{\Phi_0}\right)\right|\right)^2
        \end{aligned}
    \end{equation*}
    where the inequality follows from taking the imaginary part.
    
    To account for the higher-order terms, we count the number of ordered tuples of Hamiltonian Paulis from $\mathcal S_{\mathrm{H,eff}}$ whose product is proportional to $P_r$. For the $q$th order, this number is at most $M^{q-1}$. Therefore, using Eq.~\eqref{eq:hnd-bell-sampling-beta},
    \begin{equation*}
        p^{\mathrm H}_{(\sigma P_r)}
        \ge
        \frac{1}{|S|}\left(|\operatorname{Re}(\widetilde h_r)| t-\sum_{q\ge2}\frac{(t\beta)^q}{q!}M^{q-1}\right)^2 .
    \end{equation*}

    Set $t\coloneq 2^jT$. Using $\beta\le 1.04\cdot 2^{-j}$ and $|\operatorname{Re}(\widetilde h_r)|\ge 0.96\cdot 2^{-(j+1)}$, we get
    \begin{equation*}
        p^{\mathrm H}_{(\sigma P_r)}
        \ge
        \frac{1}{|S|}\left(0.96\cdot\frac{T}{2}-\frac{1}{M}\sum_{q\ge2}\frac{(1.04\cdot TM)^q}{q!}\right)^2 
        \ge
        \frac{1}{|S|M^2}\left(1.52TM-e^{1.04TM}+1\right)^2 .
    \end{equation*}
    For $TM\in[0.2,0.52]$, this gives $p^{\mathrm H}_{(\sigma P_r)}\ge \frac{1}{200|S|M^2}$. Therefore, using Eq.~\eqref{eq:hnd-prob-sample-Pr-mod-stabilizers-unraveling},
    \begin{equation}
    \label{eq:sample-stab-under-hamilt}
        \Pr(\text{sample $P_r$ modulo stabilizers}) \ge \tr\left(\sum_{s\in S} O_{sP_r}\,\mathcal E_t(\rho_{\mathrm{code}})\right)
        = \sum_{\sigma \in S}p^{\mathrm H}_{(\sigma P_r)}
        \ge \frac{1}{200M^2}.
    \end{equation}

    The calculation of $\Pr(\text{sample a stabilizer})$ proceeds identically except with the replacement $P_r\to I$ in Eq.~\eqref{eq:prob-sigma-Pr}. In this case the identity coefficient
    $-\frac{i}{2}k_0 I$ contributes at first order to the stabilizer amplitude. Hence, for any $\sigma\in S$ and $TM\in[0.2,0.52]$ we obtain
    \begin{equation*}
        p^{\mathrm H}_{(\sigma)}
        \ge
        \frac{1}{|S|}
        \left(
            1-\frac{k_0t}{2}
            -
            \frac{1}{M}\sum_{q\ge2}\frac{(1.04\cdot TM)^q}{q!}
        \right)^2 .
    \end{equation*}

        Using $t=2^jT$ and $k_0\le K_\infty\le 2^{-j}/50$, for
    $TM\in[0.2,0.52]$ we obtain
    \begin{equation*}
        p^{\mathrm H}_{(\sigma)}
        \ge
        \frac{1}{|S|}
        \left(
            1-\frac{1}{M}
            \left[
                \frac{TM}{100}
                +e^{1.04TM}-1-1.04TM
            \right]
        \right)^2
        \ge
        \frac{1}{|S|}
        \left(1-\frac{1}{2M}\right),
    \end{equation*}
    where the last inequality follows since
    $x/100+e^{1.04x}-1-1.04x<1/5$ for $x\in[0.2,0.52]$.

    Therefore, using Eq.~\eqref{eq:hnd-prob-sample-Pr-mod-stabilizers-unraveling} and summing over all stabilizers, we get,
    \begin{equation*}
        \Pr(\text{sample a stabilizer})
        \ge 
        \tr\left(\sum_{s\in S} O_s\,\mathcal E_t(\rho_{\mathrm{code}})\right)
        =
        \sum_{\sigma\in S}p^{\mathrm H}_{(\sigma)}
        \ge
        1-\frac{1}{2M}.
    \end{equation*}

    Since $t = 2^j/(2M)$ corresponds to $TM = 0.5 \in [0.2,0.52]$, this concludes the proof.
\end{proof}

The following corollary is useful for reducing the ancilla cost of the candidate identification in Appendix~\ref{sec:appendix-hamilt-struct-sampling}.
\begin{corollary}[Chi-matrix entries lower bounds]
    In the setting of \Cref{lem:bell-sampling-of-eff-hamilt-struct}, consider the chi-matrix representation of $e^{\LindbladEff t}$:
    \begin{equation*}
        e^{\LindbladEff t}(\rho)
        =
        \sum_{P_m, P_w\in G_{\mathrm{eff}}}
        \chi_{mw}(t) P_m \rho P_w .
    \end{equation*}
    In this representation, the sum over $G_{\mathrm{eff}}$ is understood modulo global phases, with one Hermitian representative chosen for each phase class; we choose the representative of the target phase class to be $P_r$.
    
    Then:
    \begin{equation*}
        \begin{aligned}
            \chi_{rr}(t)&=\Pr(\text{sample $P_r$ modulo stabilizers})
            \ge
            \frac{1}{200M^2},\\
             \chi_{00}(t) &=\Pr(\text{sample a stabilizer}) \ge 1-\frac{1}{2M}.
        \end{aligned}
    \end{equation*}
\end{corollary}

\begin{proof}
    Fix $\sigma\in S$ and write
    $Z_r\coloneq \sigma P_r$ for the corresponding representative of the coset $SP_r$.
    Using the chi-matrix representation of $e^{\LindbladEff t}$ and the definition of the Bell-measurements projector $O_P$, we get
    \begin{equation*}
        \begin{aligned}
            \Pr(\text{sample $Z_r$})
            \coloneq \tr(O_{Z_r} e^{\LindbladEff t}(\rho_\mathrm{code})) &= |S| \sum_{P_m,P_w\in G_{\mathrm{eff}}} \chi_{mw}(t)
            \bra{\Phi_0} (Z_r^\dagger P_m\Pi \otimes I_{n'}) \ket{\Phi_0}\bra{\Phi_0} (\Pi P_w Z_r \otimes I_{n'}) \ket{\Phi_0}  = \\
            &= \frac{1}{|S|} \sum_{P_m\in G_{\mathrm{eff}}} \chi_{mm}(t)
            \bigg|\bra{\Phi_0} (P_r P_m\otimes I_{n'}) \ket{\Phi_0}\bigg|^2 
            = \frac{\chi_{rr}(t)}{|S|} .
        \end{aligned}
    \end{equation*}
    where the second equality follows from $P_r,P_m \in G_{\mathrm{eff}}$, $\Pi = \frac{1}{|S|}\sum_{s\in S} s$, $\Pi \sigma = \sigma \Pi = \Pi$ and the assumption that no nontrivial product of elements of $\mathcal S_{\mathrm{eff}}$ is proportional to a stabilizer. The last equality follows from the orthogonality of Bell states.  Therefore, combining the probabilities across all representatives of the coset $SP_r$,
    \begin{equation} \label{eq:HND-prob-sample-Pr-modulo-stabilizers}
        \Pr(\text{sample $P_r$ modulo stabilizers})
        =
        \tr\left(\sum_{\sigma\in S} O_{\sigma P_r}\,\widetilde{\rho}\right)
        =
        \sum_{\sigma\in S}\Pr(\text{sample $\sigma P_r$})
        =
        \chi_{rr}(t).
    \end{equation}
    Similarly, by setting $P_r = I_{n'}$, we get $\Pr(\text{sample a stabilizer}) = \chi_{00}(t)$.

    Combining the expressions for diagonal $\chi$ matrix entries above with \Cref{lem:bell-sampling-of-eff-hamilt-struct} proves the remark.
\end{proof}

%% file: appendix/HDD-barrier.tex
\section{Optimality of the \texorpdfstring{\HNDAbbrev{}}{HDD} Boundary and Relation to HNLS}
\label{sec:hdd-boundary}

As shown in \cref{cor:ansatz-free-hnd}, combining our dissipator structure learner
(\cref{alg:appendix-diss-struct-learning}) with the Hamiltonian learner (\cref{alg:hamilt-hier-learning}) achieves
ansatz-free Heisenberg-limited learning of every Hamiltonian Pauli
coefficient $h_r$ whose Pauli $P_r$ satisfies the following \HND{}
(\HNDAbbrev{}) condition
\begin{equation}
\label{eq:hnd-condition}
    P_r
    \notin
    \{I\}\cup\FullDissStruct = \{I\} \cup \DiagDissStruct
    \cup
    \bigl\{
        \Herm(P_\ell P_m):
        P_\ell,P_m\in\DiagDissStruct,\,
        P_\ell\neq P_m
    \bigr\}.
\end{equation}
We denote the set of nontrivial Hamiltonian Pauli terms satisfying this condition by
\begin{equation*}
    \HNDStruct
    \coloneq
    \HamiltStruct\setminus\FullDissStruct.
\end{equation*}
Thus, the \HNDAbbrev{} condition is sufficient for Heisenberg-limited
learning even when the structures of both the Hamiltonian and
dissipator are unknown. In this section, we ask whether the \HNDAbbrev{} condition is also
necessary for ansatz-free Hamiltonian learning.

This is not implied by the standard Hamiltonian-not-in-Lindblad-span
(HNLS) criterion~\cite{zhou2018achieving,demkowicz2017qecmetrology,gorecki2020optimal,abbasgholinejad2026error}. When the dissipator is known exactly, HNLS can allow
Heisenberg-limited estimation of Hamiltonian coefficients whose Pauli
operators lie inside $\FullDissStruct$. This leaves an apparent gap
between the known-dissipator and ansatz-free black-box settings.

We prove that this gap is fundamental. In particular,
\cref{thm:hdd-local-nonextendability} shows that uniform ansatz-free
Hamiltonian learning outside the \HNDAbbrev{} sector is
standard-quantum-limited, requiring
$T=\Omega(\varepsilon^{-2})$.

For this section, we let $A=(a_{kj})\succeq0$ and write the Lindbladian as
\begin{equation}
\label{eq:hdd-boundary-lindbladian}
    \Lindblad_{H,A}(\rho)
    =
    -i\sum_{P_k\in\HamiltStruct}
    h_k\comm{P_k}{\rho}
    +
    \sum_{P_k,P_j\in\DiagDissStruct}
    a_{kj}
    \left(
        P_k\rho P_j
        -
        \frac{1}{2}\acomm{P_jP_k}{\rho}
    \right).
\end{equation}

\begin{theorem}[Optimality of the \HNDAbbrev{} boundary]
\label{thm:hdd-local-nonextendability}
Let $\Lindblad_0=\Lindblad_{H_0,A_0}$ be an $n$-qubit Lindbladian with dissipator structure $\DiagDissStruct$ and sparsity $M_0 = |\HamiltStruct| + |\DiagDissStruct|^2$. Normalize its coefficient scale as $\Lambda_0 := \max\{\max_k |h_k|,\max_{k,j}|a_{kj}|\}=1$. Suppose the learner knows $\DiagDissStruct$ and is promised that the unknown Lindbladian satisfies $M\le 2M_0$ and $\Lambda\le 2$, but is given neither the Kossakowski matrix nor the Hamiltonian structure or coefficients.

Then, for every nonidentity $P_r\in\FullDissStruct$ and all sufficiently
small $\varepsilon>0$, learning its Hamiltonian coefficient $h_r$ to
additive accuracy $\varepsilon$ with success probability at least $2/3$,
uniformly over a neighborhood of $\Lindblad_0$ within this promised
model class, requires worst-case total black-box evolution-time budget
\begin{equation*}
    T
    =
    \Omega\left(\frac{1}{\varepsilon^2}\right).
\end{equation*}
The lower bound holds for arbitrary finite-depth adaptive protocols
with noiseless ancillas and quantum memory, intermediate quantum operations
and measurements, and adaptively chosen query durations.
\end{theorem}
\begin{proof}
    See Appendix~\ref{sec:hdd-optimality-proof}.
\end{proof}

\begin{remark}[Range of $\varepsilon$]
The restriction to sufficiently small $\varepsilon$ is not a fundamental precision limit; it only ensures that the nearby hard deformation used in the lower bound remains within the promised model class. The precise condition depends on the particular ansatz-free model; for the promises above, an explicit admissible range is derived in Appendix~\ref{sec:deformation-instantiation}.
\end{remark}

To prove this lower bound, we consider the simpler task of distinguishing the given Lindbladian $\Lindblad_0$ from a nearby hypothesis
$\Lindblad_\theta$ along a one-parameter joint Hamiltonian--dissipator
deformation we describe below.

This section is organized as follows. The same organization provides
an outline of the proof:
\begin{itemize}
    \item \textbf{Relation to HNLS.}
    In Appendix~\ref{sec:hnls-vs-hdd}, we review the known HNLS no-go results in the setting where the dissipator is known and compare them with the \HNDAbbrev{} criterion.

    \item \textbf{Hard deformation and single-query distinguishability growth.}
    In Appendix~\ref{sec:hdd-deformation-single-query}, we choose the deformation so that the first-order distinguishability between
    $\Lindblad_0$ and $\Lindblad_\theta$ is canceled. For arbitrary joint
    system--ancilla states arising under the two hypotheses, we show that a
    black-box query of duration $t$ can decrease their root fidelity by at most
    $\mathcal O(\theta^2 t)$. Thus, a single query generates statistical
    distinguishability only at second order in the deformation parameter.

    \item \textbf{Instantiation of the hard Pauli deformation.}
    In Appendix~\ref{sec:deformation-instantiation}, we instantiate this deformation for every target Pauli
    $P_r\in\FullDissStruct$ and prove that, for sufficiently small $\theta$, it
    remains within the promised ansatz-free model class. In particular, the
    diagonal dissipator support and Kossakowski rank are preserved, and, where
    applicable, the functional-HNLS condition can be preserved along the hard
    family.

    \item \textbf{Adaptive distinguishability bound.}
    In Appendix~\ref{sec:adaptive-distinguish}, we represent an arbitrary finite-depth
    adaptive learning protocol as a decision tree and telescope the single-query
    fidelity bound over the measurement history. This yields a distinguishability
    bound in terms of a deterministic worst-case total black-box evolution-time
    budget $T$, while allowing arbitrary quantum memory, intermediate controls
    and measurements, and adaptive query durations.

    \item \textbf{Putting it together.}
    In Appendix~\ref{sec:hdd-optimality-proof}, we compare $\Lindblad_0$ and
    $\Lindblad_{4\varepsilon}$ for $\varepsilon$ sufficiently small that both
    hypotheses remain within the promised model class. Since their target
    Hamiltonian coefficients differ by $4\varepsilon$, any
    $\varepsilon$-accurate learner with constant success probability must
    distinguish the two hypotheses with constant bias. The adaptive
    distinguishability bound therefore implies
    \begin{equation*}
        T=\Omega\left(\frac{1}{\varepsilon^2}\right),
    \end{equation*}
    ruling out Heisenberg-limited learning beyond the \HNDAbbrev{} boundary.
\end{itemize}

\subsection{Gap between HNLS and \texorpdfstring{\HNDAbbrev{}}{HDD}} 
\label{sec:hnls-vs-hdd}

Heisenberg-limited Hamiltonian estimation is generally fragile to
Markovian noise~\cite{huelga1997improvement,escher2011general,demkowicz2012elusive,demkowicz2014using,smirne2016noisyfreqest,zhou2018achieving,demkowicz2017qecmetrology,gorecki2020optimal}.
While noiseless parameter estimation permits an error scaling as
$1/T$, generic Markovian noise reduces the asymptotic scaling to the
standard quantum limit $1/\sqrt{T}$. Quantum error correction can
restore Heisenberg scaling only when the signal Hamiltonian is
sufficiently distinguishable from the noise.

Zhou \emph{et al.}~\cite{zhou2018achieving} formulated the
Hamiltonian-not-in-Lindblad-span (HNLS) criterion for estimating a
single Hamiltonian parameter in the presence of known Markovian
dissipation. Consider a Hamiltonian $H=\omega G$ and known jump
operators $\{J_\mu\}$. For Hermitian $G$, the HNLS condition can
equivalently be written using the real Hermitian Lindblad span
\begin{equation}
\label{eq:real-lindblad-span}
    \mathsf S_L(A)
    \coloneq
    \operatorname{span}_{\mathbb R}
    \left\{
        I,\,
        (J_\mu)_h,\,
        (J_\mu)_a,\,
        (J_\mu^\dagger J_\nu)_h,\,
        (J_\mu^\dagger J_\nu)_a
    \right\}_{\mu,\nu},
\end{equation}
where
\begin{equation*}
    X_h=\frac{X+X^\dagger}{2},
    \qquad
    X_a=\frac{X-X^\dagger}{2i}.
\end{equation*}
For Hamiltonian $H = \omega G$, the HNLS condition is
\begin{equation}
\label{eq:hnls-condition}
    G\notin\mathsf S_L(A).
\end{equation}
With noiseless ancillas and arbitrarily fast control, this condition is
necessary and sufficient for Heisenberg-limited estimation of
$\omega$~\cite{zhou2018achieving,demkowicz2017qecmetrology}.

Note that our terminology differs slightly from that used in the
metrology literature. We use the Lindbladian $\Lindblad$ to denote the
full generator, including both its Hamiltonian and dissipative parts,
whereas the HNLS literature uses ``Lindblad'' primarily in referring to
the dissipative operators $J_\mu$, conventionally called Lindblad jump
operators, and to the corresponding Lindblad span.

The HNLS criterion was later extended to multiparameter
estimation~\cite{gorecki2020optimal}. Most relevant here is the
\emph{functional HNLS} condition of Abbasgholinejad
\emph{et al.}~\cite{abbasgholinejad2026error}, which characterizes
Heisenberg-limited estimation of a function of several unknown
Hamiltonian parameters. Specializing their result to an otherwise
unrestricted Pauli Hamiltonian
\begin{equation*}
    H
    =
    \sum_{P_k\in\PnHerm\setminus\{I\}}
    h_k P_k
\end{equation*}
with a fully known dissipator, the condition for learning the bare
coefficient $h_r$ is
\begin{equation}
\label{eq:functional-hnls-pauli}
    P_r\perp\mathsf S_L(A).
\end{equation}
Thus functional HNLS is stronger than ordinary HNLS: it requires every
Lindblad-span nuisance direction to have zero $P_r$ coefficient, rather
than merely requiring $P_r\notin\mathsf S_L(A)$.

For comparison with the \HNDAbbrev{} condition, define
\begin{align*}
    \HNLSStruct(A)
    &\coloneq
    \left\{
        P_r\in\HamiltStruct:
        P_r\notin\mathsf S_L(A)
    \right\},\\
    \FunctionalHNLSStruct(A)
    &\coloneq
    \left\{
        P_r\in\HamiltStruct:
        P_r\perp\mathsf S_L(A)
    \right\}.
\end{align*}
Every jump operator $J_\mu$ is supported on $\DiagDissStruct$, while
every $J_\mu^\dagger J_\nu$ is supported on the identity and pairwise
products of dissipator Paulis. Hence
\begin{equation*}
    \mathsf S_L(A)
    \subseteq
    \operatorname{span}_{\mathbb R}
    \bigl(
        \{I\}\cup\FullDissStruct
    \bigr),
\end{equation*}
and Pauli orthogonality gives the hierarchy
\begin{equation}
\label{eq:hdd-functional-hnls-hierarchy}
    \HNDStruct
    \subseteq
    \FunctionalHNLSStruct(A)
    \subseteq
    \HNLSStruct(A).
\end{equation}
Thus, exact knowledge of the dissipator can make additional Hamiltonian
coefficients inside $\FullDissStruct$ Heisenberg-learnable.

To better understand this hierarchy, it is illustrative to consider a single Hermitian jump operator
\begin{equation*}
    J=X+Y.
\end{equation*}
Since $J^\dagger J=2I$,
\begin{equation*}
    \mathsf S_L
    =
    \operatorname{span}_{\mathbb R}\{I,X+Y\}.
\end{equation*}
Hence $Z\perp\mathsf S_L$, so the bare coefficient $\alpha_Z$ in
\begin{equation*}
    H=\alpha_X X+\alpha_Y Y+\alpha_Z Z
\end{equation*}
satisfies functional HNLS. However,
$\DiagDissStruct=\{X,Y\}$ and
\begin{equation*}
    \FullDissStruct=\{X,Y,Z\},
\end{equation*}
so $Z\notin\HNDStruct$. This gives a simple example where functional
HNLS permits Heisenberg-limited learning with a known dissipator, while
the \HNDAbbrev{} condition does not. Moreover, the second inclusion in
\cref{eq:hdd-functional-hnls-hierarchy} can also be strict: $X$ obeys
ordinary HNLS for this example but not functional HNLS.

\subsection{Hard Lindbladian deformation and single-query distinguishability growth} 
\label{sec:hdd-deformation-single-query}

Recall that our goal is to construct a nearby Lindbladian family
$\{\Lindblad_\theta\}_\theta$ through $\Lindblad_0$ whose members remain
difficult to distinguish even under arbitrary adaptive experiments.

An adaptive experiment can be represented as a decision tree, whose nodes
are labeled by the preceding measurement outcomes. At each node, the learner
holds a joint state of the system and ancillas, queries the unknown
evolution for some duration $t$, and may subsequently apply parameter-independent
controls before the next measurement. We first study a single such node and
bound how much distinguishability between two neighboring Lindbladian
hypotheses can be generated by one black-box query. The resulting bound will
later be composed over the full decision tree.

Our choice of hard deformation is motivated by a quantity that appears in
general bounds for noisy quantum metrology. For a differentiable Lindblad
family with Hamiltonian $H_\theta$ and jump operators $J_{\mu,\theta}$,
Refs.~\cite{catana2015fisher,guta2017information,kurdzialek2023using,das2025universal}
identify the effective coherent tangent
\begin{equation*}
    B_\theta
    :=
    \partial_\theta H_\theta
    -
    \frac{i}{2}
    \sum_\mu
    \left[
        (\partial_\theta J_{\mu,\theta}^\dagger)J_{\mu,\theta}
        -
        J_{\mu,\theta}^\dagger(\partial_\theta J_{\mu,\theta})
    \right].
\end{equation*}
In the metrological setting, this operator appears in bounds on the quantum
Fisher information (QFI) and captures the effective coherent tangent responsible for
quadratic-in-time QFI growth. Motivated by this criterion, we impose
$B_\theta=0$ and construct explicit deformations
$\partial_\theta J_{\mu,\theta}$ that realize it. For our distinguishing task, however, rather than working with quantum Fisher
information, we directly use the resulting cancellation to control the change of the
root fidelity between neighboring hypotheses. This yields the following construction:

\begin{definition}[Compensated Lindbladian deformation]
\label{def:compensated-deformation}
Let $\Lindblad_0$ admit a jump representation with Hamiltonian $H_0$
and jump operators $\{J_\mu\}_\mu$. Given operators $\{E_\mu\}_\mu$,
define
\begin{equation*}
    J_{\mu,\theta}
    :=
    J_\mu+\theta E_\mu,
    \qquad
    H_\theta
    :=
    H_0+\theta G,
\end{equation*}
where $G := \frac{i}{2}\sum_\mu \left( E_\mu^\dagger J_\mu - J_\mu^\dagger E_\mu \right).$ We call the resulting family $\{\Lindblad_\theta\}_\theta$ a
\emph{compensated Lindbladian deformation}.
\end{definition}

For positive operators $\rho,\sigma$, we write
$f(\rho,\sigma):=\|\sqrt{\rho}\sqrt{\sigma}\|_1$ for their root
fidelity, without requiring normalization.

\begin{lemma}[Single-query root-fidelity bound]
\label{lem:hdd-single-query-fidelity}
Let $\{\Lindblad_\theta\}_\theta$ be a compensated Lindbladian
deformation as in \cref{def:compensated-deformation}, and suppose
\begin{equation*}
    \sum_\mu E_\mu^\dagger E_\mu
    =
    \zeta I_S
\end{equation*}
for some $\zeta\ge0$.
Write $\Phi_\theta^t:=e^{t\Lindblad_\theta}$.
For any finite-dimensional ancillary memory $R$, any positive
operators $\rho,\sigma$ on $S\otimes R$, and any
$\theta,\theta'\in\mathbb{R}$ and $t\ge0$, we have
\begin{equation*}
    f\!\left(
        (\Phi_\theta^t\otimes I_R)(\rho),
        (\Phi_{\theta'}^t\otimes I_R)(\sigma)
    \right)
    \ge
    e^{-\zeta(\theta-\theta')^2t/2}
    f(\rho,\sigma).
\end{equation*}
\end{lemma}

\begin{proof}
Set $\Delta:=\theta-\theta'$. For positive operators
$A,B$, we use the block-positivity characterization of root fidelity~\cite[Theorem~3.17]{watrous2018theory}: 
\begin{equation*}
    f(A,B)
    =
    \max\left\{
        |\operatorname{tr}Y|:
        \begin{pmatrix}
            A & Y\\
            Y^\dagger & B
        \end{pmatrix}
        \succeq0
    \right\}.
\end{equation*}
Let $X_0$ be the maximizing operator with its phase adjusted so that
$\operatorname{tr}X_0=f(\rho,\sigma)$.

Following~\cite{molmer2015hypothesis}, to relate the evolutions under $\mathcal L_\theta$ and  $\mathcal L_{\theta'}$, we introduce an auxiliary flag qubit $F$ and the positive block operator
\begin{equation*}
    W(0)
    :=
    \begin{pmatrix}
        \rho & X_0\\
        X_0^\dagger & \sigma
    \end{pmatrix}_F
    \succeq0.
\end{equation*}
The flag is only a proof device and is not available to the learner. Next, define a Lindbladian
$\widetilde{\Lindblad}$ on $F\otimes S$ with Hamiltonian and jump
operators
\begin{equation*}
    \begin{aligned}
        \widetilde H
        &:=
        |0\rangle\langle0|_F\otimes H_\theta
        +
        |1\rangle\langle1|_F\otimes H_{\theta'},\\
        \widetilde J_\mu
        &:=
        |0\rangle\langle0|_F\otimes J_{\mu,\theta}
        +
        |1\rangle\langle1|_F\otimes J_{\mu,\theta'}.
    \end{aligned}
\end{equation*}
Evolving $W(0)$ under
$e^{t\widetilde{\Lindblad}}\otimes I_R$ gives
\begin{equation*}
    W(t)
    =
    \begin{pmatrix}
        \rho_t & X(t)\\
        X(t)^\dagger & \sigma_t
    \end{pmatrix}_F
    \succeq0,
\end{equation*}
where
$\rho_t=(\Phi_\theta^t\otimes I_R)(\rho)$ and
$\sigma_t=(\Phi_{\theta'}^t\otimes I_R)(\sigma)$.
Positivity follows from complete positivity of the enlarged evolution.

Suppressing identity factors on $R$, the off-diagonal block satisfies
\begin{equation*}
        \frac{dX}{dt}
        ={}
        -i(H_\theta X-XH_{\theta'})
        +
        \sum_\mu J_{\mu,\theta}XJ_{\mu,\theta'}^\dagger -
        \frac12\sum_\mu
        \left(
            J_{\mu,\theta}^\dagger J_{\mu,\theta}X
            +
            XJ_{\mu,\theta'}^\dagger J_{\mu,\theta'}
        \right).
\end{equation*}
Taking the trace and using cyclicity,
\begin{equation*}
    \frac{d}{dt}\operatorname{tr}X(t)
    =
    \operatorname{tr}\!\left[
        (K_{\theta,\theta'}\otimes I_R)X(t)
    \right],
\end{equation*}
where
\begin{equation*}
    \begin{aligned}
        K_{\theta,\theta'}
        :={}&
        -i(H_\theta-H_{\theta'})
        +
        \sum_\mu
        \left(
            J_{\mu,\theta'}^\dagger J_{\mu,\theta}
            -
            \frac12J_{\mu,\theta}^\dagger J_{\mu,\theta}
            -
            \frac12J_{\mu,\theta'}^\dagger J_{\mu,\theta'}
        \right)\\
        ={}&
        -i\Delta G
        +
        \frac{\Delta}{2}\sum_\mu
        \left(
            J_\mu^\dagger E_\mu-E_\mu^\dagger J_\mu
        \right)
        -
        \frac{\Delta^2}{2}\sum_\mu E_\mu^\dagger E_\mu = {} -\frac{\zeta\Delta^2}{2} I_S.
    \end{aligned}
\end{equation*}
The terms linear in $\Delta$ cancel exactly by the definition of $G$.
Consequently,
\begin{equation*}
    \operatorname{tr}X(t)
    =
    e^{-\zeta\Delta^2t/2}\operatorname{tr}X_0
    =
    e^{-\zeta\Delta^2t/2}f(\rho,\sigma).
\end{equation*}
Applying the block-positivity characterization to $W(t)$ yields
\begin{equation*}
    f(\rho_t,\sigma_t)
    \ge
    |\operatorname{tr}X(t)|
    =
    e^{-\zeta\Delta^2t/2}f(\rho,\sigma),
\end{equation*}
which proves the claim.
\end{proof}

Any subsequent parameter-independent CPTP map $\Psi$ satisfies
$f(\Psi(\rho_t),\Psi(\sigma_t))\ge f(\rho_t,\sigma_t)$ by
monotonicity of root fidelity~\cite[Theorem~3.27]{watrous2018theory}.
Thus the same lower bound remains valid after controls, resets,
or the addition or discarding of ancillas.

\subsection{Instantiating the Lindbladian deformation}
\label{sec:deformation-instantiation}

We now choose explicit perturbations $E_\mu$ from \cref{def:compensated-deformation} so that the target Hamiltonian
coefficient changes at unit speed, while the resulting family satisfies
the promises of \cref{thm:hdd-local-nonextendability}.
Fix $\Lindblad_0$ as in that theorem and choose a traceless jump
representation $J_\mu=\sum_{P_k\in\DiagDissStruct}\ell_{\mu k}P_k$
with $a_{kj}:=(A_0)_{kj}=\sum_\mu\ell_{\mu k}\ell_{\mu j}^*$.
In particular, $0<a_{kk}\le1$ for $P_k\in\DiagDissStruct$.
For any operator $X$, write $[X]_{P_k}:=2^{-n}\operatorname{tr}(P_kX)$.
Hamiltonian supports and coefficient scales refer only to nonidentity
Pauli coefficients.

We first treat Hamiltonian targets that arise as products of dissipator
Paulis but do not themselves belong to $\DiagDissStruct$, i.e.,
$P_r\in\FullDissStruct\setminus\DiagDissStruct$. Choose distinct
$P_\ell,P_m\in\DiagDissStruct$ such that
$P_\ell P_m=\omega P_r$, with $\omega\in\{\pm1,\pm i\}$, and label them
so that $a_{mm}\ge a_{\ell\ell}$. Since the compensated Hamiltonian
tangent is
\begin{equation*}
    G=\frac{i}{2}\sum_\mu
    \left(E_\mu^\dagger J_\mu-J_\mu^\dagger E_\mu\right),
\end{equation*}
a natural way to generate the target Pauli $P_r$ is to choose
$E_\mu$ along $P_\ell$, so that it interferes with the existing
$P_m$ component of $J_\mu$ and produces
$P_\ell P_m\propto P_r$. The scalar prefactor is then chosen so that
$[G]_{P_r}=1$, giving $h_r(\theta)=h_r(0)+\theta$, while
$\sum_\mu E_\mu^\dagger E_\mu$ is proportional to the identity as
required by \cref{lem:hdd-single-query-fidelity}.

\begin{lemma}[Pair-product deformation]
\label{lem:hdd-pair-deformation}
Let $P_r\in\FullDissStruct\setminus\DiagDissStruct$ be nonidentity.
Choose distinct $P_\ell,P_m\in\DiagDissStruct$ such that
$P_\ell P_m=\omega P_r$, where $\omega\in\{\pm1,\pm i\}$, and,
relabeling them if necessary such that $a_{mm}\ge a_{\ell\ell}$.
Set $E_\mu:=i\omega\ell_{\mu m}P_\ell/a_{mm}$ in
\cref{def:compensated-deformation}.
The resulting deformation family satisfies
$h_r(\theta)=h_r(0)+\theta$ and
$\sum_\mu E_\mu^\dagger E_\mu=\zeta I$, with $\zeta=1/a_{mm}$.

For every real $\theta$, $A_\theta$ remains a valid Kossakowski matrix
with the same rank as $A_0$. Moreover, whenever
\begin{equation*}
    |\theta|
    <
    \theta_{\mathrm{pair}}
    :=
    (\sqrt{2}-1)\sqrt{a_{\ell\ell}a_{mm}},
\end{equation*}
the deformed dissipator stucture remains $\DiagDissStruct$, and the
sparsity and coefficient scale satisfy $M_\theta\le2M_0$ and
$\Lambda_\theta\le2$.
If $P_r$ satisfies the functional-HNLS condition
$P_r\perp\mathsf S_L(A_0)$ before deformation, then
$P_r\perp\mathsf S_L(A_\theta)$ throughout this interval.
\end{lemma}

\begin{proof}
Since $P_\ell^\dagger P_\ell=I$, we have
$\sum_\mu E_\mu^\dagger E_\mu
=a_{mm}^{-2}\sum_\mu|\ell_{\mu m}|^2I=a_{mm}^{-1}I$.
Substituting the perturbations into the definition of $G$ gives
\begin{equation*}
    G
    =
    \frac{1}{2a_{mm}}
    \sum_{P_k\in\DiagDissStruct}
    \left(
        \omega^*a_{km}P_\ell P_k
        +
        \omega a_{mk}P_kP_\ell
    \right).
\end{equation*}
Because $P_mP_\ell=\omega^*P_r$, the summand with $k=m$ equals
$P_r$. No other summand contributes to its coefficient: multiplication
by $P_\ell$ is injective on Pauli labels modulo phase. Thus
$[G]_{P_r}=1$, proving $h_r(\theta)=h_r(0)+\theta$.

The same injectivity shows that $G$ has at most
$|\DiagDissStruct|-1$ nonidentity Pauli coefficients, since $k=\ell$
contributes only a scalar multiple of $I$. The coefficient associated
with $P_\ell P_k$ has magnitude at most
\begin{equation*}
    \frac{|a_{km}|}{a_{mm}}
    \le
    \sqrt{\frac{a_{kk}}{a_{mm}}}
    \le
    \frac{1}{\sqrt{a_{mm}}},
\end{equation*}
where the first inequality follows from $A_0\succeq0$.
Consequently, every nonidentity Hamiltonian coefficient satisfies
$|h_k(\theta)|\le1+|\theta|/\sqrt{a_{mm}}$.

Only the $P_\ell$ coefficient of each jump changes:
$\ell_{\mu\ell}(\theta)
=\ell_{\mu\ell}+i\omega\theta\ell_{\mu m}/a_{mm}$.
Writing $e_k$ for the Pauli-coordinate unit vectors, this gives
\begin{equation*}
    A_\theta
    =
    \left(I+\frac{i\omega\theta}{a_{mm}}e_\ell e_m^\dagger\right)
    A_0
    \left(I+\frac{i\omega\theta}{a_{mm}}e_\ell e_m^\dagger\right)^\dagger.
\end{equation*}
This congruence preserves positivity. Since $\ell\neq m$,
$(e_\ell e_m^\dagger)^2=0$, so the congruence matrix is invertible
for every real $\theta$. Hence
$\operatorname{rank}A_\theta=\operatorname{rank}A_0$.

All jumps remain supported on $\DiagDissStruct$, and every diagonal
entry other than $(A_\theta)_{\ell\ell}$ is unchanged. Collecting
$\ell_{\mu\ell}(\theta)
=\ell_{\mu\ell}+i\omega\theta\ell_{\mu m}/a_{mm}$ into vectors indexed
by $\mu$ and applying the triangle and reverse triangle inequalities gives
\begin{equation*}
    \sqrt{a_{\ell\ell}}-\frac{|\theta|}{\sqrt{a_{mm}}}
    \le
    \sqrt{(A_\theta)_{\ell\ell}}
    \le
    \sqrt{a_{\ell\ell}}+\frac{|\theta|}{\sqrt{a_{mm}}}.
\end{equation*}
For $|\theta|<\theta_{\mathrm{pair}}$, we have
$|\theta|/\sqrt{a_{mm}}<(\sqrt2-1)\sqrt{a_{\ell\ell}}$, and hence
\begin{equation*}
    (2-\sqrt2)\sqrt{a_{\ell\ell}}
    <
    \sqrt{(A_\theta)_{\ell\ell}}
    <
    \sqrt{2a_{\ell\ell}}.
\end{equation*}
The lower bound proves preservation of the dissipator structure
$\DiagDissStruct$. Since $a_{\ell\ell}\le1$, the upper bound gives
$(A_\theta)_{\ell\ell}<2$, while all other diagonal entries remain at
most $1$. Positivity therefore implies
$|(A_\theta)_{kj}|\le\sqrt{(A_\theta)_{kk}(A_\theta)_{jj}}\le2$.
Similarly,
$|h_k(\theta)|\le1+|\theta|/\sqrt{a_{mm}}<\sqrt2<2$, so
$\Lambda_\theta\le2$.
Finally, since $G$ contains at most $|\DiagDissStruct|-1$ additional
nonidentity Pauli coefficients,
$M_\theta\le M_0+|\DiagDissStruct|-1\le2M_0$.

It remains to verify preservation of the functional-HNLS condition
from Eq.~\eqref{eq:functional-hnls-pauli}. Suppose
$P_r\perp\mathsf S_L(A_0)$. Since
$P_r\notin\DiagDissStruct$, every deformed jump has zero $P_r$
coefficient. Moreover,
$[J_\mu^\dagger J_\nu]_{P_r}=0$ by the initial orthogonality
assumption, while $E_\mu^\dagger E_\nu$ is proportional to $I$.
The linear correction also vanishes:
\begin{equation*}
    [E_\mu^\dagger J_\nu+J_\mu^\dagger E_\nu]_{P_r}
    =
    -\frac{i\ell_{\mu m}^*\ell_{\nu m}}{a_{mm}}
    +
    \frac{i\ell_{\mu m}^*\ell_{\nu m}}{a_{mm}}
    =
    0.
\end{equation*}
Thus $[J_{\mu,\theta}^\dagger J_{\nu,\theta}]_{P_r}=0$ for every
$\mu,\nu$, so every generator of $\mathsf S_L(A_\theta)$ remains
orthogonal to $P_r$. Hence $P_r\perp\mathsf S_L(A_\theta)$, as claimed.
\end{proof}

For targets already in $\DiagDissStruct$, the physical dissipator
can be kept fixed. The required compensation is realized through
identity shifts of the jump operators.

\begin{lemma}[Direct-support deformation]
\label{lem:hdd-direct-deformation}
Let $P_r\in\DiagDissStruct$ and set
$E_\mu:=i\ell_{\mu r}I/(2a_{rr})$ in
\cref{def:compensated-deformation}.
The resulting family has the canonical Pauli representation
\begin{equation*}
    \Lindblad_\theta
    =
    \Lindblad_{H_0+\theta K_r,A_0},
    \qquad
    K_r
    :=
    \sum_{P_k\in\DiagDissStruct}
    \frac{\operatorname{Re}a_{kr}}{a_{rr}}P_k.
\end{equation*}
It satisfies $h_r(\theta)=h_r(0)+\theta$ and
$\sum_\mu E_\mu^\dagger E_\mu=\zeta I$, with
$\zeta=1/(4a_{rr})$.

The dissipator, its support, and its rank remain fixed for every real
$\theta$. The sparsity obeys $M_\theta\le2M_0$ for every $\theta$,
and $\Lambda_\theta\le2$ whenever
$|\theta|<\theta_{\mathrm{dir}}:=\sqrt{a_{rr}}$.
In this case, the target already fails functional HNLS:
$P_r\not\perp\mathsf S_L(A_0)$.
\end{lemma}

\begin{proof}
The scalar perturbations give
$\sum_\mu E_\mu^\dagger E_\mu
=(4a_{rr}^2)^{-1}\sum_\mu|\ell_{\mu r}|^2I
=I/(4a_{rr})$.
The compensated Hamiltonian tangent is
\begin{equation*}
    G
    =
    \frac{1}{4a_{rr}}
    \sum_\mu
    \left(
        \ell_{\mu r}^*J_\mu+\ell_{\mu r}J_\mu^\dagger
    \right)
    =
    \frac{K_r}{2}.
\end{equation*}
Importantly, this is the Hamiltonian tangent in the shifted jump
representation, not yet in the canonical Pauli representation.

To relate the two representations, write
$\mathcal D[J](\rho):=J\rho J^\dagger
-\frac12\{J^\dagger J,\rho\}$.
For any scalar $c$, direct expansion gives
\begin{equation*}
    \mathcal D[J+cI](\rho)
    =
    \mathcal D[J](\rho)
    -
    i\left[
        \frac{i}{2}(c^*J-cJ^\dagger),\rho
    \right].
\end{equation*}
Applying this identity to each shift $\theta E_\mu$ shows that
the shifted jumps contribute the additional Hamiltonian $\theta G$.
Together with the explicit Hamiltonian $H_0+\theta G$, they therefore
represent $\Lindblad_{H_0+2\theta G,A_0}
=\Lindblad_{H_0+\theta K_r,A_0}$.
Since $[K_r]_{P_r}=1$, the physical target coefficient changes at
unit speed.

The dissipator is unchanged, so its positivity, support, and rank
are preserved. Because $K_r$ is supported on $\DiagDissStruct$,
we have $M_\theta\le M_0+|\DiagDissStruct|\le2M_0$.
Moreover, positivity of $A_0$ gives
\begin{equation*}
    |[K_r]_{P_k}|
    =
    \frac{|\operatorname{Re}a_{kr}|}{a_{rr}}
    \le
    \sqrt{\frac{a_{kk}}{a_{rr}}}
    \le
    \frac{1}{\sqrt{a_{rr}}}.
\end{equation*}
Hence $|h_k(\theta)|\le1+|\theta|/\sqrt{a_{rr}}\le2$
throughout the stated interval. All Kossakowski coefficients remain
bounded by $1$, proving $\Lambda_\theta\le2$.

Finally, $a_{rr}>0$ implies that $\ell_{\mu r}\neq0$ for some $\mu$.
At least one of $(J_\mu)_h$ and $(J_\mu)_a$ then has a nonzero
$P_r$ coefficient. Since both belong to $\mathsf S_L(A_0)$,
we obtain $P_r\not\perp\mathsf S_L(A_0)$.
\end{proof}

\subsection{Adaptive distinguishability bound}
\label{sec:adaptive-distinguish}

Following the learning-tree framework for adaptive quantum learning
protocols~\cite{chen2022exponential}, we represent the learner as a rooted
decision tree. Rather than reducing the analysis to the classical
distribution over measurement histories, we retain the learner's quantum
memory explicitly by associating to each node a subnormalized joint
system--memory state, whose trace is the probability of reaching that
node. This allows us to propagate root fidelity through the tree and
telescope the single-query bound from
\cref{lem:hdd-single-query-fidelity} directly over the adaptive protocol.

The root corresponds to the empty measurement history. Each subsequent
node $v$ corresponds to a complete classical transcript of the outcomes
observed so far, and is associated under hypothesis $\theta$ with a
positive subnormalized state $\rho_v^\theta$ of the system plus retained ancillas. Its trace
$\operatorname{tr}\rho_v^\theta$ is the probability of reaching $v$.
At a nonterminal node, the protocol chooses, as a function only of the
preceding transcript, a query duration $t_v$ and any
parameter-independent controls and ancilla operations. It then applies
the black-box evolution $\Phi_\theta^{t_v}$ to the system, leaving the
retained memory untouched, followed by a parameter-independent quantum
instrument $\{\mathcal M_{v\to u}\}_{u\in\operatorname{ch}(v)}$ whose
outcome determines the child node. Thus, absorbing the pre-query controls
into the state at $v$, the child branch states take the form
\begin{equation}
\label{eq:adaptive-tree-branch-state}
    \rho_u^\theta
    =
    \mathcal M_{v\to u}
    \left[
        \left(\Phi_\theta^{t_v}\otimes\operatorname{id}\right)
        (\rho_v^\theta)
    \right],
    \qquad
    u\in\operatorname{ch}(v),
\end{equation}
where each $\mathcal M_{v\to u}$ is completely positive and trace
nonincreasing and $\sum_u\mathcal M_{v\to u}$ is trace preserving.
The same tree, query durations, and conditional operations are used under
both hypotheses; only the branch states and hence the probabilities of
reaching the nodes may differ.

Retaining the complete classical transcript, the terminal state under
hypothesis $\theta$ is the classical--quantum state
\begin{equation}
\label{eq:adaptive-tree-output-state}
    \rho_\theta^{\mathrm{out}}
    =
    \bigoplus_{\ell\in\operatorname{leaf}(\mathcal T)}
    \rho_\ell^\theta .
\end{equation}
Thus the leaf probabilities are already encoded as
$\operatorname{tr}\rho_\ell^\theta$, while the blocks retain any quantum
memory available to the learner. Any final classical estimate is obtained
from this state by a parameter-independent quantum channel.

With this definition of the learning tree, we now turn to the proof.
We first record how root fidelity behaves when a measurement outcome is
retained.

\begin{lemma}[Root fidelity under recorded instruments]
\label{lem:hdd-recorded-instrument}
Let $\{\mathcal M_y\}_y$ be a quantum instrument, so that each
$\mathcal M_y$ is completely positive and trace nonincreasing, and
$\sum_y\mathcal M_y$ is trace preserving. For any positive operators
$\rho,\sigma$,
\begin{equation*}
    f\!\left(
        \bigoplus_y\mathcal M_y(\rho),
        \bigoplus_y\mathcal M_y(\sigma)
    \right)
    =
    \sum_y f\!\left(\mathcal M_y(\rho),\mathcal M_y(\sigma)\right)
    \ge
    f(\rho,\sigma).
\end{equation*}
The direct sums encode the classical outcome, while the quantum output
of each $\mathcal M_y$ retains the corresponding post-measurement memory.
The instrument may also include controls, resets, and the addition or
discarding of ancillas.
\end{lemma}

\begin{proof}
The recorded map
$\mathcal M_{\mathrm{rec}}(X):=\bigoplus_y\mathcal M_y(X)$ is CPTP, so
the inequality follows from monotonicity of root
fidelity~\cite[Theorem~3.27]{watrous2018theory}.
The equality follows because square roots, operator products, and the
trace norm respect direct sums.
\end{proof}

Crucially, the branch operators are not renormalized: their traces
already encode the different probabilities of reaching each outcome under
the two hypotheses. Thus the lemma controls the sum of the branch
fidelities, rather than the fidelities of the normalized conditional
states.

\begin{lemma}[Adaptive distinguishability under a worst-case time budget]
\label{lem:hdd-adaptive-distinguishability}
Let $\{\Lindblad_\theta\}_\theta$ satisfy the assumptions of
\cref{lem:hdd-single-query-fidelity}, with constant $\zeta$.
Consider any finite-depth adaptive protocol initialized in the same
normalized state under both hypotheses, with arbitrary noiseless quantum
memory, intermediate quantum operations and measurements, and
history-dependent query durations. Suppose that along every terminal
transcript $y$, the total black-box evolution time satisfies $\sum_{v\in\operatorname{path}(y)} t_v \le T,$ where the sum is over query nodes and $t_v$ is the duration chosen at
node $v$.

Then the final states, including the complete classical transcript, satisfy
\begin{equation*}
    f\!\left(
        \rho_\theta^{\mathrm{out}},
        \rho_{\theta'}^{\mathrm{out}}
    \right)
    \ge
    e^{-\zeta(\theta-\theta')^2T/2}.
\end{equation*}
Consequently, writing
$D_{\mathrm{tr}}(\rho,\sigma):=\frac12\|\rho-\sigma\|_1$,
\begin{equation*}
    D_{\mathrm{tr}}\!\left(
        \rho_\theta^{\mathrm{out}},
        \rho_{\theta'}^{\mathrm{out}}
    \right)
    \le
    \sqrt{1-e^{-\zeta(\theta-\theta')^2T}}
    \le
    |\theta-\theta'|\sqrt{\zeta T}.
\end{equation*}
\end{lemma}

\begin{proof}
Set $c:=\zeta(\theta-\theta')^2/2$. At a node $v$, let
$\rho_v,\sigma_v$ denote the subnormalized joint system--memory states
under the two hypotheses immediately before the next query. Their traces
are the respective probabilities of reaching $v$.

The query duration $t_v$ is determined by the preceding transcript and is
therefore the same under both hypotheses at node $v$. After the query, all
parameter-independent processing and measurement can be represented by
the quantum instrument whose outcomes label the children of $v$.
Applying
\cref{lem:hdd-single-query-fidelity,lem:hdd-recorded-instrument}
therefore gives
\begin{equation*}
    \sum_{u\in\operatorname{ch}(v)}
    f(\rho_u,\sigma_u)
    \ge
    e^{-ct_v} f(\rho_v,\sigma_v),
\end{equation*}
where $\operatorname{ch}(v)$ denotes the children of $v$.

Let $s_v$ be the accumulated black-box evolution time upon arrival at
$v$. Every child $u$ of $v$ satisfies $s_u=s_v+t_v$, and hence
\begin{equation*}
    \sum_{u\in\operatorname{ch}(v)}
    e^{cs_u}f(\rho_u,\sigma_u)
    \ge
    e^{cs_v}f(\rho_v,\sigma_v).
\end{equation*}
Thus replacing a node by its children cannot decrease the sum of these
weighted fidelities. Since the learning tree has finite depth, iterating
this inequality from the root to the terminal nodes gives
\begin{equation*}
    1
    \le
    \sum_{y\ \mathrm{terminal}}
        e^{cs_y}f(\rho_y,\sigma_y)
    \le
    e^{cT}
    \sum_{y\ \mathrm{terminal}}f(\rho_y,\sigma_y)
    =
    e^{cT}
    f\!\left(
        \rho_\theta^{\mathrm{out}},
        \rho_{\theta'}^{\mathrm{out}}
    \right).
\end{equation*}
Here the root has accumulated time zero and fidelity one, every terminal
history satisfies $s_y\le T$, and the final equality follows from the
direct-sum decomposition over the complete classical transcript.
Therefore
\begin{equation*}
    f\!\left(
        \rho_\theta^{\mathrm{out}},
        \rho_{\theta'}^{\mathrm{out}}
    \right)
    \ge
    e^{-cT}
    =
    e^{-\zeta(\theta-\theta')^2T/2}.
\end{equation*}

Finally, the output states are normalized, so the Fuchs--van de Graaf
inequality~\cite[Theorem~3.33]{watrous2018theory} gives
$D_{\mathrm{tr}}\le\sqrt{1-f^2}$. Substituting the fidelity bound and
using $1-e^{-x}\le x$ yields
\begin{equation*}
    D_{\mathrm{tr}}\!\left(
        \rho_\theta^{\mathrm{out}},
        \rho_{\theta'}^{\mathrm{out}}
    \right)
    \le
    \sqrt{1-e^{-\zeta(\theta-\theta')^2T}}
    \le
    |\theta-\theta'|\sqrt{\zeta T}.
\end{equation*}
Any final parameter-independent processing, including discarding part of
the transcript or quantum memory, can only increase root fidelity, so the
same bounds apply to the learner's actual output.
\end{proof}

\subsection{Optimality proof for the \texorpdfstring{\HNDAbbrev{}}{HDD} boundary}
\label{sec:hdd-optimality-proof}

We complete the proof by converting an accurate Hamiltonian coefficient
estimator into a test distinguishing the two hard hypotheses.

\begin{restatethm}{thm:hdd-local-nonextendability}[Optimality of the \HNDAbbrev{} boundary]
Let $\Lindblad_0=\Lindblad_{H_0,A_0}$ be an $n$-qubit Lindbladian
with dissipator structure $\DiagDissStruct$ and sparsity
$M_0=|\HamiltStruct|+|\DiagDissStruct|^2$.
Normalize its coefficient scale as
$\Lambda_0:=\max\{\max_k|h_k|,\max_{k,j}|a_{kj}|\}=1$.
Suppose the learner knows $\DiagDissStruct$ and is promised that the
unknown Lindbladian satisfies $M\le2M_0$ and $\Lambda\le2$, but is
given neither the Kossakowski matrix nor the Hamiltonian structure
or coefficients.

Then, for every nonidentity $P_r\in\FullDissStruct$ and all sufficiently
small $\varepsilon>0$, learning its Hamiltonian coefficient $h_r$ to
additive accuracy $\varepsilon$ with success probability at least $2/3$,
uniformly over a fixed neighborhood of $\Lindblad_0$ within this
promised model class, requires worst-case total black-box evolution-time
budget
\begin{equation*}
    T=\Omega\left(\frac{1}{\varepsilon^2}\right).
\end{equation*}
The lower bound holds for arbitrary finite-depth adaptive protocols
with noiseless ancillas and quantum memory, intermediate quantum
operations and measurements, and adaptively chosen query durations.
\end{restatethm}

\begin{proof}
Fix a nonidentity $P_r\in\FullDissStruct$ and a relative neighborhood
$\mathcal U$ of $\Lindblad_0$ within the promised model class,
independent of $\varepsilon$.
By \cref{lem:hdd-pair-deformation,lem:hdd-direct-deformation}, choose
the corresponding compensated family with
$h_r(\theta)=h_r(0)+\theta$.
Its single-query constant is $\zeta=1/a_{mm}$ in the pair-product
case ($P_r \propto P_\ell P_m$ for distinct $P_\ell,P_m \in \DiagDissStruct$) and $\zeta=1/(4a_{rr})$ in the direct-support case.
In either case, $\zeta$ is strictly positive, finite, and independent
of $\varepsilon$.

The explicit promises hold for both $\Lindblad_0$ and
$\Lindblad_{4\varepsilon}$ whenever
$4\varepsilon<\theta_{\mathrm{pair}}$ or
$4\varepsilon<\theta_{\mathrm{dir}}$, respectively.
By continuity of the family through $\Lindblad_0$, both hypotheses
also lie in $\mathcal U$ for all sufficiently small $\varepsilon$.
They therefore have the same known dissipator structure and satisfy
the same supplied sparsity and coefficient-scale bounds.

Suppose a learner with worst-case evolution-time budget $T$ outputs
an estimator $\widehat h_r$ that is $\varepsilon$-accurate with
probability at least $2/3$ throughout $\mathcal U$.
Thresholding $\widehat h_r$ at $h_r(0)+2\varepsilon$ distinguishes
$\Lindblad_0$ from $\Lindblad_{4\varepsilon}$ with success probability
at least $2/3$ under each hypothesis. Therefore, with equal prior probability on the two hypotheses, the
induced binary test succeeds with probability at least $2/3$.
By the Holevo--Helstrom~\cite{holevo1973statistical,helstrom1969quantum} theorem~\cite[Theorem~3.4]{watrous2018theory},
this requires $D_{\mathrm{tr}}\!\left(\rho_0^{\mathrm{out}},\rho_{4\varepsilon}^{\mathrm{out}}\right)\ge \frac{1}{3}.$ Combining this with
\cref{lem:hdd-adaptive-distinguishability} gives

\begin{equation*}
    \frac13
    \le
    D_{\mathrm{tr}}\!\left(
        \rho_0^{\mathrm{out}},
        \rho_{4\varepsilon}^{\mathrm{out}}
    \right)
    \le
    \sqrt{1-e^{-16\zeta\varepsilon^2T}}.
\end{equation*}
Rearranging yields
\begin{equation*}
    T
    \ge
    \frac{\ln(9/8)}{16\zeta\varepsilon^2}
    =
    \Omega\left(\frac{1}{\varepsilon^2}\right),
\end{equation*}
where the implicit constant depends on the fixed deformation but
not on $\varepsilon$. This proves the claim.
\end{proof}

\begin{remark}[Relation to the finite-resolution Heisenberg guarantee]
\label{rem:hdd-lower-bound-finite-resolution}
At first sight, our finite-resolution algorithm
(\cref{cor:ansatz-free-hnd-balanced-tail}) may appear to contradict
\cref{thm:hdd-local-nonextendability}, since it can learn some Hamiltonian
coefficients inside the full dissipator footprint at the Heisenberg limit.
There are two ways in which the finite-resolution guarantee avoids the
asymptotic lower-bound obstruction.

First, the target may lie entirely in the weak dissipator footprint. This
includes a direct-support target $P_r\in\DiagDissStruct$ with
$a_{rr}<\eta$, or a pair-product target $P_r=P_\ell P_m$ for which both
$a_{\ell\ell},a_{mm}<\eta$. After plugging in $\zeta$, the corresponding distinguishability bounds
scale as
\begin{equation*}
    T=\Omega\left(\frac{a_{rr}}{\varepsilon^2}\right)
    \qquad\text{or}\qquad
    T=\Omega\left(\frac{a_{mm}}{\varepsilon^2}\right),
\end{equation*}
respectively. Since the finite-resolution learner takes
$\eta=O(\varepsilon)$ and the relevant dissipator rate is smaller than
$\eta$, these bounds do not impose a scaling stronger than
$1/\varepsilon$ and therefore do not preclude Heisenberg-limited
learning.

Second, consider a pair-product target $P_r=P_\ell P_m$ in which
$P_\ell$ is unresolved but $P_m$ is stronger. This is precisely the
regime in which the finite-resolution algorithm uses the balanced
Kossakowski-tail assumption. If both endpoints of the hard deformation
remained in this promised class while $P_\ell$ stayed unresolved, then
positivity and \cref{ass:balanced-kossakowski-tails} would give
$|(A_0)_{\ell m}|,|(A_\theta)_{\ell m}|\le\kappa\eta$. However,
\cref{lem:hdd-pair-deformation} gives
\begin{equation*}
    (A_\theta)_{\ell m}
    =
    (A_0)_{\ell m}+i\omega\theta,
\end{equation*}
so necessarily $|\theta|\le2\kappa\eta$. The algorithm chooses
$\eta\le\varepsilon/(25\kappa M_D)$, and hence an
$\varepsilon$-separated hard deformation---in particular the
$\theta=4\varepsilon$ alternative used in
\cref{thm:hdd-local-nonextendability}---cannot remain inside the
balanced-tail class. If instead $P_\ell$ becomes resolved along the
deformation, then $P_r$ enters the learned dissipator footprint and is no
longer covered by the finite-resolution Heisenberg guarantee.
\end{remark}

\begin{remark}[Necessity of weak--strong Kossakowski regularity]
\label{rem:necessity-tail-regularity}
The preceding argument also shows why some additional regularity
assumption, such as the balanced Kossakowski-tail condition in
\cref{ass:balanced-kossakowski-tails}, is necessary for finite-resolution
Heisenberg learning of weak--strong footprint terms. Consider
$P_r=P_\ell P_m$, where $P_\ell$ is unresolved but $P_m$ is stronger.
If the promised model class also contains the $\theta=4\varepsilon$
alternative along the compensated deformation of
\cref{lem:hdd-pair-deformation}, then the same two-point argument gives
\begin{equation*}
    T
    =
    \Omega\left(\frac{a_{mm}}{\varepsilon^2}\right).
\end{equation*}
Thus, unless $a_{mm}=O(\varepsilon)$, such a model class cannot admit a
uniform $O(1/\varepsilon)$ learning guarantee.

Consequently, Heisenberg-limited learning of weak--strong
finite-resolution footprint terms requires an additional promise that
excludes these $O(\varepsilon)$ hard deformations. Balanced Kossakowski
tails provide one such promise by constraining coherences between weak
and strong dissipator directions. The particular condition in
\cref{ass:balanced-kossakowski-tails} is not itself necessary; other
structural assumptions or side information could exclude the same hard
directions.
\end{remark}

%% file: appendix/dissipator_coefficients_learning.tex
\section{Specified Lindbladian Coefficient Learning}
\label{sec:appendix-dissipator-hid-coeff-learning}

Recall the unknown $n$-qubit Lindbladian:
\begin{equation}\label{eq:lindblad-eq-diss-coeff}
    \Lindblad(\rho) = -i\sum_{P_k\in \HamiltStruct} h_k \comm{P_k}{\rho} + \sum_{P_k,P_j \in \DiagDissStruct} a_{kj}\left(P_k \rho P_j - \tfrac12\acomm{P_j P_k}{\rho}\right).
\end{equation}

Appendix~\ref{sec:appendix_dissipator_structure_learning} describes a learner for
$\EtaDiagDissStruct \coloneq \{P_k: a_{kk} \ge \eta\}$, the dissipator structure at scale $\eta$. Appendix~\ref{sec:appendix-hamilt-learning} then learns both the structure and the coefficients of the Hamiltonian outside the learned dissipator footprint, $\HamiltStruct \setminus \EstEtaFullDissStruct$, to accuracy $\varepsilon$. It remains to learn the dissipator coefficients and the \HD{} coefficients to accuracy $\varepsilon$. In particular, this can be achieved by learning all Hamiltonian coefficients of Paulis in $\HamiltStruct \cap \EstEtaFullDissStruct$ and all Kossakowski coefficients $a_{\ell m}$ with $P_\ell,P_m\in\EstEtaDiagDissStruct$, for the choices of $\eta$ specified in Appendix~\ref{sec:appendix-removing-known-dissipator-assumption}. This section solves a generalized version of this task.

In this section, we describe a standard-quantum-limited learner (\Cref{alg:appendix-specified-lindblad-coeff-learning}) for specified Lindbladian coefficients that works for arbitrary locality and does not require linear-system inversion. To do this, we run the classical shadows protocol~\cite{huang2020shadows} on Choi states of the short-time Lindbladian evolution. Our main contribution is an explicit construction of $2n$-qubit low-rank observables whose first derivatives directly equal the desired coefficient values. We then measure these observables at different time nodes and use Chebyshev interpolation, similarly to \cite{gu2024practical, caro2024learning}, to estimate their first derivatives and hence the Lindbladian coefficients to accuracy $\varepsilon$. Compared with the Hamiltonian learner of \cite{caro2024learning}, which also runs classical shadows protocol on Choi states, this gives a black-box evolution-time improvement from $\SoftBigO{n \|H\|_\infty^3/\varepsilon^4}$ to $\SoftBigO{\Lambda_L/\varepsilon^2}$, where $\Lambda_L \ge \|\Lindblad\|_\diamond$. Our approach parallels shadow process tomography~\cite{levy2024qpt, kunjummen2023qpt}; however, by constructing explicit observables, it avoids the need to construct a full-rank  linear system and hence avoids the associated numerical conditioning and inversion. It also matches the $\varepsilon$-scaling of the pseudo-Choi-state-based Hamiltonian learner of \cite{castaneda2025hamiltonian}, inviting further study of the comparison between these two resources.

\subsection{Operator content of the Lindbladian Choi state}
\label{sec:appendix-operator-content-of-lindblad-choi-state}

We work on $\mathcal H_L\otimes \mathcal H_R$ with $n$ Bell pairs,
\begin{equation*}
    \ket{\Phi_+}=\frac{1}{\sqrt 2}(\ket{00}+\ket{11}),\qquad
    \rho_0=\bigl(\ket{\Phi_+}\bra{\Phi_+}\bigr)^{\otimes n}.
\end{equation*}
Unless stated otherwise, operators act on the left register only, i.e. $A\equiv A\otimes I_R$.

Expanding the Lindbladian Choi state
$\mathcal E_t(\rho_0)\coloneqq (e^{t\Lindblad}\otimes I_R)(\rho_0)$ to first order in time gives
\begin{equation}
\label{eq:appendix-t-choi-state}
\begin{aligned}
    \mathcal E_t(\rho_0)
    &=
    \rho_0+t\Lindblad(\rho_0)+\BigO{t^2} \\
    &=
    \rho_0\left(1-\sum_{P_m\in\DiagDissStruct}a_{mm}t\right)
    -\sum_{P_k\in\HamiltStruct}ih_k t\,\comm{P_k}{\rho_0}
    +\sum_{P_m\in\DiagDissStruct}a_{mm}t\,P_m\rho_0P_m \\
    &\quad
    +\sum_{\substack{P_l,P_m\in\DiagDissStruct\\ l>m}}
    \Re(a_{lm})t\left(P_l\rho_0P_m+P_m\rho_0P_l-\frac{1}{2}\acomm{\acomm{P_m}{P_l}}{\rho_0}\right) \\
    &\quad
    +\sum_{\substack{P_l,P_m\in\DiagDissStruct\\ l>m}}
    i\Im(a_{lm})t\left(P_l\rho_0P_m-P_m\rho_0P_l-\frac{1}{2}\acomm{\comm{P_m}{P_l}}{\rho_0}\right)
    +\BigO{t^2}.
\end{aligned}
\end{equation}
In this expansion, we used the Hermiticity of the Kossakowski matrix, $a_{ml}=a_{lm}^*$, to collect the off-diagonal dissipator terms into real and imaginary parts. This motivates the following definition of observables:

\begin{definition}[Lindbladian Choi state observables]
\label{def:appendix-choi-observables}
Given the maximally entangled state $\rho_0=\bigl(\ket{\Phi_+}\bra{\Phi_+}\bigr)^{\otimes n}$ and the Lindbladian superoperator in the Pauli basis
\begin{equation*}
    \Lindblad(\rho)=\Ham(\rho)+\Diss(\rho)=-\sum_{P_k\in\HamiltStruct}ih_k\comm{P_k}{\rho}+\sum_{P_l,P_m\in\DiagDissStruct}a_{lm}\left(P_l\rho P_m-\frac{1}{2}\acomm{P_mP_l}{\rho}\right),
\end{equation*}
we define the following Hilbert--Schmidt orthonormal set of Hermitian operators for $k,m,l\in\{1,2,\dots,4^n-1\}$:
\begin{align}
    O_k &\coloneq -\frac{i}{\sqrt{2}}\comm{P_k}{\rho_0}, \label{eq:appendix-ok-operator}\\
    V_m &\coloneq P_m\rho_0P_m, \label{eq:appendix-vm-operator}\\
    R_{lm} &\coloneq \frac{1}{\sqrt{2}}\left(P_l\rho_0P_m+P_m\rho_0P_l\right), \qquad l>m, \label{eq:appendix-rlm-operator}\\
    Q_{lm} &\coloneq \frac{i}{\sqrt{2}}\left(P_l\rho_0P_m-P_m\rho_0P_l\right), \qquad l>m. \label{eq:appendix-qlm-operator}
\end{align}
Here $\rho_0$ is a $2n$-qubit state, while the Pauli operators $P_k$ are $n$-qubit Paulis acting nontrivially only on the left register.
\end{definition}

\begin{proof}[Proof of Hilbert--Schmidt orthonormality]
Let $\rho_0=\ket{\Phi_0}\bra{\Phi_0}$, and for $k\ge1$ set $\ket{\Phi_k}=(P_k\otimes I_R)\ket{\Phi_0}$. Then
\begin{equation*}
    \braket{\Phi_0}{\Phi_k}=0,\qquad \braket{\Phi_k}{\Phi_{k'}}=\delta_{kk'}.
\end{equation*}
Moreover,
\begin{equation*}
    P_\ell\rho_0P_m=\ket{\Phi_\ell}\bra{\Phi_m},\qquad \comm{P_k}{\rho_0}=\ket{\Phi_k}\bra{\Phi_0}-\ket{\Phi_0}\bra{\Phi_k}.
\end{equation*}
Using
\begin{equation*}
    \left\langle |a\rangle\!\langle b|,\ |c\rangle\!\langle d| \right\rangle_{\!HS}=\langle a|c\rangle\langle d|b\rangle,
\end{equation*}
we compute
\begin{align*}
    \langle V_m,V_{m'}\rangle_{\!HS}
    &= \langle \Phi_m|\Phi_{m'}\rangle\langle \Phi_{m'}|\Phi_m\rangle=\delta_{mm'},\\
    \langle O_k,O_{k'}\rangle_{\!HS}
    &= \frac{1}{2}\left(\langle \Phi_{k'}|\Phi_k\rangle\langle \Phi_0|\Phi_0\rangle+\langle \Phi_0|\Phi_0\rangle\langle \Phi_k|\Phi_{k'}\rangle\right)=\delta_{kk'},\\
    \langle O_k,V_m\rangle_{\!HS}
    &=0,\\
    \langle R_{lm},R_{l'm'}\rangle_{\!HS}
    &=\frac{1}{2}\left(\delta_{ll'}\delta_{mm'}+\delta_{lm'}\delta_{l'm}+\delta_{ml'}\delta_{m'l}+\delta_{mm'}\delta_{l'l}\right)=\delta_{ll'}\delta_{mm'},\\
    \langle Q_{lm},Q_{l'm'}\rangle_{\!HS}
    &=\delta_{ll'}\delta_{mm'},\\
    \langle R_{lm},Q_{l'm'}\rangle_{\!HS}
    &=0,\\
    \langle R_{lm},V_p\rangle_{\!HS}
    &=\frac{1}{\sqrt2}\left(\delta_{mp}\delta_{pl}+\delta_{lp}\delta_{pm}\right)=0,\\
    \langle Q_{lm},V_p\rangle_{\!HS}
    &=0,\\
    \langle O_k,R_{lm}\rangle_{\!HS}
    &=\langle O_k,Q_{lm}\rangle_{\!HS}=0.
\end{align*}
Here the swapped-pair contribution in $\langle R_{lm},R_{l'm'}\rangle_{\!HS}$ is ruled out by the ordering convention $l>m$ and $l'>m'$, and the remaining cross terms vanish because they contain an overlap of the form $\langle\Phi_0|\Phi_p\rangle$ with $p\ge1$. Thus $\{O_k\}\cup\{V_m\}\cup\{R_{lm}\}\cup\{Q_{lm}\}$ is Hilbert--Schmidt orthonormal.
\end{proof}

Importantly, the first derivatives of these observables at $t=0$ directly give the Lindbladian coefficients.

\begin{lemma}[Derivatives of Lindbladian Choi state observables]
\label{lem:appendix-derivatives-of-lindblad-choi-state-observables}
Let $h_k$ and $a_{lm}$ be the Hamiltonian coefficients and Kossakowski matrix elements of the Lindbladian generator in the Pauli basis. Then the first derivatives at $t=0$ of the observables from \Cref{def:appendix-choi-observables} satisfy
\begin{align}
    \frac{d}{dt}\tr\bigl(O_k\,\mathcal E_t(\rho_0)\bigr)\Big|_{t=0} &= \sqrt{2}\,h_k, \label{eq:appendix-tr-ok}\\
    \frac{d}{dt}\tr\bigl(V_m\,\mathcal E_t(\rho_0)\bigr)\Big|_{t=0} &= a_{mm}, \label{eq:appendix-tr-vm}\\
    \frac{d}{dt}\tr\bigl(R_{lm}\,\mathcal E_t(\rho_0)\bigr)\Big|_{t=0} &= \sqrt{2}\,\Re(a_{lm}), \qquad l>m, \label{eq:appendix-tr-rlm}\\
    \frac{d}{dt}\tr\bigl(Q_{lm}\,\mathcal E_t(\rho_0)\bigr)\Big|_{t=0} &= \sqrt{2}\,\Im(a_{lm}), \qquad l>m. \label{eq:appendix-tr-qlm}
\end{align}
\end{lemma}

\begin{proof}
For any time-independent observable $O$,
\begin{equation}\label{eq:appendix-observable-derivative}
   \frac{d}{dt}\tr\bigl(O\,\mathcal E_t(\rho_0)\bigr)\Big|_{t=0}=\tr\bigl(O\,\Lindblad(\rho_0)\bigr),
\end{equation}
so only the $t$-linear part matters. Let $\ket{\Phi_0}=\ket{\Phi_+}^{\otimes n}$, $\rho_0=\ket{\Phi_0}\bra{\Phi_0}$, and $\ket{\Phi_k}=P_k\ket{\Phi_0}$.

Using $\comm{P_k}{\rho_0}=i\sqrt2\,O_k$ and, for $l>m$,
\begin{equation*}
    P_l\rho_0P_m+P_m\rho_0P_l=\sqrt2\,R_{lm},\qquad P_l\rho_0P_m-P_m\rho_0P_l=-i\sqrt2\,Q_{lm},
\end{equation*}
the $t$-linear part of Eq.~\eqref{eq:appendix-t-choi-state} gives
\begin{equation}\label{eq:Lrho0-decomp}
    \Lindblad(\rho_0)=\sqrt2\sum_k h_kO_k+\sum_m a_{mm}V_m+\sqrt2\sum_{l>m}\bigl(\Re(a_{lm})R_{lm}+\Im(a_{lm})Q_{lm}\bigr)+\Xi,
\end{equation}
where
\begin{equation*}
    \Xi\coloneq -\left(\sum_m a_{mm}\right)\rho_0-\frac{1}{2}\sum_{l>m}\Re(a_{lm})\acomm{\acomm{P_m}{P_l}}{\rho_0}-\frac{i}{2}\sum_{l>m}\Im(a_{lm})\acomm{\comm{P_m}{P_l}}{\rho_0}.
\end{equation*}

We now check that $\Xi$ is orthogonal to all observables in \Cref{def:appendix-choi-observables}. For any operator $X$ on the left register,
\begin{equation*}
    \acomm{X}{\rho_0}=X\rho_0+\rho_0X=\ket{X\Phi_0}\bra{\Phi_0}+\ket{\Phi_0}\bra{X\Phi_0}.
\end{equation*}
Since $V_m$, $R_{lm}$, and $Q_{lm}$ are linear combinations of $\ket{\Phi_r}\bra{\Phi_s}$ with $r,s\ge1$, we have
\begin{align}
    \tr(V_m\rho_0)&=\tr(R_{lm}\rho_0)=\tr(Q_{lm}\rho_0)=0, \label{eq:appendix-overlap-of-dissipator-w-leftovers}\\
    \tr\bigl(V_m\acomm{X}{\rho_0}\bigr)&=\tr\bigl(R_{lm}\acomm{X}{\rho_0}\bigr)=\tr\bigl(Q_{lm}\acomm{X}{\rho_0}\bigr)=0.
\end{align}
It remains to check the Hamiltonian observables. Write $O_k=-\frac{i}{\sqrt2}\comm{P_k}{\rho_0}$ and let $X$ be a nonidentity Pauli string, standing for either $\comm{P_m}{P_l}$ or $\acomm{P_m}{P_l}$ when the corresponding expression is nonzero. Using the Hermiticity of $O_k$,
\begin{equation*}
    \tr\bigl(O_k\acomm{X}{\rho_0}\bigr)=-\frac{i}{\sqrt2}\tr\bigl(\comm{P_k}{\rho_0}(X\rho_0+\rho_0X)\bigr).
\end{equation*}
Expanding and using $\rho_0^2=\rho_0$ gives
\begin{equation*}
    \tr\bigl(\comm{P_k}{\rho_0}(X\rho_0+\rho_0X)\bigr)=\tr(P_k\rho_0X\rho_0)+\tr(P_k\rho_0X)-\tr(P_kX\rho_0)-\tr(\rho_0P_k\rho_0X).
\end{equation*}
Using $\rho_0 A\rho_0=\bra{\Phi_0}A\ket{\Phi_0}\rho_0$, the first and last terms vanish because $\bra{\Phi_0}P_k\ket{\Phi_0}=0$. The remaining terms are
\begin{equation*}
    \tr(P_k\rho_0X)-\tr(P_kX\rho_0)=\bra{\Phi_0}XP_k\ket{\Phi_0}-\bra{\Phi_0}P_kX\ket{\Phi_0}.
\end{equation*}
Each term can be nonzero only if $P_kX$ is proportional to the identity, in which case the two terms cancel. Hence $\tr\bigl(O_k\acomm{\comm{P_m}{P_l}}{\rho_0}\bigr)=\tr\bigl(O_k\acomm{\acomm{P_m}{P_l}}{\rho_0}\bigr)=0$. Also, $\tr(O_k\rho_0)=0$. Therefore $O_k$ is Hilbert--Schmidt orthogonal to every term in $\Xi$.

Taking the Hilbert--Schmidt inner product of Eq.~\eqref{eq:Lrho0-decomp} with each observable and using the orthonormality from \Cref{def:appendix-choi-observables}, together with the orthogonality to $\Xi$ above, gives
\begin{align*}
    \tr\bigl(O_k\Lindblad(\rho_0)\bigr)&=\sqrt2\,h_k,\\
    \tr\bigl(V_m\Lindblad(\rho_0)\bigr)&=a_{mm},\\
    \tr\bigl(R_{lm}\Lindblad(\rho_0)\bigr)&=\sqrt2\,\Re(a_{lm}),\qquad l>m,\\
    \tr\bigl(Q_{lm}\Lindblad(\rho_0)\bigr)&=\sqrt2\,\Im(a_{lm}),\qquad l>m.
\end{align*}
Combining these identities with Eq.~\eqref{eq:appendix-observable-derivative} yields Eq.~\eqref{eq:appendix-tr-ok}, Eq.~\eqref{eq:appendix-tr-vm}, Eq.~\eqref{eq:appendix-tr-rlm}, and Eq.~\eqref{eq:appendix-tr-qlm}.
\end{proof}

\subsection{Review of classical shadows}\label{sec:appendix_review_classical_shadows}

The key subroutine we use for learning the Lindbladian coefficients is the classical shadows protocol \citep{huang2020shadows}. The goal of the classical shadows is to estimate properties of a quantum state $\rho$ using a small number of single-copy measurements. The key step in the classical shadows algorithm involves creating a classical representation $\hat{\rho}$ of $\rho$ by applying random operations and performing measurements in the computational basis.

The version of classical shadows we consider requires applying a random Clifford operator $U_i \in \mathrm{Cl}(2^n)$ to the state $\rho$ in each of $N$ rounds. After applying $U_i$, one measures each qubit in the computational basis. The outcome is recorded as a bit string $\ket{b_i}$ of length $n$, where the $j$th bit of $\ket{b_i}$ records the measurement result of the $j$th qubit (i.e., $\ket{b_i}_j = 0$ corresponds to measuring $\ket{0}$). The corresponding classical snapshot is $\sigma_i = U_i^\dagger \ket{b_i}\bra{b_i} U_i$. The classical shadow for random Clifford measurements is then defined as:

\begin{definition}[Classical Shadow (From Clifford Measurements)]
\label{def:appendix-classical-shadow}
Given $N$ copies of an $n$-qubit quantum state $\rho$, the classical shadows procedure based on random Clifford measurements returns a classical shadow of $\rho$ which is of the form
\begin{equation*}
   \hat{\rho} = \{ \hat{\rho}_i \mid i \in \mathbb{Z}_N \} 
\end{equation*}

where
\begin{equation*}
    \hat{\rho}_i = (2^n + 1)(U_i^\dagger \ket{b_i}\bra{b_i} U_i) - \mathbb{I}
\end{equation*}
The hat notation signifies that $\hat{\rho}$ is an estimator.
\end{definition}

The snapshots $\sigma_i$ are stabilizer states and can be prepared efficiently using only Clifford operations starting from $\ket{0}^{\otimes n}$. Consequently, each snapshot requires only $\mathcal{O}(n^2)$ classical bits to store~\cite{aaranson2004stabilizers}. The full classical shadow of $\rho$ consists of $N$ such snapshots, requiring $N$ copies of $\rho$.

We copy the informal version of the theorem here for completeness:

\begin{theorem}[Classical Shadows \cite{huang2020shadows}]\label{thm:appendix-classical-shadows}
Classical shadows of size $N$ suffice to predict $M$ arbitrary linear target functions 
$\tr(O_1 \rho), \ldots, \tr(O_M \rho)$ up to additive error $\epsilon_s$ with probability at least $1-\delta$ given that
\begin{equation*}
N \geq \text{(order)} \, \log(M/\delta) \, \max_i \norm{O_i}_{\text{shadow}}^2 / \epsilon_s^2 .
\end{equation*}
The definition of the norm $\norm{O_i}_{\text{shadow}}$ depends on the ensemble of unitary 
transformations used to create the classical shadow.
\end{theorem}

For the Clifford ensemble we use, Huang et al. \cite{huang2020shadows} showed that $\norm{O_i}_{\text{shadow}}^2 \le \tr(O_i^2).$ Hence, we can estimate $M$ observables of constant Hilbert-Schmidt norm using $N = \BigO{\log(M/\delta) / \epsilon_s^2}$ copies of $\rho$.

\subsection{Standard-Quantum-Limited coefficient learning} 
\label{sec:appendix-ancilla_assisted_coeff_learning}

We now formulate the coefficient-learning task for arbitrary specified Lindbladian coefficients. Let
$\mathcal Z_H\subseteq \PnHerm\setminus\{I\}$ be the set of Hamiltonian Paulis whose coefficients we wish to estimate, and let
$\mathcal Z_D\subseteq\{(P_l,P_m):P_l,P_m\in\PnHerm\setminus\{I\},\,l\ge m\}$
be the set of Kossakowski matrix entries whose coefficients we wish to estimate. We write
$\mathcal Z_D^{\mathrm{diag}}\coloneq\{P_m:(P_m,P_m)\in\mathcal Z_D\}$ and
$\mathcal Z_D^{\mathrm{off}}\coloneq\{(P_l,P_m)\in\mathcal Z_D:l>m\}$, and define the corresponding observable count
\begin{equation*}
    N_{\mathrm{obs}}\coloneq |\mathcal Z_H|+|\mathcal Z_D^{\mathrm{diag}}|+2|\mathcal Z_D^{\mathrm{off}}|.
\end{equation*}
The factor of $2$ appears because each off-diagonal Kossakowski entry is reconstructed from separate observables for its real and imaginary parts. The algorithm below constructs the observables $O_k$, $V_m$, $R_{lm}$, and $Q_{lm}$ associated with these specified sets, estimates their first derivatives from Choi-state classical shadows at Chebyshev time nodes, and converts the derivative estimates directly into the requested coefficients using \Cref{lem:appendix-derivatives-of-lindblad-choi-state-observables}.

\begin{algorithm}[H]
    \caption{Specified Lindbladian coefficient learning via Choi-state classical shadows}
    \label{alg:appendix-specified-lindblad-coeff-learning}
    \DontPrintSemicolon
    \SetKwInOut{Input}{Inputs}
    \SetKwInOut{Output}{Output}
    \SetKwBlock{Schedule}{Schedule}{}
    \Input{Hamiltonian coefficient set $\mathcal Z_H\subseteq\PnHerm\setminus\{I\}$; \\
    Kossakowski coefficient set $\mathcal Z_D\subseteq\{(P_l,P_m):P_l,P_m\in\PnHerm\setminus\{I\},\,l\ge m\}$; \\
    accuracy and confidence parameters $\epsilon,\delta_{\mathrm{coeff}}\in(0,1)$; \\ 
    
    upper bound $\Lambda_L\ge\|\Lindblad\|_\diamond$; access to $e^{t\Lindblad}$; $n$ noiseless ancillas.}
    
    \Output{Estimates $\{\widehat h_k:P_k\in\mathcal Z_H\}$ and $\{\widehat a_{lm}:(P_l,P_m)\in\mathcal Z_D\}$ to accuracy $\epsilon$ with probability at least $1-\delta_{\mathrm{coeff}}$.}

    \Schedule{
      $\tau_{\max}\gets \Theta(1/\Lambda_L)$,\quad
      $r\gets \Theta(\log(\Lambda_L/\epsilon))$,\quad
      $\epsilon_s\gets \Theta(\epsilon/(\Lambda_L r^3))$,\quad
      $\delta_s\gets \Theta(\delta_{\mathrm{coeff}}/(r+1))$\;
      $\{t_q\}_{q=1}^{r+1}\gets$ Chebyshev--Gauss nodes on $(0,\tau_{\max})$;\quad
      $\{\alpha_q\}_{q=1}^{r+1}\gets$ first-derivative interpolation weights\;
      $\mathcal O_{\mathrm{tar}}\gets
      \{O_k:P_k\in\mathcal Z_H\}\cup
      \{V_m:(P_m,P_m)\in\mathcal Z_D\}\cup
      \{R_{lm},Q_{lm}:(P_l,P_m)\in\mathcal Z_D,\ l>m\}$\;
    }

    \For{$q\gets 1$ \KwTo $r+1$}{
        Prepare copies of $\rho(t_q)\gets(e^{t_q\Lindblad}\otimes I_R)(\rho_0)$ \tcp*{prepare Choi states}
        Estimate all $\{\tr(O\rho(t_q)):O\in\mathcal O_{\mathrm{tar}}\}$ to accuracy $\epsilon_s$ using global Clifford classical shadows with failure probability $\delta_s$\;
    }

    \ForEach{$O\in\mathcal O_{\mathrm{tar}}$}{
        $\widehat{\mu}^{(1)}_O\gets\sum_{q=1}^{r+1}\alpha_q\,\widehat{\tr(O\rho(t_q))}$\tcp*{first-derivative estimation via Chebyshev interpolation}
    }

    \ForEach{$P_k\in\mathcal Z_H$}{
        $\widehat h_k\gets \widehat{\mu}^{(1)}_{O_k}/\sqrt2$\;
    }
    \ForEach{$(P_m,P_m)\in\mathcal Z_D$}{
        $\widehat a_{mm}\gets \widehat{\mu}^{(1)}_{V_m}$\;
    }
    \ForEach{$(P_l,P_m)\in\mathcal Z_D$ with $l>m$}{
        $\widehat a_{lm}\gets\bigl(\widehat{\mu}^{(1)}_{R_{lm}}+i\widehat{\mu}^{(1)}_{Q_{lm}}\bigr)/\sqrt2$\;
    }

    \Return{$\{\widehat h_k:P_k\in\mathcal Z_H\}$ and $\{\widehat a_{lm}:(P_l,P_m)\in\mathcal Z_D\}$}.
\end{algorithm}

\begin{theorem}[Specified Lindbladian coefficient learning via Choi-state classical shadows]
\label{thm:appendix-lindblad-coefficients_learning_shadows}
Let $\Lindblad$ be an $n$-qubit Lindbladian and assume $\Lambda_L\ge \max\{1,\|\Lindblad\|_\diamond\}$. Assume black-box access to $e^{t\Lindblad}$. Fix coefficient index sets
$\mathcal Z_H\subseteq\PnHerm\setminus\{I\}$ and
$\mathcal Z_D\subseteq\{(P_l,P_m):P_l,P_m\in\PnHerm\setminus\{I\},\,l\ge m\}$.
Let
\begin{equation*}
    N_{\mathrm{obs}}\coloneq |\mathcal Z_H|+|\mathcal Z_D^{\mathrm{diag}}|+2|\mathcal Z_D^{\mathrm{off}}|,
\end{equation*}
where $\mathcal Z_D^{\mathrm{diag}}\coloneq\{P_m:(P_m,P_m)\in\mathcal Z_D\}$ and
$\mathcal Z_D^{\mathrm{off}}\coloneq\{(P_l,P_m)\in\mathcal Z_D:l>m\}$. Fix $\epsilon,\delta\in(0,1)$. Then running \Cref{alg:appendix-specified-lindblad-coeff-learning} returns coefficient estimates
$\{\widehat h_k:P_k\in\mathcal Z_H\}$ and
$\{\widehat a_{lm}:(P_l,P_m)\in\mathcal Z_D\}$ satisfying:

\begin{enumerate}[leftmargin=*, label=\arabic*.]
\item \textit{(Accuracy)} With probability $\ge1-\delta$,
\begin{equation*}
    |\widehat h_k-h_k|\le \epsilon,\qquad |\widehat a_{lm}-a_{lm}|\le \epsilon
\end{equation*}
for all $P_k\in\mathcal Z_H$ and all $(P_l,P_m)\in\mathcal Z_D$.

\item \textit{(Total evolution time)} It applies $e^{\Lindblad t}$ for total evolution time
\begin{equation*}
    t_{\mathrm{tot}}=\BigO{\frac{\Lambda_L}{\epsilon^2}\log\!\left(\frac{\Lambda_L}{\epsilon}\right)^7\log\!\left(\frac{N_{\mathrm{obs}}\log(\Lambda_L/\epsilon)}{\delta}\right)}.
\end{equation*}

\item \textit{(Minimum evolution time)} It only ever applies $e^{\Lindblad t_q}$ for $t_q$  such that 
\begin{equation*}
    t_q = \Omega\!\left(\frac{1}{\Lambda_L\log(\Lambda_L/\epsilon)^2}\right) \qq{and}
    \min_{1\le q\le r}|t_{q+1}-t_q|
    =\Omega\!\left(\frac{1}{\Lambda_L\log(\Lambda_L/\epsilon)^2}\right).
\end{equation*}

\item \textit{(Number of experiments)} It uses
\begin{equation*}
    N_{\exp}=\BigO{\frac{\Lambda_L^2}{\epsilon^2}\log\!\left(\frac{\Lambda_L}{\epsilon}\right)^7\log\!\left(\frac{N_{\mathrm{obs}}\log(\Lambda_L/\epsilon)}{\delta}\right)}
\end{equation*}
quantum circuits, each preparing $n$ Bell pairs, evolving one register under $e^{\Lindblad t}$ for some $t\in(0,\tau_{\max})$, applying global random Clifford measurements on the resulting $2n$-qubit Choi state, and measuring in the computational basis.

\item \textit{(Classical overhead)} A direct implementation using a phase-sensitive Clifford simulator has classical post-processing cost
\begin{equation*}
    \SoftBigO{N_{\exp}(n^3 +N_{\mathrm{obs}}n^2)},
\end{equation*}
where $N_{\exp}$ is the number of Choi-state classical-shadow snapshots in item~$(4)$.
\end{enumerate}
\end{theorem}

\begin{proof}
If $N_{\mathrm{obs}}=0$, the algorithm returns empty estimate lists and the claim is trivial, so assume $N_{\mathrm{obs}}\ge1$. Throughout this proof, we repeatedly use the classical shadows protocol from \Cref{thm:appendix-classical-shadows} applied to the $e^{\Lindblad t}$ Choi state and the unit-HS-norm $2n$-qubit observables from \Cref{def:appendix-choi-observables}. The key observation we rely on is that the first derivatives of these observables are exactly the coefficients we wish to learn, as shown in \Cref{lem:appendix-derivatives-of-lindblad-choi-state-observables}. To estimate the first derivatives based on the observables at different time nodes, we use Chebyshev polynomial interpolation from \Cref{thm:total-error-noisy}.

We first form the target observable set
\begin{equation*}
    \mathcal O_{\mathrm{tar}}\coloneq \{O_k:P_k\in\mathcal Z_H\}\cup\{V_m:(P_m,P_m)\in\mathcal Z_D\}\cup\{R_{lm},Q_{lm}:(P_l,P_m)\in\mathcal Z_D,\ l>m\}.
\end{equation*}
By construction, $|\mathcal O_{\mathrm{tar}}|=N_{\mathrm{obs}}$. For $O\in\mathcal O_{\mathrm{tar}}$, define
\begin{equation*}
    f_O(t)\coloneq \tr\bigl(O\,(e^{t\Lindblad}\otimes I_R)(\rho_0)\bigr).
\end{equation*}
Since \(\Lambda_L\ge\|\Lindblad\|_\diamond\), the functions
\(f_O(t)\) satisfy the derivative-growth assumption required by
\Cref{cor:first_chebyshev_derivative_parameters}.  Indeed, fix
\(O\in\mathcal O_{\mathrm{tar}}\) and write
\[
    \rho_t\coloneq (e^{t\Lindblad}\otimes I_R)(\rho_0).
\]
Since \(e^{t\Lindblad}\) is CPTP for \(t\ge0\), \(\rho_t\) is a density
operator and hence \(\norm{\rho_t}_1=1\).  For every integer
\(\ell\ge1\),
\begin{align*}
    f_O^{(\ell)}(t)
    &=
    \tr\!\left(
        O\,(\Lindblad^\ell\otimes I_R)(\rho_t)
    \right),
\end{align*}
and therefore, by Hölder's inequality for Schatten norms,
\begin{equation}
    |f_O^{(\ell)}(t)| \le \norm{O}_\infty \norm{(\Lindblad^\ell\otimes I_R)(\rho_t)}_1 \le \norm{O}_\infty \norm{\Lindblad^\ell}_\diamond \norm{\rho_t}_1 \le \norm{O}_\infty \norm{\Lindblad}_\diamond^\ell .
\end{equation}
All observables in \(\mathcal O_{\mathrm{tar}}\) have
\(\norm{O}_\infty\le1\), so $|f_O^{(\ell)}(t)| \le \Lambda_L^\ell$ for all $\ell\ge1$.

We now choose the Chebyshev--Gauss nodes
\(\{t_q\}_{q=1}^{r+1}\subset(0,\tau_{\max})\) with the parameters from
\Cref{cor:first_chebyshev_derivative_parameters}, applied with
\(B=1\) and \(\Lambda=\Lambda_L\). By
\Cref{thm:total-error-noisy} and \Cref{cor:first_chebyshev_derivative_parameters},
estimating each observable value \(f_O(t_q)\) to accuracy \(\epsilon_s\)
suffices to estimate \(f_O'(0)\) to accuracy \(\epsilon\).  Thus, for
each \(O\in\mathcal O_{\mathrm{tar}}\), the Chebyshev estimator
\begin{equation*}
    \widehat f_O^{(1)}(0)
    \coloneq
    \sum_{q=1}^{r+1}\alpha_q\,\widehat f_O(t_q)
\end{equation*}
satisfies
\begin{equation*}
    |\widehat f_O^{(1)}(0)-f_O'(0)|\le\epsilon
\end{equation*}
on the event that all nodewise estimates obey
\(|\widehat f_O(t_q)-f_O(t_q)|\le\epsilon_s\).

At each Chebyshev node, we apply global Clifford classical shadows to the Choi state $(e^{t_q\Lindblad}\otimes I_R)(\rho_0)$ and estimate all observables in $\mathcal O_{\mathrm{tar}}$. By \Cref{thm:appendix-classical-shadows}, estimating all $N_{\mathrm{obs}}$ observables at one node to accuracy $\epsilon_s$ with failure probability at most $\delta_s=\delta/(r+1)$ requires
\begin{equation*}
    N_{\mathrm{node}}=\BigO{\frac{1}{\epsilon_s^2}\log\!\left(\frac{N_{\mathrm{obs}}}{\delta_s}\right)}=\BigO{\frac{\Lambda_L^2 r^6}{\epsilon^2}\log\!\left(\frac{N_{\mathrm{obs}}r}{\delta}\right)}
\end{equation*}
Choi-state copies. A union bound over the $r+1$ Chebyshev nodes guarantees that all observable values, and hence all first-derivative estimates, are simultaneously accurate with probability at least $1-\delta$.

We then convert derivatives to coefficients using \Cref{lem:appendix-derivatives-of-lindblad-choi-state-observables}. For each $P_k\in\mathcal Z_H$, set
\begin{equation*}
    \widehat h_k\coloneq \frac{1}{\sqrt2}\widehat f_{O_k}^{(1)}(0).
\end{equation*}
For each diagonal Kossakowski entry $(P_m,P_m)\in\mathcal Z_D$, set
\begin{equation*}
    \widehat a_{mm}\coloneq \widehat f_{V_m}^{(1)}(0).
\end{equation*}
For each off-diagonal Kossakowski entry $(P_l,P_m)\in\mathcal Z_D$ with $l>m$, set
\begin{equation*}
    \widehat a_{lm}\coloneq \frac{\widehat f_{R_{lm}}^{(1)}(0)+i\,\widehat f_{Q_{lm}}^{(1)}(0)}{\sqrt2}.
\end{equation*}
Using Eq.~\eqref{eq:appendix-tr-ok}, Eq.~\eqref{eq:appendix-tr-vm}, Eq.~\eqref{eq:appendix-tr-rlm}, and Eq.~\eqref{eq:appendix-tr-qlm}, and decreasing the hidden constant in $\epsilon_s$ by an absolute factor to account for the real and imaginary parts of off-diagonal entries, we obtain
\begin{equation*}
    |\widehat h_k-h_k|\le\epsilon,\qquad |\widehat a_{lm}-a_{lm}|\le\epsilon
\end{equation*}
for all specified coefficients. This proves item~$(1)$.

The number of Choi-state experiments is
\begin{equation*}
    N_{\exp}=(r+1)N_{\mathrm{node}}=\BigO{\frac{\Lambda_L^2}{\epsilon^2}\log\!\left(\frac{\Lambda_L}{\epsilon}\right)^7\log\!\left(\frac{N_{\mathrm{obs}}\log(\Lambda_L/\epsilon)}{\delta}\right)},
\end{equation*}
which proves item~$(4)$. Since every experiment uses evolution time at most $\tau_{\max}=\Theta(1/\Lambda_L)$, the total black-box evolution time is
\begin{equation*}
    t_{\mathrm{tot}}\le N_{\exp}\tau_{\max}=\BigO{\frac{\Lambda_L}{\epsilon^2}\log\!\left(\frac{\Lambda_L}{\epsilon}\right)^7\log\!\left(\frac{N_{\mathrm{obs}}\log(\Lambda_L/\epsilon)}{\delta}\right)},
\end{equation*}
which proves item~$(2)$. Finally, \((3)\) is satisfied by \Cref{lem:appendix-cheb-time-resolution}.

The classical overhead in item~$(5)$ is dominated by evaluating the observables in $\mathcal O_{\mathrm{tar}}$ on the classical-shadow snapshots from \Cref{def:appendix-classical-shadow}. Since $O_k,V_m,R_{lm},Q_{lm}$ are rank-one or rank-two operators in the Bell basis, evaluating them requires phase-sensitive stabilizer amplitudes. We use a phase-sensitive Clifford simulator such as \cite{Bravyi2019simulationofquantum}. Equivalently, let $C_{\mathrm{Bell}}$ be the fixed Clifford mapping computational-basis states to the Bell basis. For each $2n$-qubit snapshot, we first apply $C_{\mathrm{Bell}}^\dagger$ classically, so that overlaps with Bell states reduce to computational-basis amplitudes. This fixed Bell-basis change uses $\BigO{n}$ Hadamards and CNOTs; in the phase-sensitive simulator, the Hadamard updates dominate and cost $\BigO{n^3}$ per snapshot. After this preprocessing, each observable evaluation only requires computing the relevant computational-basis stabilizer amplitudes, costing $\SoftBigO{n^2}$ per observable. Thus a direct implementation has post-processing cost
\begin{equation*}
    \SoftBigO{N_{\exp}n^3+N_{\exp}N_{\mathrm{obs}}n^2}.
\end{equation*}
\end{proof}

%% file: appendix/interface-noise.tex
\section{Robustness to interface noise}\label{sec:interface-noise}

Most previous work on quantum Hamiltonian learning has restricted analysis of implementation noise to the state-preparation-and-measurement (SPAM) model, arguing that $O(\epsilon^{-1})$ Heisenberg scaling can be retained in the presence of errors at these stages. As has been made transparent in the metrological setting \cite{zhou2018achieving,demkowicz2017qecmetrology}, however, SPAM errors are not the primary obstacle for Heisenberg-limited quantum sensing: instead, ambient errors can relegate a sensing protocol to the SQL unless error-correction or other error mitigation strategies can be leveraged to distinguish signal from noise. Our work completely characterizes the power of quantum error-correction in this regard, as we provide the first algorithm to simultaneously learn the entire ambient dissipator while learning the superposed Hamiltonian in the Heisenberg limit.

However, \cite{cotler2026noisylearning} illustrates that error-correction of the ambient dissipator does not eliminate all barriers to learning an underlying Hamiltonian. In particular, their work points out that even while errors within a quantum device can be protected by quantum error-correction, the logical evolution by an ambient Lindbladian acts at the physical level, and absent additional structure, the coupling between signal and quantum computer cannot be assumed to be noiseless. This is laid bare in our algorithm: we allow probe qubits to evolve under the signal, then perform control including quantum error-correction and physical Pauli twirling on those \textit{physical} degrees of freedom. In this section, we contend with these errors, demonstrating the inevitable tradeoff between Heisenberg-limited learning and the native error rate of quantum hardware. 

For concreteness, we assume that the noisy interface between our hardware and the environment is modeled by a depolarizing channel described by the dissipator
\begin{equation*}
    \mathcal{L}_{\textnormal{dep}}^\nu(\rho) = \nu\sum_{j=1}^n \sum_{\sigma\in\{X, Y, Z\}}(\sigma^{(j)}\rho \sigma^{(j)} - \rho) . 
\end{equation*}
Here, $\nu$ is a noise strength parameter governed by how noise-prone the hardware is, e.g. the relaxation times of the hardware qubits. While we work with depolarizing noise as it is a well-studied and standard noise model, we note that the following discussion can be easily extended to other noise models. Beyond parametrizing the noise strength within the generator, note that this Lindbladian will cause a non-infinitesimal error channel with error terms of various Pauli weights. The severity of these errors is governed by how long the channel evolves, which is in turn controlled by the speed of our quantum control. Operationally, let us suppose that each of our control steps (including twirling and quantum error correction) takes at most time $t_c$. Then $(\nu, t_c)$ are the two parameters governing the quality of our quantum device, and will control how well we are able to perform Heisenberg-limited Hamiltonian learning in the presence of implementation error.

\begin{definition}[Single-qubit depolarizing channel] The single-qubit depolarizing channel with strength $\lambda\in [0, 1]$ is
\begin{equation*}
    \mathcal{D}_\lambda(\rho) = (1-\lambda)\rho + \lambda\, \frac{I}{2}
\end{equation*}
for any single-qubit density matrix $\rho$. The depolarizing channel is self-adjoint with respect to the Hilbert-Schmidt inner product, and on a general 2-by-2 matrix $A$, (the adjoint of) $\mathcal{D}$ acts as
\begin{equation*}
    \mathcal{D}_\lambda(A) = (1-\lambda)A + \lambda\,\frac{\tr(A)\,I}{2}\,.
\end{equation*}
\end{definition}
Note that $\exp(\mathcal{L}_{\textnormal{dep}}^\nu t_c)$ and $\mathcal D_\lambda^{\otimes n}$ are equivalent when $\lambda = 1 - \exp(-4\nu t_c)$. The following Lemma will capture the core effect of interface noise within our algorithm.

\begin{lemma} \label{lemma:channel_noise_bd}
    Let $\Psi$ be any completely positive trace-nonincreasing map acting on $n$ qubits, and $0\leq p\leq 1$. Moreover, let $\Phi$ be a completely positive trace-nonincreasing map of the form
    \begin{equation*}
        \Phi = (1-p)\Psi + p\Xi \ ,
    \end{equation*}
    where $\Xi$ is any completely positive trace-nonincreasing map. Then for any integer $N >0$,
    \begin{equation*}
        |\Psi^N - \Phi^N|_\diamond \leq 2pN \ .
    \end{equation*}
\end{lemma}

\begin{proof}
    By telescoping,
    \begin{equation*}
    \Psi^N-\Phi^N
    =
    \sum_{i=0}^{N-1}
    \Psi^{N-i-1}\circ(\Psi-\Phi)\circ \Phi^i \ .
    \end{equation*}
    Applying the triangle inequality and submultiplicativity of the diamond norm,
    \begin{equation*}
        |\Psi^N-\Phi^N|_\diamond
        \leq
        \sum_{i=0}^{N-1}
        |\Psi^{N-i-1}|_\diamond
        |\Psi-\Phi|_\diamond
        |\Phi^i|_\diamond .
    \end{equation*}
    Since $\Psi$ and $\Phi$ are completely positive and trace-nonincreasing, their diamond norms are at most one, and hence
    \begin{equation*}
    |\Psi^N-\Phi^N|_\diamond
    \leq
    N|\Psi-\Phi|_\diamond .
    \end{equation*}
    By definition of $\Phi$, $\Psi-\Phi = p(\Psi-\Xi)$ and thus
    \begin{equation*}
    |\Psi-\Phi|_\diamond
    =
    p|\Psi-\Xi|_\diamond
    \leq
    p\bigl(|\Psi|_\diamond+|\Xi|_\diamond\bigr)
    \leq 2p,
    \end{equation*}
    where we used that $\Psi$ and $\Xi$ are CPTNI. Combining the previous two displays completes the proof
    \begin{equation*}
    |\Psi^N-\Phi^N|_\diamond \leq 2pN \ .
    \end{equation*}
\end{proof}

Now recall that our algorithm interacts with the signal Lindbladian in three ways according to Eq.~\eqref{eq:QEC-reshaping}, Eq.~\eqref{eq:logical-twirling-QEC-reshaping}, and Eq.~\eqref{eq:physical-twirling-QEC-reshaping}. We define their noisy-interface versions here, using simplified indices $1,2,3$ to denote the different maps:

\begin{itemize}
    \item \textbf{QEC reshaping map:} $A_1
        \coloneq
        \RMap \circ \bigl(e^{\Lindblad \tau}\otimes I_{n'-n}\bigr) \circ \mathcal C_\Pi$ becomes
        \begin{equation*}
             A_1'
        \coloneq
        \RMap \circ e^{\mathcal{L}_{\textnormal{dep}}^\nu t_c}\circ \bigl(e^{\Lindblad \tau}\otimes I_{n'-n}\bigr) \circ e^{\mathcal{L}_{\textnormal{dep}}^\nu t_c}\circ \mathcal C_\Pi
        \end{equation*}
    \item \textbf{Physical twirling + partial QEC reshaping:} $
        A_2
        \coloneq
        \RMap \circ \frac{1}{4^n}\sum_{P\in \PnHerm}
        \bigl(\mathcal U_P\otimes I_{n'-n}\bigr)\circ
        \bigl(e^{\Lindblad \tau}\otimes I_{n'-n}\bigr)\circ 
        \bigl(\mathcal U_P\otimes I_{n'-n}\bigr)\circ \mathcal C_\Pi$ becomes
        \begin{equation*}
            A_2'
        \coloneq
        \RMap \circ \frac{1}{4^n}\sum_{P\in \PnHerm}
        \bigl(\mathcal U_P\otimes I_{n'-n}\bigr)\circ e^{\mathcal{L}_{\textnormal{dep}}^\nu t_c}\circ
        \bigl(e^{\Lindblad \tau}\otimes I_{n'-n}\bigr)\circ e^{\mathcal{L}_{\textnormal{dep}}^\nu t_c}\circ
        \bigl(\mathcal U_P\otimes I_{n'-n}\bigr)\circ \mathcal C_\Pi
        \end{equation*}
    \item \textbf{QEC reshaping + partial logical twirling:} $A_3
        \coloneq
        \frac{1}{|\overline{\mathcal K}_{\overline P_a}|}
        \sum_{\overline Q\in \overline{\mathcal K}_{\overline P_a}}
        \Bigl(\mathcal U_{\overline Q}\circ \RMap \circ \bigl(e^{\Lindblad \tau}\otimes I_{n'-n}\bigr)\circ \mathcal U_{\overline Q}\Bigr)\circ \mathcal C_\Pi$ becomes
        \begin{equation*}
            A_3'
        \coloneq
        \frac{1}{|\overline{\mathcal K}_{\overline P_a}|}
        \sum_{\overline Q\in \overline{\mathcal K}_{\overline P_a}}
        \Bigl(\mathcal U_{\overline Q}\circ \RMap \circ e^{\mathcal{L}_{\textnormal{dep}}^\nu t_c}\circ \bigl(e^{\Lindblad \tau}\otimes I_{n'-n}\bigr)\circ e^{\mathcal{L}_{\textnormal{dep}}^\nu t_c}\circ \mathcal U_{\overline Q}\Bigr)\circ \mathcal C_\Pi
        \end{equation*}
\end{itemize}
Note that $e^{\mathcal{L}_{\textnormal{dep}}^\nu t_c}$ acts only on the $n$ ``probe'' qubits as we may assume the remaining $n'- n$ qubits can be protected by quantum error-correction within the device. Next, we see that all three channels can be treated identically in the noisy case. 

\begin{lemma} \label{lemma:channel_err_bd_As}
For each map $A_i, i\in {1,2,3}$, and every integer $N>0$,
\begin{equation*}
|A_i^N-A_i'^N|_\diamond
\leq
2Np.
\end{equation*}
where $p \le \left(1-e^{-8\nu t_c n}\right)$.
\end{lemma}

\begin{proof}
Using $e^{\mathcal{L}_{\textnormal{dep}}^\nu t_c}=\mathcal D_\lambda^{\otimes n}$ with $\lambda=1-e^{-4\nu t_c}$, a single interface-noise layer acts trivially on all $n$ probe qubits with probability $(1-\lambda)^n$. Since each $A_i'$ contains two such layers, the probability that the composition of both layers acts trivially is at least 
\begin{equation*}
(1-\lambda)^{2n}=e^{-8\nu t_c n}.
\end{equation*}
On this branch, the noisy map coincides with $A_i$. All remaining branches define some completely positive trace-nonincreasing map $\Xi_i$. Hence
\begin{equation*}
A_i'=(1-p)A_i+p\Xi_i,
\qquad
p\le1-(1-\lambda)^{2n}=1-e^{-8\nu t_c n}.
\end{equation*}
Applying \Cref{lemma:channel_noise_bd} with $\Psi=A_i$ and $\Phi=A_i'$ gives
\begin{equation*}
|A_i^N-A_i'^N|_\diamond
\leq 2pN
\le
2N\left(1-e^{-8\nu t_c n}\right).
\end{equation*}
The averaging in $A_2$ and $A_3$ does not affect the argument, since the same convex decomposition is preserved under linear combinations.
\end{proof}

We are now prepared to state the effect of interface noise on the Heisenberg-limited sensitivity of our Hamiltonian learning algorithm.
\begin{theorem} \label{thm:interface-noise}
    In the presence of $(\nu, t_c)$ interface depolarizing noise, \Cref{alg:hamilt-hier-learning} succeeds in total evolution time $t = O(M^2/\epsilon)$ when the following condition is satisfied:
    \begin{equation*}
        \epsilon > c_1M^2\sqrt{\nu n t_c}.
    \end{equation*}
\end{theorem}
where $c_1$ is a universal constant.
\begin{proof}
    From \Cref{lem:N-step-reshaping-error} , we have 
    \begin{equation*}   \big\|\, A_\tau^{\,N} - e^{\,t\LindbladEff}\,\big\|_\diamond
    \ \le\ \frac{t^2\,\|\Lindblad\|_\diamond^2}{N}, \qquad
    \big\|\,\widetilde A_\tau^{\,N} - e^{\,t\widetilde\LindbladEff}\,\big\|_\diamond
    \ \le\ \frac{t^2\,\|\Lindblad\|_\diamond^2}{N}\, .
    \end{equation*}
    By Lemmas \ref{lemma:channel_noise_bd}, \ref{lemma:channel_err_bd_As} and the triangle inequality, we thus have 
    \begin{equation*}   \big\|\, A_\tau'^{\,N} - e^{\,t\LindbladEff}\,\big\|_\diamond
    \ \le\ \frac{t^2\,\|\Lindblad\|_\diamond^2}{N} + 2pN, \qquad
    \big\|\,\widetilde A_\tau'^{\,N} - e^{\,t\widetilde\LindbladEff}\,\big\|_\diamond
    \ \le\ \frac{t^2\,\|\Lindblad\|_\diamond^2}{N}\, +2pN.
    \end{equation*}
    Under these bounds, the error does not strictly decrease as $N$ is increased. Minimizing the RHS gives 
    \begin{equation*}
        N_{\rm opt} = \frac{t\,\|\Lindblad\|_\diamond}{\sqrt{2p}}
    \end{equation*}
    In both cases, we thus obtain minimum $N$-round channel error bounded by $2\sqrt{2} t\|\Lindblad\|_\diamond\sqrt{p}$. 

    The \HNDAbbrev{} Hamiltonian learner sensitivity to the interface noise is bottle-necked by the candidate identification step, i.e. \Cref{alg:appendix-round-j-hamilt-struct-learning}. By Eq.~\eqref{eq:hamilt-struct-sampling-prob-bell-sampling-ideal}, we want the interface error to be at most $\BigO{1/M^2}$. Therefore, 
    \begin{equation*}
        t\|\Lindblad\|_\diamond\sqrt{p} <\BigO{1/M^2}\ .
    \end{equation*}
    Hence, under the longest evolution time $t = \BigO{(M\epsilon)^{-1}}$ of \Cref{alg:appendix-round-j-hamilt-struct-learning}, we can only do so when $\epsilon > c M^2\sqrt{p}$. Noting that for small $\nu, t_c$, we have $p_{\rm int}=O(\nu q t_c)$, which concludes the proof of the theorem.
\end{proof}

%% file: appendix/reducing_ancillas.tex
\section{Reducing the ancilla cost of candidate identification}\label{sec:appendix-reducing-the-ancilla-cost}

 The dissipator structure learner \Cref{alg:appendix-diss-struct-learning} and the \HNDAbbrev{} Hamiltonian learner \Cref{alg:hamilt-hier-learning} use Bell sampling for candidate identification in \Cref{alg:appendix-round-j-diss-struct-sampling,alg:appendix-round-j-hamilt-struct-learning}. In this section, we show that Bell sampling can be replaced by population recovery, following the construction of \citet{hu2025ansatz}. This substitution reduces the ancilla cost from $n+\BigO{\log M}$ to $\BigO{\log M}$ at the cost of increasing the total evolution time by $\BigO{M^2 \log n}$ as stated in \Cref{thm:ancilla-reduced-hamilt-learner}.

Before invoking the population-recovery theorem of \citet{flammia2021pauli}, we first record the identities that make this replacement possible. The key observation is that the Bell-sampling probabilities coincide with the diagonal $\chi$-rates of the effective logical channel; see Eq.~\eqref{eq:prob-sample-Qr-modulo-stabilizers} and Eq.~\eqref{eq:HND-prob-sample-Pr-modulo-stabilizers}. Let $\overline{\mathcal P}=N(S)/S$ denote the logical Pauli group on the code space $\Pi\mathcal H_{n'}$ of a stabilizer code with stabilizer group $S$. We write the effective logical channel (Eq.~\eqref{eq:Leff-qec-reshaping} and Eq.~\eqref{eq:Leff-physical-twirl-qec-reshaping}) in the logical Pauli basis as
\begin{equation}\label{eq:Leff-logical-chi-matrix-representation}
    e^{\LindbladEff t}(\rho_c)
    =
    \sum_{m,k}
        \overline{\chi}_{mk}(t)\,
        \overline Q_m \rho_c \overline Q_k ,
\end{equation}
where $\rho_c\in\mathfrak A_c$ is an arbitrary code state and the sum ranges over logical Paulis $\overline Q_m,\overline Q_k\in\overline{\mathcal P}$. By Eq.~\eqref{eq:prob-sample-Qr-modulo-stabilizers} and Eq.~\eqref{eq:HND-prob-sample-Pr-modulo-stabilizers}:
\begin{equation*}
        \Pr(\text{sample $Q_r$ modulo stabilizers}) = \Pr(\text{sample $\overline{Q}_r$})
        =
        \overline{\chi}_{rr}(t).
\end{equation*}

For dissipator-structure identification, Eq.~\eqref{eq:diss-struct-sampling-prob-bell-sampling} gives
\begin{equation} \label{eq:diss-logical-chi-rate}
\overline{\chi}_{rr}
=
\Pr(\text{sample $\overline Q_r$})
=
\Pr(\text{sample $Q_r$ modulo stabilizers})
\ge
\frac{1}{4e \, 2^j\,\widehat{\Gamma}_j}
\ge
\frac{1}{4eM_D}.
\end{equation}
The final inequality uses the sparsity bound and the round-$j$ hierarchy, $\widehat{\Gamma}_j\le 2^{-j}M_D$, matching the main-text exposition. The same reasoning applies to the desparsified version.

Similarly, for \HNDAbbrev{} candidate identification, Eq.~\eqref{eq:hamilt-struct-sampling-prob-bell-sampling} gives
\begin{equation}\label{eq:hamilt-logical-chi-rate}
\overline{\chi}_{rr}
=
\Pr(\text{sample $\overline Q_r$})
=
\Pr(\text{sample $P_r$ modulo stabilizers}\mid A_{P_r}\wedge E_{\mathrm{tar}})
\ge
\frac{1}{300M^2}.
\end{equation}
Thus, in both hierarchical candidate-identification routines, the relevant logical Pauli rates are lower bounded by sparsity-dependent constants, independent of the final target accuracy $\varepsilon$ and threshold $\eta$. Consequently, population recovery can replace Bell sampling without changing the $\varepsilon$- or $\eta$-dependence of the guarantees in \Cref{alg:hamilt-hier-learning,alg:appendix-diss-struct-learning}.

We now import the Pauli channel estimation algorithm by \cite{flammia2021pauli}.

\begin{theorem}[Learning Pauli error rates via population recovery~\cite{flammia2021pauli}]
\label{theorem:population_recovery}
Let $\mathcal{E}$ be an $n$-qubit channel with diagonal Pauli $\chi$-rates $\{\chi_{ii}\}$.
For any $0<\epsilon,\delta<1$, there exists an ancilla-free procedure that uses
\begin{equation}
\label{eq:poprec_samples}
m \;=\; \mathcal{O}\!\left(\epsilon^{-2}\log\frac{n}{\epsilon\delta}\right)
\end{equation}
applications of $\mathcal{E}$ and prepares/measures only product single-qubit Pauli eigenstates.
With probability at least $1-\delta$, it outputs a hypothesis $\widehat\chi$ such that
\begin{equation}
\label{eq:poprec_accuracy}
\max_i |\widehat\chi_{ii}-\chi_{ii}|
\;\le\; \epsilon.
\end{equation}

The hypothesis $\widehat\chi$ is supported on a set $\widehat S\subseteq\{0,\dots,4^n-1\}$
of size $|\widehat S|\;\le\;\frac{4}{\epsilon}$ (i.e., $\widehat\chi_{ii}=0$ for $i\notin \widehat S$)

The classical post-processing runs in time $\mathcal{O}\!\left(mn/\epsilon\right) = \widetilde{O}(n/\epsilon^3)$, where $\widetilde{O}(\cdot)$ omits the $\operatorname{polylog}()\,$ factors in $(n,1/\epsilon,1/\delta)$. 
\end{theorem}

The same statement applies to the logical $\chi$-matrix representation in Eq.~\eqref{eq:Leff-logical-chi-matrix-representation}: one simply runs population recovery using logical Pauli eigenstate preparations and logical Pauli-basis measurements. 

Therefore, for dissipator candidate identification, \Cref{theorem:population_recovery} with $\epsilon = \BigO{1/M_D}$ from Eq.~\eqref{eq:diss-logical-chi-rate} implies that
\begin{equation*}
m_{\mathrm{diss,PR}}=\SoftBigO{M_D^2\log n}
\end{equation*}
population-recovery experiments suffice for candidate identification, compared with
$m_{\mathrm{diss,Bell}}=\SoftBigO{M_D}$ Bell-sampling experiments. Thus replacing logical Bell sampling by logical population recovery incurs an additional factor $M_D\log n$ in the dissipator-structure learner of \Cref{thm:appendix-hier-diss-struct-learning} (in the desparsified version, the corresponding factor is $2^j\widehat{\Gamma}_j\log n$). This increases to the total evolution time of the full dissipator structure learner from \Cref{thm:appendix-hier-diss-struct-learning} to
\begin{equation}\label{eq:diss-tot-time-pop-recovery}
t_{\mathrm{diss,PR}}
=
\SoftBigO{\frac{M_D\log n}{\eta}},
\end{equation}
while the ancilla cost is reduced from $n+\BigO{\log M}$ to $\BigO{\log M}$.

Similarly, for Hamiltonian candidate identification, \Cref{theorem:population_recovery} with $\epsilon = \BigO{1/M^2}$ from Eq.~\eqref{eq:hamilt-logical-chi-rate} implies that:
\begin{equation*}
m_{\mathrm{Hamilt,PR}}=\SoftBigO{M^4\log n}
\end{equation*}
population-recovery experiments suffice for candidate identification, compared with
$m_{\mathrm{Hamilt,Bell}}=\SoftBigO{M^2}$ Bell-sampling experiments. Therefore replacing logical Bell sampling by logical population recovery incurs an additional factor $M^2\log n$ in the Hamiltonian structure learner of \Cref{thm:appendix-round-j-hamilt-struct-learn}. This increases to the total evolution time of the \HNDAbbrev{} Hamiltonian learner from \Cref{thm:appendix-hier-hamilt_learning} to
\begin{equation}\label{eq:Hamilt-tot-time-pop-recovery}
t_{\mathrm{Hamilt,PR}}
=
\SoftBigO{\frac{M^4\log n}{\varepsilon}},
\end{equation}
while the ancilla cost is reduced from $n+\BigO{\log M}$ to $\BigO{\log M}$.

Combining the population-recovery total evolution times Eq.~\eqref{eq:Hamilt-tot-time-pop-recovery} and Eq.~\eqref{eq:diss-tot-time-pop-recovery} with $\eta = \varepsilon/M_D^2$ from \Cref{cor:ansatz-free-hnd-balanced-tail} yields an ancilla-reduced version of \Cref{res:informal-main}:
\begin{theorem}[Ancilla-reduced Ansatz-free \HNDAbbrev{} Hamiltonian Learning] \label{thm:ancilla-reduced-hamilt-learner}
    There exists a quantum algorithm that, under the balanced Kossakowski tail condition (\Cref{ass:balanced-kossakowski-tails} with $\kappa\le M_D$), outputs an estimator of the true \HND{} coefficients such that with high probability:
    \begin{equation*}
        |\widehat{h_k}-h_k| \le \varepsilon, \qquad \forall P_k \in \EpsHNDStruct
    \end{equation*}
    The total evolution time under $e^{\Lindblad t}$ is Heisenberg-limited:
    \begin{equation}
        t_{\mathrm{tot}} = \SoftBigO{\frac{M^{4}\log n}{\epsilon}}.
    \end{equation}
    The algorithm employs only Clifford operations, uses $\BigO{\log M}$ noiseless ancillary qubits, and has classical processing cost $\SoftBigO{\Poly{n,M,1/\epsilon}}$.
\end{theorem}

This replacement, however, does not improve the ancilla cost of \Cref{res:ansatz-free-lindblad}, since the full Lindbladian learner remains bottlenecked by the SQL coefficient-learning stage of Appendix~\ref{sec:appendix-dissipator-hid-coeff-learning}, which uses $n$ ancillary qubits to form the Choi state. For local or sparsely interacting Lindbladians, SQL coefficient learning can be performed efficiently without ancillas ~\cite{ivashkov2026ansatzfreelindblad,stilck2024efficient}. Extending such ancilla-reduced SQL Lindbladian learning to arbitrary-locality generators remains an open question.

%% file: bib/main.bib
@article{boulant2003robust,
  title={Robust method for estimating the Lindblad operators of a dissipative quantum process from measurements of the density operator at multiple time points},
  author={Boulant, Nicolas and Havel, Timothy F and Pravia, Marco A and Cory, David G},
  journal={Physical Review A},
  volume={67},
  number={4},
  pages={042322},
  year={2003},
  publisher={APS}
}

@article{bairey2020learning,
  title={Learning the dynamics of open quantum systems from their steady states},
  author={Bairey, Eyal and Guo, Chu and Poletti, Dario and Lindner, Netanel H and Arad, Itai},
  journal={New Journal of Physics},
  volume={22},
  number={3},
  pages={032001},
  year={2020},
  publisher={IOP Publishing}
}

@article{samach2022lindblad,
  title={Lindblad tomography of a superconducting quantum processor},
  author={Samach, Gabriel O and Greene, Ami and Borregaard, Johannes and Christandl, Matthias and Barreto, Joseph and Kim, David K and McNally, Christopher M and Melville, Alexander and Niedzielski, Bethany M and Sung, Youngkyu and others},
  journal={Physical Review Applied},
  volume={18},
  number={6},
  pages={064056},
  year={2022},
  publisher={APS}
}

@article{wolf2008assessing,
  title={Assessing non-Markovian quantum dynamics},
  author={Wolf, Michael Marc and Eisert, J and Cubitt, Toby S and Cirac, J Ignacio},
  journal={Physical review letters},
  volume={101},
  number={15},
  pages={150402},
  year={2008},
  publisher={APS}
}

@article{baumgartner2008analysis,
doi = {10.1088/1751-8113/41/39/395303},
url = {https://doi.org/10.1088/1751-8113/41/39/395303},
year = {2008},
month = {sep},
publisher = {},
volume = {41},
number = {39},
pages = {395303},
author = {Baumgartner, Bernhard and Narnhofer, Heide},
title = {Analysis of quantum semigroups with GKS–Lindblad generators: {II}. General},
journal = {Journal of Physics A: Mathematical and Theoretical}
}

@article{onorati2023fitting1,
  title={Fitting quantum noise models to tomography data},
  author={Onorati, Emilio and Kohler, Tamara and Cubitt, Toby S},
  journal={Quantum},
  volume={7},
  pages={1197},
  year={2023},
  publisher={Verein zur F{\"o}rderung des Open Access Publizierens in den Quantenwissenschaften}
}

@article{onorati2023fitting2,
  title={Fitting time-dependent Markovian dynamics to noisy quantum channels},
  author={Onorati, Emilio and Kohler, Tamara and Cubitt, Toby S},
  journal={arXiv preprint arXiv:2303.08936},
  year={2023}
}

@article{liu2025robust,
  title={Robust Lindbladian Estimation for Quantum Dynamics},
  author={Liu, Yinchen and Seddon, James R and Kohler, Tamara and Onorati, Emilio and Cubitt, Toby S},
  journal={arXiv preprint arXiv:2507.07912},
  year={2025}
}

@article{dobrynin2024compressed,
  title={Compressed-sensing Lindbladian quantum tomography with trapped ions},
  author={Dobrynin, Dmitrii and Cardarelli, Lorenzo and M{\"u}ller, Markus and Bermudez, Alejandro},
  journal={Quantum Science and Technology},
  year={2024}
}

@article{wang2017experimental,
  title={Experimental quantum Hamiltonian learning},
  author={Wang, Jianwei and Paesani, Stefano and Santagati, Raffaele and Knauer, Sebastian and Gentile, Antonio A and Wiebe, Nathan and Petruzzella, Maurangelo and O’brien, Jeremy L and Rarity, John G and Laing, Anthony and others},
  journal={Nature Physics},
  volume={13},
  number={6},
  pages={551--555},
  year={2017},
  publisher={Nature Publishing Group UK London}
}

@article{stilck2024efficient,
  title={Efficient and robust estimation of many-qubit Hamiltonians},
  author={Stilck Fran{\c{c}}a, Daniel and Markovich, Liubov A and Dobrovitski, VV and Werner, Albert H and Borregaard, Johannes},
  journal={Nature Communications},
  volume={15},
  number={1},
  pages={311},
  year={2024},
  publisher={Nature Publishing Group UK London}
}

@misc{franca2025learning,
      title={Learning and certification of local time-dependent quantum dynamics and noise}, 
      author={Daniel Stilck França and Tim Möbus and Cambyse Rouzé and Albert H. Werner},
      year={2025},
      eprint={2510.08500},
      archivePrefix={arXiv},
      primaryClass={quant-ph},
      url={https://arxiv.org/abs/2510.08500}, 
}

@article{montana2025efficiently,
  title={Efficiently learning non-Markovian noise in many-body quantum simulators},
  author={Monta{\~n}{\`a}-L{\'o}pez, Jordi A and Elben, Andreas and Choi, Joonhee and Trivedi, Rahul},
  journal={arXiv preprint arXiv:2511.16772},
  year={2025}
}

@article{olsacher2025hamiltonian,
  title={Hamiltonian and Liouvillian learning in weakly-dissipative quantum many-body systems},
  author={Olsacher, Tobias and Kraft, Tristan and Kokail, Christian and Kraus, Barbara and Zoller, Peter},
  journal={Quantum Science and Technology},
  volume={10},
  number={1},
  pages={015065},
  year={2025},
  publisher={IOP Publishing}
}

@misc{ivashkov2026ansatzfreelindblad,
      title={Ansatz-Free Learning of Lindbladian Dynamics In Situ}, 
      author={Petr Ivashkov and Nikita Romanov and Weiyuan Gong and Andi Gu and Hong-Ye Hu and Susanne F. Yelin},
      year={2026},
      eprint={2603.05492},
      archivePrefix={arXiv},
      primaryClass={quant-ph},
      url={https://arxiv.org/abs/2603.05492}, 
}

@misc{kraft2025bounded,
      title={Bounded-Error Quantum Simulation via Hamiltonian and Lindbladian Learning}, 
      author={Tristan Kraft and Manoj K. Joshi and William Lam and Tobias Olsacher and Florian Kranzl and Johannes Franke and Lata Kh Joshi and Rainer Blatt and Augusto Smerzi and Daniel Stilck França and Benoît Vermersch and Barbara Kraus and Christian F. Roos and Peter Zoller},
      year={2025},
      eprint={2511.23392},
      archivePrefix={arXiv},
      primaryClass={quant-ph},
      url={https://arxiv.org/abs/2511.23392}, 
}

@article{pastori2022characterization,
  title={Characterization and verification of Trotterized digital quantum simulation via Hamiltonian and Liouvillian learning},
  author={Pastori, Lorenzo and Olsacher, Tobias and Kokail, Christian and Zoller, Peter},
  journal={PRX Quantum},
  volume={3},
  number={3},
  pages={030324},
  year={2022},
  publisher={APS},
  doi={10.1103/PRXQuantum.3.030324}
}

@misc{berg2025large,
      title={Large-scale Lindblad learning from time-series data}, 
      author={Ewout van den Berg and Brad Mitchell and Ken Xuan Wei and Moein Malekakhlagh},
      year={2025},
      eprint={2512.08165},
      archivePrefix={arXiv},
      primaryClass={quant-ph},
      url={https://arxiv.org/abs/2512.08165}, 
}

@article{cai2026optimal,
  title={Optimal detection of dissipation in Lindbladian dynamics},
  author={Cai, Yiyi},
  journal={arXiv preprint arXiv:2603.17736},
  year={2026}
}

@article{temme2010chi,
  title={The $\chi^2$-divergence and mixing times of quantum Markov processes},
  author={Temme, Kristan and Kastoryano, Michael James and Ruskai, Mary Beth and Wolf, Michael Marc and Verstraete, Frank},
  journal={Journal of Mathematical Physics},
  volume={51},
  number={12},
  year={2010},
  publisher={AIP Publishing},
  doi={10.1063/1.3511335}
}

@article{birke2026demonstrating,
  title={Demonstrating and Benchmarking Classical Shadows for Lindblad Tomography},
  author={Birke, Rune Thinggaard and Severin, Johann Bock and Marciniak, Malthe A and Hogedal, Emil and Nylander, Andreas and Ahmad, Irshad and Osman, Amr and Bizn{\'a}rov{\'a}, Janka and Rommel, Marcus and Roudsari, Anita Fadavi and others},
  journal={arXiv preprint arXiv:2602.14694},
  year={2026}
}

@misc{gong2026multiparameter,
      title={Robust multiparameter estimation using quantum scrambling}, 
      author={Wenjie Gong and Bingtian Ye and Daniel Mark and Soonwon Choi},
      year={2026},
      eprint={2601.23283},
      archivePrefix={arXiv},
      primaryClass={quant-ph},
      url={https://arxiv.org/abs/2601.23283}, 
}

@misc{sinha2026lindblad,
  doi = {10.7907/HPRR-FQ68},
  url = {https://resolver.caltech.edu/CaltechTHESIS:04142026-214248334},
  author = {Sinha, Savar Dayal},
  title = {Efficient and SPAM-Robust Ansatz-Free Lindbladian Learning},
  publisher = {California Institute of Technology},
  year = {2026},
  copyright = {No commercial reproduction, distribution, display or performance rights in this work are provided.}
}

@misc{arad2026nearoptimallearninglocallindbladians,
      title={Near-Optimal Learning of Local Lindbladians}, 
      author={Itai Arad and Zhili Chen and Naixu Guo and Patrick Rebentrost and Zhan Yu},
      year={2026},
      eprint={2606.20535},
      archivePrefix={arXiv},
      primaryClass={quant-ph},
      url={https://arxiv.org/abs/2606.20535}, 
}

@misc{mobus2026robuststructurelearningklocal,
      title={Robust Structure Learning of $k$-local Lindbladians}, 
      author={Tim Möbus and Thiago Bergamaschi and Daniel Stilck França and Cambyse Rouzé},
      year={2026},
      eprint={2606.23652},
      archivePrefix={arXiv},
      primaryClass={quant-ph},
      url={https://arxiv.org/abs/2606.23652}, 
}

@misc{lewis2026learningstructureopenquantum,
      title={Learning the structure of open quantum systems}, 
      author={Laura Lewis and Ewin Tang and John Wright},
      year={2026},
      eprint={2606.30358},
      archivePrefix={arXiv},
      primaryClass={quant-ph},
      url={https://arxiv.org/abs/2606.30358}, 
}

@misc{chen2026learning,
      title={Learning Arbitrary Lindbladians from Time Evolution}, 
      author={Zhili Chen and Zhan Yu},
      year={2026},
      eprint={2607.28610},
      archivePrefix={arXiv},
      primaryClass={quant-ph},
      url={https://arxiv.org/abs/2607.28610}, 
}

@misc{song2026efficient,
      title={Efficient Lindbladian Learning from Constant-Time Pauli Responses}, 
      author={Jiaxing Song and Yukun Zhang and Xiao Yuan and Yusen Wu},
      year={2026},
      eprint={2607.25795},
      archivePrefix={arXiv},
      primaryClass={quant-ph},
      url={https://arxiv.org/abs/2607.25795}, 
}

@misc{zhou2026characterizingarbitrarylindbladiandynamics,
      title={Characterizing Arbitrary Lindbladian Dynamics with a Few Pauli Measurements}, 
      author={Taiqi Zhou and Weiyuan Gong},
      year={2026},
      eprint={2607.23044},
      archivePrefix={arXiv},
      primaryClass={quant-ph},
      url={https://arxiv.org/abs/2607.23044}, 
}

@article{wiebe2014hamiltonian,
  title = {Hamiltonian Learning and Certification Using Quantum Resources},
  author = {Wiebe, Nathan and Granade, Christopher and Ferrie, Christopher and Cory, D. G.},
  journal = {Phys. Rev. Lett.},
  volume = {112},
  issue = {19},
  pages = {190501},
  numpages = {5},
  year = {2014},
  month = {May},
  publisher = {American Physical Society},
  doi = {10.1103/PhysRevLett.112.190501},
  url = {https://link.aps.org/doi/10.1103/PhysRevLett.112.190501}
}

@article{wiebe2014quantum,
  title = {Quantum Hamiltonian learning using imperfect quantum resources},
  author = {Wiebe, Nathan and Granade, Christopher and Ferrie, Christopher and Cory, David},
  journal = {Phys. Rev. A},
  volume = {89},
  issue = {4},
  pages = {042314},
  numpages = {16},
  year = {2014},
  month = {Apr},
  publisher = {American Physical Society},
  doi = {10.1103/PhysRevA.89.042314},
  url = {https://link.aps.org/doi/10.1103/PhysRevA.89.042314}
}

@misc{he2026optimal,
      title={Optimal classical shadow estimation of unitary channels at Heisenberg limit}, 
      author={Entong He and Zihao Li and Noam Scully and Sisi Zhou and Yuxiang Yang},
      year={2026},
      eprint={2606.13638},
      archivePrefix={arXiv},
      primaryClass={quant-ph},
      url={https://arxiv.org/abs/2606.13638}, 
}

@misc{shin2026mintime,
      title={Heisenberg-limited Hamiltonian learning without short-time control}, 
      author={Myeongjin Shin and Junseo Lee and Changhun Oh},
      year={2026},
      eprint={2604.27838},
      archivePrefix={arXiv},
      primaryClass={quant-ph},
      url={https://arxiv.org/abs/2604.27838}, 
}

@misc{depradenne2026longtimes,
      title={Learning Hamiltonians at Long Times}, 
      author={Constantin Cedillo Vayson de Pradenne and Jordan Cotler and Hsin-Yuan Huang},
      year={2026},
      eprint={2606.05690},
      archivePrefix={arXiv},
      primaryClass={quant-ph},
      url={https://arxiv.org/abs/2606.05690}, 
}

@article{hangleiter2024robustly,
  title={Robustly learning the Hamiltonian dynamics of a superconducting quantum processor},
  author={Hangleiter, Dominik and Roth, Ingo and Fuksa, Jon{\'a}{\v{s}} and Eisert, Jens and Roushan, Pedram},
  journal={Nature Communications},
  volume={15},
  number={1},
  pages={9595},
  year={2024},
  publisher={Nature Publishing Group UK London}
}

@article{valenti2019hamiltonian,
  title={Hamiltonian learning for quantum error correction},
  author={Valenti, Agnes and van Nieuwenburg, Evert and Huber, Sebastian and Greplova, Eliska},
  journal={Physical Review Research},
  volume={1},
  number={3},
  pages={033092},
  year={2019},
  publisher={APS}
}

@article{hu2025ansatz,
  title={Ansatz-free Hamiltonian learning with Heisenberg-limited scaling},
  author={Hu, Hong-Ye and Ma, Muzhou and Gong, Weiyuan and Ye, Qi and Tong, Yu and Flammia, Steven T and Yelin, Susanne F},
  journal={arXiv preprint arXiv:2502.11900},
  year={2025}
}

@article{dutkiewicz2024advantage,
  title={The advantage of quantum control in many-body Hamiltonian learning},
  author={Dutkiewicz, Alicja and O'Brien, Thomas E and Schuster, Thomas},
  journal={Quantum},
  volume={8},
  pages={1537},
  year={2024},
  publisher={Verein zur F{\"o}rderung des Open Access Publizierens in den Quantenwissenschaften}
}

@article{anshu2021sample,
  title={Sample-efficient learning of interacting quantum systems},
  author={Anshu, Anurag and Arunachalam, Srinivasan and Kuwahara, Tomotaka and Soleimanifar, Mehdi},
  journal={Nature Physics},
  volume={17},
  number={8},
  pages={931--935},
  year={2021},
  publisher={Nature Publishing Group UK London}
}

@inproceedings{bakshi2024structure,
  title={Structure learning of Hamiltonians from real-time evolution},
  author={Bakshi, Ainesh and Liu, Allen and Moitra, Ankur and Tang, Ewin},
  booktitle={2024 IEEE 65th Annual Symposium on Foundations of Computer Science (FOCS)},
  pages={1037--1050},
  year={2024},
  organization={IEEE}
}

@article{caro2024learning,
  title={Learning quantum processes and Hamiltonians via the Pauli transfer matrix},
  author={Caro, Matthias C},
  journal={ACM Transactions on Quantum Computing},
  volume={5},
  number={2},
  pages={1--53},
  year={2024},
  publisher={ACM New York, NY}
}

@article{gu2024practical,
  title={Practical Hamiltonian learning with unitary dynamics and Gibbs states},
  author={Gu, Andi and Cincio, Lukasz and Coles, Patrick J},
  journal={Nature Communications},
  volume={15},
  number={1},
  pages={312},
  year={2024},
  publisher={Nature Publishing Group UK London}
}

@inproceedings{zhao2025learning,
  title={Learning the structure of any Hamiltonian from minimal assumptions},
  author={Zhao, Andrew},
  booktitle={Proceedings of the 57th Annual ACM Symposium on Theory of Computing},
  pages={1201--1211},
  year={2025}
}

@misc{sinha2025improvedhamiltonianlearningsparsity,
      title={Improved Hamiltonian learning and sparsity testing through Bell sampling}, 
      author={Savar D. Sinha and Yu Tong},
      year={2025},
      eprint={2509.07937},
      archivePrefix={arXiv},
      primaryClass={quant-ph},
      url={https://arxiv.org/abs/2509.07937}, 
}

@article{haah2024learning,
  title={Learning quantum Hamiltonians from high-temperature Gibbs states and real-time evolutions},
  author={Haah, Jeongwan and Kothari, Robin and Tang, Ewin},
  journal={Nature Physics},
  volume={20},
  number={6},
  pages={1027--1031},
  year={2024},
  publisher={Nature Publishing Group UK London}
}

@article{huang2023learning,
  title={Learning many-body Hamiltonians with Heisenberg-limited scaling},
  author={Huang, Hsin-Yuan and Tong, Yu and Fang, Di and Su, Yuan},
  journal={Physical Review Letters},
  volume={130},
  number={20},
  pages={200403},
  year={2023},
  publisher={APS}
}

@article{li2024heisenberg,
  title={Heisenberg-limited Hamiltonian learning for interacting bosons},
  author={Li, Haoya and Tong, Yu and Gefen, Tuvia and Ni, Hongkang and Ying, Lexing},
  journal={npj Quantum Information},
  volume={10},
  number={1},
  pages={83},
  year={2024},
  publisher={Nature Publishing Group UK London}
}

@article{ma2024learning,
  title={Learning $ k $-body Hamiltonians via compressed sensing},
  author={Ma, Muzhou and Flammia, Steven T and Preskill, John and Tong, Yu},
  journal={arXiv preprint arXiv:2410.18928},
  year={2024}
}

@article{mirani2024learning,
  title={Learning interacting fermionic Hamiltonians at the Heisenberg limit},
  author={Mirani, Arjun and Hayden, Patrick},
  journal={Physical Review A},
  volume={110},
  number={6},
  pages={062421},
  year={2024},
  publisher={APS}
}

@article{ni2024quantum,
  title={Quantum hamiltonian learning for the fermi-hubbard model},
  author={Ni, Hongkang and Li, Haoya and Ying, Lexing},
  journal={Acta Applicandae Mathematicae},
  volume={191},
  number={1},
  pages={2},
  year={2024},
  publisher={Springer}
}

@article{castaneda2025hamiltonian,
  title={Hamiltonian learning via shadow tomography of pseudo-choi states},
  author={Castaneda, Juan and Wiebe, Nathan},
  journal={Quantum},
  volume={9},
  pages={1700},
  year={2025},
  publisher={Verein zur F{\"o}rderung des Open Access Publizierens in den Quantenwissenschaften}
}

@article{zubida2021optimal,
  title={Optimal short-time measurements for Hamiltonian learning},
  author={Zubida, Assaf and Yitzhaki, Elad and Lindner, Netanel H and Bairey, Eyal},
  journal={arXiv preprint arXiv:2108.08824},
  year={2021}
}

@article{yu2023robust,
  title={Robust and efficient Hamiltonian learning},
  author={Yu, Wenjun and Sun, Jinzhao and Han, Zeyao and Yuan, Xiao},
  journal={Quantum},
  volume={7},
  pages={1045},
  year={2023},
  publisher={Verein zur F{\"o}rderung des Open Access Publizierens in den Quantenwissenschaften}
}

@article{evans2019scalable,
  title={Scalable bayesian hamiltonian learning},
  author={Evans, Tim J and Harper, Robin and Flammia, Steven T},
  journal={arXiv preprint arXiv:1912.07636},
  year={2019}
}

@article{bairey2019learning,
  title={Learning a local Hamiltonian from local measurements},
  author={Bairey, Eyal and Arad, Itai and Lindner, Netanel H},
  journal={Physical review letters},
  volume={122},
  number={2},
  pages={020504},
  year={2019},
  publisher={APS}
}

@article{qi2019determining,
  title={Determining a local Hamiltonian from a single eigenstate},
  author={Qi, Xiao-Liang and Ranard, Daniel},
  journal={Quantum},
  volume={3},
  pages={159},
  year={2019},
  publisher={Verein zur F{\"o}rderung des Open Access Publizierens in den Quantenwissenschaften}
}

@article{rouze2024learning,
  title={Learning quantum many-body systems from a few copies},
  author={Rouz{\'e}, Cambyse and Fran{\c{c}}a, Daniel Stilck},
  journal={Quantum},
  volume={8},
  pages={1319},
  year={2024},
  publisher={Verein zur F{\"o}rderung des Open Access Publizierens in den Quantenwissenschaften}
}

@inproceedings{bakshi2024learning,
  title={Learning quantum Hamiltonians at any temperature in polynomial time},
  author={Bakshi, Ainesh and Liu, Allen and Moitra, Ankur and Tang, Ewin},
  booktitle={Proceedings of the 56th Annual ACM Symposium on Theory of Computing},
  pages={1470--1477},
  year={2024}
}

@misc{chen2025quantumprobetomography,
      title={Quantum Probe Tomography}, 
      author={Sitan Chen and Jordan Cotler and Hsin-Yuan Huang},
      year={2025},
      eprint={2510.08499},
      archivePrefix={arXiv},
      primaryClass={quant-ph},
      url={https://arxiv.org/abs/2510.08499}, 
}

@article{flammia2021pauli,
  title={Pauli error estimation via population recovery},
  author={Flammia, Steven T and O'Donnell, Ryan},
  journal={Quantum},
  volume={5},
  pages={549},
  year={2021},
  publisher={Verein zur F{\"o}rderung des Open Access Publizierens in den Quantenwissenschaften}
}

@article{cubitt2012complexity,
  title={The complexity of relating quantum channels to master equations},
  author={Cubitt, Toby S and Eisert, Jens and Wolf, Michael M},
  journal={Communications in Mathematical Physics},
  volume={310},
  pages={383--418},
  year={2012},
  publisher={Springer}
}

@article{cotler2026noisylearning,
  title={Noisy quantum learning theory},
  author={Cotler, Jordan and Gong, Weiyuan and Kannan, Ishaan},
  journal={Nature Communications},
  year={2026},
  publisher={Nature Publishing Group UK London},
  doi = {10.1038/s41467-026-73693-x}
}

@misc{kannan2026exponential,
      title={Exponential speedups in fault-tolerant processing of quantum experiments}, 
      author={Ishaan Kannan and Harald Putterman and Jordan Cotler},
      year={2026},
      eprint={2605.02057},
      archivePrefix={arXiv},
      primaryClass={quant-ph},
      url={https://arxiv.org/abs/2605.02057}, 
}

@article{braunstein1994statistical,
  title = {Statistical distance and the geometry of quantum states},
  author = {Braunstein, Samuel L. and Caves, Carlton M.},
  journal = {Phys. Rev. Lett.},
  volume = {72},
  issue = {22},
  pages = {3439--3443},
  numpages = {0},
  year = {1994},
  month = {May},
  publisher = {American Physical Society},
  doi = {10.1103/PhysRevLett.72.3439},
  url = {https://link.aps.org/doi/10.1103/PhysRevLett.72.3439}
}

@article{kitaev1995quantum,
  title={Quantum measurements and the Abelian stabilizer problem},
  author={Kitaev, A Yu},
  journal={arXiv preprint quant-ph/9511026},
  year={1995}
}

@article{huelga1997improvement,
  title={Improvement of frequency standards with quantum entanglement},
  author={Huelga, Susanna F and Macchiavello, Chiara and Pellizzari, Thomas and Ekert, Artur K and Plenio, Martin B and Cirac, J Ignacio},
  journal={Physical Review Letters},
  volume={79},
  number={20},
  pages={3865},
  year={1997},
  publisher={APS},
  doi = {10.1103/PhysRevLett.79.3865},
}

@article{giovannetti2006quantum,
title = {Quantum Metrology},
  author = {Giovannetti, Vittorio and Lloyd, Seth and Maccone, Lorenzo},
  journal = {Phys. Rev. Lett.},
  volume = {96},
  issue = {1},
  pages = {010401},
  numpages = {4},
  year = {2006},
  month = {Jan},
  publisher = {American Physical Society},
  doi = {10.1103/PhysRevLett.96.010401},
  url = {https://link.aps.org/doi/10.1103/PhysRevLett.96.010401}
}

@article{higgins2007entanglement,
  title={Entanglement-free Heisenberg-limited phase estimation},
  author={Higgins, Brendon L and Berry, Dominic W and Bartlett, Stephen D and Wiseman, Howard M and Pryde, Geoff J},
  journal={Nature},
  volume={450},
  number={7168},
  pages={393--396},
  year={2007},
  publisher={Nature Publishing Group UK London}
}

@article{escher2011general,
  title={General framework for estimating the ultimate precision limit in noisy quantum-enhanced metrology},
  author={Escher, BM and de Matos Filho, Ruynet Lima and Davidovich, Luiz},
  journal={Nature Physics},
  volume={7},
  number={5},
  pages={406--411},
  year={2011},
  publisher={Nature Publishing Group UK London}
}

@article{demkowicz2014using,
  title={Using entanglement against noise in quantum metrology},
  author={Demkowicz-Dobrza{\'n}ski, Rafal and Maccone, Lorenzo},
  journal={Physical review letters},
  volume={113},
  number={25},
  pages={250801},
  year={2014},
  publisher={
  APS}
}

@article{kimmel2015robust,
  title={Robust calibration of a universal single-qubit gate set via robust phase estimation},
  author={Kimmel, Shelby and Low, Guang Hao and Yoder, Theodore J},
  journal={Physical Review A},
  volume={92},
  number={6},
  pages={062315},
  year={2015},
  publisher={APS}
}

@article{smirne2016noisyfreqest,
  title = {Ultimate Precision Limits for Noisy Frequency Estimation},
  author = {Smirne, Andrea and Ko\l{}ody\ifmmode \acute{n}\else \'{n}\fi{}ski, Jan and Huelga, Susana F. and Demkowicz-Dobrza\ifmmode \acute{n}\else \'{n}\fi{}ski, Rafa\l{}},
  journal = {Phys. Rev. Lett.},
  volume = {116},
  issue = {12},
  pages = {120801},
  numpages = {6},
  year = {2016},
  month = {Mar},
  publisher = {American Physical Society},
  doi = {10.1103/PhysRevLett.116.120801},
  url = {https://link.aps.org/doi/10.1103/PhysRevLett.116.120801}
}

@article{zhou2018achieving,
  title={Achieving the Heisenberg limit in quantum metrology using quantum error correction},
  author={Zhou, Sisi and Zhang, Mengzhen and Preskill, John and Jiang, Liang},
  journal={Nature communications},
  volume={9},
  number={1},
  pages={78},
  year={2018},
  publisher={Nature Publishing Group UK London}
}

@article{demkowicz2017qecmetrology,
  title = {Adaptive Quantum Metrology under General Markovian Noise},
  author = {Demkowicz-Dobrza\ifmmode \acute{n}\else \'{n}\fi{}ski, Rafa\l{} and Czajkowski, Jan and Sekatski, Pavel},
  journal = {Phys. Rev. X},
  volume = {7},
  issue = {4},
  pages = {041009},
  numpages = {15},
  year = {2017},
  month = {Oct},
  publisher = {American Physical Society},
  doi = {10.1103/PhysRevX.7.041009},
  url = {https://link.aps.org/doi/10.1103/PhysRevX.7.041009}
}

@article{demkowicz2012elusive,
  title={The elusive Heisenberg limit in quantum-enhanced metrology},
  author={Demkowicz-Dobrza{\'n}ski, Rafa{\l} and Ko{\l}ody{\'n}ski, Jan and Gu{\c{t}}{\u{a}}, M{\u{a}}d{\u{a}}lin},
  journal={Nature communications},
  volume={3},
  number={1},
  pages={1063},
  year={2012},
  publisher={Nature Publishing Group UK London}
}

@article{gorecki2020optimal,
  title={Optimal probes and error-correction schemes in multi-parameter quantum metrology},
  author={G{\'o}recki, Wojciech and Zhou, Sisi and Jiang, Liang and Demkowicz-Dobrza{\'n}ski, Rafa{\l}},
  journal={Quantum},
  volume={4},
  pages={288},
  year={2020},
  publisher={Verein zur F{\"o}rderung des Open Access Publizierens in den Quantenwissenschaften}
}

@misc{abbasgholinejad2026error,
      title={Error-corrected function estimation advantage in multiparameter Hamiltonians}, 
      author={Erfan Abbasgholinejad and Lorcán O. Conlon and Sean R. Muleady and Jacob Bringewatt and Ali Fahimniya and Yu-Xin Wang and Alexey V. Gorshkov},
      year={2026},
      eprint={2609.16226},
      archivePrefix={arXiv},
      primaryClass={quant-ph},
      url={https://arxiv.org/abs/2609.16226}, 
}

@article{catana2015fisher,
doi = {10.1088/1751-8113/48/36/365301},
url = {https://doi.org/10.1088/1751-8113/48/36/365301},
year = {2015},
month = {aug},
publisher = {IOP Publishing},
volume = {48},
number = {36},
pages = {365301},
author = {Catana, Catalin and Bouten, Luc and Guţă, Mădălin},
title = {Fisher informations and local asymptotic normality for continuous-time quantum Markov processes},
journal = {Journal of Physics A: Mathematical and Theoretical}
}

@article{guta2017information,
    author = {Guta, Madalin and Kiukas, Jukka},
    title = {Information geometry and local asymptotic normality for multi-parameter estimation of quantum Markov dynamics},
    journal = {Journal of Mathematical Physics},
    volume = {58},
    number = {5},
    pages = {052201},
    year = {2017},
    month = {05},
    issn = {0022-2488},
    doi = {10.1063/1.4982958},
    url = {https://doi.org/10.1063/1.4982958}
}

@article{kurdzialek2023using,
  title = {Using Adaptiveness and Causal Superpositions Against Noise in Quantum Metrology},
  author = {Kurdzia{\l{}}ek, Stanis{\l{}}aw and G{\'o}recki, Wojciech and Albarelli, Francesco and Demkowicz-Dobrza{\ifmmode \acute{n}\else \'{n}\fi{}}ski, Rafa{\l{}}},
  journal = {Phys. Rev. Lett.},
  volume = {131},
  issue = {9},
  pages = {090801},
  numpages = {7},
  year = {2023},
  month = {Aug},
  publisher = {American Physical Society},
  doi = {10.1103/PhysRevLett.131.090801},
  url = {https://link.aps.org/doi/10.1103/PhysRevLett.131.090801}
}

@article{das2025universal,
  title = {Universal time scalings of sensitivity in Markovian quantum metrology},
  author = {Das, Arpan and G{\'o}recki, Wojciech and Demkowicz-Dobrza{\ifmmode \acute{n}\else \'{n}\fi{}}ski, Rafa{\l{}}},
  journal = {Phys. Rev. A},
  volume = {111},
  issue = {2},
  pages = {L020403},
  numpages = {6},
  year = {2025},
  month = {Feb},
  publisher = {American Physical Society},
  doi = {10.1103/PhysRevA.111.L020403},
  url = {https://link.aps.org/doi/10.1103/PhysRevA.111.L020403}
}

@article{molmer2015hypothesis,
  title = {Hypothesis Testing with Open Quantum Systems},
  author = {M{\o{}}lmer, Klaus},
  journal = {Phys. Rev. Lett.},
  volume = {114},
  issue = {4},
  pages = {040401},
  numpages = {5},
  year = {2015},
  month = {Jan},
  publisher = {American Physical Society},
  doi = {10.1103/PhysRevLett.114.040401},
  url = {https://link.aps.org/doi/10.1103/PhysRevLett.114.040401}
}

@article{monras2007optimal,
  title = {Optimal Quantum Estimation of Loss in Bosonic Channels},
  author = {Monras, Alex and Paris, Matteo G. A.},
  journal = {Phys. Rev. Lett.},
  volume = {98},
  issue = {16},
  pages = {160401},
  numpages = {4},
  year = {2007},
  month = {Apr},
  publisher = {American Physical Society},
  doi = {10.1103/PhysRevLett.98.160401},
  url = {https://link.aps.org/doi/10.1103/PhysRevLett.98.160401}
}

@article{benatti2014dissipative,
  title={Dissipative quantum metrology in manybody systems of identical particles},
  author={Benatti, Fabio and Alipour, Sahar and Rezakhani, Alireza T},
  journal={New Journal of Physics},
  volume={16},
  number={1},
  pages={015023},
  year={2014},
  publisher={IOP Publishing},
  doi={10.1088/1367-2630/16/1/015023}
}

@article{beau2017nonlinear,
  title = {Nonlinear Quantum Metrology of Many-Body Open Systems},
  author = {Beau, M. and del Campo, A.},
  journal = {Phys. Rev. Lett.},
  volume = {119},
  issue = {1},
  pages = {010403},
  numpages = {6},
  year = {2017},
  month = {Jul},
  publisher = {American Physical Society},
  doi = {10.1103/PhysRevLett.119.010403},
  url = {https://link.aps.org/doi/10.1103/PhysRevLett.119.010403}
}

@article{alvarez2011measuring,
  title = {Measuring the Spectrum of Colored Noise by Dynamical Decoupling},
  author = {\'Alvarez, Gonzalo A. and Suter, Dieter},
  journal = {Phys. Rev. Lett.},
  volume = {107},
  issue = {23},
  pages = {230501},
  numpages = {5},
  year = {2011},
  month = {Nov},
  publisher = {American Physical Society},
  doi = {10.1103/PhysRevLett.107.230501},
  url = {https://link.aps.org/doi/10.1103/PhysRevLett.107.230501}
}

@article{norris2016qubit,
  title = {Qubit Noise Spectroscopy for Non-Gaussian Dephasing Environments},
  author = {Norris, Leigh M. and Paz-Silva, Gerardo A. and Viola, Lorenza},
  journal = {Phys. Rev. Lett.},
  volume = {116},
  issue = {15},
  pages = {150503},
  numpages = {5},
  year = {2016},
  month = {Apr},
  publisher = {American Physical Society},
  doi = {10.1103/PhysRevLett.116.150503},
  url = {https://link.aps.org/doi/10.1103/PhysRevLett.116.150503}
}

@article{pazsilva2017multiqubit,
  title = {Multiqubit spectroscopy of Gaussian quantum noise},
  author = {Paz-Silva, Gerardo A. and Norris, Leigh M. and Viola, Lorenza},
  journal = {Phys. Rev. A},
  volume = {95},
  issue = {2},
  pages = {022121},
  numpages = {26},
  year = {2017},
  month = {Feb},
  publisher = {American Physical Society},
  doi = {10.1103/PhysRevA.95.022121},
  url = {https://link.aps.org/doi/10.1103/PhysRevA.95.022121}
}

@article{szankowski2017environmental,
  title={Environmental noise spectroscopy with qubits subjected to dynamical decoupling},
  author={Sza{\'n}kowski, Piotr and Ramon, Guy and Krzywda, Jan and Kwiatkowski, Damian and Cywi{\'n}ski, {\L}ukasz},
  journal={Journal of Physics: Condensed Matter},
  volume={29},
  number={33},
  pages={333001},
  year={2017},
  publisher={IOP Publishing},
  doi={10.1088/1361-648X/aa7648}
}

@article{gardner2025lindblad,
  title = {Lindblad estimation with fast and precise quantum control},
  author = {Gardner, James W. and Haine, Simon A. and Hope, Joseph J. and Chen, Yanbei and Gefen, Tuvia},
  journal = {Phys. Rev. Appl.},
  volume = {24},
  issue = {4},
  pages = {044055},
  numpages = {26},
  year = {2025},
  month = {Oct},
  publisher = {American Physical Society},
  doi = {10.1103/6yzb-43rs},
  url = {https://link.aps.org/doi/10.1103/6yzb-43rs}
}

@misc{wang2024exponential,
      title={Exponential entanglement advantage in sensing correlated noise}, 
      author={Yu-Xin Wang and Jacob Bringewatt and Alireza Seif and Anthony J. Brady and Changhun Oh and Alexey V. Gorshkov},
      year={2024},
      eprint={2410.05878},
      archivePrefix={arXiv},
      primaryClass={quant-ph},
      url={https://arxiv.org/abs/2410.05878}, 
}

@article{brady2026correlated,
  title = {Correlated Noise Estimation with Quantum Sensor Networks},
  author = {Brady, Anthony J. and Wang, Yu-Xin and Albert, Victor V. and Gorshkov, Alexey V. and Zhuang, Quntao},
  journal = {Phys. Rev. Lett.},
  volume = {136},
  issue = {8},
  pages = {080803},
  numpages = {9},
  year = {2026},
  month = {Feb},
  publisher = {American Physical Society},
  doi = {10.1103/sl32-jn82},
  url = {https://link.aps.org/doi/10.1103/sl32-jn82}
}

@misc{brady2026precision,
      title={Precision Limits of Multiparameter Markovian-Noise Metrology}, 
      author={Anthony J. Brady and Yu-Xin Wang and Luis Pedro García-Pintos and Alexey V. Gorshkov},
      year={2026},
      eprint={2604.14298},
      archivePrefix={arXiv},
      primaryClass={quant-ph},
      url={https://arxiv.org/abs/2604.14298}, 
}

@article{chuang1997prescription,
  title={Prescription for experimental determination of the dynamics of a quantum black box},
  author={Chuang, Isaac L},
  journal={Journal of Modern Optics},
  volume={44},
  number={11-12},
  pages={2455--2467},
  year={1997},
  publisher={Taylor \& Francis},
  doi = {10.1080/09500349708231894},
  url = {https://www.tandfonline.com/doi/abs/10.1080/09500349708231894},
}

@article{poyatos1997complete,
  title = {Complete Characterization of a Quantum Process: The Two-Bit Quantum Gate},
  author = {Poyatos, J. F. and Cirac, J. I. and Zoller, P.},
  journal = {Phys. Rev. Lett.},
  volume = {78},
  issue = {2},
  pages = {390--393},
  numpages = {0},
  year = {1997},
  month = {Jan},
  publisher = {American Physical Society},
  doi = {10.1103/PhysRevLett.78.390},
  url = {https://link.aps.org/doi/10.1103/PhysRevLett.78.390}
}

@article{knill2008randomized,
  title = {Randomized benchmarking of quantum gates},
  author = {Knill, E. and Leibfried, D. and Reichle, R. and Britton, J. and Blakestad, R. B. and Jost, J. D. and Langer, C. and Ozeri, R. and Seidelin, S. and Wineland, D. J.},
  journal = {Phys. Rev. A},
  volume = {77},
  issue = {1},
  pages = {012307},
  numpages = {7},
  year = {2008},
  month = {Jan},
  publisher = {American Physical Society},
  doi = {10.1103/PhysRevA.77.012307},
  url = {https://link.aps.org/doi/10.1103/PhysRevA.77.012307}
}

@article{erhard2019characterizing,
  title={Characterizing large-scale quantum computers via cycle benchmarking},
  author={Erhard, Alexander and Wallman, Joel J and Postler, Lukas and Meth, Michael and Stricker, Roman and Martinez, Esteban A and Schindler, Philipp and Monz, Thomas and Emerson, Joseph and Blatt, Rainer},
  journal={Nature communications},
  volume={10},
  number={1},
  pages={5347},
  year={2019},
  publisher={Nature Publishing Group UK London},
  doi          = {10.1038/s41467-019-13068-7},
  url          = {https://doi.org/10.1038/s41467-019-13068-7}
}

@article{emerson2005scalable,
doi = {10.1088/1464-4266/7/10/021},
url = {https://doi.org/10.1088/1464-4266/7/10/021},
year = {2005},
month = {sep},
publisher = {},
volume = {7},
number = {10},
pages = {S347},
author = {Emerson, Joseph and Alicki, Robert and Życzkowski, Karol},
title = {Scalable noise estimation with random unitary operators},
journal = {Journal of Optics B: Quantum and Semiclassical Optics},
}

@article{emerson2007symmetrized,
  title={Symmetrized characterization of noisy quantum processes},
  author={Emerson, Joseph and Silva, Marcus and Moussa, Osama and Ryan, Colm and Laforest, Martin and Baugh, Jonathan and Cory, David G and Laflamme, Raymond},
  journal={Science},
  volume={317},
  number={5846},
  pages={1893--1896},
  year={2007},
  publisher={American Association for the Advancement of Science}
}

@article{magesan2011scalable,
  title = {Scalable and Robust Randomized Benchmarking of Quantum Processes},
  author = {Magesan, Easwar and Gambetta, J. M. and Emerson, Joseph},
  journal = {Phys. Rev. Lett.},
  volume = {106},
  issue = {18},
  pages = {180504},
  numpages = {4},
  year = {2011},
  month = {May},
  publisher = {American Physical Society},
  doi = {10.1103/PhysRevLett.106.180504},
  url = {https://link.aps.org/doi/10.1103/PhysRevLett.106.180504}
}

@inproceedings{kempe2004complexity,
  title={The complexity of the local Hamiltonian problem},
  author={Kempe, Julia and Kitaev, Alexei and Regev, Oded},
  booktitle={International Conference on Foundations of Software Technology and Theoretical Computer Science},
  pages={372--383},
  year={2004},
  organization={Springer}
}

@article{brandao2019finite,
  title={Finite correlation length implies efficient preparation of quantum thermal states},
  author={Brandao, Fernando GSL and Kastoryano, Michael J},
  journal={Communications in Mathematical Physics},
  volume={365},
  number={1},
  pages={1--16},
  year={2019},
  publisher={Springer}
}

@article{harper2021fast,
  title={Fast estimation of sparse quantum noise},
  author={Harper, Robin and Yu, Wenjun and Flammia, Steven T},
  journal={PRX Quantum},
  volume={2},
  number={1},
  pages={010322},
  year={2021},
  publisher={APS}
}

@article{da2011practical,
  title={Practical characterization of quantum devices without tomography},
  author={da Silva, Marcus P and Landon-Cardinal, Olivier and Poulin, David},
  journal={Physical Review Letters},
  volume={107},
  number={21},
  pages={210404},
  year={2011},
  publisher={APS}
}

@misc{kuszmaul2025multiplicative,
      title={The Multiplicative Version of Azuma's Inequality, with an Application to Contention Analysis}, 
      author={William Kuszmaul and Qi Qi},
      year={2025},
      eprint={2102.05077},
      archivePrefix={arXiv},
      primaryClass={cs.DS},
      url={https://arxiv.org/abs/2102.05077}, 
}

@article{freedman1975martingales,
 ISSN = {00911798, 2168894X},
 URL = {http://www.jstor.org/stable/2959268},
 author = {David A. Freedman},
 journal = {The Annals of Probability},
 number = {1},
 pages = {100--118},
 publisher = {Institute of Mathematical Statistics},
 title = {On Tail Probabilities for Martingales},
 urldate = {2026-04-01},
 volume = {3},
 year = {1975},
 doi = {10.1214/aop/1176996452}
}

@article{atalaya2021continuous,
  title = {Continuous quantum error correction for evolution under time-dependent Hamiltonians},
  author = {Atalaya, J. and Zhang, S. and Niu, M. Y. and Babakhani, A. and Chan, H. C. H. and Epstein, J. M. and Whaley, K. B.},
  journal = {Phys. Rev. A},
  volume = {103},
  issue = {4},
  pages = {042406},
  numpages = {21},
  year = {2021},
  month = {Apr},
  publisher = {American Physical Society},
  doi = {10.1103/PhysRevA.103.042406},
  url = {https://link.aps.org/doi/10.1103/PhysRevA.103.042406}
}

@article{holevo1973statistical,
  title={Statistical decision theory for quantum systems},
  author={Holevo, Alexander S},
  journal={Journal of multivariate analysis},
  volume={3},
  number={4},
  pages={337--394},
  year={1973},
  publisher={Elsevier},
  doi = {10.1016/0047-259X(73)90028-6}
}

@article{helstrom1969quantum,
  title={Quantum detection and estimation theory},
  author={Helstrom, Carl W},
  journal={Journal of Statistical Physics},
  volume={1},
  number={2},
  pages={231--252},
  year={1969},
  publisher={Springer},
  doi={10.1007/BF01007479}
}

@article{flammia2012quantum,
doi = {10.1088/1367-2630/14/9/095022},
url = {https://doi.org/10.1088/1367-2630/14/9/095022},
year = {2012},
month = {sep},
publisher = {IOP Publishing},
volume = {14},
number = {9},
pages = {095022},
author = {Flammia, Steven T and Gross, David and Liu, Yi-Kai and Eisert, Jens},
title = {Quantum tomography via compressed sensing: error bounds, sample complexity and efficient estimators},
journal = {New Journal of Physics}
}

@article{Kliesch2019guaranteedrecovery,
  doi = {10.22331/q-2019-08-12-171},
  url = {https://doi.org/10.22331/q-2019-08-12-171},
  title = {Guaranteed recovery of quantum processes from few measurements},
  author = {Kliesch, Martin and Kueng, Richard and Eisert, Jens and Gross, David},
  journal = {{Quantum}},
  issn = {2521-327X},
  publisher = {{Verein zur F{\"{o}}rderung des Open Access Publizierens in den Quantenwissenschaften}},
  volume = {3},
  pages = {171},
  month = aug,
  year = {2019}
}

@book{nielsen2010quantum,
  title={Quantum Computation and Quantum Information: 10th Anniversary Edition},
  author={Nielsen, Michael A and Chuang, Isaac L},
  year={2010},
  place={Cambridge},
  publisher={Cambridge university press}
}

@book{majidy2024building,
  title={Building quantum computers: a practical introduction},
  author={Majidy, Shayan and Wilson, Christopher and Laflamme, Raymond},
  year={2024},
  publisher={Cambridge University Press}
}

@article{gottesman2024surviving,
  title={Surviving as a quantum computer in a classical world},
  author={Gottesman, Daniel},
  journal={Textbook manuscript preprint},
  volume={8},
  number={8.1},
  pages={8--2},
  year={2024}
}

@book{o2014boolean,
  title={Analysis of boolean functions},
  author={O'Donnell, Ryan},
  year={2014},
  publisher={Cambridge University Press}
}

@book{watrous2018theory,
  title={The theory of quantum information},
  author={Watrous, John and others},
  volume={1},
  year={2018},
  publisher={Cambridge university press Cambridge}
}

@article{lindblad1976generators,
  title={On the generators of quantum dynamical semigroups},
  author={Lindblad, Goran},
  journal={Communications in mathematical physics},
  volume={48},
  pages={119--130},
  year={1976},
  publisher={Springer}
}

@article{gorini1976completely,
  title={Completely positive dynamical semigroups of N-level systems},
  author={Gorini, Vittorio and Kossakowski, Andrzej and Sudarshan, Ennackal Chandy George},
  journal={Journal of Mathematical Physics},
  volume={17},
  number={5},
  pages={821--825},
  year={1976},
  publisher={American Institute of Physics}
}

@article{Chruscinski2017history,
issn = {1230-1612},
journal = {Open systems \& information dynamics},
pages = {1740001},
volume = {24},
publisher = {World Scientific Publishing Company},
number = {3},
year = {2017},
title = {A Brief History of the GKLS Equation},
copyright = {2017, World Scientific Publishing Company},
address = {Singapore},
author = {Chruściński, Dariusz and Pascazio, Saverio},
}

@article{huang2020shadows,
  author  = {Hsin-Yuan Huang and Richard Kueng and John Preskill},
  title   = {Predicting Many Properties of a Quantum System from Very Few Measurements},
  journal = {Nature Physics},
  volume  = {16},
  number  = {10},
  pages   = {1050--1057},
  year    = {2020},
  month   = {6},
  issn    = {1745-2481},
  doi     = {10.1038/s41567-020-0932-7},
  url     = {https://doi.org/10.1038/s41567-020-0932-7}
}

@article{levy2024qpt,
  title = {Classical shadows for quantum process tomography on near-term quantum computers},
  author = {Levy, Ryan and Luo, Di and Clark, Bryan K.},
  journal = {Phys. Rev. Res.},
  volume = {6},
  issue = {1},
  pages = {013029},
  numpages = {18},
  year = {2024},
  month = {1},
  publisher = {American Physical Society},
  doi = {10.1103/PhysRevResearch.6.013029},
  url = {https://link.aps.org/doi/10.1103/PhysRevResearch.6.013029}
}

@article{kunjummen2023qpt,
  title = {Shadow process tomography of quantum channels},
  author = {Kunjummen, Jonathan and Tran, Minh C. and Carney, Daniel and Taylor, Jacob M.},
  journal = {Phys. Rev. A},
  volume = {107},
  issue = {4},
  pages = {042403},
  numpages = {12},
  year = {2023},
  month = {4},
  publisher = {American Physical Society},
  doi = {10.1103/PhysRevA.107.042403},
  url = {https://link.aps.org/doi/10.1103/PhysRevA.107.042403}
}

@misc{franceschetto2025hamiltonian,
      title={Hamiltonian learning via quantum Zeno effect}, 
      author={Giacomo Franceschetto and Egle Pagliaro and Luciano Pereira and Leonardo Zambrano and Antonio Acín},
      year={2025},
      eprint={2509.15713},
      archivePrefix={arXiv},
      primaryClass={quant-ph},
      url={https://arxiv.org/abs/2509.15713}, 
}

@article{Becker2021QZEOpen,
  author  = {Becker, Simon and Datta, Nilanjana and Salzmann, Robert},
  title   = {Quantum Zeno Effect in Open Quantum Systems},
  journal = {Annales Henri Poincar{\'e}},
  year    = {2021},
  volume  = {22},
  number  = {11},
  pages   = {3795--3840},
  doi     = {10.1007/s00023-021-01075-8},
  url     = {https://doi.org/10.1007/s00023-021-01075-8}
}

@article{Burgarth2020QZD,
  author  = {Burgarth, Daniel and Facchi, Paolo and Nakazato, Hiromichi and Pascazio, Saverio and Yuasa, Kazuya},
  title   = {Quantum Zeno Dynamics from General Quantum Operations},
  journal = {Quantum},
  year    = {2020},
  volume  = {4},
  pages   = {289},
  month   = jul,
  doi     = {10.22331/q-2020-07-06-289},
  url     = {https://doi.org/10.22331/q-2020-07-06-289}
}

@article{Mobus2019QZEGeneralized,
  author  = {M{\"o}bus, Tim and Wolf, Michael M.},
  title   = {Quantum Zeno Effect Generalized},
  journal = {Journal of Mathematical Physics},
  year    = {2019},
  volume  = {60},
  number  = {5},
  pages   = {052201},
  doi     = {10.1063/1.5090912},
  url     = {https://doi.org/10.1063/1.5090912}
}

@article{PazSilva2012ZenoQC,
  author  = {Paz-Silva, G. A. and Rezakhani, A. T. and Dominy, J. M. and Lidar, D. A.},
  title   = {Zeno Effect for Quantum Computation and Control},
  journal = {Physical Review Letters},
  year    = {2012},
  volume  = {108},
  number  = {8},
  pages   = {080501},
  doi     = {10.1103/PhysRevLett.108.080501},
  url     = {https://doi.org/10.1103/PhysRevLett.108.080501}
}

@article{Brun2006EntanglementQEC,
  author  = {Todd A. Brun and Igor Devetak and Min-Hsiu Hsieh},
  title   = {Correcting Quantum Errors with Entanglement},
  journal = {Science},
  year    = {2006},
  volume  = {314},
  number  = {5798},
  pages   = {436--439},
  doi     = {10.1126/science.1131563}
}

@article{Wilde2008OptimalEntanglementQEC,
  author  = {Mark M. Wilde and Todd A. Brun},
  title   = {Optimal entanglement formulas for entanglement-assisted quantum coding},
  journal = {Physical Review A},
  year    = {2008},
  volume  = {77},
  number  = {6},
  pages   = {064302},
  doi     = {10.1103/PhysRevA.77.064302}
}

@article{knill1997theory,
  title={Theory of quantum error-correcting codes},
  author={Knill, Emanuel and Laflamme, Raymond},
  journal={Physical Review A},
  volume={55},
  number={2},
  pages={900},
  year={1997},
  publisher={APS}
}

@book{gottesman1997stabilizer,
  title={Stabilizer codes and quantum error correction},
  author={Gottesman, Daniel},
  year={1997},
  publisher={California Institute of Technology}
}

@article{kessler2014quantum,
  title={Quantum error correction for metrology},
  author={Kessler, Eric M and Lovchinsky, Igor and Sushkov, Alexander O and Lukin, Mikhail D},
  journal={Physical review letters},
  volume={112},
  number={15},
  pages={150802},
  year={2014},
  publisher={APS}
}

@article{zhou2020optimal,
  title={Optimal approximate quantum error correction for quantum metrology},
  author={Zhou, Sisi and Jiang, Liang},
  journal={Physical Review Research},
  volume={2},
  number={1},
  pages={013235},
  year={2020},
  publisher={APS}
}

@article{dur2014improved,
  title={Improved quantum metrology using quantum error correction},
  author={D{\"u}r, Wolfgang and Skotiniotis, Michalis and Froewis, Florian and Kraus, Barbara},
  journal={Physical Review Letters},
  volume={112},
  number={8},
  pages={080801},
  year={2014},
  publisher={APS}
}

@article{marti2015qecenhanced,
  title = {Quantum Error-Correction-Enhanced Magnetometer Overcoming the Limit Imposed by Relaxation},
  author = {Herrera-Mart\'{\i}, David A. and Gefen, Tuvia and Aharonov, Dorit and Katz, Nadav and Retzker, Alex},
  journal = {Phys. Rev. Lett.},
  volume = {115},
  issue = {20},
  pages = {200501},
  numpages = {5},
  year = {2015},
  month = {Nov},
  publisher = {American Physical Society},
  doi = {10.1103/PhysRevLett.115.200501},
  url = {https://link.aps.org/doi/10.1103/PhysRevLett.115.200501}
}

@article{arrad2014sensingqec,
  title = {Increasing Sensing Resolution with Error Correction},
  author = {Arrad, G. and Vinkler, Y. and Aharonov, D. and Retzker, A.},
  journal = {Phys. Rev. Lett.},
  volume = {112},
  issue = {15},
  pages = {150801},
  numpages = {6},
  year = {2014},
  month = {Apr},
  publisher = {American Physical Society},
  doi = {10.1103/PhysRevLett.112.150801},
  url = {https://link.aps.org/doi/10.1103/PhysRevLett.112.150801}
}

@article{reiter2017dissipative,
  title={Dissipative quantum error correction and application to quantum sensing with trapped ions},
  author={Reiter, Florentin and S{\o}rensen, Anders S{\o}ndberg and Zoller, Peter and Muschik, Christine A},
  journal={Nature communications},
  volume={8},
  number={1},
  pages={1822},
  year={2017},
  publisher={Nature Publishing Group UK London},
  doi={10.1038/s41467-017-01895-5}
}

@article{layden2018spatial,
  title={Spatial noise filtering through error correction for quantum sensing},
  author={Layden, David and Cappellaro, Paola},
  journal={npj Quantum Information},
  volume={4},
  number={1},
  pages={30},
  year={2018},
  publisher={Nature Publishing Group UK London},
  doi = {10.1038/s41534-018-0082-2}
}

@article{layden2019ancillafree,
  title = {Ancilla-Free Quantum Error Correction Codes for Quantum Metrology},
  author = {Layden, David and Zhou, Sisi and Cappellaro, Paola and Jiang, Liang},
  journal = {Phys. Rev. Lett.},
  volume = {122},
  issue = {4},
  pages = {040502},
  numpages = {6},
  year = {2019},
  month = {Jan},
  publisher = {American Physical Society},
  doi = {10.1103/PhysRevLett.122.040502},
  url = {https://link.aps.org/doi/10.1103/PhysRevLett.122.040502}
}

@article{zhou2024ancillafree,
  title = {Achieving metrological limits using ancilla-free quantum error-correcting codes},
  author = {Zhou, Sisi and Manes, Argyris Giannisis and Jiang, Liang},
  journal = {Phys. Rev. A},
  volume = {109},
  issue = {4},
  pages = {042406},
  numpages = {33},
  year = {2024},
  month = {Apr},
  publisher = {American Physical Society},
  doi = {10.1103/PhysRevA.109.042406},
  url = {https://link.aps.org/doi/10.1103/PhysRevA.109.042406}
}

@misc{qiao2026distributed,
      title={Distributed estimation of many-body Hamiltonians via punctured surface code}, 
      author={Linmu Qiao and Zhichun Ouyang and Sisi Zhou},
      year={2026},
      eprint={2605.11092},
      archivePrefix={arXiv},
      primaryClass={quant-ph},
      url={https://arxiv.org/abs/2605.11092}, 
}

@article{antu2025stabilizercodes,
  doi = {10.22331/q-2025-06-05-1766},
  url = {https://doi.org/10.22331/q-2025-06-05-1766},
  title = {Stabilizer codes for {H}eisenberg-limited many-body {H}amiltonian estimation},
  author = {Antu, Santanu Bosu and Zhou, Sisi},
  journal = {{Quantum}},
  issn = {2521-327X},
  publisher = {{Verein zur F{\"{o}}rderung des Open Access Publizierens in den Quantenwissenschaften}},
  volume = {9},
  pages = {1766},
  month = jun,
  year = {2025}
}

@misc{marrero2026encoded,
      title={Encoded Quantum Signal Processing for Heisenberg-Limited Metrology}, 
      author={Carlos Ortiz Marrero and Rui Jie Tang and Nathan Wiebe},
      year={2026},
      eprint={2603.22798},
      archivePrefix={arXiv},
      primaryClass={quant-ph},
      url={https://arxiv.org/abs/2603.22798}, 
}

@article{shettell2021practical,
  title={Practical limits of error correction for quantum metrology},
  author={Shettell, Nathan and Munro, William J and Markham, Damian and Nemoto, Kae},
  journal={New Journal of Physics},
  volume={23},
  number={4},
  pages={043038},
  year={2021},
  publisher={IOP Publishing},
  doi={10.1088/1367-2630/abf533}
}

@misc{sahu2026achieving,
      title={Achieving the Heisenberg limit using fault-tolerant quantum error correction}, 
      author={Himanshu Sahu and Qian Xu and Sisi Zhou},
      year={2026},
      eprint={2601.05457},
      archivePrefix={arXiv},
      primaryClass={quant-ph},
      url={https://arxiv.org/abs/2601.05457}, 
}

@article{kapourniotis2019faulttolerant,
  title = {Fault-tolerant quantum metrology},
  author = {Kapourniotis, Theodoros and Datta, Animesh},
  journal = {Phys. Rev. A},
  volume = {100},
  issue = {2},
  pages = {022335},
  numpages = {15},
  year = {2019},
  month = {Aug},
  publisher = {American Physical Society},
  doi = {10.1103/PhysRevA.100.022335},
  url = {https://link.aps.org/doi/10.1103/PhysRevA.100.022335}
}

@article{aaranson2004stabilizers,
  author  = {Scott Aaronson and Daniel Gottesman},
  title   = {Improved Simulation of Stabilizer Circuits},
  journal = {Physical Review A},
  volume  = {70},
  number  = {5},
  year    = {2004},
  month   = {11},
  doi     = {10.1103/PhysRevA.70.052328},
  url     = {https://doi.org/10.1103/PhysRevA.70.052328}
}

@article{Bravyi2019simulationofquantum,
  doi = {10.22331/q-2019-09-02-181},
  url = {https://doi.org/10.22331/q-2019-09-02-181},
  title = {Simulation of quantum circuits by low-rank stabilizer decompositions},
  author = {Bravyi, Sergey and Browne, Dan and Calpin, Padraic and Campbell, Earl and Gosset, David and Howard, Mark},
  journal = {{Quantum}},
  issn = {2521-327X},
  publisher = {{Verein zur F{\"{o}}rderung des Open Access Publizierens in den Quantenwissenschaften}},
  volume = {3},
  pages = {181},
  month = sep,
  year = {2019}
}

@inproceedings{chen2022exponential,
  title={Exponential separations between learning with and without quantum memory},
  author={Chen, Sitan and Cotler, Jordan and Huang, Hsin-Yuan and Li, Jerry},
  booktitle={2021 IEEE 62nd Annual Symposium on Foundations of Computer Science (FOCS)},
  pages={574--585},
  year={2022},
  organization={IEEE},
  doi={10.1109/FOCS52979.2021.00063}
}

@article{abbas2025nearly,
  title={Nearly optimal algorithms to learn sparse quantum Hamiltonians in physically motivated distances},
  author={Abbas, Amira and Cerrato, Nunzia and Guti{\'e}rrez, Francisco Escudero and Grinko, Dmitry and Mele, Francesco Anna and Sinha, Pulkit},
  journal={arXiv:2509.09813},
  year={2025},
  url={https://arxiv.org/abs/2509.09813}
}

@article{zheng2026efficient,
  title={Efficient learning of logical noise from syndrome data},
  author={Zheng, Han and Chu, Chia-Tung and Chen, Senrui and Manes, Argyris Giannisis and Lee, Su-un and Zhou, Sisi and Jiang, Liang},
  journal={arXiv preprint arXiv:2601.22286},
  year={2026}
}

@article{xiao2026insitu,
  title={In-situ benchmarking of fault-tolerant quantum circuits. I. Clifford circuits},
  author={Xiao, Xiao and Hangleiter, Dominik and Bluvstein, Dolev and Lukin, Mikhail D and Gullans, Michael J},
  journal={arXiv preprint arXiv:2601.21472},
  year={2026}
}

@article{wagner2023learning,
  title={Learning logical pauli noise in quantum error correction},
  author={Wagner, Thomas and Kampermann, Hermann and Bru{\ss}, Dagmar and Kliesch, Martin},
  journal={Physical review letters},
  volume={130},
  number={20},
  pages={200601},
  year={2023},
  publisher={APS}
}

@article{wagner2022pauli,
  title={Pauli channels can be estimated from syndrome measurements in quantum error correction},
  author={Wagner, Thomas and Kampermann, Hermann and Bru{\ss}, Dagmar and Kliesch, Martin},
  journal={Quantum},
  volume={6},
  pages={809},
  year={2022},
  publisher={Verein zur F{\"o}rderung des Open Access Publizierens in den Quantenwissenschaften}
}

@article{wagner2021optimal,
  title={Optimal noise estimation from syndrome statistics of quantum codes},
  author={Wagner, Thomas and Kampermann, Hermann and Bru{\ss}, Dagmar and Kliesch, Martin},
  journal={Physical review research},
  volume={3},
  number={1},
  pages={013292},
  year={2021},
  publisher={APS}
}

@article{remm2026experimentally,
  title={Experimentally informed decoding of stabilizer codes based on syndrome correlations},
  author={Remm, Ants and Lacroix, Nathan and B{\"o}deker, Lukas and Genois, Elie and Hellings, Christoph and Swiadek, Fran{\c{c}}ois and Norris, Graham J and Eichler, Christopher and Blais, Alexandre and M{\"u}ller, Markus and others},
  journal={Physical Review Research},
  volume={8},
  number={1},
  pages={013044},
  year={2026},
  publisher={APS}
}

@misc{takou2025estimating,
      title={Estimating decoding graphs and hypergraphs of memory QEC experiments}, 
      author={Evangelia Takou and Kenneth R. Brown},
      year={2025},
      eprint={2504.20212},
      archivePrefix={arXiv},
      primaryClass={quant-ph},
      url={https://arxiv.org/abs/2504.20212}, 
}

@article{blume2025estimating,
  title={Estimating detector error models from syndrome data},
  author={Blume-Kohout, Robin and Young, Kevin},
  journal={arXiv preprint arXiv:2504.14643},
  year={2025}
}

@article{spitz2018adaptive,
  title={Adaptive Weight Estimator for Quantum Error Correction in a Time-Dependent Environment},
  author={Spitz, Stephen T and Tarasinski, Brian and Beenakker, Carlo WJ and O'Brien, Thomas E},
  journal={Advanced Quantum Technologies},
  volume={1},
  number={1},
  pages={1800012},
  year={2018},
  publisher={Wiley Online Library}
}

@article{fowler2014scalable,
  title={Scalable extraction of error models from the output of error detection circuits},
  author={Fowler, Austin G and Sank, D and Kelly, J and Barends, R and Martinis, John M},
  journal={arXiv preprint arXiv:1405.1454},
  year={2014}
}

@misc{combes2014insitu,
      title={In-situ characterization of quantum devices with error correction}, 
      author={Joshua Combes and Christopher Ferrie and Chris Cesare and Markus Tiersch and G. J. Milburn and Hans J. Briegel and Carlton M. Caves},
      year={2014},
      eprint={1405.5656},
      archivePrefix={arXiv},
      primaryClass={quant-ph},
      url={https://arxiv.org/abs/1405.5656}, 
}

@article{fletcher2008channel,
  title={Channel-adapted quantum error correction for the amplitude damping channel},
  author={Fletcher, Andrew S and Shor, Peter W and Win, Moe Z},
  journal={IEEE Transactions on Information Theory},
  volume={54},
  number={12},
  pages={5705--5718},
  year={2008},
  publisher={IEEE},
  doi={10.1109/TIT.2008.2006458}
}

@article{aliferis2008fault,
  title = {Fault-tolerant quantum computation against biased noise},
  author = {Aliferis, Panos and Preskill, John},
  journal = {Phys. Rev. A},
  volume = {78},
  issue = {5},
  pages = {052331},
  numpages = {9},
  year = {2008},
  month = {Nov},
  publisher = {American Physical Society},
  doi = {10.1103/PhysRevA.78.052331},
  url = {https://link.aps.org/doi/10.1103/PhysRevA.78.052331}
}

@article{bonilla2021xzzx,
  title={The XZZX surface code},
  author={Bonilla Ataides, J Pablo and Tuckett, David K and Bartlett, Stephen D and Flammia, Steven T and Brown, Benjamin J},
  journal={Nature communications},
  volume={12},
  number={1},
  pages={2172},
  year={2021},
  publisher={Nature Publishing Group UK London},
  doi={10.1038/s41467-021-22274-1}
}

@article{tuckett2019tailoring,
  title = {Tailoring Surface Codes for Highly Biased Noise},
  author = {Tuckett, David K. and Darmawan, Andrew S. and Chubb, Christopher T. and Bravyi, Sergey and Bartlett, Stephen D. and Flammia, Steven T.},
  journal = {Phys. Rev. X},
  volume = {9},
  issue = {4},
  pages = {041031},
  numpages = {22},
  year = {2019},
  month = {Nov},
  publisher = {American Physical Society},
  doi = {10.1103/PhysRevX.9.041031},
  url = {https://link.aps.org/doi/10.1103/PhysRevX.9.041031}
}

@article{chuang1997bosonic,
  title = {Bosonic quantum codes for amplitude damping},
  author = {Chuang, Isaac L. and Leung, Debbie W. and Yamamoto, Yoshihisa},
  journal = {Phys. Rev. A},
  volume = {56},
  issue = {2},
  pages = {1114--1125},
  numpages = {0},
  year = {1997},
  month = {Aug},
  publisher = {American Physical Society},
  doi = {10.1103/PhysRevA.56.1114},
  url = {https://link.aps.org/doi/10.1103/PhysRevA.56.1114}
}

@article{wu2025bias,
  title={Bias-tailored single-shot quantum LDPC codes},
  author={Wu, Shixin and Brun, Todd A and Lidar, Daniel A},
  journal={arXiv preprint arXiv:2507.02239},
  year={2025},  
  url = {https://arxiv.org/abs/2507.02239}
}

@article{kuehnke2025hardware,
  title={Hardware-tailored logical Clifford circuits for stabilizer codes},
  author={Kuehnke, Eric J and Levi, Kyano and Roffe, Joschka and Eisert, Jens and Miller, Daniel},
  journal={arXiv preprint arXiv:2505.20261},
  year={2025},
  url={https://arxiv.org/abs/2505.20261}
}
